\documentclass[12pt,a4paper]{article}

\usepackage[T1]{fontenc}
\usepackage[verbose=true]{microtype}

\usepackage{amsmath}
\usepackage{amssymb}
\usepackage{txfonts}        

\usepackage{qcircuit}
\let\1\relax                
\usepackage{quantum}

\usepackage{authblk}
\usepackage[perpage]{footmisc}
\usepackage[inline]{enumitem}
\usepackage{tocloft}

\usepackage{array}
\usepackage{longtable}
\usepackage{multirow}
\usepackage{rotating}

\usepackage{calc}
\usepackage{epsfig}
\usepackage{symbols}

\usepackage{color}
\usepackage{verbatim}

\usepackage{algorithm}
\usepackage{algpseudocode}
\newcommand\Input[1]{\State{{\bf input:} #1}}
\renewcommand\Call[1]{\State{{\bf call } \textsc{#1}}}

\usepackage[backend=biber,style=alphabetic]{biblatex}
\usepackage[colorlinks,linktocpage,hypertexnames=false,pdfpagelabels]{hyperref}
\usepackage[amsmath,thmmarks,hyperref]{ntheorem}
\usepackage[capitalise]{cleveref}

\newtheorem{theorem}{Theorem}
\newtheorem{lemma}[theorem]{Lemma}
\newtheorem{corollary}[theorem]{Corollary}
\newtheorem{proposition}[theorem]{Proposition}
\newtheorem{definition}[theorem]{Definition}

\theoremstyle{break}

\theoremstyle{nonumberplain}
\theorembodyfont{\normalfont}
\theoremsymbol{$\Box$}
\newtheorem{proof}{Proof}

\theoremstyle{nonumberbreak}
\theorembodyfont{\normalfont}
\theoremsymbol{$\Box$}

\theoremstyle{nonumberplain}
\makeatletter
\newtheorem{rep@theorem}[theorem]{\noexpand\rep@title}
\newcommand{\newreptheorem}[2]{%
  \newenvironment{rep#1}[1]{%
    \def\rep@title{#2~\ref{##1}~(restated)}%
    \begin{rep@theorem}}%
    {\end{rep@theorem}}}
\makeatother
\newreptheorem{theorem}{Theorem}

\crefname{equation}{}{}
\crefrangelabelformat{equation}{\textup{(#3#1#4)}--\textup{(#5#2#6)}}
\crefname{term}{}{}
\creflabelformat{term}{(#2#1#3)}
\crefname{figure}{Figure}{Figure}
\crefname{remark}{Remark}{Remarks}
\crefname{case}{case}{cases}
\crefname{type}{type}{type}
\crefname{condition}{condition}{conditions}
\crefname{part}{part}{parts}

  {\par\noindent\ignorespacesafterend}
  {\par\noindent\ignorespacesafterend}
  {\par\noindent\ignorespacesafterend}

\makeatletter
\renewcommand\section{%
  \@startsection{section}{1}{0pt}%
  {-\baselineskip}{.2\baselineskip}%
  {\normalfont\Large\bfseries\raggedright}}
\renewcommand\subsection{%
  \@startsection{subsection}{2}{0pt}%
  {-\baselineskip}{.2\baselineskip}%
  {\normalfont\large\bfseries\raggedright}}
\renewcommand\subsubsection{%
  \@startsection{subsubsection}{3}{0pt}%
  {-\baselineskip}{.2\baselineskip}%
  {\normalfont\normalsize\bfseries\raggedright}}

\renewcommand\paragraph{%
  \@startsection{paragraph}{4}{0pt}%
  {3.25ex\@plus 1ex\@minus .2ex}{-1em}%
  {\normalfont\normalsize\bfseries\itshape}}
\makeatother

\newcommand{\affilcr}{\protect\\}

\bibliography{spectral-gap}
\DefineBibliographyStrings{english}{in = {}}

\makeatletter
\def\thm@@thmline@name#1#2#3#4{%
  \@dottedtocline{-2}{0em}{2.3em}%
  {\makebox[\widesttheorem][l]{#1 \protect\numberline{#2}}#3}%
  {#4}}
\@ifpackageloaded{hyperref}{
  \def\thm@@thmline@name#1#2#3#4#5{%
    \ifx\\#5\\%
      \@dottedtocline{-2}{0em}{2.3em}%
      {\makebox[\widesttheorem][l]{#1 \protect\numberline{#2}}#3}%
      {#4}
    \else
      \ifHy@linktocpage\relax\relax
        \@dottedtocline{-2}{0em}{2.3em}%
        {\makebox[\widesttheorem][l]{#1 \protect\numberline{#2}}#3}%
        {\hyper@linkstart{link}{#5}{#4}\hyper@linkend}%
      \else
        \@dottedtocline{-2}{0em}{2.3em}%
        {\hyper@linkstart{link}{#5}%
          {\makebox[\widesttheorem][l]{#1 \protect\numberline{#2}}#3}%
          \hyper@linkend}%
        {#4}%
      \fi
    \fi}
}
\makeatother

\newlength\widesttheorem
\AtBeginDocument{\settowidth{\widesttheorem}{Proposition 99\quad}}

\theoremlisttype{allname}

\makeatletter
\def\nobreakhline{%
  \noalign{\ifnum0=`}\fi
    \penalty\@M
    \futurelet\@let@token\LT@@nobreakhline}
\def\LT@@nobreakhline{%
  \ifx\@let@token\hline
    \global\let\@gtempa\@gobble
    \gdef\LT@sep{\penalty\@M\vskip\doublerulesep}
  \else
    \global\let\@gtempa\@empty
    \gdef\LT@sep{\penalty\@M\vskip-\arrayrulewidth}
  \fi
  \ifnum0=`{\fi}%
  \multispan\LT@cols
     \unskip\leaders\hrule\@height\arrayrulewidth\hfill\cr
  \noalign{\LT@sep}%
  \multispan\LT@cols
     \unskip\leaders\hrule\@height\arrayrulewidth\hfill\cr
  \noalign{\penalty\@M}%
  \@gtempa}
\makeatother

\newcommand{\keyword}[1]{\emph{#1}}

\DeclareMathOperator{\rem}{frac}
\newcommand{\wc}{{}\cdot{}}
\newcommand{\TM}[1]{\text{\textsc{#1}}}

\newcommand{\lambdamin}{\lambda_{\mathrm{min}}}
\newcommand{\Sbr}{{S_{\!\text{br}}}}
\newcommand{\hrow}{h_{\text{row}}}

\newcommand{\Htrans}{H_{\text{trans}}}
\newcommand{\Hpenalty}{H_{\text{pen}}}
\newcommand{\Hinit}{H_{\text{init}}}
\newcommand{\Htranspen}{H_{\text{trans,pen}}}
\newcommand{\Hpath}{H_{\text{path}}}
\newcommand{\htrans}{h_{\text{trans}}}
\newcommand{\hpenalty}{h_{\text{pen}}}
\newcommand{\hinit}{h_{\text{init}}}
\newcommand{\htranspen}{h_{\text{trans,pen}}}

\newcounter{tmpcounter}

\begin{document}


\title{Undecidability of the Spectral Gap}

\author[1,2]{Toby Cubitt\thanks{t.cubitt@ucl.ac.uk}}
\affil[1]{Department of Computer Science, University College London,\affilcr
  Gower Street, London WC1E 6EA, United Kingdom}
\affil[2]{DAMTP, University of Cambridge,
  Centre for Mathematical Sciences,\affilcr
  Wilberforce Road, Cambridge CB3 0WA, United Kingdom}
\author[3,4]{David Perez-Garcia\thanks{dperezga@ucm.es}}
\affil[3]{Departamento de An\'alisis Matem\'atico and IMI,
  Facultad de CC Matem\'aticas,\affilcr
  Universidad Complutense de Madrid,
  Plaza de Ciencias 3, 28040 Madrid, Spain}
\affil[4]{ICMAT, C/Nicol\'as Cabrera, Campus de Cantoblanco, 28049 Madrid}
\author[5,6]{Michael M.~Wolf\thanks{m.wolf@tum.de}}
\affil[5]{Department of Mathematics, Technische Universit\"at M\"unchen,
  85748 Garching, Germany}
\affil[6]{Munich Center for Quantum Science and Technology (MCQST), Germany}

\date{\today}

\maketitle

\begin{abstract}
We construct families of translationally-invariant, nearest-neighbour Hamiltonians on a 2D square lattice of $d$-level quantum systems ($d$ constant), for which determining whether the system is gapped or gapless is an undecidable problem. This is true even with the promise that each Hamiltonian is either gapped or gapless in the strongest sense: it is promised to either have continuous spectrum above the ground state in the thermodynamic limit, or its spectral gap is lower-bounded by a constant. Moreover, this constant can be taken equal to the operator norm of the local operator that generates the Hamiltonian (the local interaction strength). The result still holds true if one restricts to arbitrarily small quantum perturbations of classical Hamiltonians. The proof combines a robustness analysis of Robinson's aperiodic tiling, together with tools from quantum information theory: the quantum phase estimation algorithm and the history state technique mapping Quantum Turing Machines to Hamiltonians.
\end{abstract}

\clearpage

\tableofcontents
\clearpage



\section{Introduction}\label{sec:introduction}

\subsection{Definitions and notation}
\label{sec:preliminaries}

Let us start by defining the setup we will be considering, which is the standard framework in mathematical physics for describing (finite) \keyword{spin lattice models}. We denote by $\Lambda(L):=\{1,\ldots,L\}^2$  the set of vertices (referred to as \keyword{sites}) of a square lattice of side $L\in\mathbb{N}$, with $L\ge 2$. ${\cE}\subset\Lambda(L)\times\Lambda(L)$ will denote the set of directed edges of the square lattice, oriented so that $(i,j)\in\cal E$ implies that $j$ lies north or east of $i$. We will consider both \emph{periodic} and \emph{open} boundary conditions. In the first case, the outer rows and columns are also connected along the same direction, so that $\Lambda(L)$ becomes a square lattice on a torus. In the second case, these connections are not present.

We associate to each site $i\in\Lambda(L)$ a Hilbert space $\cH^{(i)}\simeq\C^d$, and to any subset $S\subseteq\Lambda(L)$ the tensor product $\bigotimes_{i\in S} \cH^{(i)}$. The \keyword{interactions} between neighbouring pairs $(i,j)\in\cal E$ are given by Hermitian operators $h^{(i,j)}\in\cB(\cH^{(i)}\otimes\cH^{(j)})$. In addition, we may assign an \keyword{on-site interaction} given by a Hermitian matrix $h_1^{(k)}\in\cB(\cH^{(k)})$ to each site $k\in\Lambda(L)$.

We will restrict in this paper to models that are built up from such nearest-neighbour and possibly on-site terms in a translational invariant way. That is, when identifying the isomorphic Hilbert spaces on which they act, $h_1^{(k)}=h_1^{(l)}$ for all $k,l\in\Lambda(L)$, and $h^{(i',j')}=h^{(i,j)}$ whenever there is a $v\in\mathbb{Z}^2$ such that $(i',j')=(i+v,j+v)$. The total Hamiltonian for the system is then defined as
\begin{equation}
  H^{\Lambda(L)}:=\sum_{(i,j)\in\mathcal{E}} h^{(i,j)}+\sum_{k\in\Lambda(L)} h_1^{(k)}.
\end{equation}
It is fully described for any system size $L$ by three Hermitian matrices: a $d\times d$ matrix $h_1$ and two $d^2\times d^2$ matrices $h_{\text{row}}$ and $h_{\text{col}}$, which define the interactions between neighbouring sites within any row and column respectively.
Hence, it may alternatively be written as
\begin{equation}\label{eq:H-Lambda-from-h-row-col}
  H^{\Lambda(L)} = \sum_{\text{rows}}\sum_{c} h_{\text{row}}^{(c,c+1)} + \sum_{\text{columns}}\sum_{r} h_{\text{col}}^{(r,r+1)} + \sum_{i\in\Lambda(L)} h_1^{(i)}.
\end{equation}
$\max\{\|h_{\text{row}}\|, \|h_{\text{col}}\|, \|h_1\|\}$ is called the \keyword{local interaction strength} of the Hamiltonian and can be normalised to be $1$.

The set of eigenvalues, or \keyword{energy levels}, of the Hamiltonian $H^{\Lambda(L)}$ will be denoted by $\spec H^{\Lambda(L)} := \{\lambda_0 (H^{\Lambda(L)}),\lambda_1(H^{\Lambda(L)}),\dots\}$. When the Hamiltonian is clear from context, we will sometimes omit it and just write $\{\lambda_0,\lambda_1,\dots\}$. They are always assumed to be listed in non-decreasing order $\lambda_0\leq\lambda_1\leq\ldots$. The smallest eigenvalue $\lambda_0(H^{\Lambda(L)})$ is called the \emph{ground state energy} and the corresponding eigenvectors \emph{ground states}.  The \textit{ground state energy density} is defined as
\begin{equation}
E_{\rho}:=\lim_{L\rightarrow \infty} E_\rho(L), \quad \text{where } \; E_\rho(L):=\frac{\lambda_0(H^{\Lambda(L)})}{L^2}\;.
\end{equation}
It is not difficult to show \cite{Cubitt15a} that this limit is well defined.

$H^{\Lambda(L)}$ is called \emph{frustration-free} if its ground state energy is zero while all $h^{(i,j)},h_1^{(k)}$ are positive semi-definite. That is, a ground state of a frustration-free Hamiltonian minimises the energy of each interaction term individually. $H^{\Lambda(L)}$ is called \textit{classical} if its defining interactions $h^{(i,j)},h_1^{(k)}$ are diagonal in a given product basis (e.g. the canonical one).

We can define now the main quantity under study: the \emph{spectral gap}
\begin{equation}
  \Delta(H^{\Lambda(L)}):= \lambda_1(H^{\Lambda(L)})-\lambda_0(H^{\Lambda(L)}).
\end{equation}

In this paper we are considering the behaviour of $\Delta(H^{\Lambda(L)})$ in the thermodynamic limit, that is, when $L\rightarrow \infty$. For that, we introduce the following definitions:

\begin{definition}[Gapped]\label{def:gapped}
  We say that a family $\{H^{\Lambda(L)}\}$ of Hamiltonians, as described above, characterises a gapped system if there is a constant $\gamma>0$ and a system size $L_0$ such that for all $L>L_0$, $\lambda_0(H^{\Lambda(L)})$ is non-degenerate and $\Delta(H^{\Lambda(L)})\geq\gamma$. In this case, we say that \emph{the spectral gap is at least $\gamma$}.
\end{definition}

\begin{definition}[Gapless]\label{def:gapless}
  We say that a family $\{H^{\Lambda(L)}\}$ of Hamiltonians, as described above, characterises a gapless system if there is a constant $c>0$ such that for all $\varepsilon>0$ there is an $L_0\in\mathbb{N}$ so that for all $L>L_0$ any point in $[\lambda_0(H^{\Lambda(L)}),\lambda_0(H^{\Lambda(L)})+c]$ is within distance $\varepsilon$ from $\spec H^{\Lambda(L)}$.
\end{definition}

Note that gapped is not defined as the negation of gapless; there are systems that fall into neither class. The reason for choosing such strong definitions is to deliberately avoid ambiguous cases (such as systems with degenerate ground states). Our constructions will allow us to use these strong definitions, because we are able to guarantee that each instance falls into one of the two classes. Indeed, we could further strengthen the definition of \keyword{gapless} without changing our undecidability results or their proofs, below, by demanding that $c=c(L)$ grows with $L$ so that $\lim_{L\rightarrow\infty}c(L)=\infty$.

The ideas and techniques for the proof will mainly come from quantum information theory. We therefore use notation standard to that field, such as Dirac's notation for linear algebra operations. A summary of all the relevant standard notation can be found, for example, in Chapter~2 of the classic book of Nielsen and Chuang \cite{Nielsen+Chuang}.


\subsection{Main results}
\label{sec:main-results}

For each natural number $n$, we define $\varphi=\varphi(n)$ to be the rational number whose binary fraction expansion consists of the digits of $n$ in reverse order after the decimal point. We also fix throughout a specific Universal Turing Machine, denoted by UTM.

\begin{theorem}[Main theorem]\label{thm:promise}
  For any given universal Turing Machine UTM, we can construct explicitly a dimension $d$, $d^2\times d^2$ matrices $A,A',B,C,D,D'$ a $d\times d$ diagonal projector $\Pi$ and a rational number $\beta$ which can be as small as desired, with the following properties:
  \begin{enumerate}
  \item $A$ is diagonal with entries in $\Z$.
  \item $A'$ is Hermitian with entries in $ \Z+ \frac{1}{\sqrt{2}}\Z$,
  \item $B,C$ have integer entries,
  \item $D$ is diagonal with entries in $\Z$,
  \item $D'$ is hermitian with entries in $\Z$.
  \end{enumerate}
  For each natural number $n$, define:
  \begin{align*}
    &\begin{aligned}
      &h_1(n)=\alpha(n)\Pi, &\qquad&  \\
      &h_{\text{col}}(n)=D +\beta D', &\qquad& \text{independent of $n$}
    \end{aligned}\\
    &h_{\text{row}}(n)=A + \beta\left(A'+e^{i\pi\varphi} B + e^{-i\pi\varphi} B^\dg + e^{i\pi2^{-\abs{\varphi}}} C + e^{-i\pi2^{-\abs{\varphi}}} C^\dg\right),
  \end{align*}
  where $\alpha(n)\le \beta$ is an algebraic number computable from $n$. Then:
  \begin{enumerate}
  \item  The local interaction strength is bounded by~1, i.e.\ \linebreak $\max(\norm{h_1(n)}, \norm{h_{\text{row}}(n)}, \norm{h_{\text{col}}(n)}) \leq 1$.
  \item If UTM halts on input $n$, then the associated family of Hamiltonians $\{H^{\Lambda(L)}(n)\}$ is gapped in the strong sense of \cref{def:gapped} and, moreover, the gap $\gamma\ge 1$.
  \item If UTM does not halt on input $n$, then the associated family of Hamiltonians $\{H^{\Lambda(L)}(n)\}$ is gapless in the strong sense of \cref{def:gapless}.
  \end{enumerate}
\end{theorem}

\noindent Some comments are in order:
\begin{itemize}
\item Using the classic result of \textcite{Turing} that the Halting Problem for UTM on input $n$ is undecidable, we can conclude that \textit{the spectral gap problem is also undecidable}.
\item The interaction terms in the Hamiltonians given in the Theorem are\linebreak $\beta$-perturbations of the classical Hamiltonian given by $h_{\text{row}}=A,h_{\text{col}}=D$. Since $\beta$ can be taken as small as desired, all interactions appearing in the Theorem are arbitrarily small quantum perturbations of a classical system.
\item The only dependency on $n$ in the interactions appears in some \textit{prefactors} to a set of fixed interactions.
\item In the gapped case, the size of the gap is larger than or equal to the local interaction strength, hence is \textit{as large as possible}.
\item It is straightforward (if tedious) to extract an explicit value for $d$ in \cref{thm:promise,thm:axiomatic} from the construction described in this paper.
\item By exploiting the well known connection between (algorithmic) undecidability and (axiomatic) independence (see e.g. \cite{Poonen}) one immediately obtains the following corollary:
\end{itemize}

\begin{corollary}[Axiomatic independence of the spectral gap]\label{thm:axiomatic}
  Let $d\in\mathbb{N}$ be a sufficiently large constant. For any consistent formal system with a recursive set of axioms, there exists a translationally invariant nearest-neighbour Hamiltonian on a 2D lattice with local dimension $d$ and algebraic entries for which neither the presence nor the absence of a spectral gap is provable from the axioms.
\end{corollary}

Moreover, as a key intermediate step in the proof of \cref{thm:promise}, we will prove that the ground state energy density problem is also undecidable, a result clearly interesting on its own.

\begin{theorem}[Undecidability of g.s.\ energy density]
  \label{thm:gs_density}
  Let $d\in\mathbb{N}$ be sufficiently large but fixed. We can explicitly construct a one-parameter family of translationally invariant nearest-neighbour Hamiltonians on a 2D square lattice with open boundary conditions, local Hilbert space dimension $d$, algebraic matrix entries, and local interaction strengths bounded by~1 for which determining whether $E_\rho=0$ or $E_\rho>0$ is an undecidable problem.
\end{theorem}

A short version of this paper -- including a statement of the main result, discussion of its implications, an outline of the main ideas behind the proof, together with a sketch of the argument -- was published recently in Nature \cite{Cubitt15a}. We encourage the reader to consult it in order to gain some high-level intuition about the full, rigorous proof given in this work.

\subsection{Relation to physics}
\label{sec:conclusions}
As already discussed in \cite{Cubitt15a}, this result has a number of implications for condensed matter and mathematical physics, that we briefly recall here.

Quantum spin lattice models are ubiquitous in mathematical physics. They play a central role in condensed matter physics, in quantum computation and even in high energy physics, where one route to formalising the continuum is to consider quantum many body systems on a lattice and later sending the lattice spacing to zero. In all these contexts, one often starts from a description of the microscopic interactions that govern the system. The question is then how to infer the relevant observable macroscopic properties that emerge as the lattice size (number of particles) tends to infinity.

The spectral gap is amongst the most important properties of a quantum many-body Hamiltonian. It is intimately related to the physics of a quantum many-body system. Quantum phase transitions can only occur at critical points where the gap vanishes. The intuitive reason is that the spectral gap ``protects'' the ground state properties of the system against small perturbations, since an energy of the order of the gap must be pumped into the system to transition to a different state. Formalising this intuition is a major open question. Recently Bravyi, Hastings and Michalakis \cite{BravyiHastingsMichalakis} and Michalakis and Pytel \cite{Michalakis} proved this for the particular case of frustration-free systems satisfying certain additional conditions, by proving that the spectral gap of these systems is stable under arbitrary local perturbations throughout the system.

The low-temperature behaviour of a system is also governed by the spectral gap: gapped systems exhibit ``non-critical'' behaviour, with low-energy excitations that behave as massive particles \cite{Haegeman13}, preventing long-range correlations \cite{Hastings_exp-decay, Nachtergaele06}; gapless systems exhibit ``critical'' behaviour, with low-energy excitations that behave as massless particles, allowing long-range correlations. This implies that ground states of gapped systems are somehow less complex. That intuition has been formalised in the \keyword{area law conjecture} \cite{Eisert10}, which has been proven for 1D spin systems \cite{Hastings_area-law}, and in 2D if certain additional hypothesis on the spectrum are satisfied \cite{Hastings07}. This in turn translates into better algorithms for computing properties of such systems \cite{Orus14}. A paradigmatic example of this line of research is the recent proof \cite{Vidick,Vidick2}, based on an improved version of the 1D area law theorem \cite{Arad13}, that the ground state energy density problem for gapped quantum spin systems in 1D is in the complexity class~P (i.e.\ the run-time of the algorithm scales polynomially in the system size).

Because of its central importance to many-body quantum systems, many seminal results in mathematical physics concern the spectral gap of specific systems. Examples include the Lieb-Schultz-Mattis theorem showing that the Heisenberg chain for half-integer spins is gapless \cite{LiebSchultzMattis}, extended to higher dimensional bipartite lattices by Hastings \cite{Hastings_LSM}, and the proof of a spectral gap for the 1D AKLT model \cite{AKLT}. More recently, shortly after this work first appeared on preprint servers, Bravyi and Gosset gave a complete characterisation of the spectral gap for the simplest non-trivial case of 1D chains of spin-$\tfrac{1}{2}$ particles (qubits) with frustration-free, nearest-neighbour interactions~\cite{BravyiGosset}. (The frustrated case remains open.)

The same is true of some of the most important open questions in the field. For instance, the \keyword{Haldane conjecture} \cite{Haldane}, first formulated in 1983, states that the integer-spin antiferromagnetic Heisenberg model in 1D has a non-vanishing spectral gap. Despite strong supporting evidence from numerical simulations, a proof remains elusive. The analogous question in 2D for non-bipartite lattices can be traced back to the work of Anderson in 1973 \cite{Anderson73}, where he suggested the existence of new type of materials, nowadays called topological quantum spin liquids. The existence of minerals in nature, such as herbertsmithite, whose interactions can be well approximated by the spin-$\frac{1}{2}$ antiferromagnetic Heisenberg model on the Kagome lattice -- hence are compelling candidates for systems with a topological quantum spin liquid phase \cite{Han} -- brought the problem of its spectral gap to the forefront of physics. The existence of a spectral gap in these systems is an active area of research, with techniques reducing the spectral gap problem to finite-volume criteria very recently being applied successfully (partially numerically) to certain cases such as the honeycomb lattice~\cite{Lemm_2D_AKLT,Pomata_2D_AKLT}, whilst other cases remain disputed even at the level of numerical evidence (see e.g.\ \cite{White,Iqbal14}).

Undecidability of the spectral gap has a number of immediate corollaries relevant to physics. It implies that one can write down models whose phase diagrams are so complex they are in fact uncomputable. It implies the standard approach of trying to gain insight into physics models by solving numerically for larger and larger lattice sizes necessarily fails for some systems; the system can display all the features of a gapless model, with the gap of the finite system decreasing monotonically with increasing size. Then, at some threshold size, it may suddenly switch to having a large gap. (In \cref{sec:unbounded} we also construct models exhibiting the opposite transition, from gapped to gapless.) Not only can the threshold size be arbitrarily large; the threshold size itself is an uncomputable number. In a recent paper \cite{size-driven}, we constructed very simple models with small local dimensions which exhibit this type of ``size-driven phase transition'' (without, however, the undecidability properties proven here). Our findings also imply that a result showing robustness of the spectral gap under perturbations, as for the case of frustration-free Hamiltonians~\cite{Michalakis} and for free-fermion systems~\cite{DeRoeck_fermions,Hastings_fermions}, cannot hold for general gapped systems.

Phase diagrams with infinitely many phases are known in quantum systems in connection with the quantum Hall effect, where fractal diagrams like the Hofstadter butterfly can be obtained \cite{OsadchyAvron,Hofstadter}. Since membership in many fractal sets is not decidable (when formulated in the framework of real computation; see \cite{BlumSmale}), it would be interesting to see whether quantum Hall systems could provide a real-world manifestation of our findings.

Conjectures about the spectral gap, such as the Haldane conjecture, the 2D AKLT conjecture, or the Yang-Mills mass gap conjecture in quantum field theory, implicitly assume that these questions can be answered one way or the other. Our results prove that the answer to the general spectral gap question is not determined by the axioms of mathematics. Whilst our results are still a long way from proving that any of these specific conjectures are axiomatically undecidable, they at least open the door to the possibility that these -- or similar -- questions about physical models may be provably unanswerable.

The idea that some of the most difficult open problems in physics could be mathematically proven to be impossible to solve is not new. Indeed, proving the impossibility of obtaining exact formulae for certain physical quantities in natural physical models is highlighted as one of the main open problems in mathematical physics in the list published by the International Association of Mathematical Physics in the late 90's, edited by Aizenman~\cite{Aizenmann}.
Apart from the pioneering work of Komar in this direction in 1964 \cite{Komar64}, stating undecidable properties in Quantum Field Theories, and the influential paper of Anderson entitled ``More is different'' \cite{Anderson72} in 1972, most of the initial connections between physical problems and undecidability arose in the 80s and 90s. In 1981, \cite{Pourel81} studied non-computable solutions to the wave equation. One year later, \cite{Fredkin82} found undecidable questions in models of hard frictionless balls. In 1984, Wolfram wrote the paper ``Undecidability and Intractability in Theoretical Physics'' where, motivated by the emergent complexity of very simple cellular automata, he conjectured that many natural problems in physics, and in particular in classical statistical physics, should be undecidable. Many results in this direction -- using cellular automata to prove undecidability results in physical problems -- have been shown since then (e.g \cite{Domany84} or \cite{Omohundro84}). The work of Gu, Weedbrook, Perales and Nielsen \cite{Gu-Nielsen} in 2009 found undecidable properties in the 2D classical Ising model.  Besides cellular automata, another natural connection between undecidability and physics, that we will exploit in this paper, came from tiling problems, shown to be undecidable by the works of Wang and Berger in the 60s \cite{Wang, Berger}. This idea has been exploited for instance in 1990 by Kanter \cite{Kanter90}, finding undecidable properties in anisotropic 1D Potts Hamiltonians. That same year, in a completely different direction, Moore wrote one of the most influential papers on undecidability in physics \cite{Moore90}, proving undecidability of the long-term behaviour of a particle-in-a-box problem. (See also \cite{Bennett90} for a nice commentary on that paper.)

In recent years, mainly motivated by quantum information theory and the link it established between physics and computer science, there has been a revival of interest in undecidability in quantum physics. Results in this direction have appeared in several contexts, such as measurement and control  \cite{Eisert12,Wolf11}, tensor networks \cite{Morton12, Kliesch14, Delascuevas16}, measurement-based quantum computation \cite{vandenNest08}, channel capacities \cite{Elkouss16}, or Bell inequalities \cite{Bendersky16, Slofstra16}.

A precursor of those results, due to Lloyd, appeared already in 1993 \cite{Lloyd, Lloyd94} in the early days of quantum information theory. There, it is argued that quantum computing implies certain spectral properties of quantum mechanical operators are undecidable. More precisely, the unitary evolution $U$ associated to the evolution of a computer (classical or quantum) capable of universal computation has invariant subspaces with discrete spectrum (roots of unity) and other invariant subspaces with continuous spectrum (the whole unit sphere), corresponding respectively to computations that halt and do not halt. Then, since the halting problem is undecidable, Lloyd concludes that given a quantum state associated with a program in the infinite dimensional space in which $U$ is defined, it is undecidable to know whether it has overlap with an invariant subspace having discrete spectrum or, conversely, it is supported on an invariant subspace in which the spectrum is the full unit circle. See \cite{Lloyd16, Cubitt16-Comment} for a recent discussion between the relation between Lloyd's result and the result of this paper.

In subsequent follow-up work, we have extended the undecidability results described here to 1D systems~\cite{1d_spectral_gap} and to uncomputability of phase diagrams~\cite{phase_diagrams}, and have used insights from our undecidability results to construct simple, explicit examples of the new type of ``size-driven'' phase transition implied by undecidability of the spectral gap~\cite{size-driven}.

We refer to a forthcoming review paper in collaboration with Gu and Perales \cite{review-in-preparation} for an extensive analysis and discussion of all results mentioned above, and many more, related to undecidability in physics.

\subsection{Brief overview of the proof} \label{sec:extended-overview}
An extended overview and discussion of the ideas behind our proof can be found in the Supplementary Material of \cite{Cubitt15a}. We will simply sketch here the four main steps of the proof, whose rigorous statements and proofs are given, respectively, in \cref{sec:phase-estimation,sec:local-Hamiltonians,sec:quasi-periodic,sec:put-together}.

\noindent\textbf{Step 1: write $n$ on the tape}.\\
In order to feed an arbitrary input $n\in \N$ to the UTM, and at the same time keep a uniform upper bound on the local dimension of the Hilbert space of each site in the final Hamiltonian, we need to construct a Quantum Turing Machine (QTM) with a fixed number of internal states such that, when fed with an input given by any sequence of 1's larger than the size of $n$, writes $n$ deterministically on the tape and halts. We need also to keep strict control on the time and space required for this computation. The precise result we prove along these lines is given in \cref{phase-estimation_QTM}, and  \cref{sec:phase-estimation} is devoted to proving it. The idea is to construct explicitly the QTM associated to the \textit{quantum phase estimation algorithm}. To do this, we rely heavily on results and ideas from \cite{Bernstein-Vazirani}.

\noindent\textbf{Step 2: construct, for each $n\in \N$, a 1D Hamiltonian whose ground state energy on a finite chain depends on the behaviour of the UTM on input $n$}.\\
For this, we heavily rely on the remarkable paper of Gottesman and Irani \cite{Gottesman-Irani} (see also \cite{toby-johannes-maris}), which can be seen as a milestone in a long history of papers \cite{Kitaev_book, KKR, Oliveira-Terhal, AGIK}, dating back to Feynman \cite{Feynman}, relating the ground state energy of a Hamiltonian with the time-evolution of a quantum computation. In order to achieve the desired properties of our construction, we have to modify the original construction of Gottesman and Irani, but most of the essential ideas of this step appear already in \cite{Gottesman-Irani}.

After these two steps, we have for each $n\in \N$ a 1D translationally invariant Hamiltonian whose ground state energy for a chain of size $L$ and open boundary conditions is $0$ if the UTM does not halt on $n$ in space less than $O(L)$ and time less than $O(c^L)$ (for any desired constant $c$). Otherwise, it has energy larger than $c^{-L}$. So there the difference in ground state energy depending on the halting behaviour of the UTM on $n$ vanishes exponentially fast with chain length, and is zero in the thermodynamic limit. \cref{QTM_in_local_Hamiltonian} states the precise result we obtain, and  \cref{sec:local-Hamiltonians} is devoted to its proof.

\noindent\textbf{Step 3: amplify the ground state energy difference}.\\
For this, we turn to 2D lattices, and exploit the properties of an aperiodic tiling of the 2D plane due to Robinson. The idea is to construct a Hamiltonian whose ground state mimics the tiling pattern of Robinson's tiling shown in \cref{fig:tiling} and, at the same time, places the ground state of the 1D Hamiltonian constructed in Step~2 on top of each of the 1D borders appearing in this pattern. Using the fact that there is a constant density of squares of size $4^r$ for all $r\in \N$, and making a shift in energies, we manage to show for the resulting Hamiltonian that, if the UTM halts on $n$, the ground state energy diverges to $+\infty$, whereas in the non-halting case it diverges to $-\infty$. From there, we conclude undecidability of the ground state energy density.  \cref{sec:quasi-periodic} takes care of proving all the new results we require for Robinson's tiling. The rest of the proof, together with the final step, appears in \cref{sec:put-together}.

\noindent\textbf{Step 4: from ground state energy difference to spectral gap}.\\
The final step combines the Hamiltonian $H_u$ constructed in Step 3 with two others: a trivial Hamiltonian having ground state energy~$0$ and a constant spectral gap, and a critical Hamiltonian $H_d$ having ground state energy~$0$ and a spectrum that becomes dense in $[0,\infty)$ as the system size goes to infinity. We couple these Hamiltonians in such a way that the spectrum of the resulting Hamiltonian on $\Lambda(L)$ is $$\{0\} \cup \left[\spec(H^{\Lambda(L)}_u)+\spec(H^{\Lambda(L)}_d)\right] \cup S_L,$$ with $S_L\ge \min\left\{1, 1+\lambda_0(H_u^{\Lambda(L)})\right\}$. This, together with the spectral properties of $H_u$ shown in Step~3, completes the proof.


\clearpage

\section{Unconstrained local Hilbert space dimension}
\label{sec:unbounded}

Before starting in \cref{sec:phase-estimation} on the proof of the main theorem, in this section we will describe two approaches that exploit known undecidability results for tiling and completion problems. Based on these, we can derive simpler but weaker undecidability results for the spectral gap and for other low energy properties of translationally invariant, nearest-neighbour Hamiltonians on a 2D square lattice, defined by their local interactions $\{(h_\text{row}(n),h_\text{col}(n))\}_{n\in\mathbb{N}}$. In contrast to the main theorem of this paper, however, the families of Hamiltonians that we construct here have local Hilbert space dimension $d_n$ that differs for different elements $n$ of the family, with no upper bound on $\{d_n\}_{n\in\mathbb{N}}$.

\subsection{Undecidability of the spectral gap via tiling}
\label{sec:undec_via_tiling}
As in the more sophisticated constructions that will appear later, the idea is to reduce an undecidable ground state energy problem to the spectral gap problem. If we do not constrain the local Hilbert space dimension, then this reduction can be chosen such that it directly exploits the undecidability of a tiling problem. To this end, we need two ingredients:

\subsubsection{Ingredient 1: a tiling Hamiltonian}
We start by recalling the notion of Wang tilings and tiling problems. A unit square whose edges are coloured with colours chosen from a finite set is called a \emph{Wang tile}. A finite set $\mathcal{K}$ of Wang tiles is said to \emph{tile the plane} $\mathbb{Z}^2$ if there is an assignment $\mathbb{Z}^2\rightarrow \mathcal{K}$ such that abutting edges of adjacent tiles have the same colour. (Rotations or reflections of the tiles are not allowed here.) It is a classic result of Berger~\cite{Berger} that, given any set of tiles as input, determining whether or not this set can tile the plane is undecidable.

There is an easy way to rewrite any tiling problem as a ground state energy problem for a classical Hamiltonian. If $\mathcal{K}=\{1,\ldots,K\}$ we assign a Hilbert space $\cH^{(i)}\simeq\C^K$ to each site $i$ of a square lattice, and define the local interactions via
\begin{equation}
  h_c^{(i,j)} := \sum_{(m,n)\in C^{(i,j)}} \proj[(i)]{m}\ox\proj[(j)]{n},
\end{equation}
where the set of constraints $C^{(i,j)}\subseteq\mathcal{K}\times\mathcal{K}$ includes all pairs of tiles $(m,n)$ which are incompatible when placed on adjacent sites $i$ and $j$. The overall Hamiltonian on the lattice $\Lambda(L)$ is then simply
\begin{equation}
  H_c^{\Lambda(L)} := \sum_{(i,j)\in\cE} h_c^{(i,j)}.
\end{equation}

By construction, we have $h_c^{(i,j)}\geq 0$, $H_c^{\Lambda(L)}$ is translational invariant and its spectrum is contained in $\mathbb{N}_0$. If there exists a tiling of the plane, then for open boundary conditions $\forall L: 0\in\spec{H_c^{\Lambda(L)}}$. Similarly, in the case of periodic boundary conditions, the existence of a periodic tiling implies that $0\in\spec{H_c^{\Lambda(L)}}$ holds for an unbounded sequence of $L$'s. On the other hand, if there is no tiling, then there is an $L_0$ such that
\begin{equation}
  \forall L>L_0 : \spec{H_c^{\Lambda(L)}} \geq 1.
\end{equation}
This is due to Wang's extension theorem (see e.g.\ \cite{Grunbaum+Shephard}).

\subsubsection{Ingredient 2: a gapless frustration-free Hamiltonian}
As a second ingredient we will use that there are frustration-free two-body Hamiltonians
\begin{equation}
  H_q^{\Lambda(L)}:=\sum_{(i,j)\in\cE} h_q^{(i,j)},
\end{equation}
such that for $L\rightarrow\infty$ we get $\spec{H_q^{\Lambda(L)}}\rightarrow\R_+$ in the sense that the spectrum of the finite system approaches a dense subset of $\R_+$ \cite{UncleHamiltonians}. We will denote the corresponding single site Hilbert space by $\HS_q^{(i)}\simeq\C^D$ and we can in fact choose $D=2$ for instance by assigning to each row of the lattice the XY-model with transversal field taken at a gapless point with product ground state. Note that in this case the entries of the matrices $h_q^{(i,j)}$ are rational.

\subsubsection{Reducing tiling to spectral gap}
In order to fruitfully merge the ingredients, we assign a Hilbert space $\HS^{(i)} := \HS_0^{(i)}\oplus\HS_c^{(i)}\ox\HS_q^{(i)} \simeq \C^1\oplus\C^K\ox\C^D$ to each site $i\in\Lambda(L)$. A corresponding orthonormal set of basis vectors will be denoted by $\ket{0}\in\C^1$ and $\ket{k,\alpha}:=\ket{k}\ox\ket{\alpha}\in\C^K\ox\C^D $, respectively. The Hamiltonian is then defined in terms of the two-body interactions
\begin{align}
  H^{(i,j)} := &\proj{0}^{(i)}\ox\1_{cq}^{(j)}+\1_{cq}^{(i)}\ox\proj{0}^{(j)}
              \label{eq:main1a}\\
           &+ \sum_{(m,n)\in C^{(i,j)}}
              \proj{m}^{(i)}\ox\1_q^{(i)}\ox\proj{n}^{(j)}\ox\1_q^{(j)}
              \label{eq:main2}\\
           &+ \sum_{\alpha,\beta,\gamma,\delta=1}^D\1_c^{(i)} \ox
              \ketbra{\alpha}{\beta}^{(i)} \ox \1_c^{(j)} \ox \ketbra{\delta}{\gamma}^{(j)} \; \braket{\alpha,\delta|h_q^{(i,j)}|\beta,\gamma}.
              \label{eq:main3}
\end{align}
Here $\1_c,\1_q$ and $\1_{cq}$ denote the identity operators on $\HS_c,\HS_q$ and $\HS_c\ox\HS_q$, respectively.
As before, we define the Hamiltonian on the square lattice $\Lambda(L)$ as
\begin{equation}
  H^{\Lambda(L)}:=\sum_{(i,j)\in\cE}H^{(i,j)}.
\end{equation}
\begin{theorem}[Reducing tiling to spectral gap]\label{thm:main-unbounded}
  Consider any set $\mathcal{K}$ of Wang tiles, and the corresponding family of two-body Hamiltonians $\{H^{\Lambda(L)}\}_{L}$ on square lattices $\Lambda(L)$ either with open or with periodic boundary conditions.
  \begin{enumerate}
  \item The family of Hamiltonians is frustration-free.
  \item If $\mathcal{K}$ tiles the plane $\mathbb{Z}^2$, then in the case of open boundary conditions, we have that $\spec{H^{\Lambda(L)}}\rightarrow\mathbb{R}_+$ for $L\rightarrow\infty$, i.e., there is no gap and an excitation spectrum that becomes dense in $\R_+$. Similarly, if $\mathcal{K}$ tiles any torus, then in the case of periodic boundary conditions $\spec{H^{\Lambda(L_i)}}\rightarrow\mathbb{R}_+$ for a subsequence $L_i\rightarrow\infty$.
  \item If $\mathcal{K}$ does not tile the plane, then there is an $L_0$ such that for all $L>L_0$ $H^{\Lambda(L)}$ has a unique ground state and a spectral gap of size at least one, i.e.
    \begin{equation}
      \spec{H^{\Lambda(L)}}\setminus\{0\}\geq 1.
    \end{equation}
  \end{enumerate}
\end{theorem}

\begin{proof}
  Frustration-freeness is evident since $H^{(i,j)}\geq 0$ and $H^{\Lambda(L)}\ket{0,\ldots,0} = 0$.
  In order to arrive at the other assertions, we decompose the Hamiltonian as $ H^{\Lambda(L)}=:H_0+H_c+H_q$, where the three terms on the right are defined by taking the sum over edges separately for the expressions in \cref{eq:main1a}, \cref{eq:main2} and \cref{eq:main3}, respectively.
  Let us now assign a \emph{signature} $\sigma\in\{0,\ldots,K\}^{L^2}$ to every state of our computational product basis, in the following way: $\ket{0}^{(i)}$ is assigned the signature $\sigma_i=0$, and $\ket{k,\alpha}^{(i)}$ is assigned the signature $\sigma_i=k$ irrespective of $\alpha$.
  By collecting computational basis states with the same signature we can then decompose the Hilbert space as
  \begin{equation}
    \bigotimes_{i\in\Lambda}\HS^{(i)}\simeq\bigoplus_{\sigma}\HS_\sigma.
  \end{equation}
  The Hamiltonian is block-diagonal w.r.t.\ this decomposition, i.e.\ it can be written as
  \begin{equation}
    H^{\Lambda(L)}=\bigoplus_\sigma H_\sigma,
    \quad\text{and thus}\quad
    \spec{H^{\Lambda(L)}}=\bigcup_\sigma\spec{H_\sigma}.
  \end{equation}
  In order to identify the spectra coming from different signatures we will distinguish three cases:

  \begin{enumerate}
  \item $\sigma=(0,\ldots,0)$: this yields an eigenvalue $0$ as already mentioned.

  \item $\sigma \neq (0,\ldots,0)$ but $\sigma_i=0$ for some $i$: for any unit vector $\ket{\psi_\sigma}$ with such a signature we have $\braket{\psi_\sigma|H_0|\psi_\sigma} \geq 1$. We will denote the part of the spectrum which corresponds to these signatures by $S\subset\mathbb{R}$. The only property we will use is that $S\geq 1$.

  \item $\sigma\in\{1,\ldots,K\}^{L^2}$: In the subspace spanned by all states whose signature does not contain $0$ we have that $H_0=0$ and $H_c+H_q=H_c^{\Lambda(L)}\ox\1+\1\ox H_q^{\Lambda(L)}$.
    Consequently, the spectrum stemming from this subspace equals $\spec{H_c^{\Lambda(L)}}+\spec{H_q^{\Lambda(L)}}$.
  \end{enumerate}

  Putting things together, we obtain
  \begin{equation}
    \spec{H^{\Lambda(L)}} = \{0\}\cup\Big\{\spec{H_c^{\Lambda(L)}}+\spec{H_q^{\Lambda(L)}}\Big\}\cup S,
  \end{equation}
  The equivalence between the impossibility of tiling with $\mathcal{K}$ and the existence of a spectral gap of size at least~1 now follows immediately from the properties of the ingredients $h_c$ and $h_q$. Uniqueness of the ground state in cases where no tiling exists follows from the above argument when considering the spectra as multisets rather than sets.
\end{proof}

From the undecidability of the tiling problem we now obtain the following corollary. Recall from \cref{sec:preliminaries} that a pair of Hermitian matrices $(h_\text{row},h_\text{col})$ defines a Hamiltonian $H^{\Lambda(L)}$ for each square $\Lambda(L)$  via \cref{eq:H-Lambda-from-h-row-col}, where we are now choosing open boundary conditions.\footnote{Here we make no use of an additional on-site term, so $h_1=0$.}

\begin{corollary}[Undecidability of the spectral gap for unconstrained dimension]\ \\
     Let $\epsilon >0$. There is no algorithm that, on input $(h_\text{row},h_\text{col})$ with rational entries and operator norm smaller than $1+\epsilon$, decides whether the associated family of Hamiltonians $\{H^{\Lambda(L)}\}$ describe a gapped or a gapless system (according to the definitions given in \cref{sec:preliminaries}). This holds even under the promise that one of these is true, that $H^{\Lambda(L)}$ is frustration free for all $\Lambda$, and that in the gapped case the spectral gap is at least~1.
\end{corollary}
Here, the norm bound follows from the observation that $\matnorm{H^{(i,j)}} \leq 1 + \matnorm{h_q^{(i,j)}}$ where $h_q^{(i,j)}$ can be rescaled to arbitrary small norm.

For the case of periodic boundary conditions we obtain the same statement if we replace our strong definition of `gapless' by the weaker requirement that $\exists c>0\; \forall\epsilon>0 \; \forall L_0 \; \exists L>L_0$ so that every point in $[\lambda_0(H^{\Lambda(L)}),\lambda_0(H^{\Lambda(L)})+c]$ is within distance $\epsilon$ from $\spec H^{\Lambda(L)}$. The reason for this is, that if the period of the torus does not match the required one, then a gap can be generated that, however, disappears again if we enlarge the size of the system.

We emphasise that, so far, there is no constraint on the local Hilbert space dimension. Since every instance in the above construction has a finite local Hilbert space we can, however, at least conclude axiomatic undecidability for finite local Hilbert spaces. That is, for any given consistent formal system with a recursive set of axioms, one can construct a specific instance of a Hamiltonian with properties as in the corollary and finite-dimensional local Hilbert space, such that neither the presence nor the absence of a spectral gap can be proven from the axioms.\footnote{An explicit example can, for instance, be constructed from the explicit Turing Machines of \cite{YedidiaAronson16} for which the halting problem is independent of ZFC}. However, the local dimension, whilst finite, depends on the choice of axiomatic system and there is no upper-bound on the values it can take. The main aim of this paper is to prove the much stronger \cref{thm:promise}, which shows that the spectral gap problem remains undecidable even for a fixed value of the local Hilbert space dimension. This in turn implies that for any consistent, recursive formal system, axiomatic independence holds for Hamiltonians with this fixed local dimension; the local dimension is now independent of the choice of axioms.

In the above construction, gapped cases are related to the impossibility of tiling. Since the latter always admits a proof, the axiomatically undecidable cases coming from this construction have to be gapless. The following section will provide a construction where this asymmetry is reversed.

\subsection{Undecidability of low energy properties}

\subsubsection{Reducing tile completion to a ground state energy problem}
We now modify the above construction in order to show that not only the spectral gap but many other low energy properties are undecidable as well. The idea is to invert the relation between gap and existence of tiling. In the previous construction, the existence of a gap was associated with the impossibility of a tiling. Now we want to associate it with the \emph{existence} of a tiling. A drawback is that we loose the frustration-free property. On the other hand, we do not have to rely on the undecidability of tiling (which is a rather non-trivial result \cite{Berger,Robinson}), but can exploit the simpler undecidability of the \keyword{completion problem}, already proven by Wang \cite{Wang}. In the tiling completion problem, one fixes a tile at the left-bottom corner and asks whether the first quadrant can be tiled.

Undecidability of the completion problem is relatively simple by reduction from the Halting Problem for Turing Machines (TM), and explained in full detail in \textcite[Section 4]{Robinson}. There are many different, computationally equivalent definitions of Turing Machines. (The Halting Problem is undecidable for any of these variants.) For the completion problem, the most convenient is to take a Turing Machine to be specified by a finite number of states $Q=\{q_0, q_1,\ldots \}$ with $q_0$ the initial state, a finite alphabet of tape symbols $S=\{s_0,s_1,\ldots\}$ with $s_0$ the blank symbol, together with a finite set of transition rules specified by quintuples of the form
\begin{equation*}
  (q,s,s',D,q')\in Q\times S\times S\times \{\text{left, right}\}\times Q.
\end{equation*}
The TM has a half-infinite tape of cells indexed by $\N$, and a single read/write tape head that moves along the tape. The machine starts in state $q_0$ with the head in tape cell~0. If the TM is currently in state $q$ and reads the symbol $s$ from the current tape cell, then it writes $s'$ into the current tape cell, transitions to state $q'$, and moves in the direction $D$. For each $q,s$, there must be at most one valid quintuplet.
If there is none, then the machine halts.\footnote{Note that this defines a slightly different variant of Turing Machines from those used later in \cref{sec:phase-estimation}.}

\begin{figure}[!htp]
  \centering
   \includegraphics[width=0.8\columnwidth]{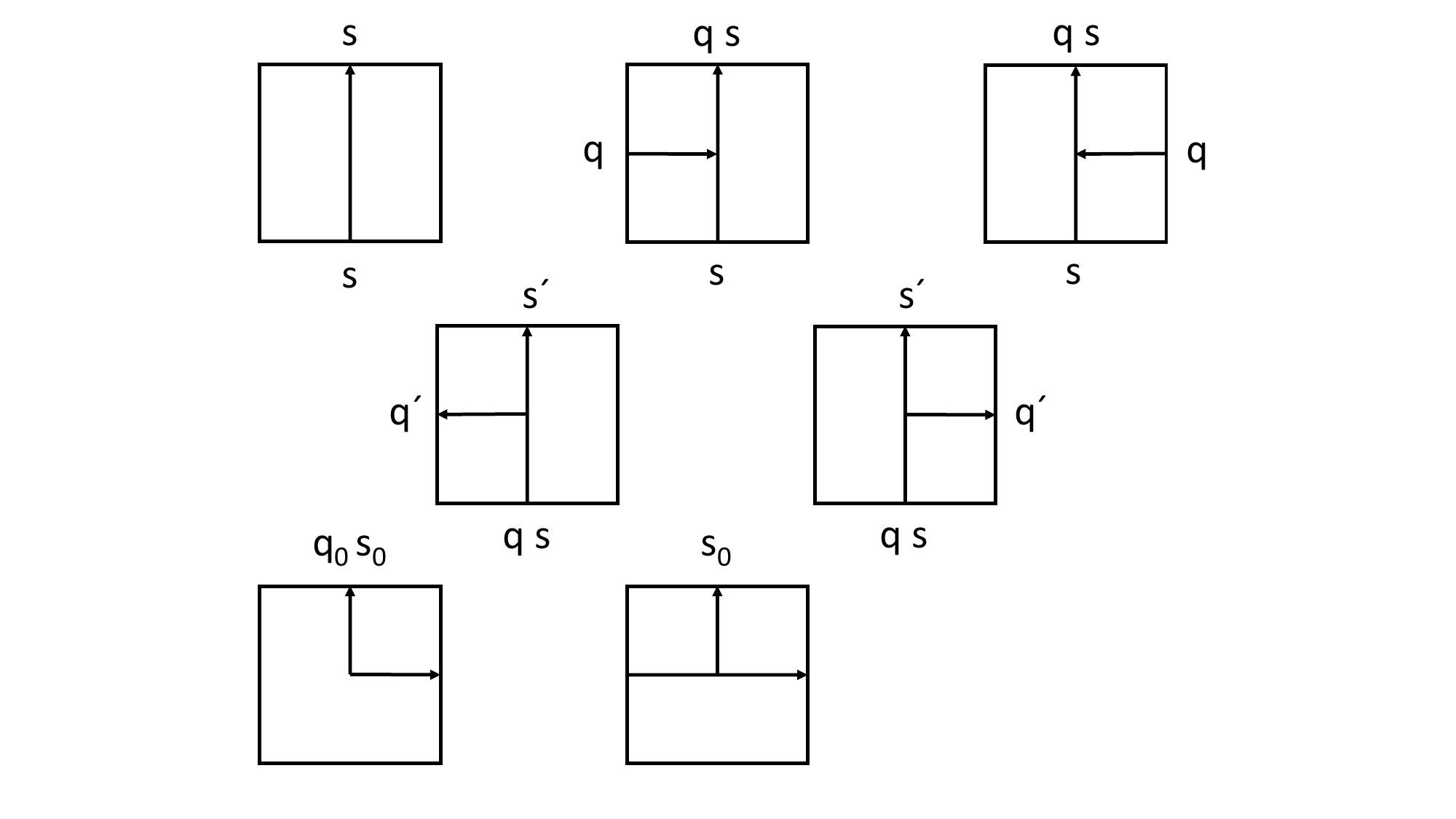}
  \caption[Undecidability of the tiling completion problem.]{Tiles for encoding a Turing machine with half-infinite tape into a completion problem. The tile in the bottom-left corner is fixed and, together with the second tile in the third row, starts the Turing machines running on the blank half-infinite tape. The tiles in the second row are the 'action tiles' that are able to change the state of the Turing machine and correspond to a left or right moving head. For each pair $(q,s)$ there is at most one action tile.}\label{fig:RobinsonTM-tiles}
\end{figure}

Following \textcite{Robinson}, we use the Wang tiles shown in \cref{fig:RobinsonTM-tiles}. For clarity, edges here are marked with labelled arrows rather than colours, and adjacent edges match if arrow heads abut arrow tails with the same label; clearly these tile markings can be identified with different colours to recover a set of standard Wang tiles. We denote the set of tiles by $\mathcal{K}$, which contains at most $K:=|\mathcal{K}|\leq 2+ (3|Q|+1)|S|$ elements.

If the tile that appears in the bottom-left corner in \cref{fig:RobinsonTM-tiles}, which we denote by $\crossURblue$, is fixed in the bottom-left corner of an $L\times L$ lattice, then any valid tiling corresponds to the evolution of the Turing Machine on the blank tape up to time $L$, where time goes up in the vertical direction and each row of tiles corresponds to the total configuration (including the tape) of the Turing machine in two consecutive instants of time (represented by the bottom and upper edges of that row of tiles). Hence,
\begin{enumerate}[label=(\arabic{*}).,ref=(\arabic{*})]
\item There exists a valid tile completion for all $L\times L$ squares if and only if the Turing Machine, when running on an initially blank tape, does not halt.
\item If such a tiling exists, then it is unique.
\end{enumerate}

Now, the completion problem can be represented as a ground state energy problem in the following straightforward way. Consider a square lattice $\Lambda(L)$ with open boundary conditions and corresponding edge set $\cE$.

Assign a weight $w_{(i,j)}(m,n)\in\mathbb{Z}$ to edge $(i,j)\in\cE$ on which a pair of tiles $(m,n)\subseteq\mathcal{K}\times\mathcal{K}$ is placed. The choice of weights is summarised in the following table. Here $\wctile$ stands for any tile different from $\crossURblue$, the unbracketed numbers define the weights for non-matching adjacent tiles, and the numbers in brackets define the weights for cases where the tiles match (i.e.\ all abutting labels and arrows match).
\begin{equation}\label{eq:corner_weights}
  \begin{array}{cc|cc}
    & & \multicolumn{2}{|c}{\text{Right tile}} \\
    & & \crossURblue & \wctile \\
    \hline
    \multirow{2}{*}{{\text{Left tile}}}
    & \crossURblue & 4 & 2\ (-1) \\
    & \wctile & 4 & 2\ \phantom{-}(0)
  \end{array}
  \qquad
  \begin{array}{cc|cc}
    & & \multicolumn{2}{|c}{\text{Tile above}} \\
    & & \crossURblue & \wctile \\
    \hline
    \multirow{2}{*}{{\text{Tile below}}}
    & \crossURblue  & 4 & 2\ (-1) \\
    & \wctile  & 4 & 2\ \phantom{-}(0)
  \end{array}
\end{equation}

\vspace{2em}

Assign a Hilbert space $\HS_u^{(i)}\simeq\mathbb{C}^{K}$ to each site $i\in\Lambda(L)$ whose computational basis states are labelled by tiles. We define a Hamiltonian $H_u^{\Lambda(L)}:=\sum_{(i,j)\in\cE}h_u^{(i,j)},$ with
\begin{equation}\label{eq:tiling-Hamiltonian}
  h_u^{(i,j)} := \sum_{m,n\in\mathcal{K}} \omega_{(i,j)}(m,n)\ \proj[(i)]{m}\ox\proj[(j)]{n},
\end{equation}
so that positive and negative weights become energy ``penalties'' or ``bonuses'' in the Hamiltonian.
By construction, the Hamiltonian is diagonal in computational basis, its eigenvectors correspond to tile configurations $\Lambda(L)\rightarrow \mathcal{K}$ and, since all weights are integers, its spectrum lies in $\mathbb{Z}$. In order to determine the ground state energy note that the tile $\crossURblue$ is the only one that can lead to negative weights. Assume first that a tile $\crossURblue$ is placed on a site different from the bottom-left corner. Then this tile will have at least one neighbour below with an arrow pointing upwards or one neighbour to the left with an arrow pointing to the right. In either case there will be an energy penalty $+4$ so that the weights involving this tile will sum up to at least $2$. If, however, $\crossURblue$ is placed in the bottom-left corner, then no such energy penalty is picked up and the tile can contribute $-2$ to the energy.

If the tiling is then completed following the evolution of the encoded Turing machine on a blank tape, then no additional penalty will occur on $\Lambda(L)$ if the Turing machine does not halt within $L$ time-steps. Hence $\lambda_0(H_u^{\Lambda(L)})=-2$, the ground state will be unique, and will be given by a product state. If the Turing machine halts, then the completion problem has no solution beyond some $L_0$ and every configuration with $\crossURblue$ at its bottom-left corner will get an additional penalty of at least $2$, leading to non-negative energy. Since there is always a configuration that achieves zero energy (e.g.\ one using only the tile types at the top-left of \cref{fig:RobinsonTM-tiles}), we obtain:

\begin{lemma}[Relating ground state energy to the halting problem]\label{TM2gsenergy1}
  Consider any Turing machine with state set $Q$ and tape alphabet $S$ running on an initially blank tape such that its head never moves left of the starting cell. There is a family of translational invariant nearest neighbour Hamiltonians $\{H_u^{\Lambda(L)}\}_L$ (specified above) on the 2D square lattice with open boundary conditions and local Hilbert space dimension at most $|S|(3|Q|+1)+2$ such that
  \begin{enumerate}
  \item\label{TMlemmaitem1} If the Turing Machine does not halt within $L$ time-steps, then $\lambda_0(H_u^{\Lambda(L)})=-2$. In this case the corresponding ground state is a product state, and is the unique eigenstate with negative energy.
  \item If the Turing machine halts, then $\exists L_0$ (given by the halting time) such that $ \forall L>L_0:\lambda_0(H_u^{\Lambda(L)})= 0$.
\end{enumerate}
\end{lemma}
Since every Turing machine can be simulated by one with a half-infinite tape, which automatically satisfies the constraint that its head never moves to the left of the origin, the above construction leads to an undecidable ground state energy problem, which we will exploit in the following.

\subsubsection{Reduction of the halting problem to arbitrary low energy properties}\label{sec:physical-implications}

Consider translational invariant Hamiltonians on square lattices with open boundary conditions. Let $H_q^{\Lambda(L)}$ describe such a family of Hamiltonians on  $L\times L$ lattices. Our aim is to prove the undecidability of essentially any low energy property displayed by this Hamiltonian in the thermodynamic limit that distinguishes it from a gapped system with unique product ground state, e.g.\ the existence of topological order, or non-vanishing correlation functions. This is implied by the following theorem:

\begin{theorem}[Relating low energy properties to the halting problem]\label{low-energy}
  Let TM be a Turing machine with state set $Q$ and alphabet $S$ running on an initially blank tape. Consider the class of translational invariant Hamiltonians on square lattices with open boundary conditions. Let $H_q^{\Lambda(L)}$ with nearest-neighbour interaction $h_q^{(i,j)}\in\cB(\HS_q^{(i)}\ox\HS_q^{(j)})$, $\HS_q^{(i)}\simeq\mathbb{C}^q$, and on-site term $h_{q1}^{(k)}\in\cB(\HS_q^{(k)})$ describe such a Hamiltonian. Assume that there is a $c\in\mathbb{R}$ such that $\lambda_0\Big(H_q^{\Lambda(L)}\Big)-cL^2\in[-\frac12,\frac12]$ for all lattice sizes $L$. Then we can construct a family of Hamiltonians $H^{\Lambda(L)}$ with the following properties:
  \begin{enumerate}
  \item $H^{\Lambda(L)}$ is translational invariant on the square lattice with open boundary conditions and local Hilbert spaces $\HS^{(i)}\simeq\mathbb{C}^d$ of dimension $d\leq 2(q+1)+|S|(3|Q|+1)$.
    \item The interactions of $H^{\Lambda(L)}$ are nearest-neighbour (possibly with on-site terms), computable and independent of $L$.
  \item\label{lowenergythm_nonhalting} If the TM does not halt within $L$ time-steps, then $H^{\Lambda(L)}$ has a non-degenerate product ground state and a spectral gap $\Delta(H^{\Lambda(L)})\geq\frac12$.
  \item \label{lowenergythm_halting} If TM halts, then $\exists L_0$ (given by the halting time) such that $\forall L>L_0$ the following identity between multisets holds in the interval $[0,\frac12)$:
    \begin{equation}
      \spec H_q^{\Lambda(L)}-\lambda_0(H_q^{\Lambda(L)})
      = \spec H^{\Lambda(L)}-\lambda_0(H^{\Lambda(L)}).
    \end{equation}
    In this case there exist isometries $V^{(i)}:\HS_q^{(i)}\rightarrow \HS^{(i)}$, $V:=\bigotimes_i V^{(i)}$ such that, for this part of the spectrum, any eigenstate $\ket{\phi_q}$ of $H_q^{\Lambda(L)}$ is mapped to the corresponding eigenstate $\ket{\phi}$ of $H^{\Lambda(L)}$ via $V\ket{\phi_q}=\ket{\phi}$.
  \end{enumerate}
\end{theorem}

Some remarks before the proof. Since, on the Turing machine level, \labelcref{lowenergythm_nonhalting} and \labelcref{lowenergythm_halting} cannot be told apart by any effective algorithm, the same has to hold for any property that distinguishes a gapped system with product ground state from the low energy sector of some frustration-free Hamiltonian. With small modifications in the proof, the nearest neighbour assumption for $h_q$ could be replaced by any fixed local interaction geometry. In this case, $H^{\Lambda(L)}$ will of course inherit this interaction geometry.

\begin{proof}
  W.l.o.g.\ $c=0$ since we can always compensate for it by adding a multiple of the identity to the on-site part of the Hamiltonian.

  We first embed the Hamiltonian given in the \namecref{low-energy} in a larger system, such that its ground state energy is shifted by $-1$. Define a translationally invariant, nearest-neighbour Hamiltonian on an auxiliary system with local Hilbert space $\HS_a^{(i)}\simeq\mathbb{C}^2$, via
  \begin{equation}
    \eta:=\sum_{(i,j)\in\cE}\eta^{(i,j)},\quad
    \eta^{(i,j)}:=-\frac12 \proj{0}^{(i)}\ox\proj{1}^{(j)}+2 \1^{(i)}\ox\proj{0}^{(j)}.
  \end{equation}
  Note that this is nothing but $\frac12 \times$ the Hamiltonian defined by \cref{eq:corner_weights} (with the number in brackets), where $\ket{0}$ equals $\crossURblue$ and $\ket{1}$ is the blank tile. Hence $\lambda_0(\eta)=-1$ and $\Delta(\eta)=1$.

  Now introduce $\HS_Q^{(i)}:=\HS_a^{(i)}\ox\HS_q^{(i)}$ and set $h_Q^{(i,j)}:=\eta^{(i,j)}\ox\1_q^{(i,j)}+\1_a^{(i,j)}\ox h_q^{(i,j)}$ acting on $\HS_Q^{(i)}\ox\HS_Q^{(j)}$. Then $H_Q^{\Lambda(L)}:=\sum_{(i,j)\in\cE} h_Q^{(i,j)}+\sum_{k\in\Lambda(L)}\1_a^{(k)}\otimes h_{q1}^{(k)}$ has the property that in the interval $\Big(-\infty,\lambda_0(H_q^{\Lambda(L)})\Big)$ we have that $$\spec H_q^{\Lambda(L)}-1=\spec H_Q^{\Lambda(L)}$$ holds as an identity between multisets (i.e.\ including multiplicities) and the corresponding eigenstates are related via a local isometric embedding.

  We now combine this with the Hamiltonian from the previous subsection by enlarging the local Hilbert space to $\HS^{(i)}:=\HS_u^{(i)}\oplus \HS_Q^{(i)}$ and defining
  \begin{equation}\label{eq:main1}
    h^{(i,j)}:=h_Q^{(i,j)}+h_u^{(i,j)}+6 \1_Q^{(i)}\ox \1_u^{(j)} + 6\1_Q^{(j)}\ox \1_u^{(i)} ,
  \end{equation}
  where $h_Q^{(i,j)}$ and $h_u^{(i,j)}$ act nontrivially on $\HS^{(i)}_Q\ox \HS_Q^{(j)}$ and $\HS_u^{(i)}\ox \HS_u^{(j)}$ respectively and $\mathbbm{1}_u$ and $\mathbbm{1}_Q$ denote the projectors onto the respective subspaces. The Hamiltonian on the square $\Lambda(L)$ is now
  \begin{equation}
    H^{\Lambda(L)}=\sum_{(i,j)\in\cE}h^{(i,j)}+\sum_{k\in\Lambda(L)}\1_a^{(k)}\otimes h_{q1}^{(k)}.
  \end{equation}

 To analyse the ground state of $H^{\Lambda(L)}$, notice that all the terms in the Hamiltonian $\eta, h_q, h_u, \1_Q\otimes\1_u$ commute with each other. Let $\tilde{H}_q = \sum_{(i,j)\in\cE} h_q^{(i,j)} + h_{q1}^{(i,j)}$, $\tilde{H}_\eta = \sum_{(i,j)\in\cE} \eta^{(i,j)} + 2\1_Q^{(i)}\ox \1_u^{(j)} + 2\1_Q^{(j)}\ox \1_u^{(i)}$, and $\tilde{H}_u = \sum_{(i,j)\in\cE} h_u^{(i,j)}+ 4 \1_Q^{(i)}\ox \1_u^{(j)} + 4\1_Q^{(j)}\ox \1_u^{(i)}$. We can then decompose $H^{\Lambda(L)}=\tilde{H}_q + \tilde{H}_u + \tilde{H}_\eta$ where the three Hamiltonians commute with each other.

 We now assign a \emph{signature} $\sigma\in\{0,1,2\}^{L^2}$ to every state of our computational product basis, so that $\ket{i}^{(j)}$ is identified with $\sigma_j\in\{0,1\}$ for all $\ket{i}\in \HS_Q^{(j)}$ of the form $\ket{\sigma_j}_a \ket{\alpha_j}_q$ and $\sigma_j=2$ for all  $\ket{i}\in \HS_u^{(j)}$. By collecting computational basis states with the same signature, we can then decompose the Hilbert space as
  \begin{equation}
    \bigotimes_{i\in\Lambda}\HS^{(i)}\simeq\bigoplus_{\sigma}\HS_\sigma.
  \end{equation}
  The Hamiltonian is block diagonal w.r.t.\ this decomposition, i.e.\ it can be written as
  \begin{equation}
    H^{\Lambda(L)}=\bigoplus_\sigma H_\sigma,
    \quad\mbox{and thus}\quad
    \spec{H^{\Lambda(L)}}=\bigcup_\sigma\spec{H_\sigma}.
  \end{equation}

  In order to identify the spectra and eigenstates coming from different signatures we will distinguish three cases:

  \paragraph{Case~1:} $\sigma=(2,\ldots,2)$. Restricted to this sector (i.e.\ the $\HS_\sigma$ component of the direct sum), the Hamiltonian acts as $H_u^{\Lambda(L)}=\sum_{(i,j)\in \cE} h^{(i,j)}_u$.  But according to \cref{TM2gsenergy1}, if up to time $L$ the TM did not halt, then $H_u^{\Lambda(L)}$ has one negative eigenvalue $-2$ whose associated eigenstate is a product. However, if $L\ge L_0$ where $L_0$ is related to the halting time, then $\spec H_u^{\Lambda(L)}\geq 0$.

  \paragraph{Case~2:} $\sigma_j=1$ for all $j\in \Lambda(L)$, except in the lowest left corner, where $\sigma_j=0$. In this sector we have the following identity between multisets:
  \begin{equation}
   \spec H_\sigma = \spec H_q^{\Lambda(L)} -1.
  \end{equation}

  \paragraph{Case~3:} Any other signature $\sigma$. It is easy to see that the energy given by $\tilde{H}_\eta$ in the sector $\HS_\sigma$ is exactly the energy given to the configuration $\sigma$ by the tiling Hamiltonian \cref{eq:tiling-Hamiltonian} with colours $\{0,1,2\}$ and weights given by the following tables from \cref{eq:corner_weights}, hence the energy is $\ge 0$.
  \begin{equation}
  \begin{array}{cc|ccc}
    & & \multicolumn{3}{c}{\text{Right tile}} \\
    & & 0 & 1 & 2 \\
    \hline
    \multirow{2}{*}{{\text{Left tile}}}
    & 0 & 2 & -\frac12 & 2\\
    & 1 & 2 &  0& 2\\
    & 2 & 2  & 2 &0
  \end{array}
  \qquad
  \begin{array}{cc|ccc}
    & & \multicolumn{3}{c}{\text{Tile above}} \\
    & & 0 & 1 & 2 \\
    \hline
    \multirow{2}{*}{{\text{Tile below}}}
    & 0 & 2 & -\frac12 & 2\\
    & 1 & 2 &  0& 2\\
    & 2 & 2  & 2 &0
  \end{array}
  \end{equation}
  Similarly, one can show that $\tilde{H}_u\ge 0$. Therefore, in this Case~3, the energy of $H^{\Lambda(L)}$ in the sector $\HS_\sigma$ is $\ge \lambda_0(H_q^{\Lambda(L)})$. \Cref{low-energy} now follows straightforwardly.
\end{proof}


\clearpage
\section{Quantum Phase Estimation Turing Machine}
\label{sec:phase-estimation}
We will see in \cref{sec:local-Hamiltonians} how to encode an arbitrary Quantum Turing Machine (QTM) into a translationally-invariant nearest-neighbour Hamiltonian, in such a way that the local Hilbert space dimension is a function only of the alphabet size and number of internal states. This will allow us to encode any specific universal (reversible) Turing Machine in a Hamiltonian with fixed local dimension. However, to prove undecidability by reduction from the Halting Problem, we will need a way to feed any desired input to this universal TM.

To this end, the main result of this section is to construct a family of quantum Turing Machines, all of which have the same alphabet size and number of internal states, but whose transition rules vary. For any given $n\in \mathbb{N}$, we can choose the transition rules such that the QTM, when started from a sequence of $N$ ones ($N$ any number larger than the number of bits $|n|$ in the binary representation of $n$) writes out the binary representation of $n$ to its tape and then halts deterministically.

Given $n$, we recall from \cref{sec:main-results} that $\varphi=\varphi(n)$ is defined as the rational number whose binary fraction expansion contains the digits of $n$ in reverse order after the decimal.\footnote{The fact that the digits are in reverse order is purely for convenience in the exposition -- it is trivial to modify the construction so that they are ordered the other way around.}

The QTM that we construct will be based on the quantum phase estimation algorithm~\cite{phase-estimation,Nielsen+Chuang}, applied to the single qubit unitary $U_\varphi=\left(\begin{smallmatrix}1&0\\0&e^{i\pi\varphi}\end{smallmatrix}\right)$, to obtain the phase $\varphi$. The input $N$ can be any upper bound on the number of binary digits in the output of the quantum phase estimation algorithm, such that $\varphi$ can be represented exactly as a binary fraction, with no approximation error. Note that the transition rules of the QTM are \emph{not} allowed to depend on $N$, though they can depend on $\varphi$ (and hence on $n$).

The important role played by $N$ can be sketched as follows. Firstly, $N$ fixes the number of output qubits to be used in the quantum phase estimation algorithm, such that an exact binary fraction representation of $\varphi$ is possible. But $N$ has a more subtle use in the last step of the quantum phase estimation algorithm, where one needs to implement an inverse quantum Fourier transform (QFT). To implement such a QFT in the standard way on the full set of output qubits, one would require quantum gates implementing phase rotations by angles that depend on $N$. But this would require $N$-dependent QTM transition rule amplitudes, which is not allowed. The solution is to use $N$ as a way to locate the least significant bit of $\varphi$, which will in turn allow the QFT to be implemented using gates that depend only on $\varphi$. (This is explained in more detail when we give the precise description analysis of the QTM, later in this section.)

This way of carrying out the QFT on a QTM comes at a cost of exponential run-time, so would not be useful if using the QFT to solve some computational problem. But for our peculiar use of the phase-estimation algorithm to generate an input string for a universal TM, which may anyway run indefinitely, this exponential run-time is not an issue. (Though it will entail a new clock construction when we come to encode the evolution of this QTM in a Hamiltonian, see \cref{sec:local-Hamiltonians}.)

More precisely, we will prove the following result (see below for the necessary basic definitions, notations, and facts about QTMs).
\begin{theorem}[Phase-estimation QTM]
  \label{phase-estimation_QTM}
  There exists a family of well-formed, normal form, unidirectional QTMs $P_n$ indexed by $n\in\N$ with the following properties:
  \begin{enumerate}
  \item \label[part]{phase-estimation_QTM:fixed_size}%
    Both the alphabet and the set of internal states are identical for all $P_n$; only the transition rules differ.

  \item \label[part]{phase-estimation_QTM:output}%
    On input $N\geq\abs{n}$ written in unary, $P_n$ behaves properly, halts deterministically after $O(\poly(N) 2^N)$ steps, uses $N+3$ space, and outputs the binary expansion of $n$ (padded to $N$ digits with leading 0's). (Here, $\abs{n}$ denotes length of the binary expansion of $n$.)

  \item \label[part]{phase-estimation_QTM:amplitudes}%
    For any $n$, and for each choice of states $p,q$, alphabet symbols $ \sigma, \tau$ and directions $D$, the transition amplitude $\delta(p,\sigma,\tau,q,D)$ is one of the elements of the set
    \begin{equation}
      \left\{0,1,\pm \frac{1}{\sqrt{2}},e^{i\pi\varphi}, e^{i\pi 2^{-\abs{\varphi}}}\right\}
    \end{equation}
    where $\varphi\in\Q$.
  \end{enumerate}

\end{theorem}
We emphasise again that the input $N$ does \emph{not} determine the output string that gets written to the tape; it only determines the number of binary digits in the output. The number represented by that output is determined (up to padding with leading zeros) by the choice of the parameter $n$ for the QTM $P_n$.

\Cref{phase-estimation_QTM} is essentially the only inherently quantum ingredient in our result, and the precise properties asserted there will be \emph{absolutely crucial} to the proof. For example, if instead of writing out the string and halting deterministically, the QTM did this only with arbitrarily high probability, our proof would not go through.

It is therefore not at all clear a priori whether QTMs fulfilling these strict requirements exist. Since our proof depends so delicately on the precise properties of these QTMs, we cannot appeal to previous results. Instead, in this section we give a detailed and explicit construction of a QTM based on the quantum phase estimation algorithm, that fulfils all the requirements of \cref{phase-estimation_QTM}.

\subsection{Quantum Turing Machinery}\label{sec:Turing_machinery}
For completeness we quote the basic definitions of Turing Machines and Quantum Turing Machines verbatim from \cite{Bernstein-Vazirani}.

\begin{definition*}[Turing Machine -- Definition~3.2
                    in~\cite{Bernstein-Vazirani}]\hfill\\
  \addtheoremline{definition}{Turing Machine}
  \label{def:TM}
  A (deterministic) Turing Machine (TM) is defined by a triplet $(\Sigma, Q, \delta)$ where $\Sigma$ is a finite alphabet with an identified blank symbol $\#$, $Q$ is a finite set of states with an identified initial state $q_0$ and final state $q_f\not = q_0$, and $\delta$ is a transition function
  \begin{equation}
    \delta: Q\times \Sigma \rightarrow \Sigma\times Q\times \{L,R\}.
  \end{equation}
  The TM has a two-way infinite tape of cells indexed by $\Z$ and a single read/write tape head that moves along the tape.
  A \keyword{configuration} of the TM is a complete description of the contents of the tape, the location of the tape head and the state $q\in Q$ of the finite control. At any time, only a finite number of the tape cells may contain non-blank symbols.

  For any configuration $c$ of the TM, the successor configuration $c'$ is defined by applying the transition function to the current state and the symbol scanned by the head, replacing them by those specified in the transition function and moving the head left (L) or right (R) according to $\delta$.

  By convention, the initial configuration satisfies the following conditions: the head is in cell $0$, called the \keyword{starting cell}, and the machine is in state $q_0$. We say that an initial configuration has \keyword{input} $x\in (\Sigma\setminus\#)^*$ if $x$ is written on the tape in positions $0,1,2,\cdots$ and all other tape cells are blank. The TM halts on input $x$ if it eventually enters the final state $q_f$. The number of steps a TM takes to halt on input $x$ is its running time on input $x$. If a TM halts, then its output is the string in $\Sigma^*$ consisting of those tape contents from the leftmost non-blank symbol to the rightmost non-blank symbol, or the empty string if the entire tape is blank. A TM is called \keyword{reversible} if each configuration has at most one predecessor.
\end{definition*}

As amplitudes in quantum mechanics can take arbitrary complex values, one has to be careful when defining quantum Turing Machines to ensure the transition amplitudes are themselves computable. Following \cite[Definition~3.2]{Bernstein-Vazirani}, we restrict the transition function to computable numbers in the standard way. Let $\tilde{\C}$ be the set consisting of $\alpha$ such that there is a deterministic classical Turing Machine $M_\alpha$ that, given input $n\in\N$, outputs the real and imaginary parts of $\alpha$ to within $2^{-n}$ in time polynomial in $n$.\footnote{We have followed \cite[Definition~3.2]{Bernstein-Vazirani} by restricting to \emph{efficiently} computable numbers in the definition of QTMs. This is not strictly necessary in our case, as we are concerned with computability rather than complexity theory here. \cite{Bernstein-Vazirani} show that this restriction is without loss of generality. In fact, in our case, the transition amplitudes of all the QTMs we construct will belong to the even more restrictive set of algebraic numbers.}

\begin{definition*}[Quantum Turing Machine -- Definition~3.2
                    in~\cite{Bernstein-Vazirani}] \hfill\newline
  \addtheoremline{definition}{Quantum Turing Machine}
  \label{def:QTM}
  A quantum Turing Machine (QTM) is defined by a triplet $(\Sigma, Q, \delta)$ where $\Sigma$ is a finite alphabet with an identified blank symbol $\#$, $Q$ is a finite set of states with an identified initial state $q_0$ and final state $q_f\neq q_0$, and $\delta$ -- the quantum transition function -- is a function
  \begin{equation}
    \delta: Q\times \Sigma \rightarrow \tilde{\C}^{\Sigma\times Q\times \{L,R\}}.
  \end{equation}
  The QTM has a two-way infinite tape of cells indexed by $\Z$ and a single read/write tape head that moves along the tape. We define configurations, initial configurations and final configurations exactly as for deterministic TMs.

  Let $\mathcal{S}$ be the inner-product space of finite complex linear combinations of configurations of $M$ with the Euclidean norm. We call each element $\phi\in \mathcal{S}$ a superposition of $M$. The QTM $M$ defines a linear operator $U_M: \mathcal{S}\rightarrow\mathcal{S}$, called the time evolution operator of $M$, as follows: if $M$ starts in configuration $c$ with current state $p$ and scanned symbol $\sigma$, then after one step $M$ will be in superposition of configurations $\psi=\sum_i\alpha_ic_i$, where each nonzero $\alpha_i$ corresponds to the amplitude $\delta(p,\sigma,\tau,q,d)$ of $\ket{\tau}\ket{q}\ket{d}$ in the transition $\delta(p,\sigma)$ and $c_i$ is the new configuration obtained by writing $\tau$, changing the internal state to $q$ and moving the head in the direction of $d$. Extending this map to the entire $\mathcal{S}$ through linearity gives the linear time evolution operator $U_M$.
\end{definition*}

Here, for convenience of programming, we will consider \emph{generalised} TMs and QTMs in which the head can also stay still ({\bf N}o-movement), as well as move {\bf L}eft or {\bf R}ight.

\begin{definition}[Generalised TM and QTM]
  A generalised TM or generalised QTM is defined exactly as a standard TM (\cref{def:TM}) or standard QTM (\cref{def:QTM}) except that the set of head movement directions is $\{L,R,N\}$ instead of just $\{L,R\}$.
\end{definition}

Following \textcite{Bernstein-Vazirani}, we define various classes of deterministic and quantum Turing Machines:

\begin{definition*}[Well-formed -- Definition~3.3
                   in~\cite{Bernstein-Vazirani}] \hfill\newline
  \addtheoremline{definition}{Well-formed QTM}
  We say that a QTM is \keyword{well-formed} if its time evolution operator is an isometry, that is, it preserves the Euclidean norm.
\end{definition*}

It is easy to see (Theorem~4.2 in~\cite{Bernstein-Vazirani}) that any reversible TM is also a well-formed QTM where the quantum transition function $\delta(p,\sigma, q,\tau, d)=1$ if $\delta(p,\sigma)=(q,\tau,d)$ for the reversible TM and $0$ otherwise.


\begin{definition*}[Normal form -- Definition~3.13
                    in~\cite{Bernstein-Vazirani}]
  \addtheoremline{definition}{Normal form QTM}
  \label{def:normal-form} \hfill\newline
  A well-formed QTM or reversible TM $M = (\Sigma,Q,\delta)$ is in \keyword{normal form} if
  \begin{equation}
    \forall\sigma\in\Sigma \quad
      \delta(q_f,\sigma) = \ket{\sigma}\ket{q_0}\ket{N}.
  \end{equation}
\end{definition*}

In \cite{Bernstein-Vazirani} the normal-form condition is $\delta(q_f,\sigma) = \ket{\sigma}\ket{q_0}\ket{R}$. However, the choice of final direction is just an arbitrary convention, since the machine already halted and never carries out those transitions. We replace $R$ with $N$ in the definition for technical reasons (see the Reversal \cref{reversal}, below).

There is a specific class of QTMs, called \keyword{unidirectional}, that play an important role in the general theory developed by \textcite{Bernstein-Vazirani}.

\begin{definition*}[Unidirectional -- Definition~3.14
                    in~\cite{Bernstein-Vazirani}]
  \addtheoremline{definition}{Unidirectional QTM}
  \label{def:unidirectional} \hfill\newline
  A QTM $M = (\Sigma,Q,\delta))$ is \keyword{unidirectional} if each state can be entered from only one direction. In other words, if $\delta(p_1, \sigma_1, \tau_1, q, d_1)$ and $\delta(p_2, \sigma_2, \tau_2, q, d_2)$ are both non-zero, then $d_1$ = $d_2$.
\end{definition*}

Note that in a unidirectional QTM, the direction component in any transition rule triple $\ket{\tau}\ket{q}\ket{D}$ is fully determined by the state $\ket{q}$. Thus we can recover the complete transition rules from the $\ket{\sigma}\ket{q}$ components alone, without the direction component. We call these the \keyword{reduced transition rules}, $\delta_r: Q\times\Sigma \mapsto \C^{Q\times\Sigma}$. (Equivalently, there exists an isometry $V:\C^Q\mapsto\C^{Q\times\{L,R,N\}}$ that maps the reduced transition rules to the original transition rules: $V\delta_r(p,\tau) = \delta(p,\tau)$.)

The following theorem gives necessary and sufficient conditions for a (partial) transition function to define a reversible Turing Machine.

\begin{theorem*}[Local well-formedness -- Thm.~B.1 and Cor.~B.3
                 in~\cite{Bernstein-Vazirani}]
  \addtheoremline{theorem}{Local well-formedness}
  \label{reversible_transitions} \hfill\newline
  A  TM $M$ is reversible iff its transition function satisfies the following conditions:
  \begin{description}[font=\normalfont\it,style=nextline]
  \item[Unidirection]
    Each state of $M$ can be entered while moving in only one direction. In other words, if $\delta(p_1,\sigma_1) = (\tau_1, q, d_1)$ and $\delta(p_2,\sigma_2) = (\tau_2, q, d_2)$ then $d_1 = d_2$.
  \item[One-to-one]
    The transition function is one-to-one when direction is ignored.
  \end{description}
  Furthermore, a partial transition function satisfying these conditions can always be completed to the transition function of a reversible TM.
\end{theorem*}

\textcite{Bernstein-Vazirani} proved that one can w.l.o.g.\ restrict to \keyword{unidirectional} QTMs. We therefore restrict the following quantum analogue of \cref{reversible_transitions} to the unidirectional case:

\begin{theorem}[Quantum local well-formedness]
  \label{QTM_transitions} \hfill\newline
  A unidirectional QTM $M=(\Sigma,Q,\delta)$ is well-formed iff its quantum transition function $\delta$ satisfies the following conditions:
  \begin{description}[font=\normalfont\it]
    \begin{subequations}
    \item[Normalisation]
      \begin{equation}
        \forall p,\sigma \in Q\times\Sigma \quad \norm{\delta(p,\sigma)} = 1.
      \end{equation}
    \item[Orthogonality]
      \begin{equation}
        \forall (p_1,\sigma_1) \neq (p_2,\sigma_2) \in Q\times\Sigma \quad
        \left\langle\delta_r(p_1,\sigma_1),\delta_r(p_2,\sigma_2)\right\rangle = 0.
      \end{equation}
    \end{subequations}
  \end{description}
  Furthermore, a partial quantum transition function satisfying these conditions can always be completed to a well-formed transition function.
\end{theorem}

\textcite{Bernstein-Vazirani} proved this result for standard QTMs, that is, where the head must move left or right in each time step. (In fact, they prove it without the restriction to unidirectional QTMs; see Theorem~5.2.2 and Lemma~5.3.4 in \cite{Bernstein-Vazirani}.) We therefore extend the proof of \cref{QTM_transitions} to generalised QTM, which is not difficult -- indeed, it is made particularly straightforward by our restriction to unidirectional machines.

\begin{proof}[of \cref{QTM_transitions}]
  Let $U$ be the time evolution operator of the QTM $M$. By definition $M$ is well-formed iff  $U$ is an isometry, or equivalently iff the columns of $U$ have unit length and are mutually orthogonal. Clearly, the normalisation condition specifies exactly that each column has unit length.

  Pairs of configurations whose tapes differ in a cell not under either of the heads, or whose tape heads are more than two cells apart, cannot yield the same configuration in a single time step. Therefore, all such pairs of columns are guaranteed to be orthogonal. The orthogonality condition imposes orthogonality of pairs of columns for configurations that differ only in that one is in state $p_1$ reading $\sigma_1$ while the other is in state $p_2$ reading $\sigma_2$.

  It remains to consider pairs of configurations whose heads are one or two cells apart, differing at most in the symbol written in these cells and in their states. However, unidirectionality implies that any update triples that share the same state must share the same direction. Thus either the state or the head location necessarily differs for all such pairs of columns, hence these too are orthogonal.

  The final claim follows straightforwardly from the fact that the normalisation and orthogonality conditions imply that a partial unidirectional reduced transition function $\delta_r$ is well-formed iff it defines an isometry on the space of states and tape symbols, and this can always be extended to a unitary. This fills in the undefined entries of $\delta_r$ by extending $\delta_r(q,\sigma)$ to an orthonormal basis for the space of states and symbols.
\end{proof}

We will often only be interested in the behaviour of a QTM (or reversible TM) on a particular subset of inputs, since the machine will only be run on those. A \keyword{proper} machine is guaranteed to behave appropriately on some subset of inputs, but not necessarily on all possible inputs.

\begin{definition}[Proper QTM]
  \label{def:proper_QTM}
  A QTM behaves \keyword{properly} (or is \emph{proper}) on a subset $\mathcal{X}$ of initial superpositions if whenever initialised in $\phi \in \mathcal{X}$, the QTM halts in a final superposition where each configuration has the tape head in the start cell\footnote{This property is called \keyword{stationarity} in \cite{Bernstein-Vazirani}}, the head never moved to the left of the starting cell, and the QTM never enters a configuration in which the head is in a superposition of different locations. We will refer to this latter property as {\it deterministic head movement}.

  Similarly, we say that a deterministic TM behaves properly (or is proper) on $\mathcal{X}\subset (\Sigma\backslash\{\#\})^*$ if the head never moves to the left of the starting cell, and it halts on every input $x\in \mathcal{X}$ with its tape head back in the starting cell.
\end{definition}
When the set $\mathcal{X}$ is clear from the context, we will omit it.

Note that, for a QTM to have deterministic head movement, it is \emph{not} sufficient that none of its transition rules $\delta(\sigma,\tau)$ produce a superposition of head directions. The head can also end up in a superposition of different locations because the tape state is in a superposition, so that two transition rules with different deterministic head movement apply in superposition.

Behaving properly is not a severe restriction on classical Turing Machines.\footnote{Nor is it a severe restriction on QTMs, but we will not need or prove this here.} In fact, given any TM, there is always an equivalent proper TM that computes the same function. One way to see this is to recall that all computable functions are computable by Turing Machines restricted to one-way infinite tapes (see any standard text book on the theory of computation, e.g.\ \textcite{Kozen_book}), and these clearly behave properly if the tape is extended to be two-way infinite. (Returning the head to the starting cell at the end of the computation poses no great difficulty.) In particular, this means that there exist proper universal Turing Machines.

Quantum Turing Machines were originally defined in \textcite{Bernstein-Vazirani} to have two-way infinite tapes. Indeed, those authors point out that there are trivial well-formed machines (such as the always-move-right machine) whose evolution would not be unitary on a one-way infinite tape, since the starting configuration would have no predecessor. However, when we come to encode our QTMs into local Hamiltonians, we will only be able to simulate tapes with a boundary.\footnote{Effectively, we can only encode quantum bounded linear automata rather than QTMs, though there is no limit on the finite tape length we can encode.} To avoid technical issues, we follow \textcite{Bernstein-Vazirani} in defining QTMs on two-way infinite tapes, but we will ensure that none of the QTMs (or reversible TMs) that we construct ever move their head before the starting cell. Thus, when encoding the QTM in a Hamiltonian, we can ignore all of the tape to the left of the starting cell.

In fact, the local Hamiltonians encoding the QTMs will only be able to simulate the evolution on a finite (but arbitrarily large) section of tape. We will therefore be interested in keeping very tight control on the space requirements of all the reversible and quantum TMs that we construct. By carefully controlling the space overhead, it will then be sufficient for our purposes to simulate the evolution of the QTM on a finite portion of tape that is essentially no longer than the input.

\subsubsection{Turing Machine Programming}
For a multi-track Turing machine with $k$ tracks and alphabet $\Sigma_i$ on track $i$, we will denote the contents of a tape cell by a tuple of symbols $[\sigma_1,\sigma_2,\ldots,\sigma_k] \in \Sigma_1\times\Sigma_2\times\dots\times\Sigma_k$ specifying the symbol written on each track. Similarly, the configuration of the tape will be specified by a tuple $[c_1,c_2,\dots,c_n] \in \Sigma_1^*\times\Sigma_2^*\times\dots\times\Sigma_k^*$, where by convention all $c_i$ are aligned to start in the same cell (which will be the starting cell unless otherwise specified). We will use $\wc$ to stand for an arbitrary track symbol. We will often describe Turing machines that act only on a subset of tracks, and leave the contents of all other tracks alone. In this case, we will only write out the states of the acted-upon tracks in the transition function; this transition function should be understood to be extended to the remaining tracks in the obvious way.

As a shorthand, a transition rule on a tuple containing a $\wc$ on one or more tracks should be understood to stand for the set of transition rules that leave the tracks marked $\wc$ unchanged, and act as indicated on the remaining tracks. We will often only specify some of the elements of a transition function when defining a reversible or quantum TM. We will call such partial transition functions ``well-formed'' if the elements that \emph{are} defined satisfy the conditions of \cref{reversible_transitions} or \cref{QTM_transitions}, since by those \lcnamecrefs{QTM_transitions} this is sufficient to guarantee that the partial transition functions can be completed to well-formed transition functions. QTM (or reversible TM) will sometimes use a finite number of auxiliary tracks. These are assumed always to start and finish in the all blank configuration.

We will have frequent recourse to the following basic TM and QTM programming primitives from \textcite{Bernstein-Vazirani}, which we slightly generalise here to account for additional properties of the resulting QTMs that will be important to us later. These primitives allow different QTMs to be combined in various ways to build up a more complex QTM.

\begin{lemma}[Subroutine Lemma] \hfill\newline
  \label{subroutine}
  Let $M_1$ be a two-track, normal form, reversible TM and $M_2$ a two-track normal form reversible TM (or well-formed, normal form, unidirectional QTM) with the same alphabet and the following properties:
  \begin{enumerate}
  \item $M_1$ is proper on initial configurations in $\mathcal{X}_1$ and $M_2$ is proper on $\mathcal{X}_2$.
  \item When started on $\mathcal{X}_1$, $M_1$ leaves the second track untouched; when started on $\mathcal{X}_2$, $M_2$ leaves the first track untouched.
  \item There is a state $q$ on $M_1$ that, when started on $\mathcal{X}_1$, can only be entered with the head in the starting cell.
  \item $\mathcal{X}_2$ contains all the output superpositions (with $q_f$ replaced by $q_0$) of $k$ consecutive executions of $M_2$ started from an initial configuration in $\mathcal{X}_2$ for all $0\le k\le r$, where $r\in \N\cup\{\infty\}$ is the maximum number of times that $q$ is entered when $M_1$ runs on input in $\mathcal{X}_1$.
  \end{enumerate}
  Then there is a normal form, reversible TM (or well-formed, normal form, unidirectional QTM) $M$ which behaves properly on $\mathcal{X}_1$ and acts as  $M_1$ except that each time it would enter state $q$, it instead runs machine $M_2$.
  \end{lemma}
\begin{proof}
  Exactly as Lemma~4.8 in \cite{Bernstein-Vazirani}.
\end{proof}

\begin{lemma}[Dovetailing Lemma]
  \label{dovetailing}
  Let $M_1$ and $M_2$ be well-formed normal form unidirectional QTMs (resp.\ normal form reversible TMs) with the same alphabet, so that $M_1$ is proper on $\mathcal{X}_1$, $M_2$ is proper on $\mathcal{X}_2$ and $\mathcal{X}_2$ contains all final superpositions (resp.\ configurations) of $M_1$ started on $\mathcal{X}_1$. Then, there is a well-formed normal form unidirectional QTM (resp.\ normal form reversible TM) $M$ which carries out the computation of $M_1$ followed by the computation of $M_2$ and that is also proper on $\mathcal{X}_1$.
  \end{lemma}
\begin{proof}
 Exactly as Lemma~4.9 in \cite{Bernstein-Vazirani}.
\end{proof}

\begin{lemma}[Reversal Lemma]
  \label{reversal}
  If $M$ is a well-formed, normal form, unidirectional QTM (resp.\ normal form reversible TM) then there is a well formed, normal form, unidirectional QTM (resp.\ normal form reversible TM) $M^\dg$ that reverses the computation of $M$ while taking two extra time steps and using the same amount of space. Moreover, if $M$ is proper on $\mathcal{X}$, then $M^\dg$ is proper on the set of final superpositions (resp.\ configurations) of $M$ started on $\mathcal{X}$.
\end{lemma}

\begin{proof}
  The proof is very similar to that of Lemma~4.12 in \cite{Bernstein-Vazirani}. We will prove it for QTMs, since reversible TMs are a special case of these. Consider an initial superposition $\ket{\phi_0}\in \mathcal{X}$, and let $\ket{\phi_1}, \cdots \ket{\phi_n}$ be the evolved sequence obtained by $\ket{\phi_{i+1}}=U_M\ket{\phi_i}$, where $\ket{\phi_n}$ is a final superposition. Since $M$ is normal form, $\ket{\phi_1},\cdots,\ket{\phi_n}$ do not have support on $q_0$.

  Let $\ket{\phi'_n}$ be the superposition obtained by replacing the state $q_f$ in $\ket{\phi_n}$ with the initial state of the new machine $q_0'$. Let $\ket{\phi_0'}$ be the superposition obtained by replacing the state $q_f$ in $\ket{\phi_0}$ with the new final state $q_f'$. We want to construct a QTM $M^\dg$ that, when started from $\ket{\phi'_n}$, halts on $\ket{\phi_0'}$ in $n+2$ steps. $M^\dagger$ will have the same alphabet and set of states as $M$, together with the new initial and final states $q_0',q_f'$. We define the transition function $\delta'$ in the following way:
  \begin{enumerate}
  \item $\delta'(q_0',\sigma)=\ket{\sigma}\ket{q_f}\ket{\bar{d}_{q_f}}$.
    \label{reversal:q0'_transition}
  \item For each $q\in Q\backslash\{q_0\}$ and each $\tau \in \Sigma$,
    \begin{equation} \delta'(q,\tau)=\sum_{p,q}\delta(p,\sigma,\tau,q,d_q)^*\ket{\sigma}\ket{p}\ket{\bar{d_p}}.
    \end{equation}
    \label{reversal:q_transition}
  \item $\delta'(q_0,\sigma)=\ket{\sigma}\ket{q_f'}\ket{d_{q_0}}$.
    \label{reversal:q0_transition}
  \item $\delta'(q_f',\sigma)=\ket{\sigma}\ket{q_0'}\ket{N}$.
    \label{reversal:qf'_transition}
  \end{enumerate}
  Here, for any state $q$, $d_q$ is the unique direction in which that state can be entered, and $\bar{d_q}$ is the opposite direction. Since $M$ is unidirectional, \cref{QTM_transitions} implies that $M^\dg$ is a well-formed, normal form, unidirectional QTM. Given a configuration $c$ in state $q$, $\pi(c)$ is defined as the configuration derived from $c$ by moving the head one step in the direction $\bar{d_q}$. $\pi$ can be extended by linearity to $\mathcal{S}$.

  Let us now analyse the behaviour of $M^\dg$ started from $\ket{\phi_n'}$. By \labelcref{reversal:q0'_transition}, $M^\dg$ maps $\ket{\phi_n'}$ in one step to $\pi(\ket{\phi_n})$. Now consider \labelcref{reversal:q_transition}. Since $M$ is normal form, it maps superposition $\ket{\psi_1}$ to superposition $\ket{\psi_2}$ with no support on state $q_0$ if and only if $\ket{\psi_1}$ has no support on state $q_f$. Denote $Q_0 = Q\backslash\{q_0\}$, $Q_f = Q\backslash\{q_f\}$. If $M$ takes a configuration $c_1$ with a state from $Q_f$ with amplitude $\alpha$ to a configuration $c_2$ (necessarily with a state in $Q_0$), then \labelcref{reversal:q_transition} ensures that $M^\dg$ takes configuration $\pi(c_2)$ to configuration $\pi(c_1)$ with amplitude $\alpha^*$. Let $S_0$ (resp.\ $S_f$) be the space of superpositions using only states in $Q_0$ (resp.\ $Q_f$). Since $M$ is well-formed, the restriction of the evolution operator $U_M$ to $S_f$ is an isometry into $S_0$. Hence $M^\dg$ implements, up to conjugation by $\pi$, the inverse of $U_M$ restricted to $U_M(S_f)$.

  As an aside, note that this implies that the evolution operator of $M$ is indeed a unitary and not just an isometry. Indeed, by \labelcref{reversal:q0'_transition,reversal:q_transition,reversal:q0_transition,reversal:qf'_transition} above, the evolution operator of $M^\dg$ is also an isometry from $S_0$ into $S_f$. Since $\pi$ is trivially a unitary on $\mathcal{S}$, this implies that $U_M$ restricted to $S_f$ is a unitary \emph{onto} $S_0$. Now, since $M$ is normal form, $U_M$ achieves any possible configuration with state $q_0$ by starting from the same configuration but with $q_0$ replaced by $q_f$. These two facts together show that $U_M$ is indeed surjective and hence a unitary.

  Returning to the proof of the \lcnamecref{reversal}, we have seen how \labelcref{reversal:q_transition} implies that $M^\dg$ executes the sequence of $n$ steps $\pi(\ket{\phi_{n-1}}),\cdots \pi(\ket{\phi_0})$. Finally, by \labelcref{reversal:qf'_transition}, in the $(n+2)$'th step, $M^\dg$ maps $\pi(\ket{\phi_0})$ to $\ket{\phi_0'}$ as desired.

  Moreover, it is trivial to see that $M^\dg$ uses exactly the same space as $M$, and behaves properly.
  \end{proof}

Note that our way of defining normal form (\cref{def:normal-form} as opposed to that in \cite{Bernstein-Vazirani}) is crucial to guarantee that the reversal machine is proper. As a corollary of the proof, we have also shown that the evolution of a well-formed, normal form unidirectional QTM is a unitary operator. (The proof of this fact for non-unidirectional QTMs is much more involved, and can be found in \cite{Bernstein-Vazirani}).

In the following sections, we give explicit constructions of the reversible and quantum Turing Machines that, when combined appropriately using \cref{subroutine,dovetailing,reversal} as described below, implement a phase-estimation QTM with the properties required to prove \cref{phase-estimation_QTM}. The formal definitions of all these Turing Machines are given by their transition rule tables, which can directly be verified to satisfy the well-formedness conditions of \cref{reversible_transitions,QTM_transitions}.

Since understanding how a TM works just from a table of transition rules is not always straightforward, we additionally give an informal pseudocode description of how each machine operates on suitable input. These pseudocode descriptions also help in verifying that the machines specified in the transition rule tables behave properly and satisfy the claimed space and run-time bounds.

Each pseudocode listing describes how the Turing Machine moves its head and reads and writes tape symbols as the algorithm proceeds, in order to implement the claimed operation. The internal state that the TM transitions into \emph{after} carrying out the corresponding line of pseudocode is indicated in the margin.\footnote{Generally, the new internal state is indicated on the line that carries out the head movement associated with the transition rule that transitions into that new state, except for transition rules where the TM head stays put.} The internal state that the TM is in at the beginning of an algorithm or subroutine is also indicated in the margin. Where no internal state is indicated in the margin, this means the internal state is left unchanged. Conditional branches in the pseudocode (i.e.\ \textbf{if}, \textbf{else if} and \textbf{else} statements) are structured to reflect how the different computational paths that can be followed by the TM during the computation, branch off and are then re-merged to ensure reversibility.

However, it should be emphasised that these pseudocode listings are not rigorous specifications of the corresponding Turing Machines; the formal, mathematically rigorous specifications are the transition rule tables.

\subsubsection{Reversible Turing Machine Toolbox}
It will be helpful to have a toolbox of reversible TMs which carry out various elementary computations. In order to satisfy the space constraints of \cref{phase-estimation_QTM}, we will be interested in keeping tight control of the space requirements of all our constructions.

\begin{lemma}[Copying machine]\label{copier}
  There is a two-track, normal form, reversible TM $\TM{Copy}$ with alphabet $\Sigma\times\Sigma$ that, on input $s$ written on the first track, behaves properly, copies the input to the second track and runs for time $2\abs{s}+1$, using $\abs{s}+1$ space.
\end{lemma}

\begin{proof}
  We simply step the head right, copying the symbol from the first track to the second. However, we defer copying the starting cell until the end of the computation, so that we can locate the starting cell again.
  In pseudocode:

  \begin{algorithm}[H]
    \caption{$\TM{Copy}$}
    \begin{algorithmic}[1]
      \Input{string $s\in(\Sigma\setminus\#)^*$ on track~1, blank track~2} \Comment{$q_0$}
      \If{read blank symbol $\#$ on both tracks}
        \State{halt} \Comment{$q_f$}
      \Else
        \State{step Right} \Comment{$q_1$}
        \While{read non-blank symbol $\sigma$ on track~1, blank symbol $\#$ on track~2}
          \State{copy symbol $\sigma$ from track~1 to track~2}
          \State{step Right}
        \EndWhile
        \State{step Left} \Comment{$q_2$}
        \While{read same non-blank symbol $\tau$ on both tracks}
          \State{step Left}
        \EndWhile
        \State{copy symbol $\sigma$ from track 1 to track 2}
        \State{halt} \Comment{$q_f$}
      \EndIf
    \end{algorithmic}
  \end{algorithm}

  It is straightforward to verify that the following normal form transition function implements $\TM{Copy}$:
  \begin{equation}
    \begin{array}{l|ccccc}
           & [\sigma,\#] & [\#,\#] & [\tau,\tau] \\
      \hline
      q_0  & ([\sigma,\#],q_1,R) & ([\#,\#],q_f,N) \\
      q_1  & ([\sigma,\sigma],q_1,R) & ([\#,\#],q_2,L) \\
      q_2  & ([\sigma,\sigma],q_f,N) & & ([\tau,\tau],q_2,L) \\
      q_f  & ([\sigma,\#],q_0,N) & ([\#,\#],q_0,N) & ([\tau,\tau],q_0,N) \\
      \multicolumn{2}{l}{\text{\footnotesize $\forall\sigma,\tau\in\Sigma\setminus\#$} \rule{0em}{1.5em}}
    \end{array}
  \end{equation}
  This partial transition function verifies the two conditions of \cref{reversible_transitions}, so it is well-formed and can be completed to give a reversible TM.
\end{proof}

\begin{lemma}[Shift-right machine]\label{shift_right}
  There exists a normal form, reversible TM $\TM{Shift}$ with alphabet $\Sigma$ that, on input $s$ behaves properly, shifts $s$ one cell to the right and runs for time $2\abs{s}+2$, using space $\abs{s}+2$.
\end{lemma}

\begin{proof}
  The Turing Machine has a separate internal state $q^\sigma$ corresponding to each tape symbol $\sigma$. It reads the symbol into this internal state, overwrites it with the preceding symbol, and steps right. In pseudocode:

  \begin{algorithm}[H]
    \caption{$\TM{Shift}$}
    \begin{algorithmic}[1]
      \Input{string $s\in(\Sigma\setminus\#)^*$} \Comment{$q_0$}
      \If{read blank symbol $\#$}
        \State{step Right} \Comment{$q_2$}
      \Else
        \State{read non-blank symbol $\sigma$}
        \State{write $\#$}
        \State{$\nu\gets\sigma$}
        \State{step Right} \Comment{$q_1^\sigma$}
        \While{read non-blank symbol $\sigma$}
          \State{write $\nu$}
          \State{$\nu\gets\sigma$}
          \State{step Right} \Comment{$q_1^\sigma$}
        \EndWhile
        \State{write $\nu$}
        \State{step Right} \Comment{$q_2$}
      \EndIf
      \State{step Left} \Comment{$q_3$}
      \While{read non-blank symbol $\sigma$}
        \State{step Left}
      \EndWhile
      \State{halt} \Comment{$q_f$}
    \end{algorithmic}
  \end{algorithm}
  The reason for including two different states $q_2$ and $q_3$ is to guarantee the unidirectionality requirement of \cref{reversible_transitions}. The state $q_2$ is entered from the right and the state $q_3$ from the left. The effect of this is that, if the input is the empty string, this Turing Machine program still steps right and then left before halting.

  It is straightforward to verify that the following normal-form transition function implements $\TM{Shift}$:
  \begin{equation}
    \begin{array}{l|ccc}
             & \# & \sigma \\
      \hline
      q_0    & (\#,q_2,R) & (\#,q_1^\sigma,R) \\
      q_1^\nu  & (\nu,q_2,R) & (\nu,q_1^\sigma,R) \\
      q_2    & (\#,q_3,L) \\
      q_3    & (\#,q_f,N) & (\sigma,q_3,L) \\
      q_f    & (\#,q_0,N) & (\sigma,q_0,N) \\
      \multicolumn{2}{l}{\text{\footnotesize $\forall\sigma,\nu\in\Sigma\setminus\#$} \rule{0em}{1.5em}}
    \end{array}
  \end{equation}
  As this partial transition function verifies the two conditions of \cref{reversible_transitions}, it is well-formed and can be completed to give a reversible TM.
\end{proof}

\begin{lemma}[Equality machine]\label{equality}
  There is a three-track, normal form, reversible TM $\TM{Eql}$ with alphabet $\Sigma\times\Sigma\times\{\#,0,1\}$ that, on input $s;t;b$, where $s$ and $t$ are arbitrary input strings and $b$ is a single bit, behaves properly and outputs $s;t;b'$, where $b'=\neg b$ if $s=t$ and $b'=b$ otherwise. Furthermore, $\TM{Eql}$ runs for time $4\abs{s}+1$ and uses $\abs{s}+1$ space.
\end{lemma}

Implementing a non-reversible equality-testing machine is trivial: just scan the head right checking if the symbols on the two input tracks match. If we reach the end of the input without encountering a non-matching pair, return to the starting cell and flip the output bit. If we encounter a non-matching pair, return to the starting cell and leave the output bit unchanged.

Doing this reversibly requires more care. The problem is that the computation  splits into two possible paths, depending on whether a non-match was encountered or not, and we must merge these two divergent computations back together again \emph{reversibly}. For example, we cannot simply halt after either flipping the output bit or leaving it unchanged, as that would give multiple transitions into the final state that write the same symbol to the starting cell.
The trick is to return to the point at which the computational paths diverged (either the first non-matching pair of symbols, or the end of the input) after setting the output bit, in order to merge the two computational paths back together, before returning to the starting cell again to halt.

To accomplish this, we use a state $q_4$ that is the only state that can transition to $q_f$ (except for the special case in which the first symbols in both input strings already differ, which is handled immediately). The state $q_1$ searches along the strings, until it either finds the end of both strings (first computational path), or it finds a point where the symbols in the two inputs differ (second computational path).

In the first path, state $q_2$ will return to the starting cell, switch the output bit and transition to $q_3$. State $q_3$ will move right again doing the same as $q_1$ did and, at the end of both strings, will transition to $q_4$. Note that we need two different states for this, due to unidirectionality: $q_2$ will be entered from the left and $q_3$ from the right.

In the second path, states $q_2'$ and $q_3'$ play the corresponding roles of $q_2$ and $q_3$, respectively. In both paths, $q_4$ will return to the starting cell and halt. Within each path there is no issue with reversibility. The key to this is that, when both computational paths are merged into state $q_4$, they do so on exactly the same input states as when the paths diverged (with $q_1$ replaced by $q_3$ or $q_3'$ depending on the path). Thus these transitions fulfil the properties required for a well-formed (partial) transition function by \cref{reversible_transitions}.

This implementation requires being able to identify the starting cell, so that we can return to it again (this is also important since we want to avoid ever moving the head before the starting cell, to ensure the TM is proper~\cref{def:proper_QTM}). If the symbols in the first cell differ, we can set the output bit immediately and halt. If they are identical, we temporarily change the symbol on the second input track to a $\#$ to mark the starting cell. We can recover the original second-track input at the end of the computation in this case, by copying over the symbol from the first track. (This is not the only way one could handle this.)

\begin{proof}[of \cref{equality}]
  In pseudocode, the Turing Machine program for $\TM{Eql}$ does the following:
  \begin{algorithm}[H]
    \caption{$\TM{Eql}$}
    \begin{algorithmic}[1]
      \Input{string $s\in(\Sigma\setminus\#)^*$ on track~1\\
        \phantom{\bf input:} string $t\in(\Sigma\setminus\#)^*$ on track~2\\
        \phantom{\bf input:} bit $b\in\{0,1\}$ on track~3} \Comment{$q_0$}
      \If{read symbols $\sigma\neq\tau$ on tracks~1 and~2}
        \State{halt}  \Comment{$q_f$}
      \Else
        \State{write $\#$ on track~2}
        \State{step Right} \Comment{$q_1$}
        \While{read non-blank symbols $\sigma=\tau$ on tracks~1 and~2}
          \State{step Right}
        \EndWhile
        \If{read blank symbols $\#$ on both tracks~1 and~2}
          \State{step Left} \Comment{$q_2$}
          \While{read non-blank symbols $\sigma=\tau$ on tracks~1 and~2}
            \State{step Left}
          \EndWhile
          \State{flip bit $b$ on track~3}
          \State{step Right} \Comment{$q_3$}
          \While{read symbols $\sigma=\tau$ on tracks~1 and~2}
            \State{step Right}
          \EndWhile
                \ElsIf{read different symbols $\sigma\neq\tau$ on tracks~1 and~2}
          \State{step Left} \Comment{$q'_2$}
          \While{read non-blank symbols $\sigma=\tau$ on tracks~1 and~2}
            \State{step Left}
          \EndWhile
          \State{step Right} \Comment{$q'_3$}
          \While{read symbols $\sigma=\tau$ on tracks~1 and~2}
            \State{step Right}
          \EndWhile
        \EndIf
          \State{step Left}\Comment{$q_4$}
          \While{read symbols $\sigma=\tau$ on tracks~1 and~2}
            \State{step Left}
          \EndWhile
         \State{copy track~1 symbol $\sigma$ to track~2}
      \State{halt} \Comment{$q_f$}
      \EndIf
       \end{algorithmic}
  \end{algorithm}

  The following normal-form transition function carries out this procedure:
  \begin{equation}
    \begin{array}{l|ccccc}
           & [\sigma,\sigma,b] & [\sigma,\tau,b] & [\sigma,\#,b] & [\#,\tau,b] \\
      \hline
      q_0  & ([\sigma,\#,b],q_1,R) & ([\sigma,\tau,b],q_f,N)
             & ([\sigma,\#,b],q_f,N) & ([\#,\tau,b],q_f,N) \\
      q_1  \\
      q_2  & & & ([\sigma,\#,\neg b],q_3,R) \\
      q_2' & & & ([\sigma,\#,b],q_3',R) \\
      q_3  \\
      q_3' \\
      q_4  & & & ([\sigma,\sigma,b],q_f,N) \\
      q_f  & ([\sigma,\sigma,b],q_0,N) & ([\sigma,\tau,b],q_0,N)
             & ([\sigma,\#,b],q_0,N) & ([\#,\tau,b],q_0,N) \\
      \multicolumn{2}{l}{}\\
           & [\sigma,\sigma,\#] & [\sigma,\tau,\#] & [\sigma,\#,\#]
           & [\#,\tau,\#] \\
      \hline
      q_1  & ([\sigma,\sigma,\#],q_1,R) & ([\sigma,\tau,\#],q_2',L)
             & ([\sigma,\#,\#],q_2',L) & ([\#,\tau,\#],q_2',L) \\
      q_2  & ([\sigma,\sigma,\#],q_2,L) \\
      q_2' & ([\sigma,\sigma,\#],q_2',L) \\
      q_3  & ([\sigma,\sigma,\#],q_3,R) \\
      q_3' & ([\sigma,\sigma,\#],q_3',R) & ([\sigma,\tau,\#],q_4,L)
             & ([\sigma,\#,\#],q_4,L) & ([\#,\tau,\#],q_4,L) \\
      q_4  & ([\sigma,\sigma,\#],q_4,L) \\
      q_f  & ([\sigma,\sigma,\#],q_0,N) & ([\sigma,\tau,\#],q_0,N)
             & ([\sigma,\#,\#],q_0,N) & ([\#,\tau,\#],q_0,N) \\
      \multicolumn{2}{l}{}\\
           & [\#,\#,\#] \\
      \cline{1-2}
      q_1  & ([\#,\#,\#],q_2,L) \\
      q_2  \\
      q_2' \\
      q_3  & ([\#,\#,\#],q_4,L) \\
      q_3' \\
      q_4  \\
      q_f  & ([\#,\#,\#],q_0,N) \\
      \multicolumn{2}{l}{\text{\footnotesize $\forall\sigma\neq\tau\in\Sigma\setminus\#, \forall b\in\{0,1\}$} \rule{0em}{1.5em}}
    \end{array}
  \end{equation}
  One can verify that this partial transition function satisfies the two conditions of \cref{reversible_transitions}, so it is well-formed and can be completed to give a reversible TM, as required.
\end{proof}

It is somewhat easier to construct reversible implementations of basic arithmetic operations if the numbers are written on the tape in little-endian order (i.e.\ least-significant bit first), as it avoids any need to shift the entire number to the right to accommodate additional digits. We adopt this convention for all the following basic arithmetic machines. We do not allow numbers to be padded with leading 0's as that would allow multiple binary representations of the same number, which is inconvenient when constructing reversible machines. Note that this means the number zero is represented by the blank string, not the string ``0''.

\begin{lemma}[Increment and decrement machines]
  \label{incrementer}
  There exist normal form, reversible TMs $\TM{Inc}$ and $\TM{Dec}$ with alphabet $\{\#,0,1\}$ that, on little-endian binary input $n$ (with $n>1$ for $\TM{Dec}$), behave properly and output $n+1$ or $n-1$ respectively. Both machines run for time $O(\log n)$ and use at most $\abs{n}+2$ space.
\end{lemma}

Incrementing a binary number on a non-reversible TM is straightforward: simply step the head along the number starting from the least significant bit, flipping 1's to 0's to propagate the carry until the first 0 or \#, then flip that 0 or \# to 1 and halt. (Little-endian order avoids any need to shift the whole input to the right to accommodate an additional digit, should one be required.) However, making this procedure reversible is more fiddly. One option of course is to use the general procedure for reversible simulation and uncomputation of a non-reversible TM due to \textcite{Bennett}, but this comes at a cost of polynomial space overhead. A more careful construction allows us to implement $\TM{Inc}$ directly, using just two additional tape cells.

\begin{proof}[of \cref{incrementer}]
  We first use the $\TM{Shift}$ machine from \cref{shift_right} to shift the entire input one cell to the right, as a convenient way of getting a $\#$ in the starting cell so that we can return to it later. We then reversibly increment the binary number written on the tape, and finish by running $\TM{Shift}^\dg$ (the reversal of the $\TM{Shift}$ machine, constructed using \cref{reversal}) to shift the output back one cell to the left.

  To implement reversibly the incrementing part of this procedure, we assign states as follows. State $q_1$ will search along the input for the first $0$ to change it to a $1$, and along the way will change all $1$s to $0$s. In this way, the carry from incrementing the first (least significant) bit is propagated along the binary string to the appropriate bit. If no such $0$ exists, a $1$ will be appended to the end of the input string by changing the blank symbol $\#$ there to a $1$.

  These two possible computation paths have to be merged back into a state $q_4$, which will then return the TM head to the starting cell and halt. To merge these two paths, which end up in states $q_2$ and $q_2'$ respectively, we add an extra state $q_3$ together with transition rules from $q_2$ and $q'_2$ into $q_3$, such that both computational paths arrive at a configuration in which the TM is in state $q_3$ and reading a $1$. Thereupon, the TM can transition into $q_4$ and complete the computation.

  In pseudocode:
  \begin{algorithm}[H]
    \caption{$\TM{Inc}$}
    \begin{algorithmic}[1]
      \Input{blank $\#$ followed by little-endian binary representation of $n\in\N$} \Comment{$q_0$}
      \State{step Right}  \Comment{$q_1$}
      \While{read 1}
        \State{write a 0}
        \State{step Right}
      \EndWhile
      \If{read 0}
              \State{write a 1}
        \State{step Right}\Comment{$q_2$}
        \State{step Left} \Comment{$q_3$}
      \ElsIf{read blank symbol $\#$}
        \State{write a 1}
        \State{step Right} \Comment{$q'_2$}
        \State{step Left} \Comment{$q_3$}
      \EndIf
        \State{step Left} \Comment{$q_4$}
      \While{read a 0}
        \State{step Left}
      \EndWhile
      \State{halt} \Comment{$q_f$}
    \end{algorithmic}
  \end{algorithm}

  \noindent The following normal-form transition function implements this:
  \begin{equation}
    \begin{array}{l|ccc}
           & \# & 0 & 1 \\
      \hline
      q_0  & (\#,q_1,R) \\
      q_1  & (1,q_2',R) & (1,q_2,R) & (0,q_1,R) \\
      q_2  & & (0,q_3,L) & (1,q_3,L) \\
      q_2' & (\#,q_3,L) \\
      q_3  & & & (1,q_4,L) \\
      q_4  & (\#,q_f,N) & (0,q_4,L) \\
      q_f  & (\#,q_0,N) & (0,q_0,N) & (1,q_0,N) \\
    \end{array}
  \end{equation}
  This partial transition function verifies the two conditions of \cref{reversible_transitions}, so it is well-formed and can be completed to give a reversible TM.

  This completes the construction of $\TM{Inc}$. To implement $\TM{Dec}$, simply construct the reversal of $\TM{Inc}$ using \cref{reversal}.
\end{proof}

We can use the $\TM{Inc}$ construction to construct a looping primitive. (This gives a slightly more general version of the Looping Lemma from \cite[Lemma~4.13]{Bernstein-Vazirani}, which also has tighter control on the space requirements.)
\begin{lemma}[Looping Lemma]\label{looping2}
  There is a two-track, normal form, reversible TM $\TM{Loop2}$ with alphabet $\{\#,0,1\}$, which has the following properties. On input $n;m$, with $n< m$ both little-endian binary numbers, $\TM{Loop2}$ behaves properly, runs for time $O((m-n)\log m)$, uses space $\abs{m}+2$ and halts with its tape unchanged. Moreover, $\TM{Loop2}$ has a special state $q$ such that on input $n;m$, it visits state $q$ exactly $m-n$ times, each time with its tape head back in the start cell.

  There is also a one-track, normal form, proper, reversible TM $\TM{Loop}$ which, on input $m\ge 1$, behaves as $\TM{Loop2}$ on input $0;m$.
\end{lemma}

\begin{proof}
  Our construction closely follows the proof of \textcite[Lemma~4.2.10]{Bernstein-Vazirani}. We will use two auxiliary tracks in addition to the two input tracks, both with alphabet $\{\#,0,1\}$ and both initially blank. 

  The core of the $\TM{Loop2}$ machine is a reversible TM $M'$ constructed out of two proper, normal form, reversible TMs $M_1$ and $M_2$.

  $M_1$ has initial and final states $q_0,q_f$, and transforms input $n;m;x;b$ into $n;m;x+1;b'$, where $b$ is a bit and $b'=\neg b$ if $x=n$ or $x=m-1$ but not both, otherwise $b'=b$.
  $M_1$ can be constructed by dovetailing together the $\TM{Eql}$ machine from \cref{equality} (with the first and third tracks as its input tracks and the fourth track as its output), then the $\TM{Inc}$ machine from \cref{incrementer} (acting only on the third track), and finally another $\TM{Eql}$ machine (this time with the second and third tracks as its input tracks and the fourth track as its output). By \cref{dovetailing} and the fact that all the constituent machines are proper, normal form and reversible, $M_1$ is proper, normal form and reversible. From \cref{incrementer,equality}, $M_1$ runs for time $O(\log m)$ and takes at most $\abs{m}+2$ space.

  $M_2$ has new initial and final states $q_\alpha,q_\omega$, with $q_0$ and $q_f$ as its only other normal (i.e.\ not initial or final) internal states. $M_2$ behaves as follows:
  \begin{enumerate}
  \item If it is in state $q_\alpha$ with $b=0$, it flips $b$ to 1 and enters state $q_0$. \label{M2_prop1}
  \item If it is in state $q_f$ with $b=0$, it enters state $q_0$. \label{M2_prop2}
  \item If it is in state $q_f$ with $b=1$, it flips $b$ to 0 and halts. \label{M2_prop3}
  \end{enumerate}

  The following normal form, partial transition function for $M_2$ is essentially the same as the corresponding construction in Lemma~4.2.6 of \textcite{Bernstein-Vazirani}, but we can simplify slightly by exploiting the fact that we are allowing generalised TMs. It acts only on the fourth track, and implements a machine $M_2$ that clearly satisfies the requirements \labelcref{M2_prop1,M2_prop2,M2_prop3}, above:
  \begin{equation}
    \begin{array}{l|ccc}
              & \# & 0 & 1 \\
      \hline
      q_\alpha  & & (1,q_0,N) \\
      q_f      & & (0,q_0,N) & (0,q_\omega,N) \\
      q_\omega  & (\#,q_\alpha,N) & (0,q_\alpha,N) & (1,q_\alpha,N)
    \end{array}
  \end{equation}
  $M_2$ is clearly normal form, and satisfies the two conditions of \cref{reversible_transitions}.

  The transition rules for $M'$ are constructed by deleting the $q_f$ to $q_0$ transition rules from $M_1$, and adding all the remaining $M_1$ rules to those of $M_2$. The initial and final states of $M'$ are $q_\alpha,q_\omega$. On input $\mbox{n;m;n;0}$, $M'$ will therefore run $M_2$ until it enters the $q_0$ state, then run $M_1$ until it halts and re-enters $M_2$. It will continue to alternate in this way between $M_2$ and $M_1$ until the former halts. Thus $M'$ goes through the following sequence of configurations:
  \begin{multline}
    \mbox{n;m;n;0} \overset{(i)}{\longrightarrow}
    \mbox{n;m;n;1} \overset{M_1}{\longrightarrow}
    \mbox{n;m;n+1;0} \overset{(ii)}{\longrightarrow}
    \mbox{n;m;n+1;0} \overset{M_1}{\longrightarrow}
    \mbox{n;m;n+2;0} \overset{(ii)}{\longrightarrow}\cdots\\
    \cdots\overset{(ii)}{\longrightarrow}
    \mbox{n;m;m-1;0} \overset{M_1}{\longrightarrow}
    \mbox{n;m;m;1} \overset{(iii)}{\longrightarrow}
    \mbox{n;m;m;0}
  \end{multline}
  It therefore runs exactly $m-n$ times and enters the state $q_f$ once in each run, so we can take $q_f$ as the special state $q$.

  To initialise the tracks, we dovetail a proper, reversible TM before $M'$ which transforms the input $n;m;\#;\#$ into the configuration $n;m;n;0$. This is easily constructed by dovetailing the $\TM{Copy}$ machine from \cref{copier} (acting on the first and third tracks) with a simple reversible TM which changes the first cell of the fourth track from $\#$ to 0 and halts. (Implementing the latter is trivial.) To return all tracks to their initial configuration at the end, we dovetail another proper, reversible TM after $M'$ which transforms the configuration $n;m;m;0$ into the final output $n;m;\#;\#$. This is easily constructed by dovetailing the reversal $\TM{Copy}^\dg$ of the copying machine from \cref{copier} (acting on the second and third tracks) with a simple reversible TM which changes the first cell of the fourth track from 0 back to $\#$. From \cref{adder}, these initialisation and reset machines run for time $O(\log n)$ and $O((m-n)\log m)$, respectively, and use $\abs{n}+1$ and $\abs{m}+1$ space.

  It remains to show that combining $M_1$ and $M_2$ as described above to form $M'$ gives a proper, normal form, reversible TM. Since $M_1$ is normal form, it has no transitions into $q_0$ or out of $q_f$ other than the ones we deleted before combining it with $M_2$ to construct $M'$. All remaining internal states of $M_1$ and $M_2$ are distinct. Thus, since $M_1$ is reversible, the transition rules for $M'$ also satisfy the two conditions of \cref{reversible_transitions}. $M'$ can therefore be completed to a normal form, reversible TM. Since $M_1$ is proper, it is easy to see that $M'$ is also proper.

  $M'$ loops on $M_1$ $m-n$ times (with constant time overhead and no additional space overhead coming from $M_2$). Each run of $M_1$ takes time $O(\log m)$, so the complete $\TM{Loop2}$ implementation takes time $O(\log m)$. None of the constituent TMs use more than $\abs{m}+2$ space, so neither does $\TM{Loop2}$. This completes the construction of $\TM{Loop2}$.

  To construct $\TM{Loop}$, we simply remove the second input track from the $\TM{Loop2}$ machine, and replace the $\TM{Eql}$ machine that acts on that track with a trivial machine that checks whether the starting cell of the third track contains the $\#$ symbol.
\end{proof}

The reversible incrementing, decrementing, and looping machines constructed above are sufficient to implement all arithmetic operations. The only ones we will need are addition and subtraction. These are now easy to construct.
\begin{lemma}[Binary adder]
  \label{adder}
  There exist two-track, normal form, reversible TMs $\TM{Add}$ and $\TM{Sub}$ with alphabet $\{\#,0,1\}^{\times 3}$, which have the following properties. On input $n;m$ with $n$ and $m$ both little-endian binary numbers and $m\ge 1$, $\TM{Add}$ behaves properly and outputs $n+m;m$. On input $n;m$ with $n> m$, $\TM{Sub}$ behaves properly and outputs $n-m;m$. Both TMs run for time $O(m\log m\log n)$ and  use $\max(\abs{n},\abs{m})+2$ space.
 \end{lemma}

\begin{proof}
  To construct $\TM{Add}$, we simply run the $\TM{Loop}$ TM of \cref{looping2} on the second track, inserting for its special state the $\TM{Inc}$ machine of \cref{incrementer} acting on the first track. Since both $\TM{Loop}$ and $\TM{Inc}$ are proper and normal form, the resulting machine is also proper and normal form. $\TM{Sub}$ is simply the reversal of $\TM{Add}$, constructed using \cref{reversal}.
\end{proof}

We will also need a reversible TM to convert input from unary representation to binary. Again, it is not difficult to construct this using our reversible incrementing TM.
\begin{lemma}[Unary to binary converter]
  \label{unary_to_binary}
  There exists a two-track, normal form, reversible TM $\TM{UtoB}$ with alphabet $\{\#,1\}\times\{\#,0,1\}$ with the following properties. On input $1^n;\#$ ($n$ written in unary on the first track), $\TM{UtoB}$ behaves properly and outputs $1^n;n$ ($n$ written in little-endian binary on the second track). Furthermore, $\TM{UtoB}$ runs for $O(n^2 \log n)$ steps and uses $n+1$ space.
\end{lemma}

\begin{proof}
  The basic idea is to step the head right along the unary track until we reach the end of the unary input string, running the $\TM{Inc}$ machine of \cref{incrementer} once each time we step right. However, $\TM{Inc}$ needs to be started with its head in the starting cell. So each time we run it, we need to temporarily mark the current head location, move the head back to the starting cell in order to run $\TM{Inc}$, then return the head to its previous location and continue stepping right.

  Consider the following normal form transition function, acting only on the unary track:
  \begin{equation}
    \begin{array}{l|ccc}
            & \# & 1 \\
      \hline
      q_0   & (\#,q_f,N) & (1,q,N) \\
      q_1   & (\#,q_4,L) & (\#,q_2,L) \\
      q_2   & (\#,q,N) & (1,q_2,L) \\
      q     & (\#,q_2',R) & (\#,q_1,R) \\
      q_2'  & (1,q_1,R) & (1,q_2',R) \\
      q_4   & (1,q_f,N) & (1,q_4,L) \\
      q_f   & (\#,q_0,N) & (1,q_0,N)
    \end{array}
  \end{equation}
  This verifies the two conditions of \cref{reversible_transitions}, so it is well-formed and defines a reversible TM. It is also easy to see that this TM is proper.

  The transition function implements a machine $M$ that steps right over the 1's on the unary track until it reaches the end of the input, at which point it moves its head back to the starting cell and halts. However, each time it steps right, it temporarily marks the current head location by overwriting the 1 on the unary track with a $\#$, returns the head to the starting cell, enters an apparently useless internal state $q$ for one time step, then returns the head back to where it was before, restores the 1 on the unary track, and continues stepping right from where it left off.

  In pseudocode:
  \begin{algorithm}[H]
    \caption{$M$}
    \begin{algorithmic}[1]
      \Input{unary representation of $n\in\N$, i.e.\ string of $n$ 1's} \Comment{$q_0$}
      \If{read blank symbol $\#$}
        \State{halt} \Comment{$q_f$}
      \Else
        \Call{$q$}{} \Comment{$q$}
        \State{write a $\#$}
        \State{step Right} \Comment{$q_1$}
        \Call{place\_marker\_and\_reset}
        \Loop
          \Call{$q$}{} \Comment{$q$}
          \State{step Right} \Comment{$q'_2$}
          \While{read a 1}
            \State{step Right}
          \EndWhile
          \State{write a 1}
          \State{step Right} \Comment{$q_1$}
          \Call{place\_marker\_and\_reset}
        \EndLoop
      \EndIf

      \State

      \Procedure{place\_marker\_and\_reset}{} \Comment{$q_1$}
        \If{read a $\#$}
          \State{step Left} \Comment{$q_4$}
          \Call{reset\_and\_halt}
        \ElsIf{read a 1}
          \State{write a $\#$}
          \State{Step Left} \Comment{$q_2$}
          \While{read a 1}
            \State{step Left}
          \EndWhile
        \EndIf
      \EndProcedure

      \State

      \Procedure{reset\_and\_halt}{} \Comment{$q_4$} \label{reset}
        \While{read a 1}
            \State{step Left}
        \EndWhile
        \State{write a 1}
        \State{halt} \Comment{$q_f$}
      \EndProcedure
    \end{algorithmic}
  \end{algorithm}

  \noindent The purpose of this apparently pointless procedure is that, by \cref{subroutine}, we can substitute the $\TM{Inc}$ machine from \cref{incrementer} for the state $q$, with $\TM{Inc}$ acting on the (initially blank) binary track.

  The above TM enters the $q$ state precisely $n$ times. Thus, when we substitute $\TM{Inc}$ for $q$, $\TM{Inc}$ will be run precisely $n$ times, leaving the number $n$ written on the binary track as required. Each iteration of $\TM{Inc}$ takes time $O(\log n)$ and at most $\abs{n}+2$ space. The step-right TM constructed above adds time overhead $O(n^2)$ and uses $n+1$ space. Neither machine ever moves its head before the starting cell. Thus the overall machine satisfies the time and space claims of the \namecref{unary_to_binary}.
\end{proof}

\subsection{Quantum phase estimation overview}
The qubits used in the quantum phase estimation circuit can be divided into two sets: the ``output'' qubits which will ultimately contain the binary fraction expansion of the phase in the black-box unitary, and the ``ancilla'' qubits on which the black-box unitary is applied. In our case, the black-box unitary will be the single-qubit unitary $U_\varphi=\left(\begin{smallmatrix}1&0\\0&e^{i\pi\varphi}\end{smallmatrix}\right)$, so the ancilla set will be a single qubit. As stated above, $\varphi$ will refer to the number whose binary decimal expansion contains the digits of $n$ in reverse order after the decimal.

The quantum phase estimation algorithm proceeds in five stages:
\begin{enumerate*}[label=(\roman{*})]
  \item A preparation stage, in which the qubits are first initialised in the $\ket{0}$ state, then Hadamard gates are applied to all the output qubits and the ancilla qubit is prepared in an eigenstate of $U_\varphi$ (in our case $\ket{1}$);
  \item The controlled-unitary stage, during which control-$U_\varphi$ operations are applied between the output qubits and the ancilla qubit;
  \item A stage in which we locate the least significant bit of the output;
  \item The quantum Fourier transform stage, in which the inverse quantum Fourier transform is applied to the output qubits;
  \item A reset stage, which resets all the auxiliary systems used during the computation to their initial configuration.
\end{enumerate*}

We will construct QTMs for each of these stages separately, and use the Dovetailing \cref{dovetailing} to chain them together. It will be useful to divide the tape into multiple tracks. 
A ``quantum track'' with alphabet $\Sigma_q=\{\#,0,1\}$ will store the qubits involved in the phase estimation algorithm. The input $1^N$ (the unitary representation of $N$) will be supplied as a string of $N$ 1's on the quantum track. All the other tracks will essentially be classical; after each stage they will be left in a single standard basis state (i.e.\ neither entangled with other tracks, nor in a superposition of basis states). These classical tracks will be used to implement the classical processing needed to control the quantum operations applied to the quantum track.

\subsection{Preparation stage}
\label{sec:preparation_stage}
We will use the first cell of the quantum track as the ancilla qubit, and the following $N$ cells as the output qubits of the quantum phase estimation algorithm.

It will be convenient to first store a copy of the input given in unary on the quantum track on an auxiliary ``input track'', but written in binary, i.e.\ to transform the configuration $1^N;\#$ of the quantum and input tracks into $1^N;N$. Running the unary-to-binary converter $\TM{UtoB}$ from \cref{unary_to_binary} on the quantum and input tracks carries out the desired transformation.

We want to initialise the $N+1$ qubits that will be used in the phase estimation, by preparing the ancilla qubit in the state $\ket{1}$ and the output qubits in the $\ket{+}$ state. The first $N$ qubits are already in the $\ket{1}$ state thanks to the input string. We therefore prepare the desired state by first temporarily flipping the state of the first qubit on the quantum track to $\ket{\#}$ to mark the starting cell, then stepping the QTM head right rotating each qubit into the $\ket{+}$ state, until we reach the first $\ket{\#}$ state which we again rotate to $\ket{+}$ to initialise the $N+1$'th qubit. We then move the head back to the start location, flip the state of the first cell of the quantum track from $\ket{\#}$ to $\ket{1}$, and halt. (The contents of all other tracks are ignored.)

In pseudocode:
\begin{algorithm}[H]
  \caption{Preparation QTM}
  \begin{algorithmic}[1]
    \Input{unary representation of $N\in\N$, i.e.\ string of $N$ 1's} \Comment{$q_0$}
    \State{apply unitary gate that rotates from $\ket{1}$ to $\ket{\#}$}
    \State{step Right} \Comment{$q_1$}
    \While{read $\ket{1}$}
      \State{apply Hadamard gate to rotate to $\ket{+}$}
      \State{step Right}
    \EndWhile
    \State{apply unitary gate that rotates from $\ket{\#}$ to $\ket{+}$}
    \State{step Right}  \Comment{$q_2$}
    \State{step Left} \Comment{$q_3$}
    \While{not read $\ket{\#}$}
      \State{Step Left}
    \EndWhile
    \State{apply unitary gate that rotates from $\ket{\#}$ to $\ket{1}$}
    \State{halt} \Comment{$q_f$}
  \end{algorithmic}
\end{algorithm}

It is straightforward to verify that the following partial transition function satisfies the conditions of \cref{QTM_transitions}, so is well-formed, and implements a proper, well-formed, normal form, unidirectional QTM that does exactly what we want.
\begin{equation}\label{eq:Hadamard_QTM}
  \begin{array}{l|ccc}
           & \# & 0 & 1 \\
    \hline
    q_0  & & & \ket{\#}\ket{q_1}\ket{R} \\
    q_1  & \tfrac{1}{\sqrt{2}}(\ket{0}+\ket{1}) \ket{q_2}\ket{R}
           & & \tfrac{1}{\sqrt{2}}(\ket{0}+\ket{1}) \ket{q_1}\ket{R} \\
    q_2  & \ket{\#}\ket{q_3}\ket{L} \\
    q_3  & \ket{1}\ket{q_f}\ket{N} & \ket{0}\ket{q_3}\ket{L}
           & \ket{1}\ket{q_3}\ket{L} \\
    q_f  & \ket{\#}\ket{q_0}\ket{N} & \ket{0}\ket{q_0}\ket{N}
           & \ket{1}\ket{q_0}\ket{N}
  \end{array}
\end{equation}

\subsection{Control-\texorpdfstring{$U_\varphi$}{Uphi} stage}
\label{sec:cU_stage}
The second stage of the phase estimation procedure is to apply control-$U_\varphi^{2^{n-1}}$ operations between the $n$'th output qubit and the ancilla qubit (see~\cref{fig:cU_circuit}). Constructing a QTM that implements this is more complex. The basic idea is to supply the QTM with an internal state $q$ that causes the $U_\varphi$ rotation to be applied to the quantum track at the current head location. We then construct classical control machinery which iterates over the output qubits, and loops on $q$ to apply $U_\varphi$ for a total of $2^{n-1}$ times.

\begin{figure}[htbp]
  \centering
  \mbox{
    \Qcircuit @C=1em @R=1em {
      \lstick{\ket{0}} & \gate{H} & \qw & \qw              & \qw                & \qw                    & \push{\makebox[2em]{$\dots$}} \qw & \ctrl{5}                  &  \rstick{\ket{0} + e^{2\pi i(2^{N-1}\varphi)}\ket{1}} \qw \\
      & \vdots \\
      \lstick{\ket{0}} & \gate{H} & \qw & \qw              & \qw                & \ctrl{3}               & \push{\makebox[2em]{$\dots$}} \qw & \qw                       &  \rstick{\ket{0} + e^{2\pi i(2^2\varphi)}\ket{1}} \qw \\
      \lstick{\ket{0}} & \gate{H} & \qw & \qw              & \ctrl{2}           & \qw                    & \push{\makebox[2em]{$\dots$}} \qw & \qw                       &  \rstick{\ket{0} + e^{2\pi i(2\varphi)}\ket{1}} \qw \\
      \lstick{\ket{0}} & \gate{H} & \qw & \ctrl{1}         & \qw                & \qw                    & \push{\makebox[2em]{$\dots$}} \qw & \qw                       & \rstick{\ket{0} + e^{2\pi i\varphi}\ket{1}} \qw \\
      \lstick{\ket{1}} & \qw      & \qw & \gate{U_\varphi} & \gate{U_\varphi^2} & \gate{U_\varphi^{2^2}} & \push{\makebox[2em]{$\dots$}} \qw & \gate{U_\varphi^{2^{N-1}}}& \rstick{\ket{1}} \qw
    }
    \hspace{4em}
  }
  \caption[Quantum phase estimation, control-phase stage.]{The preparation and control-$U_\varphi$ stages of the quantum phase estimation circuit for $\varphi$ (cf.~Fig.~5.2 in~\cite{Nielsen+Chuang}).}
  \label{fig:cU_circuit}
\end{figure}

We will use three auxiliary tracks. A ``mark track'', with alphabet $\Sigma_m=\{t,c,m_0,m_1,\#\}$, will be used to mark the position of the current control and target qubits. (The $m_{0,1}$ states will be used to temporarily store an auxiliary qubit on the mark track of the starting cell.) The other two auxiliary tracks, with alphabet $\Sigma_l=\{0,1,\#\}$, will constitute the work tapes of two different reversible looping TMs from~\cref{looping2}. 

\subsubsection{\texorpdfstring{$cU^k$}{cUk} machine}\label{sec:cU_machine}
We will need a QTM  which applies the control-$U_\varphi$ operation $k$ times. We give a construction for an arbitrary controlled single-qubit unitary $U$, as this will be useful later.

\begin{lemma}[Controlled-$U$ QTM]\label{cU_machine}
  For any single-qubit unitary $U$, there exists a well-formed normal form unidirectional QTM $cU^k$ with the following properties. The QTM has three tracks: a \emph{looping track} with alphabet $\Sigma_l=\{0,1,\#\}$, a \emph{mark track} with alphabet $\Sigma_m=\{t,c,m_0,m_1,\#\}$, and a \emph{quantum track} with alphabet $\Sigma_q=\{0,1,\#\}$. The input consists of a number $k\ge 1$ written in binary on the looping track in little-endian order, a configuration containing a single $t$ and a single $c$ within the first $n$ tape cells on the mark track (and all other cells blank), and an arbitrary $n$-qubit state on the quantum track.

  On such input, the QTM applies the control-$U$ operation $k$ times between the control and target qubits on the quantum track marked by $c$ and $t$, and then halts, having run for time $O(k n + k\log k)$, used at most $\max(n,\abs{k}) + 2$ space, behaving properly and leaving the configurations of the looping and mark tracks unchanged.
\end{lemma}

\begin{proof}
  It will be convenient to first run the $\TM{Shift}$ machine from \cref{shift_right} acting only on the mark and quantum tracks, to shift this part of the input one cell to the right. This produces a $\#$ on the mark and quantum tracks of the start cell, which we can use later to return to this cell. The remainder of the construction returns the quantum and mark tracks of the start cell to the $\#$ state, so we can run the corresponding $\TM{Shift}^\dg$ machine at the very end (where $\TM{Shift}^\dg$ is the reversal of $M_1$ constructed using \cref{reversal}) to shift these tracks back one cell to the left, so that the final output is correctly aligned. These shift operations take $O(n)$ time and use $n + 2$ space.

  We construct the core of the $cU^k$ QTM out of two simpler machines, $M_1$ and $M_2$, dovetailed together in the sequence $M_1,M_2,M_1^\dagger$ (where $M_1^\dg$ is the reversal of $M_1$). Machine $M_1$ effectively applies a CNOT gate between the qubits on the quantum track marked by $c$ and $t$ on the mark track, with $c$ as the control and $t$ as the target. However, as these qubits are not necessarily adjacent on the tape, $M_1$ must temporarily make use of an auxiliary qubit stored in the internal states of the QTM, which gets ``carried'' along with the QTM head. As we cannot leave a qubit in the internal state at the end of the computation, however, $M_1$ must reversibly erase this internal qubit after it has served its purpose, by applying a second CNOT between it and the control qubit. (See below for further details and quantum circuit diagrams.)

  $M_1$ implements this in stages: first, it scans right until it encounters a $c$ on the mark track, at which point it applies a CNOT between the quantum track and a qubit stored in its internal state. It then returns the head to the starting cell, and applies a CNOT between this internal qubit and an auxiliary qubit $m$ on the mark track (defined by the two orthogonal states $\ket{m_{0}}$ and $\ket{m_1}$, which are identified with $\ket{0}$ and $\ket{1}$ respectively). Finally, it returns to the location of the $c$ on the mark track, applies a second CNOT between the quantum track and the internal qubit, returns to the starting cell, and halts.

  In pseudocode:
  \begin{algorithm}[H]
    \caption{$M_1$}
    \begin{algorithmic}[1]
      \Input{mark track containing one $c$, one $t$, and $\#$ everywhere else;\\
        \phantom{\bf input:} quantum track containing an arbitrary quantum state.} \Comment{$q_0$}
      \State{write $m_0$ to mark track}
      \State{step Right}  \Comment{$q_1$}
      \While{not read $c$ on mark track}
        \State{step Right}
      \EndWhile
      \State{apply CNOT gate with quantum track as control, internal qubit $j$ as target}
      \State{step Left} \Comment{$q_2^j$}
      \While{read $\#$ or $t$ on mark track}
        \State{step Left}
      \EndWhile
      \State{apply CNOT with internal qubit $j$ as control, mark track qubit $m$ as target} \Comment{$q_3^j$}
      \While{not read $c$ on mark track}
        \State{step Right}
      \EndWhile
      \State{apply CNOT gate with quantum track as control, internal qubit $j$ as target}
      \State{step Left} \Comment{$q_4$}
      \While{read $\#$ or $t$ on mark track}
        \State{step Left}
      \EndWhile
      \State{halt} \Comment{$q_f$}
    \end{algorithmic}
  \end{algorithm}
  \noindent Clearly, $M_1$ is proper, runs for time at most $O(n)$ and uses at most $n+1$ space.

  The following well-formed, normal form, partial quantum transition function implements $M_1$ (which acts only on the mark and quantum tracks). Note that two of the internal states of $M_1$ comes in two varieties, indicated by a superscript $j=\{0,1\}$ and identified with $\ket{0},\ket{1}$, respectively, which are used to temporarily store a qubit in the internal state of the machine:
  \begin{equation}
    \begin{array}{l|ccccc}
            & [\#,\#] & [\#,i] & [t,\wc] & [c,j] & [m_k,\wc] \\
      \hline
      q_0   & \ket{m_0,\#}\ket{q_1}\ket{R} \\
      q_1   & & \ket{\#,i}\ket{q_1}\ket{R} & \ket{t,\wc}\ket{q_1}\ket{R}
            & \ket{c,j}\ket{q_2^j}\ket{L} \\
      q_2^j & & \ket{\#,i}\ket{q_2^j}\ket{L} & \ket{t,\wc}\ket{q_2^j}\ket{L}
            & & \ket{m_{k\oplus j}}\ket{q_3^j}\ket{R} \\
      q_3^j & & \ket{\#,i}\ket{q_3^j}\ket{R} & \ket{t,\wc}\ket{q_3^j}\ket{R}
            & \ket{c,j}\ket{q_4}\ket{L} & \\
      q_4   & & \ket{\#,i}\ket{q_4}\ket{L} & \ket{t,\wc}\ket{q_4}\ket{L}
            & & \ket{m_k}\ket{q_f}\ket{N} \\
      q_f   & \ket{\#,\#}\ket{q_0}\ket{N} & \ket{\#,i}\ket{q_0}\ket{N}
            & \ket{t,\wc}\ket{q_0}\ket{N} & \ket{c,j}\ket{q_0}\ket{N}
            & \ket{m_k,\wc}\ket{q_0}\ket{N} \\
      \multicolumn{2}{l}{\text{\footnotesize $\forall i,j,k\in\{0,1\}$} \rule{0em}{1.5em}}
    \end{array}
  \end{equation}

  Machine $M_2$ loops $k$ times, where $k$ is specified by the number written in binary on the looping track, applying a control-$U$ operation between the auxiliary qubit stored on the mark track of the start cell and the target qubit, and then halts. We construct $M_2$ using the reversible looping TM $\TM{Loop}$ from \cref{looping2}, and inserting for its special state a QTM $M'$ that is very similar to $M_1$. The $M'$ machine first applies a CNOT between the auxiliary qubit stored in the mark track of the start cell and an internal qubit. It then  scans right until it finds the $t$, and applies a single control-$U$ operation between the internal qubit and the target qubit. Finally, it moves the head back to the starting cell, and applies another CNOT between the internal qubit and the auxiliary qubit, before halting.

  In pseudocode:
  \begin{algorithm}[H]
    \caption{$M'$}
    \begin{algorithmic}[1]
      \Input{mark track containing auxiliary qubit $m$ in the first cell, \\
        \phantom{\bf input:} one $c$, one $t$ and $\#$ everywhere else.} \Comment{$q_0$}
      \State{apply CNOT gate with mark track qubit $m$ as control, internal qubit $j$ as target}
      \State{step Right}  \Comment{$q_1^j$}
      \While{not read $t$ on mark track}
        \State{step Right}
      \EndWhile
      \State{apply $cU$ gate with internal qubit $j$ as control, quantum track as target}
      \State{step Left} \Comment{$q_2^j$}
      \While{read $\#$ or $c$ on mark track}
        \State{step Left}
      \EndWhile
      \State{apply CNOT gate with mark track qubit $m$ as control, internal qubit $j$ as target}
      \State{halt} \Comment{$q_f$}
    \end{algorithmic}
  \end{algorithm}

  If $u_{ij}$ denotes the $i,j$'th matrix element of $U$, then the following well-formed,  normal form, partial quantum transition function (which acts only on the mark and quantum tracks) implements $M'$:
  \begin{equation}\label{eq:cU_QTM}
    \begin{array}{l|ccccc}
            & [m_0,\wc] & [m_1,\wc] & [\#,\wc] & [t,j] & [c,\wc] \\
      \hline
      q_0   & \ket{m_0,\wc}\ket{q_1^0}\ket{R} & \ket{m_1,\wc}\ket{q_1^1}\ket{R} \\
      q_1^0 & & & \ket{\#,\wc}\ket{q_1^0}\ket{R}
            & \ket{t,j}\ket{q_2^0}\ket{L} & \ket{c,\wc}\ket{q_1^0}\ket{R} \\
      q_1^1 & & & \ket{\#,\wc}\ket{q_1^1}\ket{R}
            & \sum_i u_{ij}\ket{t,i}\ket{q_2^1}\ket{L}
            & \ket{c,\wc}\ket{q_1^1}\ket{R} \\
      q_2^0 & \ket{m_0,\wc}\ket{q_f}\ket{N} & & \ket{\#,\wc}\ket{q_2^0}\ket{L}
            & & \ket{c,\wc}\ket{q_2^0}\ket{L} \\
      q_2^1 & & \ket{m_1,\wc}\ket{q_f}\ket{N} & \ket{\#,\wc}\ket{q_2^1}\ket{L}
            & & \ket{c,\wc}\ket{q_2^1}\ket{L} \\
      q_f   & \ket{m_0,\wc}\ket{q_0}\ket{N} & \ket{m_1,\wc}\ket{q_0}\ket{N}
            & \ket{\#,\wc}\ket{q_0}\ket{N} & \ket{t,j}\ket{q_0}\ket{N}
            & \ket{c,\wc}\ket{q_0}\ket{N} \\
      \multicolumn{2}{l}{\text{\footnotesize $\forall i,j,k\in\{0,1\}$} \rule{0em}{1.5em}}
    \end{array}
  \end{equation}

  $M'$ is proper, never alters the mark track, and, for given mark track configuration, always runs for the same number of time steps. So substituting $M'$ in the looping machine \TM{Loop} of \cref{looping2} produces a proper QTM $M_2$. Furthermore, since $M'$ takes $O(n)$ time and at most $n+1$ space, $M_2$ runs for time $O(k n + k \log k)$, uses at most $\max(n,\abs{k})+2$ space, and never moves the head before the starting cell.

  The only \emph{qubits} on which the overall QTM acts are the two qubits in the quantum tracks of the cells marked by $t$ and $c$, the auxiliary qubit stored on the mark track (which we label $m$), and the two internal qubits stored in the internal states $q_2^j$ of $M_1$ and $q_{1,2}^j$ of $M_2$ (let's label these $j_{1,2}$, respectively). Qubits $m$, $j_1$ and $j_2$ are all initially in the $\ket{0}$ state. Apart from classical processing to move the head into the correct location, $M_1$ just applies a CNOT between the $c$ and $j_1$ qubits, followed by a CNOT between the $j_1$ and $m$ qubits, and finally another CNOT between the $c$ and $j_1$ qubits:
  \[
  \Qcircuit @C=1em @R=1em {
    \lstick{c}            & \ctrl{1} & \qw      & \ctrl{1} & \qw \\
    \lstick{\ket{0}_{j_1}} & \targ    & \ctrl{1} & \targ    & \qw \\
    \lstick{\ket{0}_m} & \qw      & \targ    & \qw      & \qw \\
  }
  \]
  \vspace{.5em}

  Similarly, $M'$ applies a CNOT between the $k$ and $j_2$ qubits, a control-$U$ operation between the $j_2$ and $t$ qubits, and a final CNOT between the $k$ and $j_2$ qubits. Letting $cU$ denote the control-$U$ gate, and recalling that the $j_2$ qubit starts off in the $\ket{0}$ state, the overall effect of this is:
  \begin{equation}
    CNOT_{m,j_2}\; cU_{j_2,t}\; CNOT_{m,j_2} \ket[m,t]{\psi}\ket[j_2]{0}
    = cU\ket[m,t]{\psi}\ket[j_2]{0},
  \end{equation}
  or, as a quantum circuit diagram:
  \[
  \Qcircuit @C=1em @R=1em {
    \lstick{m} & \ctrl{1} & \qw      & \ctrl{1} & \qw \\
    \lstick{\ket{0}_{j_2}} & \targ    & \ctrl{1} & \targ    & \qw \\
    \lstick{t} & \qw      & \gate{U} & \qw      & \qw \\
  }
  \mspace{50mu} \raisebox{-1.7em}{$\equiv$} \mspace{70mu}
  \Qcircuit @C=1em @R=3em {
    \lstick{m} & \ctrl{1} & \qw \\
    \lstick{t}   & \gate{U} & \qw \\
  }
  \]
  i.e.\ $M'$ effectively applies a single control-$U$ operation between the $m$ and $t$ qubits. $M_2$ repeats the $M'$ operations $k$ times, so the overall action of $M_2$ is to apply a control-$U$ operation $k$ times between the $m$ and $t$ qubits:
  \begin{equation}
    (CNOT_{m,j_2}\; cU_{j_2,t}\; CNOT_{m,j_2})^k \ket[m,t]{\psi}\ket[j_2]{0}
    = cU^k\ket[m,t]{\psi}\ket[j_2]{0},
  \end{equation}
  or, in quantum circuit diagrams:
  \[
  \underbrace{\mbox{
  \Qcircuit @C=1em @R=1em {
    \lstick{m} & \ctrl{1} & \qw      & \ctrl{1} & \qw & \cdots & & \ctrl{1} & \qw      & \ctrl{1} & \qw \\
    \lstick{\ket{0}_{j_2}} & \targ    & \ctrl{1} & \targ    & \qw & \cdots & & \targ    & \ctrl{1} & \targ    & \qw \\
    \lstick{t} & \qw      & \gate{U} & \qw      & \qw & \cdots & & \qw      & \gate{U} & \qw      & \qw \\
  }}}_{k\text{ times}}
  \mspace{50mu} \raisebox{-1.7em}{$\equiv$} \mspace{70mu}
  \Qcircuit @C=1em @R=3em {
    \lstick{m} & \ctrl{1}   & \qw \\
    \lstick{t}   & \gate{U^k} & \qw \\
  }
  \]
  Since the CNOT gate is self-inverse, the time-reversal $M_1^\dg$ in fact applies the same sequence of CNOTs as $M_1$. So, on the control, target, mark and internal qubits, dovetailing $M_1,M_2,M_1^\dg$ carries out the operation:
  \begin{multline}
    (CNOT_{c,j_1}\; CNOT_{j_1,m}\; CNOT_{c,j_1}\; cU^k_{m,t}\;\cdot\\ CNOT_{j_1,m}\; CNOT_{c,j_1}\; CNOT_{c,j_1})
      \ket[c,t]{\psi}\ket[j_1]{0}\ket[m]{0}\\
    = cU^k_{c,t}\ket[c,t]{\psi}\ket[j_1]{0}\ket[m]{0},
  \end{multline}
  or, in quantum circuit diagrams:
  \[
  \Qcircuit @C=1em @R=1em {
    \lstick{c}              & \ctrl{1} & \qw      & \ctrl{1} & \qw      & \qw       & \qw      & \ctrl{1} & \qw      & \ctrl{1} & \qw \\
    \lstick{\ket{0}_{j_1}}   & \targ    & \ctrl{1} & \targ    & \qw      & \qw       & \qw      & \targ    & \ctrl{1} & \targ    & \qw \\
    \lstick{\ket{0}_{m}}   & \qw      & \targ    & \qw      & \ctrl{1} & \qw       & \ctrl{1} & \qw      & \targ    & \qw      & \qw \\
    \lstick{\ket{0}_{j_2}}   & \qw      & \qw      & \qw      & \targ    & \ctrl{1}  & \targ    & \qw      & \qw      & \qw      & \qw \\
    \lstick{t}              & \qw      & \qw      & \qw      & \qw      & \gate{U^k} & \qw     & \qw       & \qw       & \qw      & \qw \\
  }
  \mspace{50mu} \raisebox{-3.5em}{$\equiv$} \mspace{50mu}
  \Qcircuit @C=1em @R=6.5em {
    \lstick{c} & \ctrl{1}   & \qw \\
    \lstick{t} & \gate{U^k} & \qw \\
  }
  \]

  Thus the overall action of the entire QTM is to apply a $cU^k$ operation between the control and target qubits, as required, leaving the looping and mark tracks back in the configuration they started in. Furthermore, the QTM runs for time $O(k n + k \log k)$, uses at most $\max(n,\abs{k})+2$ space, and is well-formed, normal form, unidirectional and proper, as claimed.
\end{proof}

\subsubsection{Iterating over the control qubits}
\label{sec:cU_stage_iteration}
We want to apply a $cU_\varphi^{2^{n-1}}$ operation between the $n$'th output qubit and the ancilla qubit, for each of the $N$ output qubits. (Note that the output qubits are stored in reverse order on the quantum track starting from the 2nd cell, so the 1st output qubit is stored in the $N+1$'th cell of the quantum track and the $N$'th output qubit is stored in the 2nd cell; recall that the 1st cell holds the target qubit.) As we will need to use a second looping machine to iterate over the qubits, we refer to the looping track for the $cU^k$ machine as the ``$cU$ looping track'', and the track for the second looping machine used here as the ``outer looping track''.

We will need a reversible TM which, on suitable mark and $cU$ looping track configurations (cf.\ \cref{cU_machine}), moves the $c$ marker on the mark track one cell to the left, and doubles the value $k$ written on the $cU$ looping track. It is convenient (though slightly less efficient) to divide the implementation into two parts, dovetailed together: $\TM{Step}_c$ which shifts the $c$ marker one cell to the left, and $\TM{Dbl}$ which doubles the value on the $cU$ looping track.

In pseudocode, $\TM{Step}_c$ does the following:
\begin{algorithm}[H]
  \caption{$\TM{Step}_c$}
  \begin{algorithmic}[1]
    \Input{string $s\in\{t,c,\#\}^*$ starting with the symbol $t$, followed by zero or\\
      \phantom{\bf input:} more $\#$'s, ending with a single $c$ symbol.} \Comment{$q_0$}
    \State{step Right} \Comment{$q_1$}
    \While{read $\#$}
      \State{step Right}
    \EndWhile
    \State{write a $\#$}
    \State{step Left} \Comment{$q_2$}
    \State{write a $c$}
    \State{step Left} \Comment{$q_3$}
    \While{not read a $t$}
      \State{step Left}
    \EndWhile
    \State{halt} \Comment{$q_f$}
  \end{algorithmic}
\end{algorithm}

The following well-formed, normal form, partial transition function implements the reversible TM $\TM{Step}_c$ (which acts only on the mark track):
\begin{equation}
  \begin{array}{l|ccc}
         & \# & t & c \\
    \hline
    q_0  & & (t,q_1,R) \\
    q_1  & (\#,q_1,R) & & (\#,q_2,L) \\
    q_2  & (c,q_3,L) \\
    q_3  & (\#,q_3,L) & (t,q_f,N) \\
    q_f  & (\#,q_0,N) & (t,q_0,N) & (c,q_0,N) \\
    \multicolumn{2}{l}{\text{\footnotesize $\forall i\in\{0,1\}$} \rule{0em}{1.5em}}
  \end{array}
\end{equation}

Doubling a number in binary simply appends a 0 onto the binary representation of the number. The Turing Machine implementation of this is particularly convenient in the little-endian convention we are using, as it simply means finding the first blank symbol and appending the 0 there. We can simplify the implementation slightly and avoid the Turing Machine head ever moving before the starting cell, by taking advantage of the fact that the input in our case is always a power of 2, so, in little-endian order, always consists of a string of zero or more 0's followed by a single~1; in particular, the input is never blank. Thus we can overwrite the initial 0 or 1 with a blank symbol to mark the start of the tape, and restore the initial tape cell to 0 at the end of the algorithm.
In pseudocode:

\begin{algorithm}[H]
  \caption{$\TM{DBL}$}
  \begin{algorithmic}[1]
    \Input{binary representation of $k=2^n$ for $n\in\N$.} \Comment{$q_0$}
  \If{read a 1}
  \State{write a $\#$}
  \State{step Right} \Comment{$q_2$}
   \Else
  \State{write a $\#$}
    \State{step Right} \Comment{$q_1$}
    \While{read a 0}
    \State{step Right}
    \EndWhile
    \State{write a 0}
    \State{step Right} \Comment{$q_2$}
    \EndIf
    \State{write a 1}
    \State{step Left}  \Comment{$q_3$}
    \While{read 0}
      \State{step Left}
    \EndWhile
    \State{write a 0}
    \State{halt} \Comment{$q_f$}
\end{algorithmic}
\end{algorithm}

The following well-formed, normal form, partial transition function for $\TM{Dbl}$ (which acts only on the $cU$ looping track) does precisely this:
\begin{equation}
  \begin{array}{l|ccc}
         & \# & 0 & 1 \\
    \hline
    q_0  & & (\#,q_1,R) & (\#,q_2,R) \\
    q_1  & & (0,q_1,R) & (0,q_2,R) \\
    q_2  & (1,q_3,L) \\
    q_3  & (0,q_f,N) & (0,q_3,L) & \\
    q_f  & (\#,q_0,N) & (0,q_0,N) & (1,q_0,N) \\
  \end{array}
\end{equation}

We are now in a position to implement the complete controlled-$U_\varphi$ stage of the quantum phase estimation algorithm. We initialise the states of the auxiliary tracks, by first running a simple reversible TM that changes the $[\#,\#]$ in the first cell of the mark and $cU$ looping tracks into $[t,1]$, then returns to the starting cell and halts. (Constructing a reversible TM for this is trivial.) We dovetail this with a proper, normal form reversible TM that scans to the end of the output qubits, initialises the mark track in that cell to a $c$, then returns to the starting cell and halts.
In pseudocode:

\begin{algorithm}[H]
  \caption{controlled-$U_{\varphi}$ initialisation}
  \begin{algorithmic}[1]
    \Input{mark track containing $t$ in the first cell;\\
      \phantom{\bf input:} quantum track with ancilla and output qubits followed by $\#$s} \Comment{$q_0$}
    \State{step Right} \Comment{$q_1$}
    \While{not read $\#$ on quantum track}
      \State{step Right}
    \EndWhile
    \State{step Left}  \Comment{$q_2$}
    \State{write $c$ on the mark track}
    \State{step Left}  \Comment{$q_3$}
    \While{not read $t$ on mark track}
      \State{step Left}
    \EndWhile
    \State{halt} \Comment{$q_f$}
  \end{algorithmic}
\end{algorithm}

The following normal form, partial transition function (which acts only on the mark and quantum tracks) accomplishes this:
\begin{equation}
  \begin{array}{l|cccc}
         & [\#,\#] & [\#,i] & [t,j] \\
    \hline
    q_0  & & & ([t,j],q_1,R) \\
    q_1  & ([\#,\#],q_2,L) & ([\#,i],q_1,R) \\
    q_2  & & ([c,i],q_3,L) \\
    q_3  & & ([\#,i],q_3,L) & ([t,j],q_f,N) \\
    q_f  & ([\#,\#],q_0,N) & ([\#,i],q_0,N) & ([t,j],q_0,N) \\
    \multicolumn{2}{l}{\text{\footnotesize $\forall i,j\in\{0,1\}$} \rule{0em}{1.5em}}
  \end{array}
\end{equation}
On the mark track, this leaves a $t$ and $c$ over the ancilla qubit and the first output qubit,\footnote{Recall that the first output qubit is the last one on the quantum track} respectively, and blanks everywhere else. The configuration prepared on the $cU$ looping track corresponds to the number 1 written in binary.

Next, we copy $N$ from the input track to the outer looping track using the $\TM{Copy}$ machine from \cref{copier}. We then run a reversible looping TM from~\cref{looping2} which uses this track as its input track (so it will loop $N$ times in total). For the special state of this looping machine, we substitute a QTM which dovetails the $cU^k$ machine from \cref{cU_machine} with the $\TM{Step}_c$ and $\TM{Dbl}$ machines constructed above.

On appropriate input, the $cU^k$ machine from~\cref{cU_machine} runs for a number of steps which depends only on the classical part of the input supplied on the mark and looping tracks, and not on the quantum state in the quantum track. The $\TM{Step}_c$ and $\TM{Dbl}$ reversible TMs don't touch the quantum track at all. So as long as the mark and $cU$ looping tracks are always in a suitable configuration (cf.\ \cref{cU_machine}) before the $cU^k$ machine is run in each iteration, the outer looping machine will be proper.

Now, we already initialised the $cU$ looping track to $k=1$, above, and marked the ancilla qubit as the target and the first output qubit as the control. So the initial configuration of the mark and $cU$ looping tracks is suitable input for a $cU^k$ machine. When $cU^k$ is run in the first iteration of the outer looping machine, it applies a $cU^1 = cU$ between the first output qubit and the target qubit. $\TM{Step}_c$ and $\TM{Dbl}$ machines then run, which shifts the control marker one cell to the left onto the second output qubit and doubles $k$ to $2$, ready for the next iteration. In the second iteration, the looping machine therefore applies a $cU^2$ between the second output qubit and the ancilla, and doubles $k$ to 4. The looping machine goes through a total of $N$ iterations, each time feeding suitable input to the $cU^k$ machine to make it apply a $cU^{2^{n-1}}$ between the $n$'th output qubit and the ancilla, before halting with $N$ written on the outer looping track, a $t$ and $c$ in the first and second cell of the mark track, and $N+1$ written on the $cU$ looping track.

Thus the outer looping machine is well-formed, normal form, unidirectional and proper. It runs for a total of $N$ iterations with overhead $N \log N$ (cf. \cref{looping2}). In the $n$'th iteration it runs the $cU^k$ machine from \cref{cU_machine} once with $k=2^{n-1}$, and runs the $\TM{Step}_c$ and $\TM{Dbl}$ machines once each. The $n$'th iteration of the $cU^k$ machine takes time $O(N + 2^{n-1}\log 2^{n-1})$, and the $\TM{Step}_c$ and $\TM{Dbl}$ machines always take time $O(n)$. So the outer looping machine runs for a total time $O(N^2 2^N)$.

Finally, we can uncompute (reset) the auxiliary tracks by running a reversible TM which transforms the configuration $N;N;tc;N+1$ left on the input, outer looping, mark, and $cU$ looping tracks, into $N;\#;\#;\#$. We do this by running the reversal $\TM{Copy}^\dg$ of the copying machine from \cref{copier} on the input and outer looping tracks to erase the outer looping track, decrementing the $cU$ looping track and running $\TM{Copy}^\dg$ again to erase that track, and running a trivial reversible TM that changes $tc$ on the mark track into $\#\#$.

Putting everything together, the control-$U$ stage runs for a total time $O(N^2 2^N)$ and requires space $N+3$ (2 more cells than the number of qubits, which is $N+1$). When dovetailed with the preparation stage from \cref{sec:preparation_stage}, the overall QTM we have constructed implements the control-$U$ circuit of \cref{fig:cU_circuit}. If $\varphi_k$ denotes the $k$'th digit in the binary fraction expansion of $\varphi$, then this prepares the state~\cite{Nielsen+Chuang}
\begin{multline}
  \frac{1}{2^N}
  \left(\ket{0} + e^{2\pi i 0.\varphi_N}\ket{1}\right)
  \left(\ket{0} + e^{2\pi i 0.\varphi_{N-1}\varphi_N}\ket{1}\right) \cdots\\
  \cdots
  \left(\ket{0} + e^{2\pi i 0.\varphi_2\dots\varphi_{N-1}\varphi_N}\ket{1}\right)
  \left(\ket{0} + e^{2\pi i 0.\varphi_1\varphi_2\dots\varphi_N}\ket{1}\right)
\end{multline}
on the $N$ output qubits (ordered as they are on the quantum track). All other tracks are blank, except for the input track which still has the input $N$ written on it as a little-endian binary number.

\subsection{Locating the LSB}\label{sec:LSB_stage}
The final stage of the quantum phase estimation algorithm is to apply the inverse quantum Fourier transform to the quantum state generated by the control-$U$ stage. The structure of the quantum circuit for the QFT is rather reminiscent of that of the control-$U$ stage. It again involves applying a ``cascade'' of control-$U^{2^n}$ gates between pairs of qubits, which we already know how to implement. Only now there is a new cascade starting from each output qubit (see \cref{fig:QFT_circuit}).

However, the phase $2^{-N}$ of the control-$U_{2^{-N}}$ rotation needed in the QFT circuit depends on the total number of qubits $N$ that the QFT is being applied to. If we simply implemented the inverse QFT circuit directly on all $N$ output qubits, the entries of the transition function in our $cU^k$ machine from \cref{sec:cU_stage} would have to depend on the input $N$, which is not allowed. It is not at all obvious whether the inverse QFT circuit on $N$ qubits can be implemented \emph{exactly} when $N$ is given as input.\footnote{The universal QTM construction of \textcite{Bernstein-Vazirani} can approximate the QFT on $N$ qubits to arbitrary precision, but it does not implement it exactly.}

On the other hand, the entries of the transition function \emph{are} allowed to depend on the phase $\varphi$ that we are estimating -- indeed, they necessarily do so, since the output of the $P_n$ QTM in \cref{phase-estimation_QTM} depends on $n$ (or equivalently on $\varphi$). Rather than implementing the QFT on $N$ qubits, we take a different approach. We supply our QTM with the $U_{2^{-\abs{\varphi}}}$ gate, and implement the inverse QFT on $\abs{\varphi}$ qubits, independent of the input $N$. (I.e.\ just enough qubits to hold all the digits of the binary fraction expansion of $\varphi$.) However, for this to work, we must first identify which output qubit holds the least significant bit (LSB) of $\varphi$, so that we know which qubits to apply the $\abs{\varphi}$-qubit inverse QFT to. The role of the input $N$ is merely to provide us with an upper-bound on $\abs{\varphi}$, which allows us to identify the LSB in finite time and space.

Recall that the control-$U$ stage prepared the state~\cite{Nielsen+Chuang}
\begin{multline}\label{eq:cU_output}
  \frac{1}{2^{\frac{N}{2}}}
  \left(\ket{0} + e^{2\pi i 0.\varphi_N}\ket{1}\right)
  \left(\ket{0} + e^{2\pi i 0.\varphi_{N-1}\varphi_N}\ket{1}\right) \cdots\\
  \cdots
  \left(\ket{0} + e^{2\pi i 0.\varphi_2\dots\varphi_{N-1}\varphi_N}\ket{1}\right)
  \left(\ket{0} + e^{2\pi i 0.\varphi_1\varphi_2\dots\varphi_N}\ket{1}\right)
\end{multline}
on the $N$ output qubits (where $\varphi_k$ denotes the $k$'th digit in the binary fraction expansion of $\varphi$).

The least significant bit of $\varphi$ is the $\abs{\varphi}$'th bit, and by assumption (cf.\ \cref{phase-estimation_QTM}) $\abs{\varphi} \leq N$. So the $\abs{\varphi}$'th bit of $\varphi$ is 1 and the $\abs{\varphi}+1\dots N$'th bits are all 0. Thus the last $N-\abs{\varphi}$ output qubits are in the $\ket{+}$ state, and the $\abs{\varphi}$'th qubit, which corresponds to the least significant bit of $\varphi$, is in the $\ket{-}$ state. (Recall that the output qubits are stored on the quantum track in reverse order.)

Therefore, if we construct a QTM that steps right, applying Hadamard rotations to the output qubits in the quantum track until it obtains a $\ket{1}$ and halts, we will be able to locate the least significant bit of $\varphi$. More precisely, this will rotate the first $N-\abs{\varphi}$ output qubits into the $\ket{0}$ state, and halt with the $N-\abs{\varphi}+1$'th qubit (which corresponds to the $\abs{\varphi}$'th bit of $\varphi$) rotated into the $\ket{1}$ state. The least significant bit of $\varphi$ is therefore identified by the first $\ket{1}$ on the quantum track after this machine has finished running.

\begin{algorithm}[H]
  \caption{LSB QTM}
  \begin{algorithmic}[1]
    \Input{quantum tape containing $\ket{1}$ followed by $N$-qubit state from \cref{eq:cU_output}.} \Comment{$q_0$}
    \State{write $\ket{\#}$}
    \Repeat
      \State{step Right} \Comment{$q_1$}
      \State{apply a Hadamard gate} \Comment{$q_2$}
    \Until{read a $\ket{1}$}
    \State{step Left} \Comment{$q_3$}
    \While{read $\ket{0}$}
      \State{step Left}
    \EndWhile
    \State{halt} \Comment{$q_f$}
  \end{algorithmic}
\end{algorithm}

The following well-formed, normal form, partial quantum transition function (which acts only on the quantum track) implements a proper QTM that steps along the quantum track, applying Hadamards until it obtains a $\ket{1}$. It can be verified that it obeys the three conditions of \cref{QTM_transitions}, so can be completed to a well-formed QTM:
\begin{equation}\label{eq:LSB_QTM}
  \begin{array}{l|ccc}
         & \# & 0 & 1 \\
    \hline
    q_0  & & & \ket{\#}\ket{q_1}\ket{R} \\
    q_1  & \ket{\#}\ket{q_3}\ket{L}
           & \tfrac{1}{\sqrt{2}}(\ket{0} + \ket{1})\ket{q_2}\ket{N}
           & \tfrac{1}{\sqrt{2}}(\ket{0} - \ket{1})\ket{q_2}\ket{N} \\
    q_2  & & \ket{0}\ket{q_1}\ket{R} & \ket{1}\ket{q_3}\ket{L} \\
    q_3  & \ket{\#}\ket{q_f}\ket{N} & \ket{0}\ket{q_3}\ket{L} \\
    q_f  & \ket{\#}\ket{q_0}\ket{N} & \ket{0}\ket{q_0}\ket{N}
           & \ket{1}\ket{q_0}\ket{N} \\
  \end{array}
\end{equation}
(Note that the qubit in the starting cell of the quantum track is still in the state $\ket{1}$ from the preparation and control-$U$ stages. As we will not need this qubit again, this machine sets it to $\ket{\#}$ to mark the starting cell and leaves it in that state when it halts, for later convenience.)

The configuration written on the quantum track after this machine has finished is the $N+1$-qubit state:
\begin{multline}\label{eq:LSB}
  \ket{\#}\ket{0}\overset{N-\abs{\varphi}}{\cdots}\ket{0}\ket{1}
  \frac{1}{2^{\frac{\abs{\varphi}-1}{2}}}
  \left(\ket{0}
    + e^{2\pi i 0.\varphi_{\abs{\varphi}-1}\varphi_{\abs{\varphi}}}
    \ket{1}\right)
  \left(\ket{0}
    + e^{2\pi i 0.\varphi_{\abs{\varphi}-2}\varphi_{\abs{\varphi}-1}\varphi_{\abs{\varphi}}} \ket{1}\right)
  \cdots\\
  \cdots
  \left(\ket{0}
    + e^{2\pi i 0.\varphi_1\varphi_2\dots\varphi_{\abs{\varphi}-1}\varphi_{\abs{\varphi}}} \ket{1}\right).
\end{multline}
Note that the first $N-\abs{\varphi}+2$ cells of the quantum track are not entangled with the rest of the track.

It will be helpful for later to record the number of output bits that we are skipping, i.e.\ to compute $N-\abs{\varphi}$ and store the result on a separate auxiliary ``count track''. We can do this using a construction that is very similar to the unary-to-binary converter of \cref{unary_to_binary}. Specifically, we will construct a reversible TM $\TM{Count}$ that steps right over $\ket{0}$'s until it finds a $\ket{1}$ on the quantum track, running the $\TM{Inc}$ machine once each time it steps right. (This means first returning the head to the starting cell, running $\TM{Inc}$, then returning the head back to where it was previously.)

Consider the following partial transition function, which satisfies the conditions of \cref{reversible_transitions} so can be completed to a proper, normal form, reversible TM:
\begin{equation}\label{eq:LSB_counting_TM}
  \begin{array}{l|ccc}
         & \# & 0 & 1 \\
    \hline
    q_0  & (\#,q_1,R) \\
    q_1  & (\#,q_4,L) & (\#,q_2,L) & (1,q_4,L) \\
    q_2  & (\#,q,N) & (0,q_2,L) \\
    q    & (\#,q_3,R) \\
    q_3  & (0,q_1,R) & (0,q_3,R) \\
    q_4  & (\#,q_f,N) & (0,q_4,L) \\
    q_f  & (\#,q_0,N) & (0,q_0,N) & (1,q_0,N) \\
  \end{array}
\end{equation}
This machine effectively steps right over 0's until it encounters a 1, at which point it moves its head back to the starting cell and halts. (It assumes that the first cell is marked with a $\#$.) However, each time it is about to step right, it first temporarily marks the current head location by overwriting the 0 with a $\#$, returns the head to the starting cell, enters a special internal state $q$ for one time step, then returns the head back to where it was before, resets the quantum track to 0, and continues stepping right from where it left off. In pseudocode:

\begin{algorithm}[H]
  \caption{LSB-counting TM}
  \begin{algorithmic}[1]
    \Input{string $s\in\{0,1,\#\}^*$ consisting of a single $\#$ followed by any\\
      \phantom{\bf input:} binary string} \Comment{$q_0$}
    \State{step Right} \Comment{$q_1$}
    \While{read a 0}
      \State{write a $\#$}
      \State{step Left} \Comment{$q_2$}
      \While{read a 0}
        \State{step Left}
      \EndWhile
      \State{enter special state $q$} \Comment{$q$}
      \State{step Right} \Comment{$q_3$}
      \While{not read a $\#$}
        \State{step right}
      \EndWhile
      \State{write a 0}
      \State{step Right} \Comment{$q_1$}
    \EndWhile
    \State{step Left} \Comment{$q_4$}
    \While{read a 0}
      \State{step Left}
    \EndWhile
    \State{halt} \Comment{$q_f$}
  \end{algorithmic}
\end{algorithm}

By \cref{subroutine}, we can substitute for the state $q$ the $\TM{Inc}$ machine from \cref{incrementer} (acting on an initially blank ``count track''). If we run the above TM on the quantum track, it will step right precisely $N-\abs{\varphi}$ times before finding the first $\ket{1}$, so it will halt with $N-\abs{\varphi}$ written to the count track.

The configuration left on the tape now consists of the little-endian binary number $N$ written on the input track, the little-endian binary number $N-\abs{\varphi}$ written on the count track, and the $N+1$-qubit state from \cref{eq:LSB} written on the quantum track.

As the qubits storing the $N-\abs{\varphi}$ trailing 0's of the $N$-digit binary expansion of $\varphi$ now play no further role, it is convenient to shift the starting cell of subsequent machines to the tape cell containing the LSB. To this end, we copy the values $N$ and $N-\abs{\varphi}$ from the input and count tracks onto new auxiliary tracks. We then use a $\TM{Loop}$ machine from \cref{looping2}, which uses the original count track as its input and work track, to run a $\TM{Shift}$ machine from \cref{shift_right} acting on the tracks holding the copies of the count and input tracks. Thus the $\TM{Shift}$ machine will run a total of $N-\abs{\varphi}$ times, thereby shifting the binary strings on the copies so that they start in the $N-\abs{\varphi}$'th cell -- the one containing the LSB of $\abs{\varphi}$ in the quantum track. We then run a trivial TM that steps the head right over the initial string of $\ket{0}$s on the quantum track, until it encounters the first $\ket{1}$ identifying the LSB, and halts with the head at this LSB cell. When we refer to the input and count tracks in the following section, we mean the copies shifted to start at the LSB cell. (At the very end of the computation, we will uncompute the shifted copies of the count and input tracks and step the head left back to the original starting cell, to ensure the overall QTM remains proper.)

Now, we have been careful to ensure that the head is never moved to a location before the starting cell in any of the reversible and quantum TMs that we have constructed. Furthermore, the section of the complete tape configuration located before the LSB cell is unentangled with the rest (see \cref{eq:LSB}, and note that all tracks other than the quantum track are in classical configurations). Thus, if we start any of our TMs in the LSB cell, it acts as if its input were the section of the tape configuration located after (and including) the LSB cell. We can therefore ignore the tape configuration of the first $N-\abs{\varphi}$ cells in the following section, and restrict our attention to the following $\abs{\varphi}$ cells.

\subsection{QFT stage}\label{sec:QFT_stage}
We are now ready to apply the $\abs{\varphi}$-qubit inverse QFT to the $\abs{\varphi}$-qubit state stored on the quantum track. The inverse QFT circuit on $\abs{\varphi}$ qubits consists of cascades of $cU_{2^{-\abs{\varphi}}}^k$ gates, very reminiscent of the control-$U$ stage we have already implemented (see \cref{fig:QFT_circuit}). The construction will therefore be similar. In the following, as we have shifted the starting cell, we relabel the output qubits and refer to the LSB qubit as the 1st qubit, the last one as the $\abs{\varphi}$'th qubit.

\begin{figure}[htbp]
  \centering
  \resizebox{.85\textwidth}{!}{
    \mbox{\scriptsize\hspace{5em}
    \Qcircuit @C=.5em @R=1em {
      \lstick{\ket{0} + e^{2\pi i 0.j_1\dots j_n}\ket{1}}
        & \qw & \qw & \qw & \qw & \qw & \qw & \qw & \qw & \qw
        & \gate{U_{2^{-\abs{\varphi}}}^1}
        & \gate{U_{2^{-\abs{\varphi}}}^2}
        & \push{\makebox[2em]{$\dots$}} \qw
        & \gate{U_{2^{-\abs{\varphi}}}^{2^{\abs{\varphi}}}}
        & \gate{\makebox[\totalheightof{$U_{2^{-\abs{\varphi}}}^2$}][c]{$H
            \vphantom{U_{2^{-\abs{\varphi}}}^2}$}}
        & \rstick{\ket{j_1}} \qw \\
      \lstick{\ket{0} + e^{2\pi i 0.j_2\dots j_n}\ket{1}}
        & \qw & \qw & \qw & \qw & \qw
        & \gate{U_{2^{-\abs{\varphi}}}^2}
        & \gate{U_{2^{-\abs{\varphi}}}^4}
        & \push{\makebox[2em]{$\dots$}} \qw
        & \gate{\makebox[\totalheightof{$U_{2^{-\abs{\varphi}}}^2$}][c]{$H
            \vphantom{U_{2^{-\abs{\varphi}}}^2}$}}
        & \qw & \qw & \qw & \ctrl{-1} & \qw
        & \rstick{\ket{j_2}} \qw \\
      \vdots\\
      \lstick{\ket{0} + e^{2\pi i 0.j_{n-1}\dots j_n}\ket{1}}
        & \qw & \qw
        & \gate{U_{2^{-\abs{\varphi}}}^{2^{\abs{\varphi}}}}
        & \gate{\makebox[\totalheightof{$U_{2^{-\abs{\varphi}}}^2$}][c]{$H
            \vphantom{U_{2^{-\abs{\varphi}}}^2}$}}
        & \push{\makebox[2em]{$\dots$}} \qw
        & \qw & \ctrl{-2} & \qw & \qw & \qw & \ctrl{-3} & \qw & \qw & \qw
        & \rstick{\ket{j_{N-1}}} \qw \\
      \lstick{\ket{0} + e^{2\pi i 0.j_n}\ket{1}}
        & \qw
        & \gate{\makebox[\totalheightof{$U_{2^{-\abs{\varphi}}}^2$}][c]{$H
            \vphantom{U_{2^{-\abs{\varphi}}}^2}$}}
        & \ctrl{-1} & \qw
        & \push{\makebox[2em]{$\dots$}} \qw
        & \ctrl{-3} & \qw & \qw & \qw & \ctrl{-4} & \qw & \qw & \qw & \qw
        & \rstick{\ket{j_{N-1}}} \qw
    }}
  }
  \caption[Quantum phase estimation, inverse QFT stage.]{The inverse QFT stage of the quantum phase estimation circuit (cf.~Fig.~5.1 in~\cite{Nielsen+Chuang}).}
  \label{fig:QFT_circuit}
\end{figure}

In each cascade, we want to apply a $cU_{2^{m-n}} = (cU_{2^{-\abs{\varphi}}})^{2^{\abs{\varphi}+m-n}}$ gate between the $m$'th and $n$'th qubit ($m<n$). Once again, we will use mark and $cU$ looping tracks to hold the input to a $cU_{2^{-\abs{\varphi}}}^k$ machine from \cref{cU_machine}. However, the main loop will now consist of two nested loops: an outer loop to iterate the control qubit $m$ of the $cU_{2^{-\abs{\varphi}}}^{2^{\abs{\varphi}+m-n}}$ gate over each output qubit, and an inner loop to iterate the target qubit $n$ over qubits $m+1$ through $\abs{\varphi}$.

We first run a TM that initialises the mark and $cU$ looping tracks, so that the mark track contains the configuration $ct$ in the first two cells, and the $cU$ looping track contains a string of $\abs{\varphi}-1$ 0's followed by a 1. Note that this initial configuration of the $cU$ looping track is the little-endian binary representation of the number $2^{\abs{\varphi}-1}$. (Constructing a proper normal form, reversible TM that implements all of this is an easy exercise.)

We also use the $\TM{Inc}$ machine from \cref{incrementer} to increment the number $N-\abs{\varphi}$ stored on the count track to $N-\abs{\varphi}+1$. We then run the $\TM{Sub}$ machine from \cref{adder}, with the input and count tracks as the input tracks and the inner looping track as the output track, to write the number $N-(N-\abs{\varphi}+1) = \abs{\varphi}-1$ to the inner looping track.

\paragraph{Inner loop} The inner looping machine is very similar to the main looping machine from the control-$U$ stage. We use a $\TM{Loop}$ machine from \cref{looping2}, with the inner looping track as its input and work track. In each iteration, we first run a TM to divide the value on the $cU$ looping track by 2. (This is simply the reversal $\TM{Dbl}^\dg$ of the machine constructed in \cref{sec:cU_stage_iteration}.) We dovetail this with the $cU_{2^{-\abs{\varphi}}}^k$ machine from \cref{cU_machine}. Finally, we dovetail this with a TM that moves the target marker $t$ on the mark track one cell to the right. (This can be implemented by taking the reversal $\TM{Step}_c^\dg$ of the machine from \cref{sec:cU_stage_iteration} and replacing $c$ with $t$ in its transition rules to give a $\TM{Step}_t^\dg$ machine.)

Assume that the inner looping track is initialised to $\abs{\varphi}-m$, the control and target markers on the mark track are initially over the $m$'th and $m+1$'th qubits, and the $cU$ looping track is initialised to $2^{\abs{\varphi}}$. Then the effect of this inner looping machine is to apply $cU_{2^{m-n}}$ operations between the $m$'th and $n$'th qubit for all $m < n \leq \abs{\varphi}$ -- the cascade of $cU$'s starting from the $m$'th qubit in \cref{fig:QFT_circuit}.

This inner looping TM leaves the mark track in the configuration with a $c$ in the same cell that it started out in, a $t$ in the $\abs{\varphi}+1$'th cell (i.e.\ the cell after the last output qubit), and $2^{m-1}$ written on the $cU$ looping track. (The latter is the configuration consisting of an initial string of 0's, followed by a 1 in the cell containing the $c$ on the mark track.)

\paragraph{Outer loop} For the outer looping machine, we use a $\TM{Loop2}$ machine from \cref{looping2} running on the input and count tracks (which hold the numbers $N$ and $N-\abs{\varphi}+1$, respectively). By \cref{looping2}, this machine will therefore loop $\abs{\varphi}-1$ times.

In each iteration, we first run the inner looping machine. Note that for the first iteration, the auxiliary tracks are already initialised as assumed above for the value $m=1$. We dovetail this with a simple QTM that applies a Hadamard operation to the current control qubit (the one marked by a $c$ on the mark track).

We then run a TM that changes the 1 on the $cU$ looping track to a 0, shifts the $c$ on the mark track one cell to the right, steps right, and changes $[\#,\#]$ on the mark and $cU$ looping tracks to $[t,0]$. The machine then steps right, changing $\#$ to 0 on the $cU$ looping track as it goes, until it reaches the end of the output qubits. Whereupon it changes $\#$ on the $cU$ looping track to $1$, steps right, and changes the $t$ on the mark track to $\#$. Finally, it returns to its starting cell and halts. (Again, by now, constructing a proper, normal form, reversible TM for this is straightforward.) We dovetail this with the $\TM{Dec}$ TM, acting on the inner looping track.

The effect of all this is to reset the configuration of the auxiliary tracks, ready for the next iteration of the inner loop. The control marker $c$ is shifted to the next qubit along, qubit $m$ say, and the target marker $t$ is placed over the adjacent $m+1$'th qubit. The $cU$ looping track is reset to $2^{\abs{\varphi}-1}$, and the inner looping track is decremented to $\abs{\varphi}-m$, as required. In other words, the auxiliary tracks are initialised to the desired configuration for the new value of $m$.

Thus the outer looping machine runs the inner looping machine $\abs{\varphi}-1$ times for each of the output qubits $m=1\dots \abs{\varphi}-1$. Each time it runs, the inner looping machine applies the desired cascade of $cU$ operations between the $m$'th and $n$'th qubits for all $m < n \leq \abs{\varphi}$. The outer looping machine then applies a final Hadamard operation to the $m$'th qubit, before moving onto the next qubit. One can therefore see that the overall effect is to apply the inverse quantum Fourier Transform circuit to the $\abs{\varphi}$ output qubits.

\subsection{Reset Stage}
\label{sec:reset_stage}
To reset all the auxiliary tracks, we dovetail the outer looping TM with a sequence of TMs that uncompute all the configurations we previously prepared on these tracks.

The first of these TMs changes the $ct$ left on the mark track over the final two qubits to $\#\#$, and changes the final configuration $1;2^{\abs{\varphi}-1}$ of the inner looping and $cU$ looping tracks to the blank configuration. (Note that $2^{\abs{\varphi}-1}$ is the configuration consisting of a leading string of 0's followed by a single 1 on the final output qubit, so is straightforward to reset without any arithmetic computations.)

To reset the final configuration $N;N-\abs{\varphi}+1$ of the input and count tracks, we start by decrementing the count track to $N;N-\abs{\varphi}$ (using $\TM{Dec}$). Recall that we redefined the location of the starting cell before implementing the inverse QFT machine, so that the starting cell became the cell containing the LSB of $\varphi$. We also shifted the copies of the input and count tracks accordingly (\cref{sec:LSB_stage}). We now step the head back to the original starting cell (which is easily located as we left a $\#$ written there on the quantum track, cf.\ \cref{sec:LSB_stage}). We can then run the reversal $\TM{Shift}^\dg$ (constructed using \cref{reversal}) of the machine used in \cref{sec:LSB_stage}, to shift the copies of the input and count tracks used in the QFT stage back left again to the original starting cell, and run the reversal $\TM{Copy}^\dg$ of the copy machines to erase the copies of the input and count tracks.

To erase the original count track, which contains $N-\abs{\varphi}$, we must run the reversal $\TM{Count}^\dg$ that we constructed in \cref{sec:LSB_stage}. (The first $N-\abs{\varphi}$ qubits remain in the $\ket{0}$ state and the LSB qubit in the $\ket{1}$ state, so $\TM{Count}^\dg$ will indeed uncompute $N-\abs{\varphi}$.) Finally, to erase the original input track, we first use the $\TM{Shift}^\dg$ machine from \cref{shift_right} on the \emph{quantum} track, to shift the entire final state of the output qubits left. Note that this will work because the first qubit (the ancilla qubit of the control-$U$ stage) was set to $\#$ in \cref{sec:LSB_stage}. We can then run the reversal $\TM{UtoB}^\dg$ of the unary-to-binary machine from \cref{unary_to_binary}, acting on the quantum track and input track, to uncompute the binary conversion of the unary input encoded in the length of the qubit state. Note that the  $\TM{UtoB}$ machine only checks whether the symbols on the unary track are $\#$ or non-$\#$. So the fact that the configuration of the quantum track is no longer necessarily a string of $N$ $\ket{1}$'s does not matter; all that matters is that it is a string of $N$ non-$\ket{\#}$'s.

The end result of all this is to reset all the auxiliary tracks to the blank state, leaving the output of the inverse QFT circuit stored in the first $N$ cells of the quantum track.

\subsection{Analysis}
By dovetailing together the preparation stage (\cref{sec:preparation_stage}), control-$U_\varphi$ stage (\cref{sec:cU_stage}), inverse-QFT stage (\cref{sec:QFT_stage}) and reset stage (\cref{sec:reset_stage}), we have succeeded in constructing a family of well-formed, normal form, unidirectional quantum Turing Machines $P_n$ that behave properly on input $N\ge \abs{\varphi}$ written in unary, and implement the quantum phase estimation algorithm on $N$ qubits for phase $\varphi$. By construction, $P_n$ satisfies \cref{phase-estimation_QTM:fixed_size} of \cref{phase-estimation_QTM}.

Now, we know~\cite{Nielsen+Chuang} that the quantum phase estimation algorithm outputs the exact binary fraction expansion of the phase, as long as we run it on enough qubits to store the entire binary decimal expansion of the phase. Thus for $N\geq\abs{\varphi} = \abs{n}$, the $P_n$ writes out the binary decimal expansion of $\varphi$ to $N$ bits (in little-endian order, padding with 0's as necessary). We choose $\varphi$ to be the rational number whose binary decimal expansion contains the digits of $n$ in reverse order after the decimal, where $n$ indexes $P_n$. So our QTM $P_n$ implements the computation claimed in \cref{phase-estimation_QTM:output} of \cref{phase-estimation_QTM}.

We were careful throughout to keep tight control on the space requirements of the reversible and quantum TMs that we used to construct $P_n$. In fact, none of them used more than $N+3$ space. Finally, all steps of the computation take time $O(\poly N)$, except the $cU_\varphi^k$ and $cU_\varphi^k$ computations, which take time $O(2^N)$. Thus the overall run-time is $O(\poly(N) 2^N)$. Thus $P_n$ fulfils the space and time requirements in  \cref{phase-estimation_QTM:output} of \cref{phase-estimation_QTM}.

The last thing we must check in order to finish the proof of \cref{phase-estimation_QTM} is \cref{phase-estimation_QTM:amplitudes}. But the partial transition function defined so far for $P_n$ satisfies \cref{phase-estimation_QTM:amplitudes}. It is trivial to check that one can complete this to a full transition function, whilst still satisfying \cref{phase-estimation_QTM:amplitudes}. Indeed, by \cref{QTM_transitions} the problem is equivalent to completing an orthonormal set of vectors with coefficients in
\begin{equation}
  \mathcal{S} = \left\{
    0,1,\pm \frac{1}{\sqrt{2}},e^{i\pi\varphi}, e^{i\pi 2^{-\abs{\varphi}}}
  \right\}
\end{equation}
to a full orthonormal basis with coefficients in $\mathcal{S}$. Since any normalised vector with coefficients in $\mathcal{S}$ must be proportional to a vector in the canonical basis $\{e_j\}_j$ or of the form $\frac{1}{\sqrt{2}}(e_j\pm e_k)$, $j\not = k$, the result is immediate. This completes the proof of \cref{phase-estimation_QTM}.

Note that the QTM we have constructed only implements the computation correctly when supplied a valid upper-bound on the number of digits in the binary fraction expansion of the phase. If the input is not actually a correct upper bound, we make no claim about the behaviour of the QTM. (This will be significant later, when we come in \cref{sec:put-together} to bound ground state energies of a Hamiltonian constructed out of this QTM.)


\clearpage
\section{Encoding QTMs in local Hamiltonians}
\label{sec:local-Hamiltonians}
In \cref{sec:phase-estimation}, we obtained a QTM that implements exact quantum phase estimation deterministically, using a number of time-steps exponential in the bit-length of the phase being estimated. This gives us a way to generate any desired input to feed into a universal (reversible) Turing Machine (UTM). Dovetailing the phase estimation QTM and the UTM, we therefore obtain a family of QTMs of constant size (fixed alphabet size and number of internal states) for which the halting problem is undecidable on blank input tape.

In this section we will show how for any QTM, one can construct an associated Hamiltonian whose ground state encodes in a concrete way the evolution of the QTM. In particular, for the family of QTMs just described, it encodes the associated halting problem.
We start by distilling the main ideas behind the construction from a historical perspective, to serve as a roadmap for this section.

The idea of encoding quantum computations into ground states of Hamiltonians goes back to Feynman~\cite{Feynman}, and was developed into its modern form by Kitaev~\cite{Kitaev_book}. The key object is the computational history state, which encodes the entire history of the computation in superposition. As these history states will play an important role in this section, we define then rigorously here:

\begin{definition}[Computational history state]\label{def:history-state}
  A \keyword{computational history state} $\ket[CQ]{\psi} \in \HS_C\ox\HS_Q$ is a state of the form
  \begin{equation}
    \ket[CQ]{\psi} = \frac{1}{\sqrt{\dim \HS_C}} \sum_{t=0}^{d_C}\ket{t}\ket{\psi_t},
  \end{equation}
  where $\{\ket{t}\}$ is an orthonormal basis for $\HS_C$, and $\ket{\psi_t} = \prod_{i=1}^t U_i\ket{\psi_0}$ for some initial state $\ket{\psi_0}\in\HS_Q$ and set of unitaries $U_i\in\cB(\HS_Q)$.

  $\HS_C$ is called the \keyword{clock register} and $\HS_Q$ is called the \keyword{computational register}. If $U_t$ is the unitary transformation corresponding the $t$'th step of a quantum computation, then $\ket{\psi_t}$ is the state of the computation after $t$ steps. We say that the history state $\ket{\psi}$ \keyword{encodes} the evolution of this quantum computation.
\end{definition}

As Feynman realised, given a quantum circuit defined by the sequence of unitaries $\{U_t\}_{1\le t\le T}$, it is straightforward to write down a Hamiltonian for which the ground space is precisely spanned by the set of computational history states (\cref{def:history-state}), where one allows any initial state $\ket[Q]{\psi_0}$.

In order to obtain such a Hamiltonian, one first focuses on the clock register $\HS_C$. Let $d_C=T$, and look for a Hamiltonian that has as ground state the superposition of all clock time steps:
\begin{equation} \label{eq:superposition-clock-intro}
  \ket[C]{\psi} = \frac{1}{\sqrt{T}} \sum_{t=1}^{T}\ket{t}.
\end{equation}
This can be enforced by a standard hopping Hamiltonian:
\begin{equation}\label{eq:hoping-hamilt-clock}
 H_C=\sum_{t=1}^T \left(\ket{t}-\ket{t-1}\right) \left(\bra{t}-\bra{t-1}\right).
\end{equation}

To obtain the history state of \cref{def:history-state} from \cref{eq:superposition-clock-intro}, one applies the controlled unitary $\mathcal{U}_{CQ}=\sum_{t} \ketbra{t}{t} \otimes U_{[1\cdots t]}$ on $\ket[C]{\psi}\ket[Q]{\psi_0}$, where $U_{[1\cdots t]}$ is a shorthand notation for
$U_t U_{t-1}\cdots U_1$. At the level of the ground state, this boils down to considering the rotated Hamiltonian $H_{CQ}=\mathcal{U}_{CQ} (H_C\otimes \1_Q)\mathcal{U}_{CQ}^\dagger$, which is nothing more than
\begin{equation}\label{eq:Feynman-Hamiltonian}
  H_{CQ}= \sum_{t=1}^T \left(\ketbra{t}{t} \otimes \1 + \ketbra{t-1}{t-1} \otimes \1 - \ketbra{t}{t-1}\otimes U_t- \ketbra{t-1}{t}\otimes U_t^\dagger \right).
\end{equation}

The difficulty arises when one wants to implement this construction with a Hamiltonian that is (a)~local, (b)~one-dimensional and (c)~translational invariant. Issues (a)--(c) were addressed consecutively by \textcite{Kitaev_book}, \textcite{AGIK} and \textcite{Gottesman-Irani}.

As we already illustrated, the key ingredient to construct the \keyword{history state Hamiltonian} \cref{eq:Feynman-Hamiltonian} is the clock. An important insight from \cite{AGIK} was to define a 1D local Hamiltonian whose ground state encodes in superposition a series of sweeps of particular product states, which act as as an oscillator of a clock. (Cf.~\cref{fig:clock-oscillator}, which represents the particular clock oscillation we will use later).

\begin{figure}
  \begin{align*}
    &\sixcellshoriz{$\leftend$}{\,\arrR\,}{\blankR}{\blankR}{\blankR}{\rightend}\\[.5em]
    &\sixcellshoriz{\leftend}{\blankL}{\,\arrR\,}{\blankR}{\blankR}{\rightend}\\[.5em]
    &\sixcellshoriz{\leftend}{\blankL}{\blankL}{\,\arrR\,}{\blankR}{\rightend}\\[.5em]
    &\sixcellshoriz{\leftend}{\blankL}{\blankL}{\blankL}{\,\arrR\,}{\rightend}\\[.5em]
    &\sixcellshoriz{\leftend}{\blankL}{\blankL}{\blankL}{\,\arrL\,}{\rightend}\\[.5em]
    &\sixcellshoriz{\leftend}{\blankL}{\blankL}{\,\arrL\,}{\blankR}{\rightend}\\[.5em]
    &\sixcellshoriz{\leftend}{\blankL}{\,\arrL\,}{\blankR}{\blankR}{\rightend}\\[.5em]
    &\sixcellshoriz{\leftend}{\,\arrL\,}{\blankR}{\blankR}{\blankR}{\rightend}\\[.5em]
    &\sixcellshoriz{\leftend}{\,\arrR\,}{\blankR}{\blankR}{\blankR}{\rightend}
  \end{align*}
  \caption{Example of a clock oscillation cycle.}
  \label{fig:clock-oscillator}
\end{figure}

Now, apart from hopping terms in \cref{eq:hoping-hamilt-clock} (called \keyword{evolution} or \keyword{transition rule} terms) enforcing the evolution of the clock, one now needs to also include \keyword{penalty terms} that restrict the type of configurations that can appear in the ground state, by giving a positive energy to configurations that do not appear in the clock oscillation cycle. In the example of \cref{fig:clock-oscillator}, for instance, giving a positive energy to configurations of the form \;$\twocellshoriz{\blankR}{\arrL\,}$\; guarantees that no $\blankR$ can appear to the right of a $\arrL$ in any component of the ground state.

The type of constraints that can be enforced by penalty terms is characterised by the notion of a regular expressions, which for example (informally) allows one the express constraints of the form ``write one $\leftend$, then as many $\blankL$'s as you want, then write a $\,\arrR\,$ or a $\,\arrL\,$, then as many $\blankR$ as you want, and then a $\rightend$''. (Note that this regular expression captures precisely the states present in the oscillation cycle of \cref{fig:clock-oscillator}.) The formal definition of regular expression is recalled in \cref{def:regular-expression}, and the key result from \cite{Gottesman-Irani} connecting regular expressions with penalty terms is stated in \cref{regexp}. An important point here is that we will be able to guarantee by other means that the sites at the two ends are always in states $\leftend$ and $\rightend$, respectively. (We will enforce this using tiling constructions, described in detail \cref{sec:quasi-periodic}; the same condition is enforced in a different way in \cite{Gottesman-Irani}.) We can therefore restrict our analysis just to these ``bracketed'' configurations.

The idea of a local clock was developed further by \textcite{Gottesman-Irani} to obtain a translationally invariant clock where, analogous to a real clock, after each oscillation cycle (representing one ``tick'' of the clock), a counter gets incremented by one. In \cite{Gottesman-Irani} the clock counter counts in unary, and is represented on a separate track of the TM. For a finite chain of length $L$, this means the clock can tick for $O(L)$ steps before the counter cannot be incremented further and the clock stops ticking. In our case, we need the clock to tick for an exponentially larger number of time steps. Hence we need to count in binary rather than unary. In fact, it will be convenient to have it count in an even larger (but fixed) base $\zeta$. To achieve this, we will have to generalise the binary counter machine constructed in the previous section. This is done in \cref{sec:counter_TM_construction}.

There is an important subtlety. When one takes into account the full clock (oscillation and counter), penalty terms and transition rule terms are not enough to characterise the desired clock evolution. An example of this will be given below, immediately after the statement of \cref{counter_TM_evolve_to_illegal}.
The key idea used to solve this problem in \cite{AGIK}, also used crucially later in \cite{Gottesman-Irani}, was a more general way to guarantee for those particular constructions that all other product state configurations (except for those giving the desired clock evolution) have larger energy. Note that if a state does not follow the correct evolution, it picks up a positive energy contribution from the transition rule terms. Some of the undesired configurations pick up a positive energy contribution directly from the penalty terms, as before. There are other undesirable configurations that are not detected directly by the penalty terms. But these remaining undesirable configurations all evolve under the transition rule terms into configurations that \emph{do} pick up such an energy penalty. \textcite{AGIK} call this the \keyword{Clairvoyance Lemma}. This result may or may not hold for any given construction; the clock and Hamiltonian construction must be designed such that this property holds and the Clairvoyance Lemma proven for the specific construction at hand.

\textcite{AGIK} proved their Clairvoyance Lemma by dividing the Hilbert space of the clock into subsets of product states that are invariant under the clock evolution, i.e.~into orbits of the action of the clock Hamiltonian. If the above property holds -- that undesired states get penalties or evolve to states with penalties -- all orbits pick up energy penalties except the one corresponding to a valid clock evolution. This in turn ensures that the ground states of the full Hamiltonian are precisely the set of history states (\cref{def:history-state}). Moreover, the analysis gives explicit lower bounds for the energies of the excited states.

We will prove a version of the Clairvoyance Lemma for our construction following the same approach. We first introduce penalty terms and transition rule terms for the clock Hamiltonian. We prove (building on the key result \cref{counter_TM_evolve_to_illegal}) that all product states that do not appear in the desired clock evolution are either pick up an energy penalty from the penalty terms, or evolve to a configuration that picks up such a penalty. This allows us to prove that the clock Hamiltonian consisting of the sum over these penalty and transition rule terms has as its unique ground state the uniform superposition over the product states appearing in the desired clock evolution. This is the key ingredient in our version of the Clairvoyance Lemma (\cref{sec:local-Hamiltonian_analysis}), which shows that for the complete Hamiltonian that includes both the clock and computational registers $C$ and $Q$, the unique ground state is the computational history state of \cref{def:history-state}, and all other eigenstates have energies lower bounded by $\Omega\left(\frac{1}{T^3}\right)$.

The final insight to solve the issue (c)~related to the translational invariance property comes from \textcite{Gottesman-Irani}. There, it was shown how all the ideas of \cite{AGIK} to address the local 1D case, could be  implemented keeping the Hamiltonian translationally invariant. Indeed, this section can be seen as a self-contained analysis of a local Hamiltonian construction that generalises that of \textcite{Gottesman-Irani} for the particular purposes of this paper.

Apart from adapting the clock construction to a translational invariant setting, one of the main changes with respect to \cite{AGIK} (or to the original Feynman-Kitaev construction outlined above) in order to make the construction translationally invariant was to switch from quantum circuits to QTMs. That is, after each clock tick, a step of a QTM evolution is executed on a different track. In this case, the computational register $Q$ in \cref{def:history-state} corresponds to the whole QTM configuration (internal state, head location, and tape configuration), encoded across a number of separate tracks. However, in order to apply our version of the Clairvoyance Lemma, one needs to guarantee that the $U_t$ terms appearing in \cref{eq:Feynman-Hamiltonian} are indeed unitaries, and not partial isometries, even in the case in which the evolution corresponds to a QTM rather than a quantum circuit. In our case, this is guaranteed by restricting to a special class of QTMs, first identified in \textcite{Bernstein-Vazirani}, that has especially nice properties (well-formed~\cref{def:well-formed}, normal-form~\cref{def:normal-form} and unidirectional~\cref{def:unidirectional}, to which we also add proper~\cref{def:proper_QTM}) but that, at the same time, is general enough for our purposes.\footnote{Indeed, it is general enough to admit universal QTMs~\cite{Bernstein-Vazirani}.}
We give a more detailed overview of how these properties, together with a modification to the encoded computation, allow us to prove our version of the key Clairvoyance Lemma, in \cref{sec:local-Hamiltonian_analysis}.

The following is the main result of this section:

\begin{theorem}[Local Hamiltonian QTM encoding]
  \label{QTM_in_local_Hamiltonian} \hfill\\
  Let $\C^d = \C^C\otimes\C^Q$ be the local Hilbert space of a 1\nobreakdash-dimensional chain of length $L$, with special marker states $\ket{\sleftend},\ket{\srightend}$. Denote the orthogonal complement of $\linspan(\ket{\sleftend}, \ket{\srightend})$ in $\C^d$ by $\C^{d-2}$.

  For any well-formed unidirectional Quantum Turing Machine $M = (\Sigma,Q,\delta)$ and any constant $K>0$ \footnote{$K$ will be needed to deal with the extra space required by the QTM defined in \cref{phase-estimation_QTM}. It will be set to $K=3$ in \cref{sec:put-together} where all components of the proof are combined.}, we can construct a two-body interaction $h \in \cB(\C^d\ox\C^d)$ such that the 1\nobreakdash-dimensional, translationally-invariant, nearest-neighbour Hamiltonian $H(L)=\sum_{i=1}^{L-1} h^{(i,i+1)} \in \cB(\HS(L))$ on the chain of length $L\geq K+3$ has the following properties:
  \begin{enumerate}
  \item \label[part]{QTM_in_local_Hamiltonian:local_dim}%
    $d$ depends only on the alphabet size and number of internal states of $M$.

  \item \label[part]{QTM_in_local_Hamiltonian:FF}%
    $h \geq 0$, and the overall Hamiltonian $H(L)$ is frustration-free for all $L$.

  \item \label[part]{QTM_in_local_Hamiltonian:gs}%
    Denote $\HS(L-2) := (\C^{d-2})^{\ox L-2}$ and define $\Sbr=\linspan(\ket{\sleftend}) \ox \HS(L-2) \ox \linspan(\ket{\srightend})\subset\HS$. Then the unique ground state of $H(L)|_\Sbr$ is a computational history state encoding the evolution of $M$ on input corresponding to the unary representation of the number $L-K-3$, running on a finite tape segment of length $L-3$.

    \setcounter{tmpcounter}{\value{enumi}}
  \end{enumerate}
  Moreover, if $M$ is proper on input given by the unary representation of $L-K-3$, then:
  \begin{enumerate}\setcounter{enumi}{\value{tmpcounter}}
  \item \label[part]{QTM_in_local_Hamiltonian:halt}%
    The computational history state always encodes $\Omega(\zeta^L)$ time-steps, where $\zeta=\abs{\Sigma\times Q}$ \footnote{This choice of $\zeta$ guarantees that the QTM $M$ has enough time to halt (if it is going to halt) within this number of time-steps in the finite tape segment available.}. If $M$ halts in fewer than the number of encoded time steps, exactly one $\ket{\psi_t}$ has support on a state $\ket{\top}$ that encodes a halting state of the QTM. The remaining time steps of the evolution encoded in the history state leave $M$'s tape unaltered, and have zero overlap with $\ket{\top}$.

  \item \label[part]{QTM_in_local_Hamiltonian:out-of-tape}%
    If $M$ runs out of tape within a time $T$ less than the number of encoded time steps (i.e.\ in time-step $T+1$ it would move its head before the starting cell or beyond cell $L-3$), the computational history state only encodes the evolution of $M$ up to time $T$. The remaining steps of the evolution encoded in the computational history state leave $M$'s tape unaltered.

  \item \label[part]{QTM_in_local_Hamiltonian:explicit-form}%
    Finally, if $M$ satisfies \cref{phase-estimation_QTM:amplitudes} of \cref{phase-estimation_QTM}, then $h$ has the following form
    \begin{equation}
      h = A + (e^{i\pi\varphi}B + e^{i\pi2^{-\abs{\varphi}}} C + \hc),
    \end{equation}
    with $B,C \in \cB(\C^d\ox\C^d)$ independent of $n$ and with coefficients in $\Z$, and $A\in \cB(\C^d\ox\C^d)$ Hermitian independent of $n$ and with coefficients in $\Z+\frac{1}{\sqrt{2}}\Z$.
  \end{enumerate}
\end{theorem}

Though we will not need it for our purposes, it is not difficult to prove that the next-highest eigenstate of $H(L)|_\Sbr$ has energy $\Omega(1/T^3)$, where $T$ is the total number of time-steps encoded in the computational history state.

\subsection{Preliminaries}\label{sec:local_Hamiltonian_preliminaries}
As our construction draws heavily on \cite{Gottesman-Irani}, we will follow their notation and terminology, which we summarise here. We divide the chain into multiple tracks:
\begin{center}
  \begin{tabular}{|l|clc|r|}
    \hline
    $\leftend$ & $\cdots$ & Track 1: Clock oscillator & $\cdots$ & $\rightend$ \\
    \hline
    $\leftend$ & $\cdots$ & Track 2: Counter TM head and state & $\cdots$ & $\rightend$\\
    \hline
    $\leftend$ & $\cdots$ & Track 3: Counter TM tape & $\cdots$ & $\rightend$\\
    \hline
    $\leftend$ & $\cdots$ & Track 4: QTM head and state & $\cdots$ & $\rightend$\\
    \hline
    $\leftend$ & $\cdots$ & Track 5: QTM tape & $\cdots$ & $\rightend$\\
    \hline
    $\leftend$ & $\cdots$ & Track 6: Time-wasting tape & $\cdots$ & $\rightend$\\
    \hline
  \end{tabular}
\end{center}

\begin{figure}
  \begin{centering}
    \includegraphics[scale=.5]{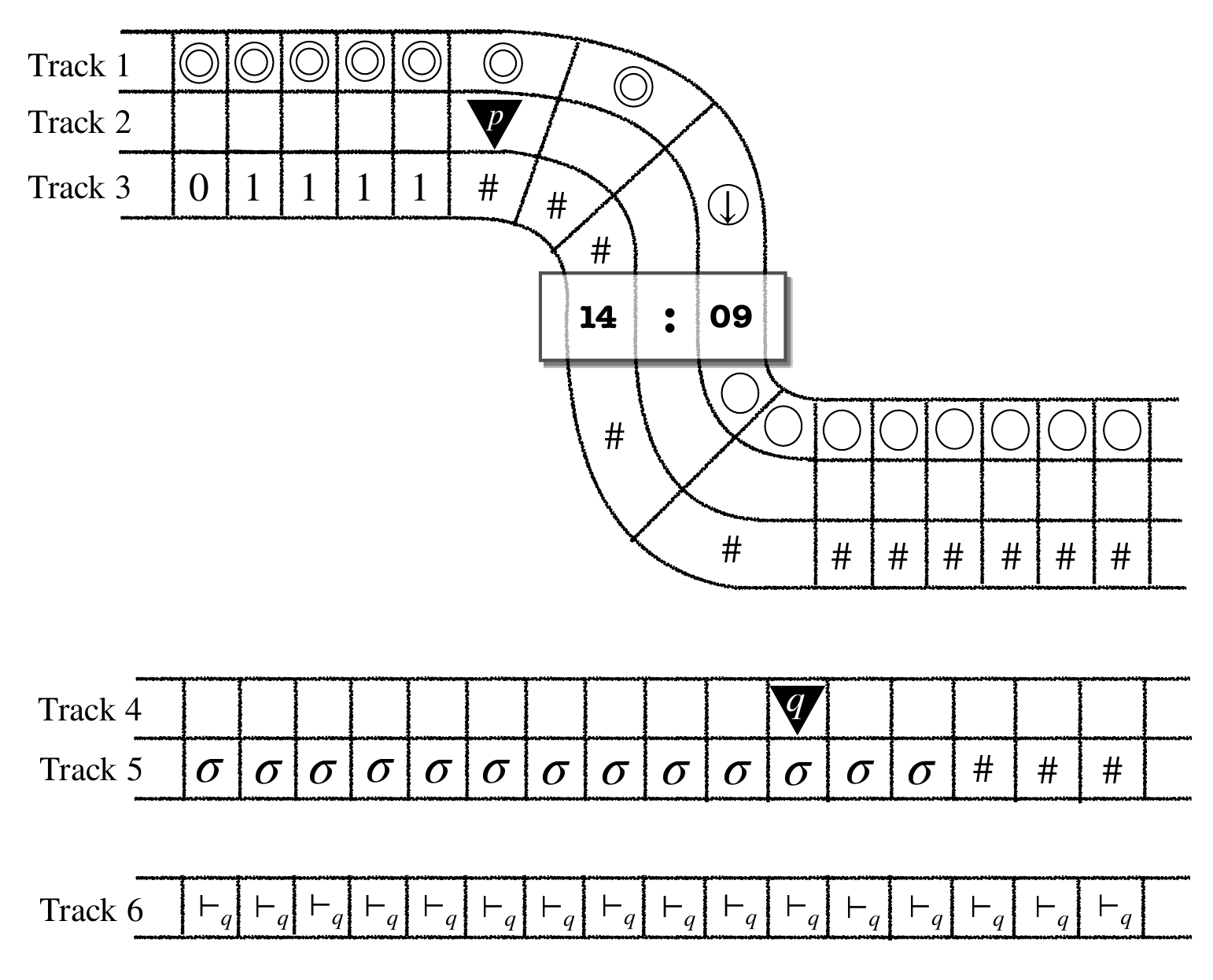}
  \end{centering}
  \caption{Cartoon illustration of the role of the different tracks. Track~1 acts as the clock's ``second hand'', repeating one full cycle for every clock increment. Track~3 acts as the ``minute hand'', getting incremented by one for every cycle of Track~1. Track~2 stores the counter Turing Machine head and internal state machinery needed to implement this incrementing. These clock tracks drive the Quantum Turing Machine in Tracks~4 and~5, Track~4 storing the QTM head and internal state, Track~5 storing the QTM tape qudits. If the QTM ever halts or runs out of space on Track~5, it switches to writing garbage on the ``time-wasting'' trap stored on Track~6 to use up the remaining time on the clock.}
  \label{fig:track-cartoon}
\end{figure}

Tracks 1-3 correspond to the clock register $C$ and tracks 4-6 to the computational register $Q$ in \cref{def:history-state} (see \cref{fig:track-cartoon} for an illustration).

The local Hilbert space at each site is the tensor product of the local Hilbert space of each of the six tracks $\HS = \otimes_{i=1}^6\HS_{i}$, where
\begin{equation}\label{eq:local_Hilbert_space}
  \begin{split}
    \HS_{1}
    := &\linspan\bigl\{\ket{s}\bigr\} \oplus
    \bigl\{\ket{\sleftend},\ket{\srightend}\bigr\}\\
    &s \in \{\arrRzero,\arrLzero,\arrLzeroi,\arrRone,\arrLone,
    \blankL,\blankR\} \text{ for } i\in\{1,\dots,K\},\\
    \HS_{2}
    := &\linspan\bigl\{\ket{p}\bigr\} \oplus
    \bigl\{\ket{\sleftend},\ket{\srightend}\bigr\}\\
    &p \in P'\cup\{\blankL,\blankR\}
    \text{ where } P:=P_L\cup P_N\cup P_R \text{ and } P':=P\cup P'_R,\\
    \HS_{3}
    := &\linspan\bigl\{\ket{\tau}\bigr\} \oplus
    \bigl\{\ket{\sleftend},\ket{\srightend}\bigr\}\\
    &\tau \in \Xi
    \text{ where } \Xi := \{\vdash,\#, 0,\ldots, \zeta-1\}
    \text{ with } \zeta = \abs{\Sigma\times Q},\\
    \HS_{4}
    := &\linspan\bigl\{\ket{q}\bigr\} \oplus
    \bigl\{\ket{\sleftend},\ket{\srightend}\bigr\}\\
    &q \in Q'\cup P\cup P'_L\cup \{r_x\}\cup \{\blankL,\blankR\}
    \text{ where } Q:=Q_L\cup Q_N\cup Q_R, Q':=Q\cup Q'_L\\
    &\text{ and } x \in Q,\\
    \HS_{5}
    := &\linspan\bigl\{\ket{\sigma}\bigr\} \oplus
    \bigl\{\ket{\sleftend},\ket{\srightend}\bigr\}\\
    &\sigma \in \Sigma,\\
    \HS_{6}
    := &\linspan\bigl\{\ket{\gamma}\bigr\} \oplus
    \bigl\{\ket{\sleftend},\ket{\srightend}\bigr\}\\
    &\gamma \in \Xi \cup \{\vdash_q\} \text{ where } q\in Q'.
  \end{split}
\end{equation}
$\Sigma$ is the tape alphabet of our given QTM $M$. $\Xi$ is the alphabet of the counter TM. $P_L,P_N,P_R$ are the sets of internal states of the counter TM that can be entered by the TM head moving left, not moving, or moving right (respectively). The states $p'\in P_R'$ duplicate the states $p\in P_R$, and $p'\in P'_L$ duplicate those in $P_L$. Similarly for the internal states $Q_L,Q_N,Q_R$ of the QTM, with $q'\in Q'_L$ duplicating the states $q\in Q_L$. Recall that by the unidirection property of reversible and quantum Turing Machines (\cref{reversible_transitions,QTM_transitions}), these sets are disjoint.

The $\blankL$ and $\blankR$ Track~2 and~4 symbols are used for cells that do not currently hold the head.\footnote{There are two blank symbols because we will need different symbols to the left and right of the head, in order to enforce the constraint that there is only one head on the track.} The role of the Track~1 states is described in \cref{sec:clock-oscillator}. That of the Track~6 ``time-wasting'' tape, as well as the role of the additional Track~4 $P\cup P'_L$ states, will become apparent in \cref{sec:unitarity}.

The marker states $\ket{\sleftend}$, $\ket{\srightend}$ appearing in  \cref{QTM_in_local_Hamiltonian} will just be the states $\ket{\sleftend} = \bigotimes_{i=1}^6\ket{\sleftend}_{\text{track}\; i}$ and  $\ket{\srightend} = \bigotimes_{i=1}^6\ket{\srightend}_{\text{track}\; i}$. When a subset of tracks $T$ is clear from the context, we will also write $\ket{\sleftend}, \ket{\srightend}$ to denote $\bigotimes_{i\in T}\ket{\sleftend}_{{\rm track}\; i}$ and $\bigotimes_{i\in T}\ket{\srightend}_{{\rm track}\; i}$, respectively.

This set of states defines a standard basis for the single-site Hilbert space $\HS$. The product states over this single-site basis then give a basis for the Hilbert space $\HS^{\ox L}$ of the chain. We call these the \keyword{standard basis states}.

We will sometimes use regular expressions to specify subsets of standard basis states. A regular expression denotes a (possibly infinite) subset of finite-length strings over a finite alphabet. Equivalently, it can be thought of as a pattern that matches all the strings in the subset and no others. Regular expressions, and the regular expression notation we will use, can be defined inductively in the standard way:
\begin{definition}[Regular expression]\label{def:regular-expression}
  Given a finite alphabet $\Sigma$, let $\Sigma^*$ denote the set of all finite-length strings of symbols from that alphabet.
  The following are regular expressions:
  \begin{itemize}
  \item The empty regular expression $\epsilon$ denotes the set containing only the empty string.
  \item Any symbol $x\in\Sigma$ denotes the singleton set $\{x\}$ containing only the string $x$.
  \end{itemize}
  Given two regular expressions $R_1,R_2$, the following are also regular expressions:
  \begin{itemize}
  \item $R_1 R_2$ denotes the set $\{xy : x\in R_1, y\in S_2\}$ of all concatenations of strings matching $R$ and $S$.
  \item The alternation $R_1|R_2$ denotes the set $R_1\cup R_2$ of strings matching either $R_1$ or $R_2$ (or both).
  \item The Kleene-star $R^*$ denotes the Kleene closure of $R$, i.e.\ the smallest set of strings $S$ such that $\epsilon, R\in S$ and $S$ is closed under concatenation.
  \item For $x_1,\dots,x_n\in\Sigma$, square brackets $[x_1,x_2,\dots,x_n] := x_1|x_2|\dots|x_n$ are a shorthand notation for alternations of symbols.
  \end{itemize}
  Parentheses $(\dots)$ are used to group subexpressions, and have the highest precedence. Square brackets have higher precedence than the Kleene-star operator, which has higher precedence than the alternation operator. Concatenation has the lowest precedence.
\end{definition}

\begin{definition}[Bracketed state]\label{def:bracketed}
  We call standard basis states that match the regular expression $\leftend\wcsymbol{}^*\rightend$ \keyword{bracketed states} (where $\wcsymbol$ stands for any single-site state other than $\leftend,\rightend$).
  We denote by $\Sbr$ the subspace spanned by the bracketed states.
\end{definition}

We will often denote configurations of multiple tracks by writing them vertically, e.g.
\begin{alignat*}{3}
  \twocellsvert{\arrL\,}{p_1}\;, &\qquad&
  \twocellshoriz{\blankL}{\arrL\,}\;, &\qquad&
  \sixcellsvert{\arrL\,}{\blankR}{\blankL}{p_1}{0}{1}
\end{alignat*}
denote, respectively, a single-site configuration of two tracks, a configuration on two neighbouring sites of one track, and a configuration on two neighbouring sites of three tracks. Since we will restrict throughout to the subspace $\Sbr$ of bracketed states, the $\leftend$ and $\rightend$ states will only ever appear at the left and right ends of the chain, across all the tracks. As a shorthand, we will therefore denote configurations involving $\leftend$ and $\rightend$ with symbols stretching vertically across all tracks, e.g.
\begin{alignat*}{2}
  \threecellsL{\arrL\,}{p_1}\;, &\qquad&
  \fourcellsR{\blankR}{p_1}{\;1\;\;}\;.
\end{alignat*}
We will also sometimes need to specify configurations in which the tracks are in any configuration \emph{other than} $\leftend,\rightend$. We denote this with a $\neg\leftend$ or $\neg\rightend$ symbol stretching vertically across all tracks, e.g.
\begin{alignat*}{2}
  \threecellsnotL{\arrL\,}{p_1}\;, &\qquad&
  \fourcellsnotR{\blankR}{p_1}{\;1\;\;}\;.
\end{alignat*}

As usual in such constructions, the two-body Hamiltonian will contain two types of terms: \keyword{penalty terms} and \keyword{transition rule terms}. Penalty terms have the form $\proj{ab}$ where $a,b$ are standard basis states. This adds a positive energy contribution to any configuration containing $a$ to the left of $b$. We call $ab$ an \keyword{illegal pair}, and denote a penalty term $\proj{ab}$ in the Hamiltonian by its corresponding illegal pair. We will sometimes also make use of single-site illegal states $a$. (Note that, even if we don't allow ourselves to use single-site penalty terms, single-site illegal states are easily implemented in terms of illegal pairs, by adding penalty terms $\proj{ax}$ and $\proj{xa}$ for all pairs $ax$ and $xa$ in which the single-site state appears.)
\begin{definition}[Legal and illegal states]\label{def:illegal}
  We call a standard basis state \keyword{legal} if it does not contain any illegal pairs, and \keyword{illegal} otherwise.
\end{definition}

By using single-site illegal states one can enforce that, on all legal states, the marker $\ket{\sleftend}$ (resp.\ $\ket{\srightend}$) can only ever appear simultaneously on all tracks. In this case, if one restricts to global bracketed states, one also gets bracketed states in each of the tracks individually. The single-site illegal states enforcing this are summarised in \cref{tbl:end-marker_illegal_states}.

{
  \centering
  \setlength{\extrarowheight}{4pt}
  \setlength{\tabcolsep}{12pt}
  \begin{longtable}{>{$}c<{$}}
    \caption{Single-site illegal states enforcing that end-marker states must appear across all tracks simultaneously.}
    \label{tbl:end-marker_illegal_states}\\*
    \nobreakhline
    \text{Track~i--j single-site illegal states, for all pairs $i\neq j$}\\*
    \nobreakhline\\*
    \twocellsvert{\leftend}{\neg\leftend}\,,\;\;
    \twocellsvert{\neg\leftend}{\leftend}\,,\;\;
    \twocellsvert{\rightend}{\neg\rightend}\,,\;\;
    \twocellsvert{\neg\rightend}{\rightend} \\*\\*
    \hline
  \end{longtable}
}

We will make repeated use of Lemma~5.2 from \cite{Gottesman-Irani}, which lets us use penalty terms to restrict to configurations matching a regular expression:
\begin{lemma}[Regexps]\label{regexp}
  For any regular expression over the standard basis of single-site states in which each state appears at most once, we can use penalty terms to ensure that any legal standard basis state for the system is a substring of a string in the regular set.
\end{lemma}
All the regular expressions we will use start with a left-bracket state $\leftend$ and end with a right-bracket $\rightend$. So whenever we apply this \namecref{regexp}, the only substrings of the regular set that are \emph{bracketed states} are in fact complete strings in the regular set, not substrings.

Transition rule terms have the form $\tfrac{1}{2}(\ket{\psi}-\ket{\varphi})(\bra{\psi}-\bra{\varphi})$, where $\ket{\psi},\ket{\varphi}$ are states on the same pair of adjacent sites. We will often take $\ket{\psi} = \ket{ab}$ and $\ket{\varphi} = \ket{cd}$ to be standard basis states. This forces any zero-energy eigenstate with amplitude on a configuration containing $ab$ to also have equal amplitude on the configuration in which that $ab$ is replaced by $cd$. Conversely, if a zero-energy eigenstate has amplitude on a configuration containing $cd$, it must also have equal amplitude on the configuration in which that $cd$ is replaced by $ab$. Thus we can think of these terms as implementing the transition rules $ab \rightarrow cd$, which we arbitrarily call the ``forwards'' direction, and the corresponding ``backwards'' transition $cd\rightarrow ab$. Following the notation in \cite{Gottesman-Irani}, we will denote transition rule Hamiltonian terms by their associated forwards transitions $ab\rightarrow cd$ or, more generally, $\ket{\psi} \rightarrow \ket{\varphi}$.

When a transition rule acts diagonally on a subset of the tracks, we will sometimes consider the \keyword{restriction} of the transition rule to those tracks. That is, for a transition rule with the general form $\ket[T]{ab}\ket[T^c]{\psi} \rightarrow \ket[T]{cd}\ket[T^c]{\varphi}$, where $T$ is some subset of the tracks, the restriction of the rule to $T$ is given by $\ket[T]{ab} \rightarrow \ket[T]{cd}$.

When we specify a Hamiltonian term only on a subset of the tracks, we implicitly mean that it acts as identity on the remaining (unspecified) tracks. We will assume throughout this section that the ground state subspace of the Hamiltonian is restricted to the subspace $\Sbr$ of bracketed states, with $\leftend$ at the left end of the chain and $\rightend$ at the right. (We show how to enforce this later, in \cref{sec:put-together}.)

\subsection{Clock Oscillator}\label{sec:clock-oscillator}
Every clock -- even one unusual enough to be encoded in superposition into the ground state of a quantum many-body Hamiltonian! -- needs some form of \keyword{oscillator} that oscillates with a fixed period (such as a pendulum), and a \keyword{counter} that is incremented after each complete oscillation of the oscillator (such as the hands on a clock). For the counter, we will use a reversible counter Turing Machine, described in \cref{sec:counter_TM}. This section describes the Track~1 clock oscillator.

The legal configurations of Track~1 are states matching the regular expression $\leftend\blankL{}^*[\arrRzero,\arrLzero,\arrLzeroi,\arrRone,\arrLone]\blankR{}^*\rightend$. By \cref{regexp}, we can enforce this using penalty terms. Independent of the configuration of any other track, a $\arrR$ arrow (either $\arrRzero$ or $\arrRone$) sweeps right along the chain until it reaches the end, whereupon it turns around and becomes a $\arrL$. (When a $\arrRzero$ reaches the end and turns around, the label on the arrow steps through the sequence $(0),(1),\dots,(K)$. This is used later on to correctly initialise the other tracks. The restriction to $L\ge K+3$ in \cref{QTM_in_local_Hamiltonian} ensures that there is enough space for such sequences of transitions.) This $\arrL$ arrow sweeps left until it reaches the beginning of the chain, at which point it turns around and becomes a $\arrR$ again. We call this entire sequence an \keyword{oscillator cycle}. (One complete cycle is illustrated in \cref{fig:clock-oscillator} for a chain of length six.) On a chain of length $L$, one complete oscillator cycle takes $2(L-2)$ steps. The following transition rules on Track~1 enforce this:
\begin{equation}
  \begin{alignedat}{3}
    \twocellshoriz{\arrRzero}{\blankR} \longrightarrow
    \twocellshoriz{\blankL}{\arrRzero}\,,
    &\qquad&
    \twocellshoriz{\arrRzero}{\rightend\,} \longrightarrow
    \twocellshoriz{\arrLzeroz}{\rightend\,}\,,
    \\[1em]
    \twocellshoriz{\blankL}{\arrLzeroi} \longrightarrow
    \twocellshoriz{\arrLzeroipo}{\blankR}\,,
    &\qquad&
    \twocellshoriz{\blankL}{\arrLzeroK} \longrightarrow
    \twocellshoriz{\arrLzero}{\blankR}\,,
    \\[1em]
    \twocellshoriz{\blankL}{\arrLzero} \longrightarrow
    \twocellshoriz{\arrLzero}{\blankR}\,,
    &\qquad&
    \twocellshoriz{\arrRone}{\blankR} \longrightarrow
    \twocellshoriz{\blankL}{\arrRone}\,,
    \\[1em]
    \twocellshoriz{\arrRone}{\rightend\,} \longrightarrow
    \twocellshoriz{\arrLone}{\rightend\,}\,,
    &\qquad&
    \twocellshoriz{\blankL}{\arrLone} \longrightarrow
    \twocellshoriz{\arrLone}{\blankR}\,.
  \end{alignedat}
\end{equation}


When a left-moving $\arrL$ arrow returns to the beginning of the chain, it turns around and becomes a right-moving $\arrR$ arrow again. However, the label $0$ on the arrow may change to a $1$, depending on the Track~2 state. The $\arrLzero$ arrow transitions to $\arrRone$ if the Track~2 state is $p_{\alpha}$ (the initial state of the counter TM). Whereas $\arrLone$ always transitions to $\arrRone$ \emph{unless} the Track~2 state is $p_{\alpha}$. The following transition rules implement this:
\begin{alignat}{2}
  \threecellsL{\arrLzero}{p_{\alpha}} \longrightarrow
    \threecellsL{\arrRone}{p_{\alpha}}\,,
  &\qquad&
  \threecellsL{\arrLone}{\neg p_{\alpha}} \longrightarrow
    \threecellsL{\arrRone}{\neg p_{\alpha}}\,,
\end{alignat}
where $\neg p_{\alpha}$ here denotes any Track~2 state \emph{other than} $p_{\alpha}$.

\subsection{Initialisation sweep}\label{sec:initialisation_sweep}
During the initial sweep of the Track~1 $\arrRzero$ from left to right and back, Track~1 contains a $\arrRzero$ or $\arrLzero$. We call this the \keyword{initialisation sweep}, and we want to use it to force Tracks~2 and~3 to be in the initial configurations
\begin{equation}\label{eq:track23_initial_configurations}
  \leftend\,p_0\,\blankR{}^*\rightend \text{ and } \leftend\#{}^*\rightend,
\end{equation}
respectively. We do this by adding illegal pairs to forbid Track~2 from being anything other than $\blankR$ when a $\arrRzero$ is over it (except at the beginning of the chain):
\begin{equation}
  \threecellsnotL{\arrRzero}{\neg\blankR}\,,
\end{equation}
and similarly to forbid Track~3 from being anything other than $\#$ when a $\arrRzero$ is over it (except at the beginning of the chain):
\begin{equation}
  \threecellsnotL{\arrRzero}{\neg\#}\,,
\end{equation}
where $\neg\wcsymbol$ again denotes anything \emph{other than} the state $\wcsymbol$.

At the beginning of the chain, we use illegal pairs to forbid Track~2 from being anything other than $p_{\alpha}$ when $\arrRzero$ is over it. We also forbid it from being $p_{\alpha}$ when a $\arrLone$ is at the beginning of the chain:
\begin{equation}
  \threecellsL{\arrRzero}{\neg p_{\alpha}}\,,
  \qquad
  \threecellsL{\arrLone}{p_{\alpha}}\,.
\end{equation}
Thus any standard basis state containing a $\arrRzero$, $\arrLzero$ or $\arrLzeroi$ on Track~1 that matches the regular expressions \cref{eq:track23_initial_configurations}, and is \emph{not} in the initial configuration on Tracks~2 and~3, will evolve (either forwards or backwards) under the oscillator transition rules into an illegal configuration, in at most $L$ steps.

If Tracks~2 and~3 are in their initial configurations, then a standard basis state containing a $\arrRzero$, $\arrLzero$ or $\arrLzeroi$ will evolve backwards until it reaches the initial configuration $\ket{\phi_0}$, the standard basis state for Tracks~1 to~3 which has the form:
\begin{equation}
  {%
    \setlength{\extrarowheight}{0pt}%
    \setlength{\tabcolsep}{0pt}%
    \arraycolsep=2pt%
    \def\arraystretch{1.2}%
    \begin{array}{|c|c|c|c|}
      \hline
      \multirow{3}{*}{$\leftend$}
      & \arrRzero & \blankR\quad\cdots\quad\blankR & \multirow{3}{*}{$\rightend$}\\
      \cline{2-3}
      & p_{\alpha} & \blankR\quad\cdots\quad\blankR & \\
      \cline{2-3}
      & \vdash & \#\quad\cdots\quad\# & \\
      \hline
    \end{array}\;.
  }
\end{equation}

\subsection{Clock Counter}\label{sec:counter_TM}

\subsubsection{Counter TM construction}\label{sec:counter_TM_construction}
We construct the counter TM using a generalisation of the binary incrementing machine $\TM{Inc}$ from \cref{incrementer}, which will increment integers written in base-$\zeta$ instead of in binary. Recall that $\zeta= \abs{\Sigma\times Q}$. Constructing a base-$\zeta$ version of $\TM{Inc}$ is largely a straightforward extension of \cref{incrementer}.

Ensuring reversibility is similar to the binary case. The TM state $q_1$ scans along the digits of the little-endian input, changing all $\zeta-1$'s to $0$'s as it goes, until it finds the first digit different from $\zeta-1$. It then increments this by one, to add the carry from all the previous $\zeta-1$s digits. If no such digit exists, a $1$ will be appended to the end of the input. The difference with the binary case is that now there are three cases (or computational paths) that we must distinguish: the case in which a digit other than $0$ is incremented by one, and two different cases in which a $1$ can be written to the tape: incrementing $0$, or writing a new $1$ in place of the first $\#$ symbol after the input. These three computational paths correspond respectively to TM states $p_2$, $p_2'$ and $p_2''$. The latter two paths are first merged into state $p_3'$. The former transitions instead into state $p_3$. Then, in a second stage, the two remaining paths are merged into state $p_4$ which, as in the binary case, is the state that then returns the head to the starting cell and halts.

In pseudocode:

\begin{algorithm}[H]
  \caption{$\TM{INC}_\zeta$}\label{alg:Inc_zeta}
  \begin{algorithmic}[1]
    \Input{string $s\in\{0,\dots,\zeta-1,\vdash,\#\}^*$ consisting of the symbol $\vdash$ followed by\\
       \phantom{\bf input:} the little-endian base-$\zeta$ representation of $n\in\N$.} \Comment{$q_0$}
    \State{step Right}  \Comment{$p_1$}
    \While{read a $\zeta-1$}
      \State{write a 0}
      \State{step Right}
    \EndWhile
    \If{read $0<i<\zeta-2$}
          \State{write $i+1$}
      \State{step Right} \Comment{$p_2$}
      \State{step Left} \Comment{$p_3$}
    \Else
    \If{read $0$}
       \State{write $1$}
       \State{step right} \Comment{$p_2'$}
     \ElsIf{read $\#$}
      \State{write a $1$}
      \State{step Right} \Comment{$p_2''$}
    \EndIf
    \State{step Left}\Comment{$p_3'$}
    \EndIf
    \State{step Left} \Comment{$p_4$}
    \While{not read a $\vdash$}
      \State{step Left}
    \EndWhile
    \State{halt} \Comment{$p_f$}
   \end{algorithmic}
\end{algorithm}

The following well-formed, normal-form, partial transition function implements a reversible TM $\TM{INC}_\zeta$ that increments any little-endian base-$\zeta$ number written on the tape with a leading ``start-of-tape'' symbol $\vdash$, and returns the head to the starting cell without ever moving the head before the starting cell:
\begin{equation}\label{eq:base-zeta_incrementer}
\begin{array}{l|cccccc}
        & \vdash & \# & 0 & 1 & i & \zeta -1\\
  \hline
  p_0   & (\vdash,p_1,R) \\
  p_1   & & (1,p_2'',R) & (1,p_2',R) & (2,p_2,R) & (i+1,p_2,R) & (0,p_1,R) \\
  p_2   & & (\#,p_3,L) & (0,p_3,L) & (1,p_3,L) & (i,p_3,L) & (\zeta-1,p_3,L) \\
  p_2'  & & & (0,p_3',L) & (1,p_3',L) & (i,p_3',L) & (\zeta-1,p_3',L) \\
  p_2'' & & (\#,p_3',L) \\
  p_3   & & & & & (i,p_4,L) & (\zeta-1,p_4,L) \\
  p_3'  & & & & (1,p_4,L) \\
  p_4   & (\vdash,p_f,N) & & (0,p_4,L) & \\
  p_f   & (\vdash,p_0,R) & (\#,p_0,R) & (0,p_0,R) & (1,p_0,R) & (i,p_0,R) & (\zeta-1,p_0,R) \\
  \multicolumn{4}{l}{\text{\footnotesize $\forall i\in\{2,\dots,\zeta-2\}$} \rule{0em}{1.5em}}
\end{array}
\end{equation}

We now modify this machine slightly in order to make it loop forever. However, we must ensure that the loop reenters the initial state $p_0$ of $\TM{Inc}_\zeta$ reversibly. To accomplish this, we introduce new initial and final states $p_\alpha$ and $p_\omega$, and modify the $p_0$ and $p_f$ transitions such that the TM reversibly transitions back into the state $p_0$ instead of halting, exploiting the fact that the symbol in the second tape cell will be blank in the first iteration, and non-blank thereafter.

In pseudocode:

\begin{algorithm}[H]
  \caption{\TM{Base-$\zeta$ Counter}}
  \begin{algorithmic}[1]
    \Input{$\vdash$ in first cell, followed by a blank tape.} \Comment{$p_\alpha$}
    \State{step Right} \Comment{$p_{-1}$}
    \State{step Left} \Comment{$p_0$}
    \Loop
   \Call{$\TM{INC}_\zeta$} \Comment{$p_f$}
   \State{step Right} \Comment{$p_f'$}
   \State{step Left}\Comment{$p_0$}
    \EndLoop
  \end{algorithmic}
\end{algorithm}

The following set of transition rules accomplishes this:
\begin{equation}\label{eq:base-zeta_counter}
\begin{array}{l|cccccc}
         & \vdash & \# & 0 & 1 & i & \zeta-1 \\
  \hline
  p_\alpha & (\vdash,p_{-1},R) \\
  p_{-1}   & & (\#,p_0,L) \\
  p_0     & (\vdash,p_1,R) \\
  p_1     & & (1,p_2'',R) & (1,p_2',R) & (2,p_2,R) & (i+1,p_2,R) & (0,p_1,R) \\
  p_2     & & (\#,p_3,L) & (0,p_3,L) & (1,p_3,L) & (i,p_3,L) & (\zeta-1,p_3,L) \\
  p_2'    & & & (0,p_3',L) & (1,p_3',L) & (i,p_3',L) & (\zeta-1,p_3',L) \\
  p_2''   & & (\#,p_3',L) \\
  p_3     & & & & & (i,p_4,L) & (\zeta-1,p_4,L) \\
  p_3'    & & & & (1,p_4,L) \\
  p_4     & (\vdash,p_f,N) & & (0,p_4,L) & \\
  p_f     & (\vdash,p_f',R) \\
  p_f'    & & & (0,p_0,L) & (1,p_0,L) & (i,p_0,L) & (\zeta-1,p_0,L) \\
  p_\omega & (\vdash,p_\alpha,R) & (\#,p_\alpha,R) & (0,p_\alpha,R) & (1,p_\alpha,R)
          & (i,p_\alpha,R) & (\zeta-1,p_\alpha,R) \\
  \multicolumn{4}{l}{\text{\footnotesize $\forall i\in\{2,\dots,\zeta-2\}$} \rule{0em}{1.5em}}
\end{array}
\end{equation}
These transition rules implement a reversible base-$\zeta$ counter TM. When started from the tape configuration consisting of a $\vdash$ symbol in the first cell followed by the all-blank tape (representing the number~0), this TM will loop indefinitely, incrementing the number written on the tape by~1 in each complete iteration. For brevity, we will refer to this particular initial tape configuration as the \keyword{standard input} to the counter TM tape.

We now declare certain configurations of the counter TM to be ``illegal''. We will choose these illegal configurations to be such that they are never entered by the counter TM started from the standard input. Later on, when we come to encode the counter TM in a local Hamiltonian, we will use penalty terms to give energy penalties to all the configurations we declare ``illegal'' here, so that the corresponding standard basis states of the spin chain are indeed illegal in the sense of \cref{def:illegal}. For now, however,  the ``illegal configurations'' simply define a particular subset $\mathcal{I}$ of the complete set of counter TM head and tape configurations.

Anticipating the later use of penalty terms, when defining illegal configurations we will often make use of the illegal pair notation introduced above, where the top row denotes the counter TM head position and internal state $p$, and the bottom row denotes the tape symbols $ab$ in the section of tape near the head:
\begin{equation}
  \fourcells{p}{\wc}{\;a\;\;}{\;b\;\;}\qquad \text { or } \qquad
  \fourcells{\wc}{p}{\;a\;\;}{\;b\;\;}\;.
\end{equation}
A single row always denotes a section of tape:
\begin{equation}
  \twocellshoriz{\;a\;\;}{\;b\;\;}\;.
\end{equation}
Certain configurations that we declare to be illegal will specify that the counter TM head is or is not located at the starting cell. Again anticipating later use of penalty terms, we introduce the following notation. We denote the starting cell by placing a $\leftend$ symbol on the left. We denote tape cells \emph{other than} the starting cell with $\neg\leftend$ on the left. For illustration, two examples of this would be:
\begin{equation}
  \threecellsL{\;p\;\;}{a}\;, \quad \threecellsnotL{\;p\;\;}{a}\;.
\end{equation}

Note that the counter TM started from the standard input never moves its head before the starting cell, never reenters its initial state $p_\alpha$, and never enters its final state $p_\omega$. We therefore declare any counter TM configuration which is in state $p_\alpha$ with the head anywhere other than the starting cell to be illegal:
\begin{equation}\label{eq:illegal_p0}
  \threecellsnotL{p_\alpha}{\wc}\;.
\end{equation}
We also declare any configuration in state $p_\omega$ to be illegal:\footnote{We could alternatively remove state $p_\omega$ from the construction entirely. We have elected to retain it, since inclusion of a halting state is necessary to conform to the definition of a TM.}
\begin{equation}\label{eq:illegal_pf}
  \fourcells{p_\omega}{\wc}{\wc}{\wc}\;.
\end{equation}
Similarly, we declare any configuration in state $p_\alpha$ with the head adjacent to a non-blank symbol to be illegal, i.e.\ any configuration containing the illegal pair
\begin{equation}\label{eq:illegal_p0+non-blank}
  \fourcells{p_\alpha}{\wc}{\wc}{\neg\#}\;.
\end{equation}

The transition table in \cref{eq:base-zeta_counter} is partial; some transitions are never used, so were not defined. For each $(p,\tau)$ for which no transition rule is defined in \cref{eq:base-zeta_counter}, we declare configurations in which the counter TM is in state $p$ and the head is reading $\tau$ to be illegal, i.e.\ any configuration containing the illegal pair
\begin{equation}\label{eq:illegal_forward_transitions}
  \twocellsvert{\;p\;\;}{\tau}\;.
\end{equation}
For each configuration $(p_L,\tau,L)$, $(p_R,\tau,R)$ or $(p_N,\tau,N)$ which is not \emph{entered} by any transition rule in \cref{eq:base-zeta_counter}, we declare the corresponding TM head and tape configurations to be illegal, i.e.\ any configurations containing one of the following illegal pairs:
\begin{equation}\label{eq:illegal_backwards_transitions}
  \fourcells{\wc}{\;p_R\;\;}{\tau}{\wc}\;,\qquad
  \fourcells{\;p_L\;\;}{\wc}{\wc}{\tau}\;,\qquad
  \fourcells{\;p_N\;\;}{\wc}{\tau}{\wc}\;.
\end{equation}

Note that during the evolution of the counter TM started from a blank tape, the head is never more than one cell to the right of a non-blank tape symbol. In fact, the only moment at which it is over a blank symbol at all is when the incrementer TM needs to carry a digit, and moves the head to the next blank cell in order to write the carry. We therefore declare configurations in which the head is more than one cell to the right of a non-blank tape symbol to be illegal, i.e.\ all configurations containing the illegal pair
\begin{equation}\label{eq:illegal_head+blank}
  \fourcells{\wc}{\;p\;\;}{\#}{\#}\;.
\end{equation}

Note also that, when started from the standard input, the counter TM never modifies the $\vdash$ in the first cell, never writes a $\vdash$ anywhere \emph{other than} the first cell, and never creates an embedded blank symbol on the tape. We therefore define all tape configurations \emph{without} a $\vdash$ in the starting cell, all tape configuration with a $\vdash$ anywhere other than the first cell, and all tape configurations with an embedded blank, to be illegal.
These illegal TM configurations correspond exactly to tape configurations matching the regular expression
\begin{equation}\label{eq:track3_regexp}
  \vdash\,[0,\dots,\zeta-1]^*\,\#^*.
\end{equation}

The little-endian numbers written by the counter TM are never padded with leading 0's, so the tape never has a~0 to the left of a~$\#$. We therefore define this combination to be illegal, i.e.\ all tape configurations containing the illegal pair
\begin{equation}\label{eq:illegal_0+blank}
  \twocellshoriz{\;0\;\;}{\;\#\;\;}\;.
\end{equation}

Later on, we will only be interested in configurations of the first $L$ cells of the tape, and will stop the counter TM just before it tries to apply a transition rule that maps out of this portion of tape. Therefore, a~0 will never appear in the $L$'th tape cell, and we declare configurations containing a~0 in this cell to be illegal:
\begin{equation}\label{eq:illegal_0+rightend}
  \twocellshoriz{\;0\;\;}{\rightend}
\end{equation}

\begin{lemma}[Evolve-to-illegal]\label{counter_TM_evolve_to_illegal}
  Any counter TM configuration that is reached starting from the standard input is legal. All other configurations with the head in position $r \in [0,L]$ are either illegal, or evolve forwards or backwards to an illegal configuration within $O(L)$ time steps in such a way that the head never leaves the portion of tape $[0,L]$.
\end{lemma}

Before delving into the technical proof, let us briefly explain why the set of configurations reached by the counter TM during its computation (or rather, the complement of this set) cannot be characterised by illegal pairs. And why, therefore, one needs to exclude many of the invalid configurations by showing that they necessarily go on to evolve into an illegal configuration.

As an illustrative example, consider the valid configurations associated with state $p_1$ located in position $5$. These are of the form:
\begin{gather*}
  \begin{array}{|c|c|c|c|c|c|}
    \multicolumn{4}{c}{} & \multicolumn{1}{c}{p_1}\\
    \hline
    \vdash & 0 & 0& 0 & \tau & \cdots \\
    \hline
  \end{array}\\
\end{gather*}
with $\tau$ any element other than $\vdash$. Any attempt to guarantee a $0$ in position $3$ in this configuration by declaring extra illegal pairs (which we recall can only be defined between nearest neighbours) would contradict the fact that, for state $p_0$, all configurations of the form:
\begin{gather*}
  \begin{array}{|c|c|c|c|c|c|}
    \multicolumn{1}{c}{p_0} & \multicolumn{4}{c}{}\\
    \hline
    \vdash & i_1 & i_2 & i_3 & i_4 & \cdots \\
    \hline
  \end{array}\\
\end{gather*}
with $i_j\in \{0,\ldots \zeta-1\}$ are valid configurations reached by the counter TM during its evolution.

\begin{proof}[\cref{counter_TM_evolve_to_illegal}]
  The first part of the \namecref{counter_TM_evolve_to_illegal} is true by construction, since the partial transition rules of \cref{eq:base-zeta_counter} implement the counter TM without using any of the undefined transitions, and the tape configurations defined to be illegal never occur when the counter TM is started from the standard input.

  Recall that the alphabet of the counter TM is $\Xi = \{\vdash,\#\}\cup\{0,\dots,\zeta-1\}$. Any tape configuration that does not start with a $\vdash$, or which contains a $\vdash$ anywhere other than the first cell, or which contains an embedded blank, or contains a~0 to the left of a~$\#$, or with a $0$ in the $L$-th cell is illegal. So legal tape configurations match the regular expression
  \begin{equation}\label{eq:legal-tape}
    \vdash(\neg \#)^*\neg[0,\#]\#^*.
  \end{equation}
  Configurations in which the head is more than one cell away from a non-blank symbol are illegal by \cref{eq:illegal_head+blank}, so all legal configurations have the head located in, or immediately adjacent to, the non-blank portion of the tape. All other configurations are illegal.

  We divide the legal configurations into separate cases, according to the internal state of the counter TM:

  \paragraph{State $p_\alpha$:}
  All configurations in state $p_\alpha$ are illegal by \cref{eq:illegal_p0}, except those with the head in the starting cell. The latter are illegal by \cref{eq:illegal_p0+non-blank} unless the tape cell to the right of the head is blank. The only such tape configuration which is legal is the standard input. Thus the only legal configuration in state $p_\alpha$ is the initial configuration of the counter TM started from the standard input.

  \paragraph{State $p_{-1}$:}
  There are no transitions out of $(p_{-1},x)$ where $x\neq\#$ in \cref{eq:base-zeta_counter}, so by \cref{eq:illegal_forward_transitions} the only legal $p_{-1}$ configurations have their head over a $\#$. Moreover, there is only one transition into $p_{-1}$ in \cref{eq:base-zeta_counter}. Hence, evolving any such configuration one step \emph{backwards} according to \cref{eq:base-zeta_counter} either enters a configuration that is illegal due to \cref{eq:illegal_forward_transitions}, or steps the head left and transitions to $p_\alpha$. But we have shown that the only legal $p_\alpha$ configuration is the standard input. Thus any configuration in state $p_{-1}$ not reachable starting from the standard input will evolve backwards to an illegal configuration in one time step.

  \paragraph{State $p_1$:}
  Every TM configuration of the form
  \begin{gather*}
    \begin{array}{|c|c|c|c|c|c|}
      \multicolumn{4}{c}{} & \multicolumn{1}{c}{p_1}\\
      \hline
      \vdash & 0 & \cdots & 0 & \tau & \cdots \\
      \hline
      \multicolumn{2}{c}{} &
        \multicolumn{4}{l}{\raisebox{.5em}{$\underbrace{\mspace{80mu}}_n$}}
    \end{array}\\
    \tau\in \Xi/_{\{\vdash\}},\;\;
    n\ge 0\;\;
  \end{gather*}
  with the rest of the tape compatible with \cref{eq:legal-tape} is reached when evolving the counter TM from the standard input; this configuration is produced whilst the counter TM is incrementing the number with little-endian base-$\zeta$ expansion
  \begin{equation*}
    \underbrace{(\zeta-1)\dots(\zeta-1)}_n \tau\cdots.
  \end{equation*}

  The remaining legal configurations have at least one tape symbol $x$ to the left of the head that is neither 0, nor $\#$, nor $\vdash$:
  \begin{gather*}
    \begin{array}{|c|c|c|c|c|c|c|c|c|c|}
      \multicolumn{8}{c}{} & \multicolumn{1}{c}{p_1}\\
      \hline
      \vdash & i_1 & \cdots & i_m & x & 0 & \cdots & 0 & \tau & \cdots \\
      \hline
      \multicolumn{5}{c}{} &
        \multicolumn{3}{l}{\raisebox{.5em}{$\underbrace{\mspace{70mu}}_n$}}
    \end{array}\\
    x \in \Xi/_{\{\vdash,\#,0\}},\;\;
    i_1,\dots,i_m \in \Xi/_{\{\vdash,\#\}},\;\;
    \tau\in \Xi/_{\{\vdash\}},\;\;
      m,n\ge 0.
  \end{gather*}
  Evolving any such configuration \emph{backwards} according to \cref{eq:base-zeta_counter} either enters into an illegal configuration due to \cref{eq:illegal_forward_transitions}, or rewinds the head left replacing 0's with $\zeta-1$'s, eventually reaching the configuration with the head immediately to the right of the $x$:
  \begin{gather*}
    \begin{array}{|c|c|c|c|c|c|c|c|c|c|}
      \multicolumn{5}{c}{} & \multicolumn{1}{c}{p_1}\\
      \hline
      \vdash & i_1 & \cdots & i_m & x & \zeta-1 & \cdots & \zeta-1 & \tau &\cdots \\
      \hline
      \multicolumn{5}{c}{} &
        \multicolumn{3}{l}{\raisebox{.5em}{$\underbrace{\mspace{130mu}}_n$}}
    \end{array}\\
    x \in \Xi/_{\{\vdash,\#,0,1\}},\;\;
    i_1,\dots,i_m \in \Xi/_{\{\vdash,\#\}},\;\;
    \tau\in \Xi/_{\{\vdash,\#\}},\;\;
       m,n\ge 0.
  \end{gather*}
  The state $p_1$ can only be entered by moving right. But the configuration $(p_1,x,R)$ is not entered by any transition rule in \cref{eq:base-zeta_counter}, so this configuration is illegal by \cref{eq:illegal_backwards_transitions}.

  \paragraph{State $p_2$:}
  The state $p_2$ is entered when incrementing any digit other than~0, $\#$ or~$\zeta-1$, so every TM configuration of the form
  \begin{gather*}
    \begin{array}{|c|c|c|c|c|c|c|}
      \multicolumn{5}{c}{} & \multicolumn{1}{c}{p_2}\\
      \hline
      \vdash & 0 & \cdots & 0 & x & \tau & \cdots  \\
      \hline
      \multicolumn{1}{c}{} &
        \multicolumn{4}{l}{\raisebox{.5em}{$\underbrace{\mspace{80mu}}_n$}}
    \end{array}\\
    x \in \Xi/_{\{\vdash,\#,0,1\}},\;\;
    \tau \in \Xi/_{\{\vdash\}},\;\;
    n\ge 0.
  \end{gather*}
  with the rest of the tape compatible with \cref{eq:legal-tape} is reached when evolving the counter TM from the standard input. This configuration is produced whilst the counter TM is incrementing the number with little-endian base-$\zeta$ expansion
  \begin{equation*}
    \underbrace{(\zeta-1)\dots(\zeta-1)}_n (x-1) \tau\cdots
  \end{equation*}

  The state $p_2$ can only be entered by moving right. Configurations $(p_2,x,R)$ where $x\in\{\vdash,\#,0,1\}$ are not entered by any transition rule in \cref{eq:base-zeta_counter}, so these configurations are illegal by \cref{eq:illegal_backwards_transitions}.

  The remaining legal configurations have at least one tape symbol $y$ to the left of the head that is neither~0, $\#$, nor~$\vdash$:
  \begin{gather*}
    \begin{array}{|c|c|c|c|c|c|c|c|c|c|c|}
      \multicolumn{9}{c}{} & \multicolumn{1}{c}{p_2}\\
      \hline
      \vdash & i_1 & \dots & i_m & y & 0 & \cdots & 0 & x & \tau &  \cdots \\
      \hline
      \multicolumn{5}{c}{} &
        \multicolumn{4}{l}{\raisebox{.5em}{$\underbrace{\mspace{80mu}}_n$}}
    \end{array}\\
    x \in \Xi/_{\{\vdash,\#,0,1\}},\;\;
    y \in \Xi/_{\{\vdash,\#,0\}},\;\;
    \tau \in \Xi/_{\{\vdash\}},\;\;
    i_1,\ldots, i_m \in \Xi/_{\{\vdash,\#\}},\;\;
    m,n \ge 0.
  \end{gather*}
  Evolving any such configuration \emph{backwards} one time step according to \cref{eq:base-zeta_counter} either produces an illegal configuration due to \cref{eq:illegal_forward_transitions}, or produces the configuration:
  \begin{gather*}
    \begin{array}{|c|c|c|c|c|c|c|c|c|c|c|}
      \multicolumn{8}{c}{} & \multicolumn{1}{c}{p_1}\\
      \hline
      \vdash & i_1 & \dots & i_m & y & 0 & \cdots & 0 & x-1 & \tau & \cdots \\
      \hline
      \multicolumn{5}{c}{} &
        \multicolumn{4}{l}{\raisebox{.5em}{$\underbrace{\mspace{80mu}}_n$}}
    \end{array}\\
    x \in \Xi/_{\{\vdash,\#,0,1\}},\;\;
    y \in \Xi/_{\{\vdash,\#,0\}},\;\;
    \tau \in \Xi/_{\{\vdash\}},\;\;
    i_1,\ldots, i_m \in \Xi/_{\{\vdash,\#\}},\;\;
    m,n \ge 0.
  \end{gather*}
  But we have already shown above that all such $p_1$ configurations evolve backwards in time to an illegal configuration.

  \paragraph{States $p_2',p_2''$:}
  The argument for states $p_2'$ and $p_2''$ is very similar to that for $p_2$, except that the legal value of $x$ is now $1$.

  \paragraph{State $p_3$:}
  The state $p_3$ is entered by stepping left from $p_2$, so every TM configuration of the form
  \begin{gather*}
    \begin{array}{|c|c|c|c|c|c|}
      \multicolumn{4}{c}{} & \multicolumn{1}{c}{p_3}\\
      \hline
      \vdash & 0 & \cdots & 0 & \tau &  \cdots \\
      \hline
      \multicolumn{1}{c}{} &
        \multicolumn{4}{l}{\raisebox{.5em}{$\underbrace{\mspace{80mu}}_n$}}
    \end{array}\\
    \tau \in \Xi/_{\{\vdash,\#,0,1\}},\;\;
    n \ge 0.
  \end{gather*}
  with the rest of the tape compatible with \cref{eq:legal-tape} is reached when evolving the counter TM from the standard input. This configuration is produced whilst the counter TM is incrementing the number with little-endian base-$\zeta$ expansion
  \begin{equation*}
    \underbrace{(\zeta-1)\dots(\zeta-1)}_n(\tau-1) \cdots
  \end{equation*}

  There is no transition out of $(p_3,x)$ where $\tau\in\{\vdash,\#,0,1\}$ in \cref{eq:base-zeta_counter}, so all such configurations are illegal by \cref{eq:illegal_forward_transitions}. The remaining legal configurations have at least one tape symbol $x$ to the left of the head that is neither~0, $\#$, nor~$\vdash$:
  \begin{gather*}
    \begin{array}{|c|c|c|c|c|c|c|c|c|c|}
      \multicolumn{8}{c}{} & \multicolumn{1}{c}{p_3}\\
      \hline
      \vdash & i_1 & \dots & i_m & x & 0 & \cdots & 0 & \tau &  \cdots \\
      \hline
      \multicolumn{5}{c}{} &
        \multicolumn{4}{l}{\raisebox{.5em}{$\underbrace{\mspace{80mu}}_n$}}
    \end{array}\\
    \tau \in \Xi/_{\{\vdash,\#,0,1\}}, x \in \Xi/_{\{\vdash,\#,0\}}\;\;
    i_1,\dots,i_m \in \Xi/_{\{\vdash,\#\}},\;\;
    m,n\ge 0.
  \end{gather*}
  Evolving any such configuration forwards one time step according to \cref{eq:base-zeta_counter} either produces an illegal configuration (for example if $n=0$), or produces the configuration:
  \begin{gather*}
    \begin{array}{|c|c|c|c|c|c|c|c|c|c|}
      \multicolumn{7}{c}{} & \multicolumn{1}{c}{p_4}\\
      \hline
      \vdash & i_1 & \dots & i_m & x & 0 & \cdots & 0 & \tau & \cdots \\
      \hline
      \multicolumn{5}{c}{} &
        \multicolumn{4}{l}{\raisebox{.5em}{$\underbrace{\mspace{80mu}}_n$}}
    \end{array}\\
    \tau \in \Xi/_{\{\vdash,\#,0,1\}}, x \in \Xi/_{\{\vdash,\#,0\}}\;\;
    i_1,\dots,i_m \in \Xi/_{\{\vdash,\#\}},\;\;
    m,n\ge 0.
  \end{gather*}
  We show below that any such $p_4$ configuration evolves to an illegal configuration.

  \paragraph{State $p_3'$:}
  The argument for state $p_3'$ is very similar to that for $p_3$, except that $\tau=1$ in this case.

  \paragraph{State $p_4$:}
  In a legal state $p_4$ configuration, the head must be over a $0$ or a $\vdash$, since there is no transition out of $(p_4,x)$ for $x\notin\{0,\vdash\}$. We first analyse the case in which it is over a $0$.

  Any TM configuration of the form
  \begin{gather*}
    \begin{array}{|c|c|c|c|c|}
      \multicolumn{3}{c}{} & \multicolumn{1}{c}{p_4}\\
      \hline
      \vdash & 0 & \cdots & 0 &   \cdots \\
      \hline
      \multicolumn{1}{c}{} &
        \multicolumn{4}{l}{\raisebox{.5em}{$\underbrace{\mspace{80mu}}_n$}}
    \end{array}\\
    n \ge 1
  \end{gather*}
  with the rest of the tape compatible with \cref{eq:legal-tape} is reached by evolving the counter TM from the standard input.

  The remaining legal configurations have at least one tape symbol $x$ to the left of the head that is neither~0, $\#$, nor~$\vdash$:
  \begin{gather*}
    \begin{array}{|c|c|c|c|c|c|c|c|c|}
      \multicolumn{7}{c}{} & \multicolumn{1}{c}{p_4}\\
      \hline
      \vdash & i_1 & \dots & i_m & x & 0 & \cdots & 0  &  \cdots \\
      \hline
      \multicolumn{5}{c}{} &
        \multicolumn{4}{l}{\raisebox{.5em}{$\underbrace{\mspace{80mu}}_n$}}
    \end{array}\\
    x \in \Xi/_{\{\vdash,\#,0\}},\;\;
    i_1,\dots,i_m \in \Xi/_{\{\vdash,\#\}},\;\;
    m\ge 0, n\ge 1
  \end{gather*}
  Evolving any such configuration according to \cref{eq:base-zeta_counter} steps the head left over the 0's until the head is over the $x$:
  \begin{gather*}
    \begin{array}{|c|c|c|c|c|c|c|c|c|}
      \multicolumn{4}{c}{} & \multicolumn{1}{c}{p_4}\\
      \hline
      \vdash & i_1 & \dots & i_m & x & 0 & \cdots & 0   & \cdots \\
      \hline
      \multicolumn{5}{c}{} &
        \multicolumn{4}{l}{\raisebox{.5em}{$\underbrace{\mspace{80mu}}_n$}}
    \end{array}\\
    x \in \Xi/_{\{\vdash,\#,0\}},\;\;
    i_1,\dots,i_m \in \Xi/_{\{\vdash,\#\}},\;\;
    m\ge 0, n\ge 1.
  \end{gather*}
  But there is no transition out of $(p_4,x)$ in \cref{eq:base-zeta_counter}, so this configuration is illegal by \cref{eq:illegal_forward_transitions}.

  We now analyse the case in which $p_4$ is over $\vdash$. Evolving one step backwards according to \cref{eq:base-zeta_counter} steps the head right and either enters an illegal configuration, transitions to state $p_4$ with the had over a $0$, or transitions to states $p_3$ or $p_3'$. In all three cases, we have already shown that all such configurations not reachable from the standard input evolve to illegal configurations.

  \paragraph{State $p_f$:}
  There are no transitions out of $(p_f,x)$ where $x\neq\vdash$ in \cref{eq:base-zeta_counter}, so the only legal $p_f$ configurations have the head over a $\vdash$. The latter can only appear in the starting cell in legal configurations. Evolving any such configuration backwards according to \cref{eq:base-zeta_counter} for one time step transitions to the state $p_4$, and we have already shown than all $p_4$ configurations not reachable from the standard input evolve to illegal configurations.

  \paragraph{State $p_f'$:}
  There are no transitions out of $(p_f',x)$ where $x\in\{\vdash,\#\}$ in \cref{eq:base-zeta_counter}, so the only legal $p_f'$ configurations have the head over a symbol in the set $\{0,\dots,\zeta-1\}$. Evolving any such configuration backwards according to \cref{eq:base-zeta_counter} for one time step transitions to the state $p_f$, and we have already shown than all $p_f$ configurations not reachable from the standard input evolve to illegal configurations.

  \paragraph{State $p_0$:}
  There are no transitions out of $(p_0,x)$ where $x\neq\vdash$ in \cref{eq:base-zeta_counter}, so the only legal $p_0$ configurations have the head over a $\vdash$. The latter can only appear at the beginning of the tape in legal configurations. Since the only transitions in \cref{eq:base-zeta_counter} into $p_0$ are from $p_1$ or $p_f'$, evolving any such configuration one step backwards according to \cref{eq:base-zeta_counter} either enters an illegal configuration, or steps the head right into state $p_{-1}$ or $p_f'$. In both cases we have already shown that configurations not reachable from the standard input evolve to illegal ones.

  \paragraph{State $p_\omega$:}
  The counter TM never enters the state $p_\omega$, and all such configurations are illegal.
\end{proof}

\subsubsection{Counter TM Hamiltonian}\label{sec:counter_TM_encoding}
We want all legal Track~2 configurations to contain a single state $p\in P'$ marking the location of the counter TM head, and blanks everywhere else. I.e.\ the Track~2 configuration should match a regular expression $\leftend\blankL{}^*p\,\blankR{}^*\rightend$ where $p\in P'$. By \cref{regexp}, we can enforce this using penalty terms. This Track~2 regular expression allows us to identify Tracks~2 and~3 with the configurations of the counter TM (with internal state $p$, its position being the head location, and Track~3 the tape). This will be explicitly or implicitly exploited in all that follows.

To encode the evolution of this counter TM in our local Hamiltonian, we encode its transition rules in transition terms. However, the transition rules will only apply when a $\arrRone$ arrow on Track~1 sweeps past the counter TM head encoded on Track~2, so that the counter TM advances exactly one step for each complete left-to-right sweep of the $\arrRone$ arrow. Since we are restricting to two-body interactions and the arrow is moving to the right, transitions in which the head moves left must be triggered when the $\arrRone$ is to the left of the TM head, in order to both move the arrow and update the head location. This way of implementing the left transitions means that
\begin{equation}\label{eq:left-move-illegal}
  \threecellsvert{\arrRone}{p}{\sigma}
\end{equation}
never occurs during the evolution of the counter TM if $\delta(p,\sigma) = (\tau,p_L,L)$. We add a penalty term to make such configurations illegal.

Transitions in which the head moves to the right or stays still will be triggered when the $\arrRone$ is on top of the head. In order to avoid two transitions in which the head moves right from being triggered during the same sweep of the $\arrRone$, we must split these latter transitions into two stages, in which we first transition into an auxiliary state $p'_R\in P'_R$, and then transition into the correct state $p_R\in P_R$ in the following time step (adapting the construction of \cite{Gottesman-Irani}). Thus, when the $\arrRone$ is on top of the head and the head will move right, we will update the tape, move the head, and step the $\arrRone$ to the right. But we transition into the auxiliary state $p'_R\in P'_R$ instead of $p_R\in P_R$. In the next step, with the $\arrRone$ now on top of the $p'_R$ state, we transition to the correct state $p_R$, and step the $\arrRone$ to the right once more so that it is no longer on top of the head. This way of carrying out the right-moving transitions means that $p'_R$ only ever appears below a $\arrRone$, so we add a penalty term to make all other $p'_R$ configurations illegal:
\begin{equation}\label{eq:p-prime-illegal}
  \twocellsvert{\neg\arrRone}{p'_R}
\end{equation}

For TM transition rules $\delta(p,\sigma) = (\tau,p_N,N)$, $\delta(p,\sigma) = (\tau,p_L,L)$ and $\delta(p,\sigma) = (\tau,p_R,R)$ (where $p\in P\backslash_{\{p_\alpha\}}$, $p_{L,N,R}\in P_{L,N,R}$ and $\sigma,\tau\in\Xi$), the following local transition terms on Tracks~1 to~3 implement the desired transformations (cells marked $\wc$ can be in any state, and are left unchanged by the transition):
\begin{subequations}
\begin{gather}
  \sixcellsvert{\arrRone}{\blankR}{p}{\blankR}{\sigma}{\wc}
    \longrightarrow \sixcellsvert{\blankL}{\arrRone}{p_N}{\blankR}{\tau}{\wc}\,,
  \qquad
  \fourcellsR{\arrRone}{p}{\sigma} \longrightarrow
    \fourcellsR{\arrLone}{p_N}{\tau}\,,
  \label{eq:counter_TM_transitions1}
  \\[1em]
  \sixcellsvert{\arrRone}{\blankR}{\blankL}{p}{\wc}{\sigma} \longrightarrow
    \sixcellsvert{\blankL}{\arrRone}{p_L}{\blankR}{\wc}{\tau}\,,
  \qquad
  \sixcellsvert{\arrRone}{\blankR}{p}{\blankR}{\sigma}{\wc} \longrightarrow
    \sixcellsvert{\blankL}{\arrRone}{\blankL}{p'_R}{\tau}{\wc}\,,
  \label{eq:counter_TM_transitions2}
  \\[1em]
  \sixcellsvert{\arrRone}{\blankR}{p'_R}{\blankR}{\wc}{\wc} \longrightarrow
    \sixcellsvert{\blankL}{\arrRone}{p_R}{\blankR}{\wc}{\wc}\,,
  \qquad
  \fourcellsR{\arrRone}{p'_R}{\wc} \longrightarrow
    \fourcellsR{\arrLone}{p_R}{\wc}\,.
  \label{eq:counter_TM_right-moving}
\end{gather}
\end{subequations}
Transitions in which the head moves left from the end of the chain are already covered by the left-moving transition in \cref{eq:counter_TM_transitions2}. Transitions in which the head would move right off the end of the chain are not implemented. (Transitions in which the head would move left off the beginning of the chain are not implemented either, but in fact the counter TM construction of \cref{sec:counter_TM} will never attempt such a transition when initialised properly. Moreover, we just declared such configurations illegal in \cref{eq:left-move-illegal}.)

When started from the blank input (which represents the number~0), the encoded counter TM will loop $\zeta^{L-3}$ times before it exceeds the $L-2$ tape space available on Track~2 (the space overhead of~1 is due to the way the incrementer TM is implemented in \cref{eq:base-zeta_incrementer}). The incrementer machine takes $\Theta(\log x)$ steps to increment the number $x$ (written on the tape in base-$\zeta$). Thus the counter TM will run for at least $\Omega(\zeta^L)$ (and at most $O( L \zeta^L)$) time-steps before it exceeds the available tape space.

The transition function for the counter TM defines a transition rule for each pair $(p,\tau)\in P\times \Xi$. So for any legal configuration containing a $p\in P$ on Track~2, exactly one of the transition rules in \cref{eq:counter_TM_transitions1,eq:counter_TM_transitions2} will apply during each $\arrRone$ sweep. Similarly, for any configuration containing a $p'_R\in P'_R$, a \cref{eq:counter_TM_right-moving} rule will apply. It is easy to verify that if any of the transition rules from \cref{eq:counter_TM_transitions1,eq:counter_TM_transitions2,eq:counter_TM_right-moving} is applied to any legal configuration, none of the \cref{eq:counter_TM_transitions1,eq:counter_TM_transitions2} rules can apply again to the resulting configuration. If the final \cref{eq:counter_TM_transitions2} rule applies, the \cref{eq:counter_TM_right-moving} rule will apply exactly once in the following step, after which none of the rules can apply again to the resulting configuration. The transition rules in \cref{eq:counter_TM_transitions1,eq:counter_TM_transitions2,eq:counter_TM_right-moving} therefore implement a single step of the TM during each left-to-right sweep of the $\arrRone$ state on Track~1, as required. We call this phase of the evolution, in which Track~1 contains a $\arrRone$ or $\arrLone$, the \keyword{computation phase}.

We also modify the clock oscillator rules from \cref{sec:clock-oscillator} involving a right-moving $\arrRone$ arrow to only apply when there is no TM head under the $\arrRone$, since the $\arrRone$ movement is already taken care of by the above transition rules when there is a TM head present:
\begin{equation}
  \fourcells{\arrRone}{\blankR}{\blankL}{\blankL}
    \longrightarrow \fourcells{\blankL}{\arrRone}{\blankL}{\blankL}\,,
  \qquad
  \fourcells{\arrRone}{\blankR}{\blankR}{\blankR}
    \longrightarrow \fourcells{\blankL}{\arrRone}{\blankR}{\blankR}\,,
  \qquad
  \threecellsR{\arrRone}{\blankR} \longrightarrow
      \threecellsR{\arrLone}{\blankR}\,.
\end{equation}

We must also enforce illegality of all the configurations declared to be illegal in \cref{sec:counter_TM_construction}. To do this, we simply introduce local penalty terms corresponding to the combinations declared illegal in \cref{eq:illegal_p0,eq:illegal_pf,eq:illegal_p0+non-blank,eq:illegal_forward_transitions,eq:illegal_backwards_transitions,eq:illegal_head+blank,eq:illegal_0+blank,eq:illegal_0+rightend}, and enforce the Track~3 regular expression of \cref{eq:track3_regexp} using penalty terms and \cref{regexp}.

The complete set of Track~1 to~3 transition rules is summarised in \cref{tbl:track123_rules}. The regular expressions on Tracks~1 to~3 enforced by penalty terms, together with all the additional illegal pairs defined so far, are summarised in \cref{tbl:track123_illegal_pairs}.

{
  \centering
  \setlength{\extrarowheight}{4pt}
  \setlength{\tabcolsep}{12pt}
  \begin{longtable}{>{$}c<{$} >{$}c<{$} >{$}c<{$}}
    \caption{All transition rules for Tracks~1 to~3.}
    \label{tbl:track123_rules}\\
    \nobreakhline
    \multicolumn{3}{c}{\text{Track~1 rules}}\\*
    \arrRzero,\arrLzero \text{ rules} & 
      & \arrLone \text{ rules} \\*
    \nobreakhline
    &&\\*
    \twocellshoriz{\arrRzero}{\blankR} \longrightarrow
      \twocellshoriz{\blankL}{\arrRzero}
    &&\phantom{\quad (*)}
      \twocellshoriz{\blankL}{\arrLone} \longrightarrow
      \twocellshoriz{\arrLone}{\blankR} \quad (*)\footnote{Transition rules marked with an $(*)$ will be replaced in \cref{sec:QTM_Hamiltonian} by a set of rules that also act on Tracks~4, 5 and 6.}
    \\*[1.5em]
    \twocellshoriz{\arrRzero}{\rightend\,} \longrightarrow
      \twocellshoriz{\arrLzeroz}{\rightend\,}
    \\*[1.5em]
    \twocellshoriz{\blankL}{\arrLzeroi} \longrightarrow
      \twocellshoriz{\arrLzeroipo}{\blankR}
    \\*[1.5em]
    \twocellshoriz{\blankL}{\arrLzeroK} \longrightarrow
      \twocellshoriz{\arrLzero}{\blankR}
    \\*[1.5em]
    &&\\
    &&\\
    \hline
    \multicolumn{3}{c}{\text{Track~1 and~2 rules}}\\*
    \arrRzero,\arrLzero \text{ rules} & \arrRone \text{ rules}
      & \arrLone \text{ rules} \\*
    \nobreakhline
    &&\\*
    \threecellsL{\arrLzero}{p_\alpha} \longrightarrow
      \threecellsL{\arrRone}{p_\alpha}
    &\fourcells{\arrRone}{\blankR}{\blankL}{\blankL}
      \longrightarrow \fourcells{\blankL}{\arrRone}{\blankL}{\blankL}
    &\phantom{\quad (*)}
      \threecellsL{\arrLone}{\neg p_\alpha} \longrightarrow
      \threecellsL{\arrRone}{\neg p_\alpha} \quad (*)
    \\*[2em]
    &\fourcells{\arrRone}{\blankR}{\blankR}{\blankR}
      \longrightarrow \fourcells{\blankL}{\arrRone}{\blankR}{\blankR}
    &\\*[2em]
    &\phantom{\quad (*)}
      \threecellsR{\arrRone}{\blankR} \longrightarrow
      \threecellsR{\arrLone}{\blankR}\quad (*)
    &\\
    &&\\
    \hline
    \multicolumn{3}{c}{\text{Track~1, 2 and~3 rules}}\\*
    \arrRzero,\arrLzero \text{ rules} & \arrRone \text{ rules}
      & \arrLone \text{ rules} \\*
    \nobreakhline
    &&\\*
    &\sixcellsvert{\arrRone}{\blankR}{\blankL}{p}{\wc}{\sigma} \longrightarrow
      \sixcellsvert{\blankL}{\arrRone}{p_L}{\blankR}{\wc}{\tau}
    &\\*[3em]
    &\sixcellsvert{\arrRone}{\blankR}{p}{\blankR}{\sigma}{\wc}
      \longrightarrow \sixcellsvert{\blankL}{\arrRone}{p_N}{\blankR}{\tau}{\wc}
    &\\*[3em]
    &\phantom{\quad (*)}
      \fourcellsR{\arrRone}{p}{\sigma} \longrightarrow
      \fourcellsR{\arrLone}{p_N}{\tau} \quad (*)
    &\\*[3em]
    &\sixcellsvert{\arrRone}{\blankR}{p}{\blankR}{\sigma}{\wc} \longrightarrow
      \sixcellsvert{\blankL}{\arrRone}{\blankL}{p'_R}{\tau}{\wc}
    &\\*[3em]
    &\sixcellsvert{\arrRone}{\blankR}{p'_R}{\blankR}{\wc}{\wc} \longrightarrow
      \sixcellsvert{\blankL}{\arrRone}{p_R}{\blankR}{\wc}{\wc}
    &\\*[3em]
    &\phantom{\quad (*)}
      \fourcellsR{\arrRone}{p'_R}{\wc} \longrightarrow
      \fourcellsR{\arrLone}{p_R}{\wc}\quad (*)\\*
    &&\\*
    \nobreakhline
  \end{longtable}
}

{
  \centering
  \setlength{\extrarowheight}{4pt}
  \setlength{\tabcolsep}{12pt}
  \begin{longtable}{>{$}c<{$} >{$}c<{$} >{$}c<{$} >{$}c<{$}}
    \caption{All illegal pairs and regular expressions enforced by illegal pairs for Tracks~1 to~3.}
    \label{tbl:track123_illegal_pairs}\\*
    \nobreakhline
    \multicolumn{4}{c}{\text{Regular expressions}} \\*
    \multicolumn{2}{c}{\text{Track~1}} & \text{Track~2} & \text{Track~3} \\*
    \nobreakhline
    \multicolumn{2}{c}{} &&\\*
    \multicolumn{2}{>{$}c<{$}}{
    \leftend\blankL{}^*[\arrRzero,\arrLzero,\arrLzeroi,\arrRone,\arrLone]
      \blankR{}^*\rightend}
    & \leftend\blankL{}^*\,p\,\blankR{}^*\rightend
    & \leftend\,\vdash\,i^*\,\#^*\rightend
    \\*
    &&&\\
    \hline
    \multicolumn{4}{c}{Illegal pairs} \\*
    \multicolumn{4}{c}{Tracks~1 and~2} \\*
    \nobreakhline
    \\*
    \multicolumn{4}{>{$}c<{$}}{
      \threecellsnotL{\arrRzero}{\neg\blankR}\,, \qquad
      \threecellsL{\arrRzero}{\neg p_\alpha}\,, \qquad
      \threecellsL{\arrLone}{p_\alpha}\,, \qquad
      \twocellsvert{\neg\arrRone}{p'_R}
    }
    \\
    &&&\\
    \hline
    \multicolumn{4}{c}{Illegal pairs} \\*
    \text{Tracks~1 and~3} & \text{Track~2} & \text{Tracks~2 and~3}
      & \text{Track~3} \\*
    \nobreakhline
    &&&\\*
    \threecellsnotL{\arrRzero}{\neg\#\;}
    & \twocellshoriz{\neg\leftend\,}{\,p_\alpha\;}
    & \fourcells{\,p_\alpha\;}{\wc}{\wc}{\neg\#}
    & \twocellshoriz{\;0\;\;}{\;\#\;\;}
    \\*[1.5em]
    &&\fourcells{\blankL}{\;p\;\;}{\;\#\;\;}{\;\#\;\;} &  \twocellshoriz{\;0\;\;}{\rightend}\\
    \\
    &&&\\
    \hline
    \multicolumn{4}{c}{Illegal pairs} \\*
    \multicolumn{4}{c}{Tracks~2 and~3 for undefined transitions in
      \cref{eq:base-zeta_counter}}\\*
    \nobreakhline
    \\*
    \multicolumn{4}{>{$}c<{$}}{
      \twocellsvert{\;p\;\;}{\tau}\,, \qquad
      \fourcells{\blankL}{\,p_R\;}{\tau}{\wc}\,, \qquad
      \fourcells{\,p_L\;}{\blankR}{\wc}{\tau}\,, \qquad
      \fourcells{\,p_N\;}{\blankR}{\tau}{\wc}
    }\\*
    \\*
    \hline
    \multicolumn{4}{c}{Illegal pairs} \\*
    \multicolumn{4}{c}{Tracks~1, 2 and~3  if $\delta(p,\sigma)=(p_L,\tau,L)$ \cref{eq:left-move-illegal}}\\*
    \nobreakhline
    \\*
    \multicolumn{4}{>{$}c<{$}}{
      \threecellsvert{\arrRone}{\;p\;\;}{\;\sigma\;\;}
    }\\*
    \\*
    \hline
  \end{longtable}
}

\subsection{Clock Hamiltonian}\label{sec:clock_Hamiltonian}
Before considering the remaining tracks, it is helpful to analyse in depth the Hamiltonian defined so far on the first three tracks. For that we start with the definition of well-formed states.


\begin{definition}[Well-formed state]
  \label{def:well-formed}
  We say that a standard basis state on Tracks~1 to~3 is \keyword{well-formed} if it is a bracketed state, its Track~1 configuration matches the regular expression $\leftend\blankL{}^*[\arrLzero,\arrLzeroi,\arrRzero,\arrLone,\arrRone]\blankR{}^*\rightend$, and its Track~2 configuration matches the regular expression $\leftend\blankL{}^*p\,\blankR{}^*\rightend$ where $p\in P'$.
\end{definition}

The penalty terms defined previously give an energy penalty of at least~1 to any standard basis state in $\Sbr$ that does not match the desired regular expressions on Tracks~1\nobreakdash--3. Thus only well-formed states can have zero energy. The following result shows that only one transition rule can apply to each well-formed state.

\begin{lemma}[Well-formed transitions]\label{well-formed}
  For any well-formed standard basis state, at most one transition rule applies in the forward direction, and at most one in the backwards direction. Furthermore, the set of well-formed states is closed under the transition rules.
\end{lemma}
\begin{proof}
  A well-formed standard basis state contains exactly one of $\arrRzero,$ $\arrLzero$, $\arrLzeroi$, $\arrRone$ or $\arrLone$ on Track~1. Thus clearly only transition rules from one of the sets in \cref{tbl:track123_rules} (the set of $\arrRzero,\arrLzero$ rules, $\arrRone$ rules, or $\arrLone$ rules) can apply in the forward direction. The left hand sides of the rules within the $\arrRzero,\arrLzero$ and $\arrLone$ sets are manifestly mutually exclusive. Since the counter TM is deterministic, there is a unique TM transition that applies at any step, so the left-hand-sides of the $\arrRone$ rules are also mutually exclusive.

  The counter TM is reversible, so there is also a unique backwards TM transition at every step. Thus the same argument applies in the backwards direction to the right hand sides of the rules in each set, with the exception of the $\arrLzero$ turning rule that changes a $\arrRone$ into a $\arrLzero$ (in the backwards direction). But this rule clearly cannot apply at the same time as the right hand side of any rule from the $\arrRone$ set, which concludes the proof of the first part of the lemma.

  It is straightforward to verify that all the transition rules in \cref{tbl:track123_rules} preserve well-formedness, implying the second part.
\end{proof}

Recall that the state $\ket{\phi_0}$ is the standard basis state with $\leftend\arrRzero\blankR{}^*\rightend$ on Track~1, $\leftend\,p_\alpha\,\blankR{}^*\rightend$ on Track~2, and $\leftend\,\vdash\,\#^*\,\rightend$ on Track~3. This corresponds to having the clock oscillator in the first state of the Track~1 sequence (see \cref{fig:clock-oscillator}), and the counter TM Tracks~1 and~2 in their initial configurations. There is no backwards transition out of $\ket{\phi_0}$, so we will refer to it as the \keyword{initial clock state}. Let $\ket{\phi_t}$ be the standard basis state obtained by applying $t$ transitions (in the forwards direction) to $\ket{\phi_0}$, which is well-defined thanks to \cref{well-formed}.

Let us consider the form of the states $\ket{\phi_t}$. Note that $\ket{\phi_0}$ contains a $\arrRzero$ on Track~1. The only transition rules that apply when Track~1 contains a $\arrRzero$, $\arrLzero$ or $\arrLzeroi$ (see \cref{tbl:track123_rules}) are those that sweep the arrow all the way to the right and back again, without affecting Tracks~2 or~3. We saw in \cref{sec:clock-oscillator} that one complete oscillator cycle takes $2(L-2)$ steps. Thus $t\leq 2(L-2)$ corresponds to the initialisation sweep, in which the state $\ket{\phi_t}$ contains a $\arrRzero$, $\arrLzero$ or $\arrLzeroi$ on Track~1, and Tracks~2 and~3 remain in their initial configurations.

At $t = 2(L-2)+1$, the $\arrLzero$ turns around and (because Track~2 contains a $p_\alpha$ at the beginning of the chain) becomes a $\arrRone$. This $\arrRone$ is now over a $p_\alpha$ on Track~2. Thus the next step implements the first step of the counter TM on Tracks~2 and~3, causing Track~2 to transition to $p_\alpha$.

Recall from \cref{sec:counter_TM} that the transition rules implement one step of the counter TM during each left-to-right sweep of the $\arrRone$ on Track~1, and then move the $\arrLone$ back to the beginning of the chain in the second half of the oscillator cycle. Note that the counter TM construction of \cref{sec:counter_TM} never transitions back to its initial state $p_\alpha$. So when the $\arrLone$ reaches the beginning of the chain, it will never find a $p_\alpha$ on Track~3. Thus it will always transition into $\arrRone$, never into $\arrRzero$ (see \cref{tbl:track123_rules}). Therefore, for all $t > 2(L-2)$ the state $\ket{\phi_t}$ contains a $\arrRone$ or $\arrLone$ on Track~1. At $t = 2(L-2)(n+1)$ the $\arrRone$ in $\ket{\phi_t}$ is back at the beginning of the chain, and Tracks~2 and~3 are in the configuration corresponding to the $n$'th step of the counter TM.

Eventually, the number written on Track~3 will be incremented until it reaches $\zeta^{L-4}$, the maximum number that can be written in base-$\zeta$ in $L-4$ digits (the $-4$ accounts for the space-overhead of~2 required by the $\TM{Inc}$ TM implementation, plus the two bracket states at the ends of the chain). When the counter TM tries to increment this number, the head will move right until it reaches the cell adjacent to the $\rightend$ at the end of the chain. At this point, it will be in an internal state that would normally step right in order to write the carry. But the site to the right of the head contains $\rightend$ instead of a blank Tape~2 cell. When the $\arrRone$ on Track~1 reaches the head at the end of on Track~2, there is no further forwards transition out of this configuration (see \cref{tbl:track123_rules}). Let $T$ be the number of transitions required to reach this configuration $\ket{\phi_T}$ starting from the initial state $\ket{\phi_0}$. We will refer to $\ket{\phi_T}$ as the \keyword{final clock state}. Since this configuration occurs when the counter TM is about to exceed $L$ tape space, which occurs after at most $O(L \zeta^L)$ steps of the TM (see \cref{sec:counter_TM}) each of which requires $2(L-2)$ transitions, we have that $T = O(L^2 \zeta^L)$. (Though this bound on the run-time will not be important for our purposes.)

The following result will be important later when we come to analyse the Hamiltonian.

\begin{lemma}[Evolve-to-illegal]\label{evolve_to_illegal}
  Evolving any $\ket{\phi_t}$ forwards or backwards in time according to the transition rules will never reach an illegal configuration. All other well-formed standard basis states will evolve either forwards or backwards to an illegal configuration after $O(L^2)$ transitions.
\end{lemma}

\begin{proof}
  As in \cref{counter_TM_evolve_to_illegal} the first part of the \namecref{evolve_to_illegal} is true by construction. For the second part, let us analyse all possible well-formed standard basis states, which we divide into two cases depending on the type of arrow state we have on Track~1:

  \paragraph{Case 1:} Track~1 contains $\arrRzero$, $\arrLzero$ or $\arrLzeroi$. If Tracks~2 and~3 are in the initial configuration, then we are in one of the $\ket{\phi_t}$ states with $0 \le t \le 2(L-2)$.

  If Tracks~2 and~3 are not in the initial configuration, then within at most $O(L)$ steps the $\arrRzero$ or $\arrLzero,\arrLzeroi$ will move forwards or backwards along the chain due to the initialisation sweep clock oscillator transition rules, until it is over a site containing the wrong initial Track~2 or Track~3 state. But this is illegal due to the Track~1\nobreakdash--3 illegal pairs from \cref{tbl:track123_illegal_pairs}.

  \paragraph{Case 2:} Track~1 contains $\arrRone$ or $\arrLone$. Recall that there is a one-to-one correspondence between configurations of the counter TM and well-formed standard basis states that do not contain a $p'_R\in P'_R$ on Track~2.

  Well-formed configurations containing a $p'_R\in P'_R$ on Track~2 evolve by a single backwards transition from \cref{eq:counter_TM_transitions2} to a configuration containing a $p\notin P'_R$ on Track~2, and evolve by a single forwards transition from \cref{eq:counter_TM_right-moving} to a configuration containing a $p_R\in P_R$. A configuration containing $p'_R\in P'_R$ is therefore reachable (resp.\ unreachable) by transition rules from the initial configuration iff the corresponding counter TM configuration with $p'_R$ replaced by the corresponding $p_R$ is reachable (resp.\ unreachable).

  For any well-formed configuration, each sweep of $\arrRone$, the transition rules implements exactly one step of the counter TM encoded in Tracks~2 and~3. The only counter TM transitions that are not implemented are those that would move the head beyond the $L$ sites between the $\leftend$ and $\rightend$. Therefore, if the configuration of Tracks~2 and~3 is not reachable from the initial configuration, then \cref{counter_TM_evolve_to_illegal} guarantees that evolving it forwards or backwards by the transition rules will reach an illegal configuration in $O(L)$ steps of the counter TM, which takes $O(L^2)$ transitions. (Note that it is crucial for this argument that to reach an illegal configuration in \cref{counter_TM_evolve_to_illegal} never requires moving the head outside the portion of tape $[0,L]$.)

  It only remains to consider well-formed states with Track~2 and~3 configurations that are reachable from the initial configuration. For the initial Tracks~2\nobreakdash--3 configuration itself, if $\arrRone$ is in the starting cell then we have the state $\ket{\phi_t}$ for $t=2(L-2)+1$. If not, either we have $\arrRone$ to the right of the head or $\arrLone$ on Track~1, with $p_\alpha$ at the start of Track~2. In both cases, evolving such a state forwards in time will reach a state with a $\arrLone$ over the $p_\alpha$ within $O(L)$ transitions, which is illegal by the second Track~1 and~2 illegal pair from \cref{tbl:track123_illegal_pairs}.

  If the Track~2\nobreakdash--3 configuration is not the initial one, then all positions of $\arrRone$ and $\arrLone$ give rise to elements in the set $\{\ket{\phi_t}\}_t$ except exactly those declared illegal in \cref{eq:left-move-illegal} and \cref{eq:p-prime-illegal}.
  This completes the proof of the \namecref{counter_TM_evolve_to_illegal}.
\end{proof}

\subsection{QTM Hamiltonian}\label{sec:QTM_Hamiltonian}
With the clock in place, encoding an arbitrary quantum Turing Machine in the Hamiltonian is now possible using ideas from \cite{Gottesman-Irani} (and is similar to the construction used in \cref{sec:counter_TM} to encode the reversible counter TM). The configurations are now quantum states, so there can be arbitrary superpositions over standard basis states. But it suffices to verify that the construction does the right thing on standard basis states; this then extends to arbitrary superpositions by linearity and well-formedness.

When we later come to analyse the spectrum, we will require our Hamiltonian to be \keyword{standard-form} in the following sense:
\begin{definition}[Standard-form Hamiltonian]
  \label{def:standard-form_H}
  We say that a Hamiltonian $H = \Htrans + \Hpenalty$ acting on a Hilbert space $\HS = (\C^C\ox\C^Q)^{\ox L} = (\C^C)^{\ox L}\ox(\C^Q)^{\ox L} =: \HS_C\ox\HS_Q$ is of \keyword{standard form} if $\Htranspen = \sum_{i=1}^{L-1} \htranspen^{(i,i+1)}$, and $\htranspen$ satisfy the following conditions:
  \begin{enumerate}
  \item $\htrans \in \cB\left((\C^C\ox\C^Q)^{\ox 2}\right)$ is a sum of transition rule terms, where all the transition rules act diagonally on $\C^C\ox\C^C$ in the following sense. Given standard basis states $a,b,c,d\in\C^C$, exactly one of the following holds:
    \begin{itemize}
    \item there is no transition from $ab$ to $cd$ at all; or
    \item $a,b,c,d\in\C^C$ and there exists a unitary $U_{abcd}$ acting on $\C^Q\ox\C^Q$ together with an orthonormal basis $\{\ket{\psi_{abcd}^i}\}_i$ for $\C^Q\ox\C^Q$, both depending only on $a,b,c,d$, such that the transition rules from $ab$ to $cd$ appearing in $\htrans$ are exactly $\ket{ab}\ket{\psi^i_{abcd}}\rightarrow \ket{cd}U_{abcd}\ket{\psi^i_{abcd}}$ for all $i$.
    \end{itemize}
    \label[part]{standard-form_H:transition_terms}
  \item $\hpenalty \in \cB\left((\C^C\ox\C^Q)^{\ox 2}\right)$ is a sum of penalty terms which act non-trivially only on $(\C^C)^{\ox 2}$ and are diagonal in the standard basis.
    \label[part]{standard-form_H:penalty_terms}
  \end{enumerate}
\end{definition}

In our case, $\C^C$ corresponds to Tracks 1--3 and $\C^Q$ to Tracks 4--6. Transitions from $ab$ to $cd$ will correspond exactly to the transitions shown for Tracks 1--3 in \cref{tbl:track123_rules}. While adding new transitions involving Tracks 4--6, we will need to remove some of the transitions in Tracks 1--3, but they will be recovered as restrictions of the new rules to those tracks. (In \cref{tbl:track123_rules}, transitions that will be replaced later are marked with an asterisk.) We will define the unitaries $U_{abcd}$ partially, and then complete them to a full unitary. All we have to take care of is, firstly, the new transition involving the rest of the tracks, when restricted to Tracks 1--3, recover exactly the transitions removed from \cref{tbl:track123_rules}; and, secondly, that orthogonality is preserved in the construction.

By the standard form,
\begin{equation}
  \htrans = \sum_{ab\rightarrow cd} (\ket{ab}-\ket{cd})(\bra{ab}-\bra{cd})
            \ox (2\1-U_{abcd}-U_{abcd}^\dagger)\;.
\end{equation}
Now, if we start with a family of QTM $P_n$ that satisfies \cref{phase-estimation_QTM:amplitudes} of \cref{phase-estimation_QTM}, it will be immediate from our construction that the partial definition of all $U_{abcd}$ will only involve elements in the set
\begin{equation}
  \mathcal{S} = \left\{0,1,\pm \frac{1}{\sqrt{2}},e^{i\pi\varphi}, e^{i\pi 2^{-\abs{\varphi}}}\right\}.
\end{equation}
As argued above, the same will hold for the full $U_{abcd}$, which gives \cref{QTM_in_local_Hamiltonian:explicit-form} in \cref{QTM_in_local_Hamiltonian}.

\subsubsection{QTM transition rules}\label{sec:QTM_transition_rules}
We first analyse how to incorporate the QTM transition rules into the Hamiltonian. We will simulate the QTM transition rules $\delta(p,\sigma) = \sum_{\tau,q,D}\delta(p,\sigma,\tau,q,D)\ket{\tau, q, D}$ during the second half of the clock oscillator cycle, when the Track~1 $\arrLone$ state sweeps from right to left. This is done by including the following transition rule terms (cf.\ \cref{eq:counter_TM_transitions1,eq:counter_TM_transitions2,eq:counter_TM_right-moving}), where $q$ is any state in $Q/\{q_f\}$ (i.e.\ we exclude all the transitions out of the QTM's final state):
\begin{subequations}\label{eq:QTM_transition_rules}
\begin{gather}
  \begin{split}
    \Ket{\,\sixcellsvert{\blankL}{\arrLone}{q}{\blankR}{\sigma}{\wc}\,}
      \longrightarrow
    &\sum_{\tau,q_R}\delta(q,\sigma,\tau,q_R,R)
      \Ket{\,\sixcellsvert{\arrLone}{\blankR}{\blankL}{q_R}{\tau}{\wc}\,}
    + \sum_{\tau,q_N}\delta(q,\sigma,\tau,q_N,N)
      \Ket{\,\sixcellsvert{\arrLone}{\blankR}{q_N}{\blankR}{\tau}{\wc}\,}\\
    &\qquad + \sum_{\tau,q_L}\delta(q,\sigma,\tau,q_L,L)
      \Ket{\,\sixcellsvert{\arrLone}{\blankR}{q'_L}{\blankR}{\tau}{\wc}\,}\,,
  \end{split} \label{eq:QTM_transition_step}
  \\[2em]
  \Ket{\,\sixcellsvert{\blankL}{\arrLone}{\blankL}{q'_L}{\wc}{\wc}\,}
    \longrightarrow \Ket{\,\sixcellsvert{\arrLone}{\blankR}{q_L}{\blankR}{\wc}{\wc}\,}\,.
    \label{eq:QTM_left-moving}
  \end{gather}
\end{subequations}
As in the implementation of the reversible counter TM (\cref{sec:counter_TM}), these transition rules implement the \textbf{R}ight-moving, \textbf{N}on-moving and \textbf{L}eft-moving transitions separately. This time, since the $\arrLone$ arrow is sweeping to the left, it is the \emph{left}-moving transitions that are implemented in two stages, first transitioning to an auxiliary state $q_L'\in Q_L'$, then in a second step transitioning to the corresponding $q_L$ state.

Note that if the QTM head ever ends up at the end of the chain, the transition rules of \cref{eq:QTM_transition_rules} can never apply, because they require the $\arrLone$ to be to the right of the QTM head. Thus the final Track~5 tape cell is not used, and the effective tape length available to the QTM for a chain of length $L+3$ is only $L$ rather than\footnote{Two cells are always used by the $\leftend,\rightend$ markers.} $L+1$. If the Track~4 QTM head ends up next to the $\rightend$ at the very end of the chain, this indicates that the head has stepped off the usable portion of the tape, regardless of the internal state.

We also include the following transition rules involving a left-moving $\arrLone$ arrow to cover the case where there is no QTM head present:
\begin{equation}\label{eq:QTM_leftarrow_sweep}
  \fourcells{\blankL}{\arrLone}{\blankL}{\blankL}
    \longrightarrow \fourcells{\arrLone}{\blankR}{\blankL}{\blankL}\,,
  \qquad
  \fourcells{\blankL}{\arrLone}{\blankR}{\blankR}
    \longrightarrow \fourcells{\arrLone}{\blankR}{\blankR}{\blankR}\,,
  \qquad
  \fourcells{\blankL}{\arrLone}{\blankL}{q}
    \longrightarrow \fourcells{\arrLone}{\blankR}{\blankL}{q}\,,
\end{equation}

\subsubsection{QTM Unitarity}\label{sec:unitarity}
With the set of transition rule terms defined so far in \cref{eq:QTM_transition_rules}, there is no transition out of the final state $q_f$ of the QTM. There is also no transition out of a QTM configuration in which the QTM head is at the end of the chain, and the next step would move the head to the right, off the end of the chain. The same is true of configurations trying to move the head left when it is located at the beginning of the chain.\footnote{Although the QTMs we will encode will in fact never do this.}

Ideally, we would like it if, when the QTM reaches one of these configurations, the clock continued ticking but the encoded QTM did nothing for the remaining time steps. However, simply omitting transitions out of these configurations causes a subtle but important issue relating to unitarity of the quantum transition rules, which will be crucial to the analysis of the resulting Hamiltonian in \cref{sec:local-Hamiltonian_analysis}.

If we complete the partial $U$ (associated with the Track 1--3 transitions $\twocellshoriz{\blankL}{\arrLone} \longrightarrow \twocellshoriz{\arrLone}{\blankR}$ ) as defined so far by \cref{eq:QTM_transition_rules,eq:QTM_qf_transition_rule,eq:QTM_leftarrow_sweep} to a unitary, this will add additional transitions out of the final state $q_f$, and also out of configurations with the head at the beginning or end of the chain. But this means that, instead of the encoded QTM doing nothing for the remaining time steps after reaching such a configuration, it will continue to evolve in some arbitrary way, depending on the particular choice of completion of $U$. On the other hand, if we omit transitions out of these configurations, the resulting $U$ will only be a partial isometry, not a full unitary.

It will be important in what follows (\cref{sec:local-Hamiltonian_analysis}) to ensure that the local update rules are extended to a unitary. We would therefore like to use up any remaining time steps in a controlled way, such that the Track~5 QTM tape is left unaltered after the QTM enters $q_f$ or runs out of tape. To this end, we deliberately add additional transitions out of the final state $q_f$, and also out of configurations with the head at the beginning or end of the chain, which cause the QTM to switch over to running some other, inconsequential computation to use up the remaining time. This time-wasting computation will use Track~6 as its tape, leaving the configuration of the Track~~5 QTM tape untouched.

The only requirement on this time-wasting computation is that it must be guaranteed not to halt or run out of tape before it has used up all the remaining time steps of the clock (otherwise we face the same unitarity issue once again). The natural choice is simply to run the same base-$\zeta$ counter computation as the clock (but running on Tracks~4 and~6). The counter TM never halts; it increments the number written on its tape forever, or -- when encoded on a finite chain -- until it runs out of tape space. Furthermore, since the clock will already have ``ticked'' for some positive number of time steps before this time-wasting counter TM starts running, the clock counter TM is guaranteed to run out of tape (and hence the clock stop ticking) first.

Transitioning from $q_f$ to the time-wasting counter TM can be accomplished simply by dovetailing the counter TM after the QTM, and encoding this dovetailed machine instead of the original QTM. Concretely, this means we must add to the Hamiltonian more transition rule terms of the form~\cref{eq:QTM_transition_rules} that encode all the transition rules of the counter TM from \cref{eq:base-zeta_counter}.\footnote{Since these are the transition rules of a classical reversible TM, they have a particularly simple form, the right hand side of each \cref{eq:QTM_transition_step} rule containing a single term.} Note that Track~6 contains additional subscripted variants $\vdash_q$ of the $\vdash$ tape symbol; the time-wasting counter TM transition rules simply ignore the subscript, and treat all of these as a $\vdash$. (I.e.\ the counter TM rules that read a $\vdash$ symbol in \cref{eq:base-zeta_counter} are duplicated for each $\vdash_q$ variant.) These time-wasting counter TM transition rules involve a disjoint set of Track~4 internal states $q\in P'$ to those $q\in Q'$ used by the QTM. Thus the left- and right-hand sides of all time-wasting counter TM \cref{eq:QTM_transition_rules} transition rules are orthogonal to those encoding the QTM transition rules.

We also add the following transition out of the final state $q_f$ of the original QTM into the initial state $p_\alpha$ of the counter TM, which dovetails the counter TM after the QTM. This transition rule acts on Tracks~1, 4 and~6:
\begin{equation}\label{eq:QTM_qf_transition_rule}
  \sixcellsvert{\blankL}{\arrLone}{q_f}{\blankR}{\vdash}{\wc}
  \longrightarrow
    \sixcellsvert{\arrLone}{\blankR}{p_\alpha}{\blankR}{\;\vdash_{q_f}}{\wc}\,.
\end{equation}

We must also add transitions that switch to running the time-wasting counter TM if the QTM runs out of tape, i.e.\ if it enters a configuration in which the head is at the beginning of the tape and would move left in the next time step, or one with the head at the end of the chain that would move right. (Though in the case of the proper machines considered in \cref{QTM_in_local_Hamiltonian:halt,QTM_in_local_Hamiltonian:out-of-tape,QTM_in_local_Hamiltonian:explicit-form} of \cref{QTM_in_local_Hamiltonian}, the former will never occur.) Moving left off the beginning of the tape is particularly easy, both because the head is already at the beginning of the chain, and because left-moving QTM transitions are implemented by first transitioning to an auxiliary $q'_L\in Q'_L$ state without moving the head. The following transition rule acting on Tracks~1, 2, 4 and~6 accomplishes this:
\begin{equation}\label{eq:QTM_leftend_transition_rule}
  \fivecellsL{\arrLone}{\neg p_\alpha}{q'_L}{\vdash} \longrightarrow
    \fivecellsL{\arrRone}{\neg p_\alpha}{p_\alpha}{\;\vdash_{q'_L}}\,.
\end{equation}
Note that, to preserve reversibility (unitarity), we must keep a record of which internal state $q'_L$ led to the QTM running out of tape. We record this in the $\vdash_q$ symbol written to the Track~6 time-wasting TM tape.

Configurations in which the head steps right off the end of the chain involve slightly more effort, as the head is at the wrong end of the chain to start running the time-wasting counter TM. From \cref{sec:QTM_Hamiltonian}, a configuration in which the Track~4 QTM head is at the very end of the chain indicates that the head has stepped right off the usable portion of the Track~5 tape. We first transition from any such configuration into an auxiliary Track~4 head reset state $r$, which moves the head all the way back to the beginning of the chain, before transitioning into the initial state of the counter TM. This requires three additional sets of transition rules.

We want the first new rule to act on Tracks~1 and~4 at the very end of the chain, and transition into the auxiliary reset state $r_q$ (depending on the state $q$ it was in immediately before). 
However, we have to make it compatible with the transitions already defined for Tracks 1--3. Hence 
we include the following transitions acting on Tracks 1,2,4 and 1,2,3,4:
\begin{equation}\label{eq:QTM_rightend_transition_rule1}
  \fourcellsR{\arrRone}{\blankR}{q} \longrightarrow
    \fourcellsR{\arrLone}{\blankR}{r_q}\, , \quad \quad
    \fivecellsR{\arrRone}{p}{\sigma}{q} \longrightarrow
    \fivecellsR{\arrLone}{p_N}{\tau}{r_q}\, ,
    \quad \quad
    \fivecellsR{\arrRone}{p'_R}{\cdot}{q} \longrightarrow
    \fivecellsR{\arrLone}{p_R}{\cdot}{r_q}\,
\end{equation}

The second new rule also acts on Tracks~1 and~4, and steps the auxiliary $r$ state to the left along with the $\arrLone$:
\begin{equation}\label{eq:QTM_rightend_transition_rule2}
  \fourcells{\blankL}{\arrLone}{\blankL}{r_q} \longrightarrow
    \fourcells{\arrLone}{\blankR}{r_q}{\blankR}\,.
\end{equation}
The third acts on on Tracks~1, 2 4 and~6 at the very beginning of the chain, and transitions into the initial configuration of the time-wasting counter TM:
\begin{equation}\label{eq:QTM_rightend_transition_rule3}
  \fivecellsL{\arrLone}{\neg p_\alpha}{r_q}{\vdash} \longrightarrow
    \fivecellsL{\arrRone}{\neg p_\alpha}{p_\alpha}{\;\vdash_q}\,.
\end{equation}
Again, to preserve reversibility, we record in the $\vdash_q$ symbol which internal state caused the QTM to run out of tape.

{
 \centering
  \setlength{\extrarowheight}{4pt}
  \setlength{\tabcolsep}{4pt}
  \begin{longtable}{>{$}c<{$}}
    \caption{All transition rules for Tracks 4, 5 and 6.}
    \label{tbl:track456_rules}\\
    \nobreakhline
    \text{Track~1 and~4 rules}\\*
    \nobreakhline
    \\*
    \fourcells{\blankL}{\arrLone}{\blankL}{\blankL}
      \longrightarrow \fourcells{\arrLone}{\blankR}{\blankL}{\blankL}\,,
    \qquad
    \fourcells{\blankL}{\arrLone}{\blankR}{\blankR}
      \longrightarrow \fourcells{\arrLone}{\blankR}{\blankR}{\blankR}\\*
    \\*
    \fourcells{\blankL}{\arrLone}{\blankL}{r_q} \longrightarrow
      \fourcells{\arrLone}{\blankR}{r_q}{\blankR}\,,
    \qquad
    \fourcells{\blankL}{\arrLone}{\blankL}{p} \longrightarrow
      \fourcells{\arrLone}{\blankR}{\blankL}{p}\\*
    \\*
    \nobreakhline
    \text{Track~1, 4 and~5 rules}\\*
    \nobreakhline
    \\*
    \begin{array}{rcl}
    \Ket{\,\sixcellsvert{\blankL}{\arrLone}{p}{\blankR}{\sigma}{\wc}\,}
      &\longrightarrow&
      \displaystyle\sum_{\tau,q_R}\delta(p,\sigma,\tau,q_R,R)
        \Ket{\,\sixcellsvert{\arrLone}{\blankR}{\blankL}{q_R}{\tau}{\wc}\,}
        \\[2em]
        &&+ \displaystyle\sum_{\tau,q_N}\delta(p,\sigma,\tau,q_N,N)
          \Ket{\,\sixcellsvert{\arrLone}{\blankR}{q_N}{\blankR}{\tau}{\wc}\,}
        \\[2em]
        &&+ \displaystyle\sum_{\tau,q_L}\delta(p,\sigma,\tau,q_L,L)
          \Ket{\,\sixcellsvert{\arrLone}{\blankR}{q'_L}{\blankR}{\tau}{\wc}\,}\\
    &&\\
    \Ket{\,\sixcellsvert{\blankL}{\arrLone}{\blankL}{q'_L}{\wc}{\wc}\,}
    &\longrightarrow& \Ket{\,\sixcellsvert{\arrLone}{\blankR}{q_L}{\blankR}{\wc}{\wc}\,}
    \end{array}
    \\\\
    \hline
    \text{Track~1, 4 and~6 rules}\\*
    \nobreakhline
    \\*
    \sixcellsvert{\blankL}{\arrLone}{q_f}{\blankR}{\vdash}{\wc}
    \longrightarrow
      \sixcellsvert{\arrLone}{\blankR}{p_\alpha}{\blankR}{\;\vdash_{q_f}}{\wc}\,,
    \qquad
    \sixcellsvert{\blankL}{\arrLone}{p}{\blankR}{\sigma}{\wc} \longrightarrow
    \sixcellsvert{\arrLone}{\blankR}{\blankL}{p_R}{\tau}{\wc}\,,
    \qquad
    \sixcellsvert{\blankL}{\arrLone}{p}{\blankR}{\sigma}{\wc}
    \longrightarrow \sixcellsvert{\arrLone}{\blankR}{p_N}{\blankR}{\tau}{\wc}\\*
    \\*
    \sixcellsvert{\blankL}{\arrLone}{p}{\blankR}{\wc}{\sigma} \longrightarrow
    \sixcellsvert{\arrLone}{\blankR}{p'_L}{\blankR}{\wc}{\tau}\,,
    \qquad
    \sixcellsvert{\blankL}{\arrLone}{\blankL}{p'_L}{\wc}{\wc} \longrightarrow
    \sixcellsvert{\arrLone}{\blankR}{p_L}{\blankR}{\wc}{\wc}
    \\\\
    \nobreakhline
    \text{Track~1, 2 and 4 rules}\\*
    \nobreakhline
    \\*
    \fourcellsR{\arrRone}{\blankR}{q} \longrightarrow
    \fourcellsR{\arrLone}{\blankR}{r_q}\,
    \\\\
    \nobreakhline
    \text{Track~1, 2, 4 and 6 rules}\\*
    \nobreakhline
    \\*
    \fivecellsL{\arrLone}{\neg p_\alpha}{q'_L}{\vdash} \longrightarrow
    \fivecellsL{\arrRone}{\neg p_\alpha}{p_\alpha}{\;\vdash_{q'_L}}\,. \quad
    \fivecellsL{\arrLone}{\neg p_\alpha}{r_q}{\vdash} \longrightarrow
    \fivecellsL{\arrRone}{\neg p_\alpha}{p_\alpha}{\;\vdash_q}\,.
    \\\\
    \nobreakhline
    \text{Track~1, 2, 3 and 4 rules}\\*
    \nobreakhline
    \\*
    \fivecellsR{\arrRone}{p}{\sigma}{q} \longrightarrow
    \fivecellsR{\arrLone}{p_N}{\tau}{r_q}\, ,
    \quad \quad
    \fivecellsR{\arrRone}{p'_R}{\cdot}{q} \longrightarrow
    \fivecellsR{\arrLone}{p_R}{\cdot}{r_q}\, \\*
    \\*
    \nobreakhline
  \end{longtable}
}

The transition rules in \cref{tbl:track456_rules} implement all the transitions that will take place for a properly initialised QTM. All of them have the desired form $\ket{ab}\ket{ij}\rightarrow \ket{cd}U_{abcd}\ket{ij}$ where $ab\rightarrow cd$ is a Track 1--3 transition and $U_{abcd}$ are partial isometries defined by their action on a subset $\ket{ij}$ of the computational basis of Tracks 4--6. We can now complete the unitaries $U_{abcd}$ at will. All we have to check is that they do indeed define a partial isometry.

The left hand sides of each transition rule in \cref{eq:QTM_transition_rules,eq:QTM_leftarrow_sweep,eq:QTM_qf_transition_rule,eq:QTM_leftend_transition_rule,eq:QTM_rightend_transition_rule1,eq:QTM_rightend_transition_rule2,eq:QTM_rightend_transition_rule3} are manifestly orthogonal. The right hand sides of all \cref{eq:QTM_left-moving,eq:QTM_leftarrow_sweep,eq:QTM_qf_transition_rule,eq:QTM_leftend_transition_rule,eq:QTM_rightend_transition_rule1,eq:QTM_rightend_transition_rule2,eq:QTM_rightend_transition_rule3} rules are clearly mutually orthogonal and normalised, and are also orthogonal to the right hand sides of all \cref{eq:QTM_transition_step} rules.

It remains to show that \cref{eq:QTM_transition_step} defines an isometry. Recall that there are two sets of \cref{eq:QTM_transition_step} terms, one encoding the QTM transition rules, and the other the time-wasting counter TM. The left- and right-hand sides of all rules in the QTM set are orthogonal to those of the counter TM set, as they involve disjoint sets  of internal states $Q'$ and $P'$ (respectively). Thus it suffices to prove that each set of transition rules considered separately defines an isometry. Since classical reversible TMs are special cases of QTMs, we prove the result for QTMs.

Let $\ket{\psi(p,\sigma)}$ denote the state on the RHS of \cref{eq:QTM_transition_step}. Then
\begin{equation}
  \begin{split}
    \norm{\psi(p,\sigma)}
    &= \sum_{\tau,q_R}\abs{\delta(p,\sigma,\tau,q_R,R)}^2
       + \sum_{\tau,q_N}\abs{\delta(p,\sigma,\tau,q_N,N)}^2\\
       &\qquad + \sum_{\tau,q_L}\abs{\delta(p,\sigma,\tau,q_L,L)}^2\\
    &= \norm{\delta(p,\sigma)}
    = 1
  \end{split}
\end{equation}
by the normalisation condition of \cref{QTM_transitions}, so the right hand sides of \cref{eq:QTM_transition_step} are normalised.

Similarly, for $(p_1,\tau_1) \neq (p_2,\tau_2)$,
\begin{equation}
  \begin{split}
    \braket{\psi(p_1,\sigma_1)|\psi(p_2,\sigma_2)}
    &= \sum_{\tau,q_R} \delta(p_1,\sigma_1,\tau,q_R,R)^*
                     \delta(p_2,\sigma_2,\tau,q_R,R)\\
      &\qquad + \sum_{\tau,q_N} \delta(p_1,\sigma_1,\tau,q_N,N)^*
                              \delta(p_2,\sigma_2,\tau,q_N,N)\\
      &\qquad\qquad + \sum_{\tau,q_L} \delta(p_1,\sigma_1,\tau,q_L,L)^*
                                    \delta(p_2,\sigma_2,\tau,q_L,L)\\
    &= \langle \delta(p_1,\tau_1), \delta(p_2,\tau_2) \rangle
    = 0
  \end{split}
\end{equation}
by the orthogonality condition of \cref{QTM_transitions}, so the right hand sides of \cref{eq:QTM_transition_step} are mutually orthogonal. Thus the QTM and time-wasting counter TM transition rules from \cref{eq:QTM_transition_step} preserve orthonormality, as required.


\subsubsection{QTM initialisation sweep}
We use penalty terms that only apply during the initialisation sweep to enforce the correct initial configurations of Tracks~4, 5 and~6. For Track~4, we use the following penalty terms, which force a single $q_0$ at the very beginning of the track:
\begin{equation}\label{eq:init-track4}
  \threecellsnotL{\arrRzero}{\neg\blankR}\,,
  \qquad
  \threecellsL{\arrRzero}{\neg q_0}\,.
\end{equation}

The $\arrLzeroi$ Track~1 states only ever occur during the initialisation sweep in the final $K$ sites at the right end of the chain. We use these to force the initial configuration of Track~5 to contain $K$ blank symbols at the very end, as required by \cref{QTM_in_local_Hamiltonian}.\footnote{These additional blank symbols will be needed later to provide for the small space overhead of the QTM constructed in \cref{phase-estimation_QTM}.}
\begin{equation}\label{eq:init-track5}
  \twocellsvert{\arrLzero}{\neg 1}\,,
  \qquad
  \twocellsvert{\arrLzeroi}{\neg \#}\,.
\end{equation}

Track~6 is forced to be in the all-blank configuration, except for the very first cell which contains a $\vdash$:
\begin{equation}\label{eq:init-track6}
  \threecellsnotL{\arrRzero}{\neg\#}\,,
  \qquad
  \threecellsL{\arrRzero}{\neg\vdash}\,.
\end{equation}

{
  \centering
  \setlength{\extrarowheight}{4pt}
  \setlength{\tabcolsep}{12pt}
  \begin{longtable}{>{$}c<{$} >{$}c<{$} >{$}c<{$}}
    \caption{Initialisation illegal pairs for Tracks~4--6.}
    \label{tbl:track45_illegal_pairs}\\
    \nobreakhline
    \multicolumn{3}{c}{Illegal pairs} \\*
    \text{Tracks~1 and~4} & \text{Tracks~1 and~5} & \text{Tracks~1 and~6} \\*
    \nobreakhline
    \\*
    \threecellsnotL{\arrRzero}{\neg\blankR}
    & \twocellsvert{\arrLzero}{\neg 1}
    & \threecellsnotL{\arrRzero}{\neg\#} \\*[2em]
    \threecellsL{\arrRzero}{\neg q_0}
    & \twocellsvert{\arrLzeroi}{\neg \#}
    & \threecellsL{\arrRzero}{\neg\vdash} \\*[2em]
    \\
    \hline
  \end{longtable}
}

\subsection{Analysis}\label{sec:local-Hamiltonian_analysis}
\Cref{well-formed,evolve_to_illegal} are the key results needed to prove the desired ground state properties of the Hamiltonian we have constructed, thanks to the ``Clairvoyance Lemma'' from \textcite[Lemma~4.2]{AGIK} (see also \cite[Lemma~5.6]{Gottesman-Irani}).\footnote{Note that our \cref{def:illegal} of ``legal'' states follows that of \cite{Gottesman-Irani}, which differs from the definition in~\cite{AGIK}.} Since the Clairvoyance \namecref{clairvoyance} has previously only been stated and proven for specific constructions, we state and prove a very general version of the \namecref{clairvoyance} here, which applies to any standard-form Hamiltonian, as defined above.

First, we give some intuition behind the argument, before giving the rigorous proof. The starting point for the proof is that the structure of standard-form Hamiltonians (\cref{def:standard-form_H}) implies that the invariant subspaces of $H$ decompose as a tensor product across the clock and quantum tracks, with the subspaces spanned states connected by transition rules.

Furthermore, these invariant subspaces come in three types: \begin{enumerate*}[label=(\arabic{*}),ref=\arabic{*}]
\item\label[type]{type_illegal} subspaces in which all states are illegal (\cref{def:illegal}), i.e.\ are in the support of one or more penalty terms;
\item\label[type]{type_legal_illegal} subspaces that contain legal and illegal states; and
\item\label[type]{type_legal} subspaces that only contain legal states, of which there can be at most one, spanned by the states corresponding to a valid evolution of the QTM.
\end{enumerate*}
We can then analyse the spectrum in each type of subspace separately, to show that the ground state energy (minimum eigenvalue) is either $>0$ or $=0$, respectively, depending on whether the QTM halts or not.

The \cref{type_illegal} invariant subspaces trivially have positive energy from the penalty terms.
\Cref{evolve_to_illegal} guarantees that any legal clock state in a \cref{type_legal_illegal} evolves to an illegal state, implying that all eigenstates in \cref{type_legal_illegal} subspaces also have positive energy.

The analysis of the \cref{type_legal} subspaces is reminiscent of Kitaev's original QMA-hardness proof for local Hamiltonians~\cite{Kitaev_book}. We first construct a unitary transformation $W$ that brings the Hamiltonian $H$ into the form $E_C\ox\1_Q$, where $E_C$ is a tridiagonal almost-Toeplitz matrix acting on the Hilbert space of the classical tracks.

This unitary is constructed by taking a product over the local unitaries corresponding to the local transition rules at each time step of the QTM. It is crucial that we have a local \emph{unitary} at each time step, and not a partial isometry, otherwise this product would not necessarily be unitary.
As seen in \cref{sec:Turing_machinery}, the fact that the QTM is well-formed, normal form and unidirectional guarantees via \cref{QTM_transitions} that the quantum transition rules define an isometry, thus can always be extended to a local unitary. Note it is \emph{not} required (nor is it the case) that the \emph{clock} transitions be extended to a full unitary; only the QTM transitions.

However, as discussed in \cref{sec:unitarity}, extending the QTM transition rules to a local unitary requires including the transitions out of the Halting state required of normal-form QTM's (\cref{def:normal-form}), as well transitions out of configurations where the QTM has run out of tape space. The time-wasting construction of \cref{sec:unitarity} ensures we still have a handle on the behaviour of the QTM after such a transition. In particular, it will never enter the Halting state if it has not already done so before running out of tape.

Having brought the Hamiltonian into the almost-diagonal form $H\simeq \E\ox\1$, we can analyse the spectrum of the \cref{type_legal} invariant subspaces. These subspaces do not contain any illegal states, so are in the kernel of all penalty terms except possibly the Halting penalty. In the absence of the Halting penalty, the minimum eigenvalue in this subspace is therefore given by that of the tridiagonal almost-Toeplitz matrix $E$, which is~0 by standard results. The associated eigenvector corresponds to the uniform superposition over all legal clock states. Inverting the unitary transformation $W$, this is exactly the history state of \cref{def:history-state}.

Therefore, considering all invariant subspaces, if the QTM never halts, the ground state is the 0-energy history state, which lives in a \cref{type_legal} subspace. If the QTM \emph{does} halt, so that the \cref{type_legal} subspace is no longer in the kernel of the Halting penalty term, the minimum eigenvalue in \cref{type_legal} can also be bounded away from~0 by standard arguments, thus the ground state energy is necessarily positive.

We now give the rigorous version of this argument.

\begin{lemma}[Invariant subspaces]
  \label{invariant_subspaces}
  Let $\Htrans$ and $\Hpenalty$ define a standard-form Hamiltonian as in \cref{def:standard-form_H}. Let $\mathcal{S}=\{S_i\}$ be a partition of the standard basis states of $\HS_C$ into minimal subsets $S_i$ that are closed under the transition rules (where a transition rule $\ket[CD]{ab}\ket{\psi} \rightarrow \ket[CD]{cd}U_{abcd}\ket{\psi}$ acts on $\HS_C$ by restriction to $(\C^C)^{\ox 2}$, i.e.\ it acts as $ab \rightarrow cd$).

  Then $\HS = \left(\bigoplus_S \mathcal{K}_{S_i}\right)\ox\HS_Q$ decomposes into invariant subspaces $\mathcal{K}_{S_i}\ox\HS_Q$ of $H = \Hpenalty + \Htrans$ where $\mathcal{K}_{S_i}$ is spanned by $S_i$.
\end{lemma}

\begin{proof}
   $\Hpenalty$ is diagonal in the standard basis by definition (\cref{standard-form_H:penalty_terms} of \cref{def:standard-form_H}), so $\mathcal{K}_{S_i}\ox\HS_Q$ are trivially invariant under $\Hpenalty$. But, by the form of the transition rule terms (\cref{standard-form_H:transition_terms} of \cref{def:standard-form_H}), the image $\Htrans\ket[C]{x}\ket[Q]{\varphi}$ of a standard basis state $\ket[C]{x}$ under $\Htrans$ has support only on standard basis states of $\HS_C$ that are reachable by transition rules from $\ket[C]{x}$. Closure of $\mathcal{K}_{S_i}\ox\HS_Q$ under $H_\text{trans}$ is then immediate from the definition of $S_i$.
\end{proof}

\begin{lemma}[Clairvoyance Lemma]\label{clairvoyance}
  Let $H = \Htrans + \Hpenalty$ be a standard-form Hamiltonian as specified in \cref{def:standard-form_H}, and let $\mathcal{K}_S$ be defined as in \cref{invariant_subspaces}. Let $\lambda_0(\mathcal{K}_S)$ denote the minimum eigenvalue of the restriction $H\vert_{\mathcal{K}_S\ox \HS_Q}$ of $H = \Htrans + \Hpenalty$ to the invariant subspace $\mathcal{K}_S\ox\HS_Q$.

  Assume that there exists a subset $\mathcal{W}$ of standard basis states for $\HS_C$ with the following properties:
  \begin{enumerate}
  \item All legal standard basis states for $\HS_C$ are contained in $\mathcal{W}$. \label[condition]{clairvoyance:legal}
  \item $\mathcal{W}$ is closed with respect to the transition rules.  \label[condition]{clairvoyance:closure}
  \item At most one transition rule applies in each direction to any state in $\mathcal{W}$. \label[condition]{clairvoyance:single_transition}
  \item For any subset $S\subseteq\mathcal{W}$ that contains only legal states, there exists at least one state to which no backwards transition applies. \label[condition]{clairvoyance:initial_state}
  \end{enumerate}

  \noindent Then each $\mathcal{K}_S$ falls into one of the following categories:
  \begin{enumerate}[label=(\arabic{*}).,ref=(\arabic{*})]
  \item $S$ contains only illegal states, and $\lambda_0(\mathcal{K}_S) \geq 1$. \label[case]{S_illegal}

  \item $S$ contains both legal and illegal states, and $\lambda_0(\mathcal{K}_S) = \Omega(1/\abs{S}^3)$.
  \label[case]{S_legal-illegal}

  \item $S$ contains only legal states, and $\lambda_0(\mathcal{K}_S) = 0$. The corresponding eigenspace is
  \begin{equation}
    \ker\left(\Htrans + \Hpenalty\right) =
    \linspan\left\{
      \frac{1}{\sqrt{\abs{S}}}\sum_{t=0}^{\abs{S}-1}\ket[C]{t}\ket[Q]{\psi_t}
    \right\}
  \end{equation}
  where $\ket[C]{t}$ are the states in $S$, $\ket{\psi_0}$ is any state in $\HS_Q$, and $\ket{\psi_t} := U_t\dots U_1\ket[Q]{\psi_0}$ where $U_t$ is the unitary on $\HS_Q$ appearing in the transition rule that takes $\ket[c]{t-1}$ to $\ket[C]{t}$. All other states in $\mathcal{K}_S\ox\HS_Q$ have energy at least $\Omega(1/\abs{S}^2)$.
  \label[case]{S_legal}
  \end{enumerate}
\end{lemma}

\noindent
To prove this, we will need Kitaev's geometrical lemma:

\begin{lemma*}[Geometrical Lemma -- Lemma~14.4 in \cite{Kitaev_book}]
  \label{Kitaev_geometrical}
  Let $A,B \geq 0$ be positive semidefinite operators such that $\ker A \cap \ker B = \{0\}$, $\lambdamin(A\vert_{\supp A}) \geq \mu$ and $\lambdamin(B\vert_{\supp B}) \geq \mu$. Then
  \begin{equation}
    A + B \geq 2\mu\sin^2\left(\frac{\theta}{2}\right)
  \end{equation}
  where
  \begin{equation}
    \cos\theta := \max_{\substack{\ket{\psi} \in \ker A\\ \ket{\varphi} \in \ker B}}\abs{\braket{\psi|\varphi}}.
  \end{equation}
\end{lemma*}\addtheoremline{lemma}{Geometrical Lemma}

\begin{proof}[of Clairvoyance \cref{clairvoyance}]\hfill\\
  \Cref{S_illegal} is trivial since $\bra[C]{x}\bra[Q]{\psi}\Hpenalty\ket[C]{x}\ket[Q]{\psi} \geq 1$ for any illegal standard basis state $\ket[C]{x}$.

  Now consider \cref{S_legal,S_legal-illegal}. By assumption, all legal standard basis states of $\HS_C$ are contained in $\mathcal{W}$, which is closed under transition rules. Thus closure of~$S$ (\cref{invariant_subspaces}) implies $S \subseteq \mathcal{W}$. Consider the directed graph of states in $\mathcal{W}$ formed by adding a directed edge between pairs of states connected by transition rules. By assumption, only one transition rule applies in each direction to any state in $\mathcal{W}$, so the graph consists of a union of disjoint paths (which could be loops in \cref{S_legal-illegal}). Minimality of $S$ (\cref{invariant_subspaces}) implies that $S$ consists of a single such connected path.

  Let $t=0,\dots,|S|-1$ denote the states in $S$ enumerated in the order induced by the directed graph. $\Htrans$ then acts on the subspace $\mathcal{K}_S\ox\HS_Q$ as
  \begin{equation}\label{eq:H_KS}
    \bigl.\Htrans\bigr\vert_{\mathcal{K}_S\ox\HS_Q}
    = \sum_{t=0}^T \frac{1}{2}\left(
        \proj{t}\ox\id + \proj{t+1}\ox\id
        - \ketbra{t+1}{t}\ox U_t - \ketbra{t}{t+1}\ox U_t^\dg
      \right),
  \end{equation}
  where $U_t$ is the unitary on $\HS_Q$ appearing in the transition rule that takes $\ket[C]{t}$ to $\ket[C]{t+1}$. $T = \abs{S}-1$ if the path in $S$ is a loop, otherwise $T = \abs{S}-2$.

  Consider a single term $H_t := \proj{t}\ox\id + \proj{t+1}\ox\id + \ketbra{t+1}{t}\ox U_t + \ketbra{t}{t+1}\ox U_t^\dg$ from \cref{eq:H_KS}. Defining the unitary $W_t = \proj{t+1}\ox U_t^\dg + (\id-\proj{t+1})\ox\id$, we have
  \begin{equation}
    W_t H_t W_t^\dg
    = \left(\ket{t}-\ket{t+1}\right)\left(\bra{t}-\bra{t+1}\right) \ox\id
    \geq 0.
  \end{equation}
  Thus, whether the path in $S$ forms a loop or not, we have\footnote{This operator has the same form as Kitaev's Hamiltonian, and the analysis from here on is similar to that in \textcite{Kitaev_book}.}
  \begin{equation}\label{eq:Kitaev_Hamiltonian}
    \begin{split}
      \bigl.\Htrans\bigr\vert_{\mathcal{K}_S\ox\HS_Q}
      &\geq \sum_{t=0}^{\abs{S}-2} \frac{1}{2}\left(
          \proj{t}\ox\id + \proj{t+1}\ox\id
          - \ketbra{t+1}{t}\ox U_t - \ketbra{t}{t+1}\ox U_t^\dg
        \right)\\
      &=: \Hpath,
    \end{split}\raisetag{1.5em}
  \end{equation}
  with equality if the path is not a loop.

  Defining the unitary
  \begin{equation}\label{eq:W_unitary}
    W = \sum_{t=0}^{\abs{S}-2} \proj{t}\ox\prod_{i=0}^t U_i^\dg + \Proj{(\abs{S}-1)}\ox\id,
  \end{equation}
  we have $\Hpath \simeq W \Hpath W^\dg = E\ox\id$ where
  \begin{equation}\label{eq:E_matrix}
    E = \begin{pmatrix}
          \frac{1}{2}  & -\frac{1}{2} &      0       &    \dots     &        &    \dots     &       0      &       0      \\
          -\frac{1}{2} &       1      & -\frac{1}{2} &    \ddots    &        &              &              &       0      \\
                0      & -\frac{1}{2} &      1       & -\frac{1}{2} &        &              &              &    \vdots    \\
             \vdots    &    \ddots    & -\frac{1}{2} &      1       & \ddots &              &              &              \\
                       &              &              &    \ddots    & \ddots &              &    \ddots    &    \vdots    \\
             \vdots    &              &              &              &        &    \ddots    &    \ddots    &       0      \\
                0      &              &              &              & \ddots &    \ddots    &       1      & -\frac{1}{2} \\
                0      &       0      &    \dots     &              & \dots  &      0       & -\frac{1}{2} & \frac{1}{2}
        \end{pmatrix}.
  \end{equation}
  The matrix $E$ is the Laplacian of the random walk on a line, and it is well known that its eigenvalues are given by $\lambda_k = 1-\cos q_k$ where $q_k = k\pi/\abs{S}$, $k=0,\dots,\abs{S}-1$, with corresponding eigenvectors $\ket{\phi_k} \propto \sum_{j=0}^{\abs{S}-1} \cos\left(q_k(j+\frac{1}{2})\right) \ket{j}$~\cite{Kitaev_book}. Thus
  \begin{equation}
    \ker \Hpath
    = \linspan\left\{
        W^\dg \frac{1}{\sqrt{\abs{S}}}\sum_{t=0}^{\abs{S}-1}\ket[C]{t}\ket[Q]{\psi_0}
      \right\}
    = \linspan\left\{
        \frac{1}{\sqrt{\abs{S}}}\sum_{t=0}^{\abs{S}-1}\ket[C]{t}\ket[Q]{\psi_t}
      \right\}
  \end{equation}
  where $\ket{\psi_t} := U_t\dots U_0 \ket{\psi_0}$ for any $\ket{\psi_0}$.

  First consider \cref{S_legal-illegal}. Take $A = \Hpath$ and $B = \Hpenalty|_{\mathcal{K}_S\ox\HS_Q}$ in \cref{Kitaev_geometrical}. We have $\lambdamin(A|_{\supp A}) = 1 - \cos q_1 = \Omega(1/\abs{S}^2)$ and $\lambdamin(B|_{\supp B}) = 1$. Furthermore, since $S$ contains at least one illegal state in \cref{S_legal-illegal}, we have $\ker A\cap \ker B = \{0\}$ and
  \begin{equation}
    \cos^2\theta
    = \max_{\ket{\psi}} \bra[C]{\phi_0}\bra[Q]{\psi}
        W^\dg\,\Pi_{\ker B}W \ket[C]{\phi_0}\ket[Q]{\psi}
    \leq 1 - 1/\abs{S},
  \end{equation}
  where $\Pi_{\ker B}$ is the projector onto $\ker B$ and we have used the fact that $B = \Hpenalty|_{\mathcal{K}_S\ox\HS_Q}$ is diagonal in the standard basis. Invoking \cref{Kitaev_geometrical}, we obtain
  \begin{equation}
    \left(\Htrans + \Hpenalty\middle)\right\vert_{\mathcal{K}\ox\HS_Q}
    \geq \Hpath + \Hpenalty\vert_{\mathcal{K}\ox\HS_Q}
    = \Omega(1/\abs{S}^3),
  \end{equation}
  thus $\lambda_0(\mathcal{K}_S) = \Omega(1/\abs{S}^3)$ as claimed.

  Finally, consider \cref{S_legal}. Since $S$ contains only legal states in this case, $\Hpenalty\vert_{\mathcal{K}_S\ox\HS_Q} = 0$. Furthermore, by \cref{clairvoyance:initial_state} the states in $S$ cannot form a loop, so $H\vert_{\mathcal{K}_S\ox\HS_Q} = \Htrans\vert_{\mathcal{K}_S\ox\HS_Q} = \Hpath \simeq E\ox\id$. The claim in the \namecref{clairvoyance} now follows from the form of the eigenvalues and kernel of $E$, given above.
\end{proof}

Note that the proof of the Clairvoyance \namecref{clairvoyance} relied crucially on unitarity of the operators $U_{abcd}$ appearing in the transition rules (see \cref{invariant_subspaces}). In particular, it is \emph{not} sufficient for the $U_{abcd}$ to be partial isometries. That would not allow us to unitarily transform $\Htrans\vert_{\mathcal{K}_S\ox\HS_Q}$ into $E\ox\1$ using the unitary $W$ in \cref{eq:W_unitary,eq:E_matrix}. Thus, in order to apply the Clairvoyance \cref{clairvoyance}, we \emph{must} complete the quantum parts of the transition rules to a full unitary. This necessitates the ``time-wasting'' construction in \cref{sec:unitarity} (or similar).

We are finally in a position to prove \cref{QTM_in_local_Hamiltonian}, the main result of this section. As commented above, let $\HS_C = (\C^C)^{\ox L}$ and $\HS_Q = (\C^Q)^{\ox L}$ be the Track~1\nobreakdash--3 and Track~4\nobreakdash--6 Hilbert spaces, respectively, for a chain of length $L$, as specified in \cref{eq:local_Hilbert_space}. Let $\htrans \in \cB((\C^C\ox\C^Q)^{\ox 2})$ be the sum of all the transition rule terms defined in \cref{tbl:track123_rules,tbl:track456_rules} (omitting those marked in \cref{tbl:track123_rules}) after completing to a full unitary so that the resulting Hamiltonian is standard-form. Let $\hpenalty \in \cB((\C^C\ox\C^Q)^{\ox 2})$ be the sum of all penalty terms defined in \cref{tbl:end-marker_illegal_states,tbl:track123_illegal_pairs}, and $\hinit \in \cB((\C^C\ox\C^Q)^{\ox 2})$ be the sum of all Track~4--6 initialisation penalty terms defined in \cref{tbl:track45_illegal_pairs}.

Define the standard-form (\cref{def:standard-form_H}) Hamiltonian $H(L) = \Htrans(L) + \Hpenalty(L)  \in \HS_C\ox\HS_Q$ on a chain of length $L$, where $\Htranspen = \sum_{i=1}^{L-1} \htranspen^{(i,i+1)}$, and similarly define $\Hinit(L) := \sum_{i=1}^{L-1} \hinit^{(i,i+1)}$.

\begin{proposition}[Unique ground state]\label{H_unique_gs}
  The unique 0-energy eigenstate of\linebreak $\bigl(\Htrans(L) + \Hpenalty(L) + \Hinit(L)\bigr)\bigl|_\Sbr\bigr.$ is the computational history state
  \begin{equation}\label{eq:0-energy_eigenstate}
    \frac{1}{\sqrt{T}}\sum_{t=0}^T \ket[C]{\phi_t}\ket[Q]{\psi_t},
  \end{equation}
  where $T = \Omega\left(L\,\zeta^L\right)$, $\ket[Q]{\psi_0}\in\HS_Q$ is the initial configuration of the QTM with the unary representation of the number $L-K-3$ as input, and $\ket[Q]{\psi_t}\in\HS_Q$ is the sequence of states produced by evolving $\ket{\psi_0}$ under the QTM and time-wasting counter TM transition rules (where each such state is duplicated $O(L)$ times in succession in the sequence).
\end{proposition}

\begin{proof}
  Let us first apply the Clairvoyance \namecref{clairvoyance} to $\bigl.\bigl(\Htrans(L) + \Hpenalty(L)\bigr)\bigr|_\Sbr$. We have to check that $\htrans$ and $\hpenalty$ fulfil the requirements of the \namecref{clairvoyance}.

  Take $\HS_C$ to be the Hilbert space of Tracks~1\nobreakdash--3, and $\HS_Q$ the Hilbert space of Tracks~4\nobreakdash--5. Define $\mathcal{W}$ to be the set of all well-formed Track~1\nobreakdash--3 standard basis states. Any state that is not well-formed violates an illegal pair enforcing a regular expression, so all legal Track~1\nobreakdash--3 states are well-formed. $\mathcal{W}$ therefore fulfils \cref{clairvoyance:legal} of \cref{clairvoyance}. By \cref{well-formed}, the set of well-formed states is closed under the transition rules and at most one transition rule applies in each direction to any well-formed state, so $\mathcal{W}$ fulfils \cref{clairvoyance:closure,clairvoyance:single_transition} of \cref{clairvoyance}. Finally, by \cref{evolve_to_illegal}, the only subset of legal Track~1\nobreakdash--3 standard basis states that is closed under the transition rules is the set of clock states $\{\ket{\phi_t}\}$, which has one state $\ket{\phi_0}$ to which no backward transition applies, and one state $\ket{\phi_T}$ to which no forward transition applies. So $\mathcal{W}$ also fulfils \cref{clairvoyance:initial_state} of \cref{clairvoyance}.

  All penalty terms are diagonal in the standard basis, as required by \cref{invariant_subspaces,clairvoyance}, and all transition rules are of the form required in \cref{clairvoyance}, by construction. $\htrans$ and $\hpenalty$ therefore fulfil all the requirements of \cref{invariant_subspaces,clairvoyance}.

  Invoking the Clairvoyance \cref{clairvoyance}, the 0-energy eigenspace of $(\Htrans + \Hpenalty)|_\Sbr$ has the form
  \begin{equation}
    \ker\left(\Htrans + \Hpenalty\middle)\right|_\Sbr =
    \linspan\left\{
      \frac{1}{\sqrt{\abs{S}}}\sum_{t=0}^{\abs{S}-1}\ket[C]{\phi_t}\ket[Q]{\psi_t}
    \right\},
  \end{equation}
  where $\ket[Q]{\psi_t} = U_t\dots U_1\ket[Q]{\psi_0}$ and $\ket[Q]{\psi_0}$ is any state of $\HS_Q$.

  Now we include the $\hinit$ terms, and show that $\ket[Q]{\psi_0}$ must be as claimed in \cref{H_unique_gs}. For $0 \leq t \leq 2L-1$, the clock states $\ket{\phi_t}$ consist of an $\arrRzero$ that sweeps from left to right and back along the chain (where for the time steps $t=L+1,\dots,L+K$ of this sweep, the arrow is in the states $\arrLzeroz,\dots,\arrLzeroK$, respectively). The transition rules that apply during the initialisation sweep act trivially on Tracks~4\nobreakdash--5, so $\ket[Q]{\psi_t} = \ket[Q]{\psi_0}$ for $t \leq 2L-1$. By construction, the $\hinit$ penalty terms from \cref{tbl:track45_illegal_pairs} give an additional energy penalty to any Track~4\nobreakdash--5 states that are not in the desired initial QTM configuration when the $\arrLzero$ sweeps past.

  Consider any Track~4\nobreakdash--5 state $\ket[Q]{\psi_0}$ that is not the desired initial QTM configuration. Then there is some $0 \leq t \leq 2L-1$ for which $\bra[C]{\phi_t}\bra[Q]{\psi_t} \Hinit \ket[C]{\phi_t}\ket[Q]{\psi_t} > 0$. Noting that the overall Hamiltonian $H \geq 0$ and the $\ket{\phi_t}$ are mutually orthogonal, this immediately implies that the unique 0-energy state is that given in \cref{eq:0-energy_eigenstate}.

  The initial state $\ket[Q]{\psi_0}$ on Track~4 matches the regular expression $\leftend\blankL{}^*q\blankR{}^*\rightend$. The transition rules $U_t$ preserve this form -- which allows us to identify Tracks~4 and~5 with the head and tape configuration of the QTM -- and by construction implement one step of the QTM or time-wasting counter TM in each $\arrLone$ sweep (all other unitaries applied to Tracks~4 and~5 during each sweep being identities).

  Finally, the clock counter TM counts up to $\Omega(\zeta^L)$ before it runs out of tape, and each step of this TM requires one complete sweep of the Track~1 arrow which takes $\Omega(L)$ steps. Thus $T = \Omega(L\,\zeta^L)$ as claimed.\footnote{This lower-bound is certainly not tight, as incrementing the base-$\zeta$ number on the tape takes more than one step of the TM, but it suffices for our purposes.}) We are done.
\end{proof}

The local Hilbert space dimension in \cref{eq:local_Hilbert_space} manifestly depends only on the alphabet size and internal states of the counter TM encoded on Tracks~2 and~3, the QTM encoded on Tracks~4 and~5, and the time-wasting counter TM encoded on Tracks~4 and~6. However, the alphabet size and internal state space of both counter TMs depends only on the base $\zeta$ we choose, and this is completely determined by the alphabet size and internal state space of the QTM ($\zeta=\abs{\Sigma\times Q}$). Thus the local Hilbert space dimension $d$ depends only on the alphabet size and number of internal states of the QTM, as claimed in \cref{QTM_in_local_Hamiltonian:local_dim} of \cref{QTM_in_local_Hamiltonian}.

Since the local interaction $h$ of our Hamiltonian is constructed by summing transition rule terms and penalty terms, and $\htrans,\hpenalty,\hinit \geq 0$, we have $h \geq 0$. But we have shown that the overall Hamiltonian has a 0-energy eigenstate, hence it is frustration-free, satisfying \cref{QTM_in_local_Hamiltonian:FF} of \cref{QTM_in_local_Hamiltonian}.

\Cref{H_unique_gs} implies that the unique ground state of $H|_\Sbr$ has the computational history state form claimed in \cref{QTM_in_local_Hamiltonian:gs} of \cref{QTM_in_local_Hamiltonian}. Moreover, since the history state encodes $\Omega(\abs{\Sigma \times Q}^L)$ complete sweeps of the Track~1 arrow, and the QTM is advanced by exactly one step in each right-to-left sweep, the computational history state encodes $\Omega(\abs{\Sigma \times Q}^L)$ steps of the computation. If the QTM halts before this, it transitions by the $q_f$ rule of \cref{eq:QTM_qf_transition_rule} to running the time-wasting counter TM, which never alters the QTM tape encoded on Track~5. Track~5 is therefore only in the state $q_f$ for one time step. Thus \cref{QTM_in_local_Hamiltonian:halt} of \cref{QTM_in_local_Hamiltonian} is also satisfied.

Now imagine that the QTM reaches a configuration in which the next step would move the head before the starting cell. (Though in the case of the proper machines considered in \cref{QTM_in_local_Hamiltonian:halt,QTM_in_local_Hamiltonian:out-of-tape,QTM_in_local_Hamiltonian:explicit-form} of \cref{QTM_in_local_Hamiltonian}, this will never occur.) Consider the corresponding $\ket{\psi_t}$ in the computational history state. The QTM transition rules do not alter Track~6, so Track~6 is still in its initial configuration. Since the QTM has deterministic head movement, at no point during the evolution is its head ever in a superposition of locations. $\ket{\psi_t}$ is therefore of the form:
\begin{equation}
  \ket{\psi_t} =
    \ket{\sleftend} \left( \sum_q
      \ket[\text{Track~4}]{q\;\blankR\dots\blankR}
      \ox\ket[\text{Track~5}]{\varphi_q}
      \ox\ket[\text{Track~6}]{\vdash_q\#\dots\#}
    \right) \ket{\srightend},
\end{equation}
where $\ket{\varphi_q}$ are unnormalised Track~5 QTM tape states, and all configurations in the superposition would step the QTM head left in the next time-step. The $q_f$ transition rule from \cref{eq:QTM_qf_transition_rule} therefore applies to all states in the superposition, and the next state must have the form:
\begin{equation}
  \ket{\psi_{t+1}} =
    \ket{\sleftend} \left( \sum_q
      \ket[\text{Track~4}]{p_\alpha\;\blankR\dots\blankR}
      \ox\ket[\text{Track~5}]{\varphi_q}
      \ox\ket[\text{Track~6}]{\vdash_q\#\dots\#}
    \right) \ket{\srightend}.
\end{equation}
Thenceforth, only time-wasting counter TM transition rules apply.

A very similar argument applies if the QTM reaches a configuration in which the head moves beyond cell $L$. Since the time-wasting counter TM transitions never alter Track~5, \cref{QTM_in_local_Hamiltonian:out-of-tape} of \cref{QTM_in_local_Hamiltonian} follows.

This concludes the proof of \cref{QTM_in_local_Hamiltonian}, and this section.


\clearpage

\section{Quasi-periodic tilings}
\label{sec:quasi-periodic}
In order to ultimately prove the required spectral properties of our final Hamiltonian, we require significantly stronger properties of quasi-periodic tilings than those used to prove tiling results. In particular, we will need strong ``rigidity'' results, which show that the quasi-periodic structure is in a sense robust to errors. As far as we are aware, these rigidity results are new. All the tiling results used in the proof of our main results are gathered in this section.

We exploit the very particular properties of a quasi-periodic tiling due to Robinson, which we briefly review in the following section. The Robinson tiling has a hierarchical geometric structure that lends itself to inductive proofs of the required rigidity results.

\subsection{Robinson's tiling}\label{sec:Robinson}
We now describe Robinson's quasi-periodic tiling. It was discovered by Robinson in 1971 \cite{Robinson} as a tool to simplify Berger's proof of the undecidability of the tiling problem \cite{Berger}.

\begin{figure}[hbtp]
  \centering
  \includegraphics[width=0.7\columnwidth]{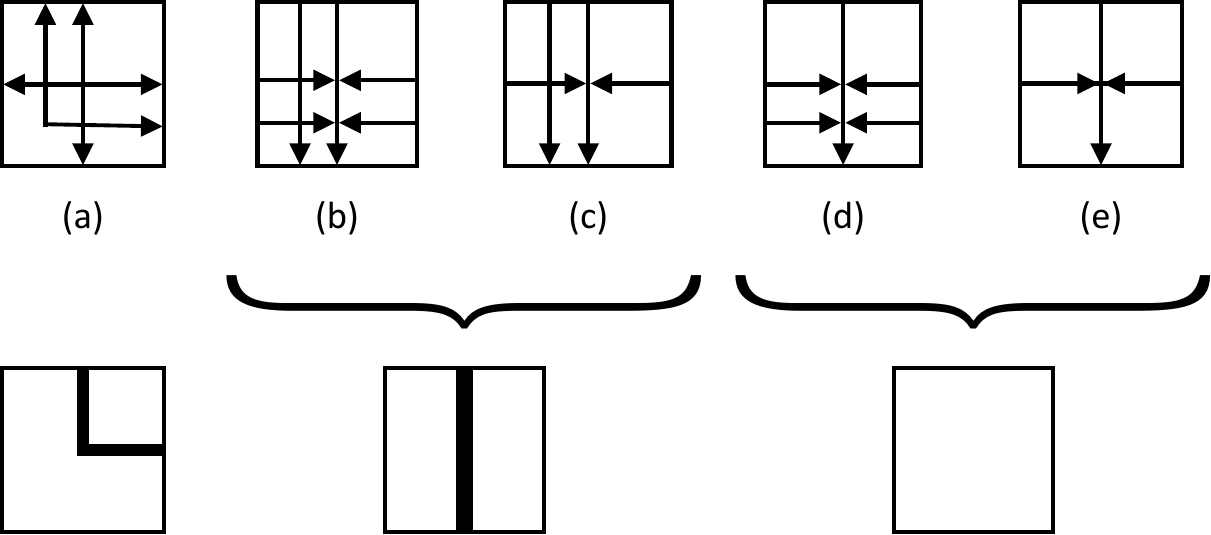}\\
  \caption[The five basic tiles of Robinson's tiling.]{The five basic tiles of Robinson's tiling (top), and a simplified schematic representation of these used in \cref{fig:tiling} (bottom).}
  \label{fig:Robinson-tiles}
\end{figure}

\begin{figure}[hbtp]
  \centering
  \includegraphics[width=0.8\columnwidth]{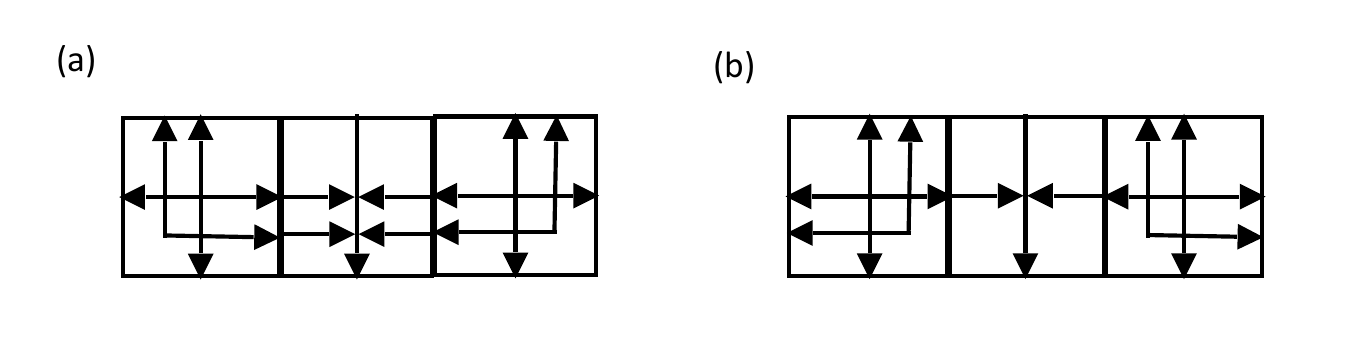}\\
  \caption{(a) Facing crosses, and (b) back-to-back crosses in Robinson's tiling.}
  \label{fig:cross-orientation}
\end{figure}

Robinson's tiling is based on the five basic tiles showed in \cref{fig:Robinson-tiles} and all rotations and reflections thereof. Tile~(a) is called a \textit{cross}, whereas tiles~(b)\nobreakdash--(e) are called \textit{arms}. In a valid tiling, arrow heads must meet arrow tails. As drawn, cross~(a) is said to face up/right. With this orientation convention, two consecutive crosses (meaning with no other cross in between them) in the same row or column are said to be facing each other, or to be back-to-back, as shown in \cref{fig:cross-orientation}.  An arm is said to point in the direction of its unique complete central arrow. Arrows that do not start or end in the mid-point of a square's side are called \textit{side arrows}.

\begin{figure}[hbtp]
  \centering
  \includegraphics[width=0.6\columnwidth]{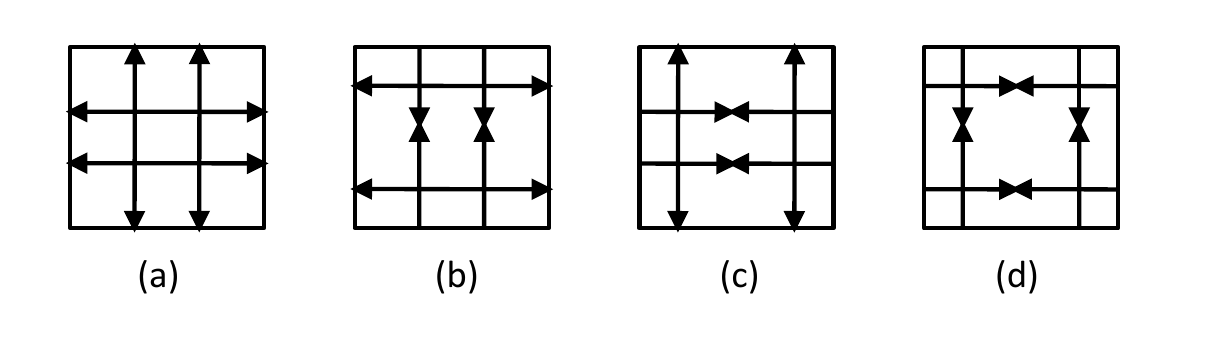}\\
  \caption{Parity markings to force the position of the crosses.}
  \label{fig:parity-markings}
\end{figure}

For each one of the five basic tiles of \cref{fig:Robinson-tiles}, we introduce two possible colours, red or green, to each of the side arrows (note that there is one tile without colour) with the following restrictions: colours must match between adjacent side arrows; at most one colour may be used horizontally and at most one colour may be used vertically; for the cross, the same colour must be used in both directions; for tile~(b) one colour must be used horizontally and the other colour vertically. Moreover, we impose the restriction that green crosses must appear in alternate positions in alternate rows (and potentially in other positions too).

The latter can be achieved by adding parity markings to the above tiles: the additional arrows showed in \cref{fig:parity-markings}, which should match properly in any valid tiling, with the following rules: parity marking~(a) is associated with green crosses, parity marking~(b) with horizontal green arrows, and~(c) with vertical green arrows. Parity marking~(d) is associated with all tiles. It is important to remark here that parity markings ``live in a different layer'' than the five basic tiles. That is, arrows from the parity markings must match only those from the parity markings and arrows from the basic tiles must match only those from the basic tiles. Note that the parity markings \cref{fig:parity-markings} already form a closed set under rotation and reflection---(c)~is the rotation of (b)---in contrast to the five basic tiles of \cref{fig:Robinson-tiles} which are extended to the full set by including all rotations and reflections.

\begin{figure}[hbtp]
  \centering
  \includegraphics[width=.7\columnwidth]{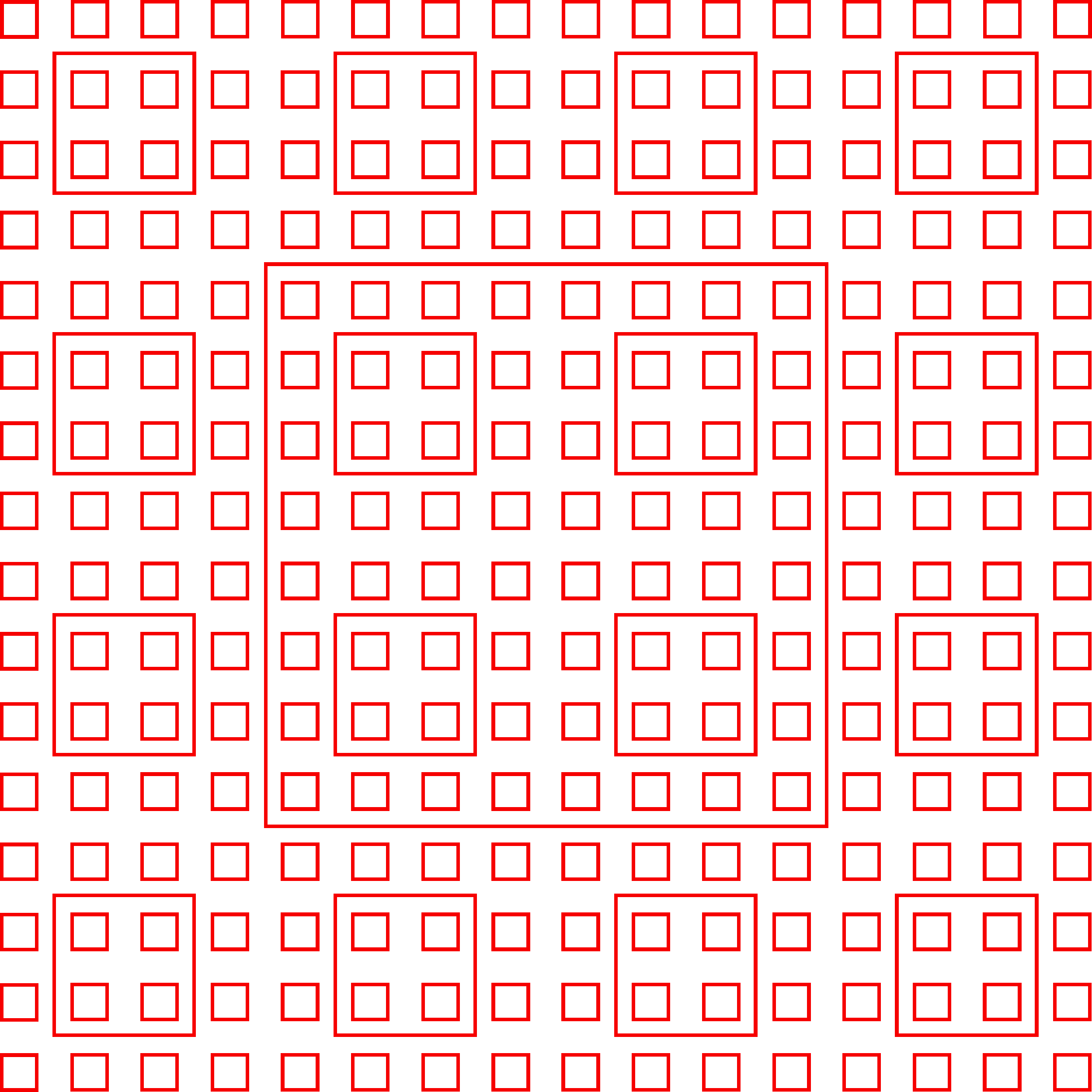}
  \caption{A possible Robinson tiling of the plane.}
  \label{fig:tiling}
\end{figure}

If we now draw only the red coloured lines, one possible tiling of the plane looks like \cref{fig:tiling}, which has the crucial (for our purposes) quasi-periodic structure consisting of interlocking squares of increasing size.

\begin{figure}[hbtp]
\centering
  \begin{tabular}{cc}
    \includegraphics[width=0.5\columnwidth]{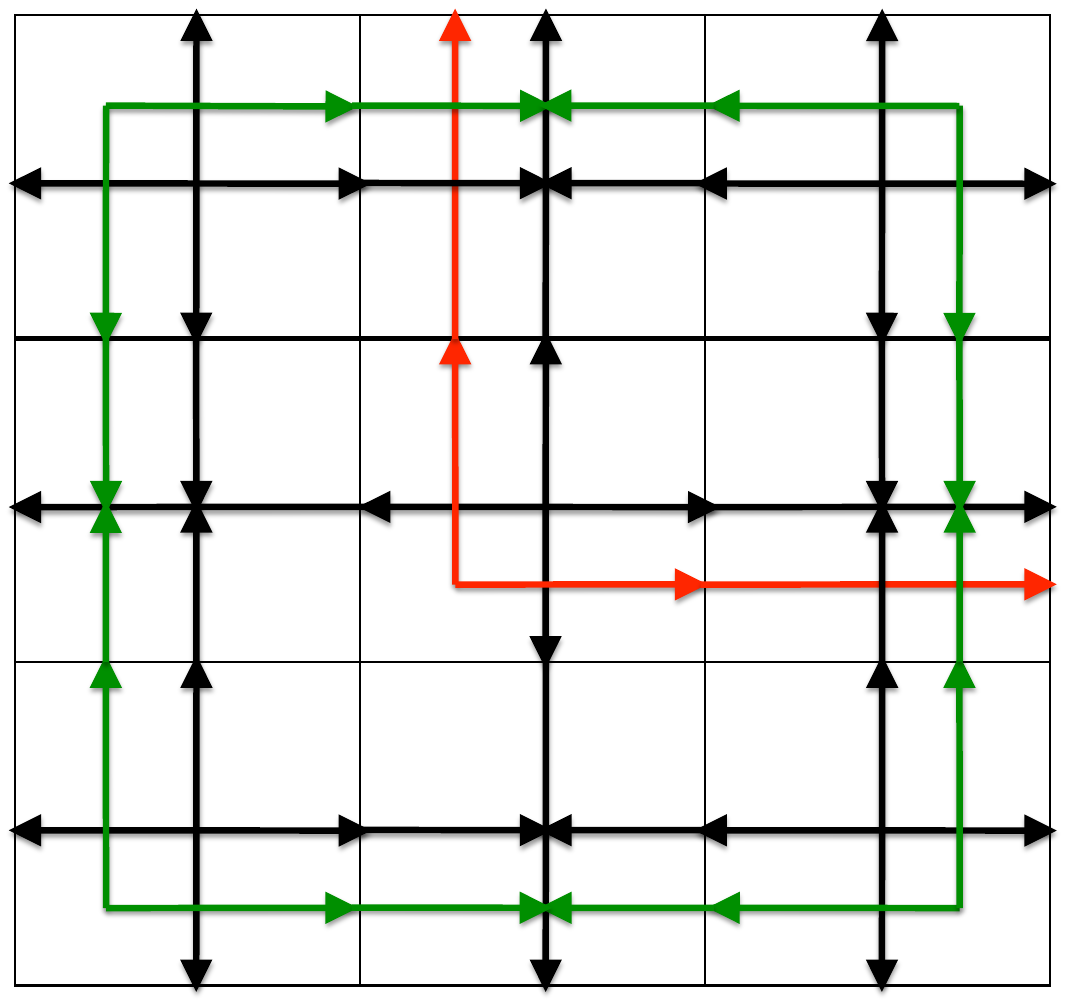} & \includegraphics[width=0.5\columnwidth]{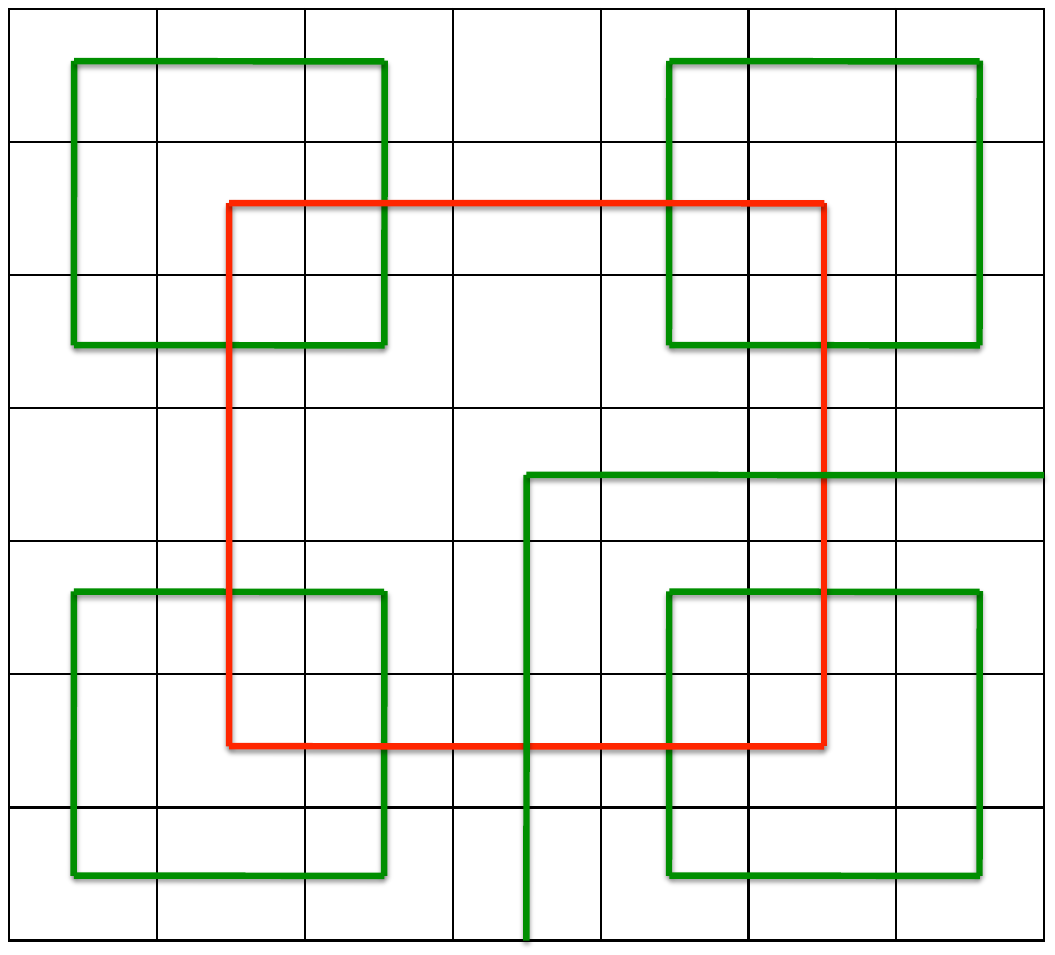}
  \end{tabular}
  \caption[Robinson tiling squares.]{\textit{Left}: A $3\times 3$ square of the Robinson tiling, with the choice of a right/up facing cross in the middle tile. \textit{Right}: A $7\times 7$ square of the Robinson tiling, with the choice of a right/down facing cross in the middle tile. (To avoid confusion, we have only drawn the coloured lines of the tiles in this second figure.)}
  \label{fig:3-square}
\end{figure}

Let us try to explain briefly (full details can be found in \cite{Robinson}) how \cref{fig:tiling} emerges, and the remaining freedom it allows in constructing valid tilings of the plane. The fact that crosses must appear in alternate rows in alternate positions, together with the form of the five basic tiles, means that any given cross completely determines the structure of the $3\times 3$ square constructed in the direction it faces. For example, assume we start with a green cross facing left/down. Then we are forced to complete the $3\times 3$ square having the cross we started as the top/right corner, as shown in \cref{fig:3-square}. Furthermore, the central tile must be a red cross, with the only freedom being the choice of direction it faces.

Once the direction of the central cross is fixed, the $7\times 7$ square obtained by extending the $3\times 3$ square in the direction this red cross faces is fixed as shown in \cref{fig:3-square}. The green cross in the central position is forced, with the only freedom again being the choice of direction it faces. Continuing this procedure gives a tiling similar to that shown in \cref{fig:tiling}. It consists of a quasi-periodic structure of squares, formed by a continuous line of a single colour going through all the tiles on the perimeter of the square. These are called \keyword{borders} and have sizes $4^n$ (which---following Robinson's terminology---means that the distance between two of the facing crosses delimiting the border is $4^n+1$) repeated with period $2^{2n+1}$, for all $n\in \N$.

\begin{figure}[hbtp]
\centering
    \includegraphics[width=0.45\columnwidth]{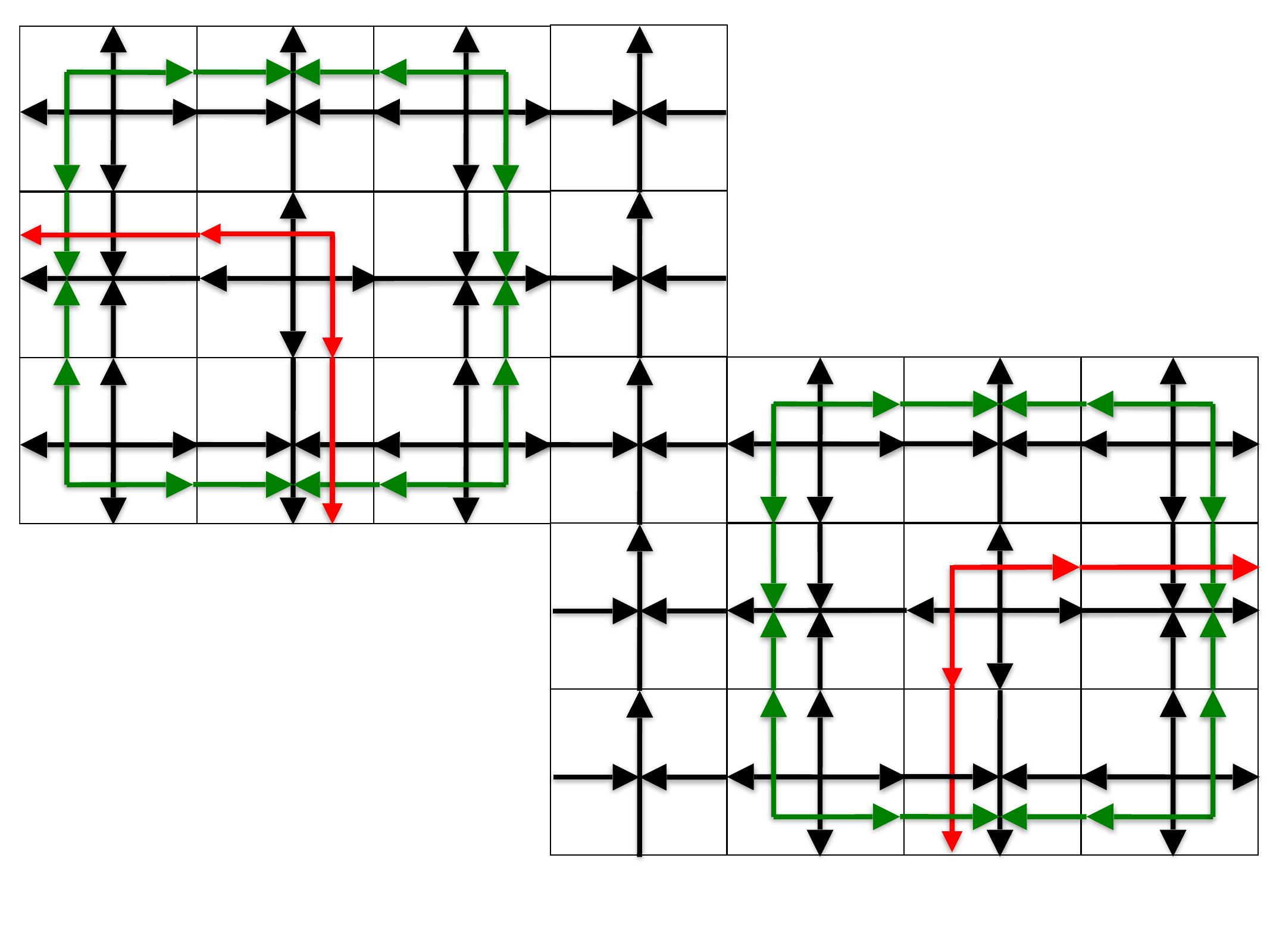}   \includegraphics[width=0.45\columnwidth]{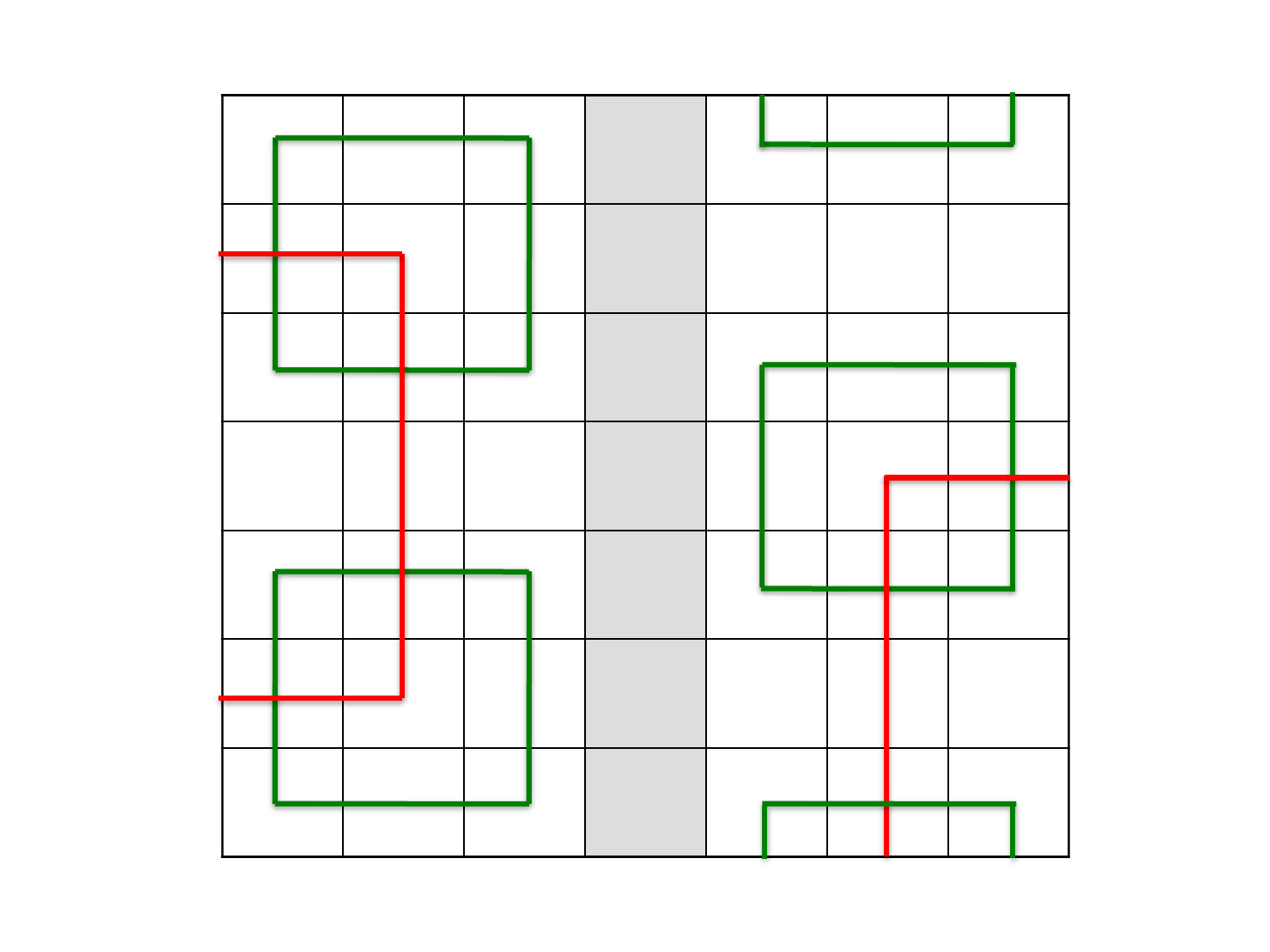}
  \caption{Left: an example of a fault line, in which the patterns are shifted. Right: a schematic way, with the fault line in grey, to complete the figure in the right to a $7\times7$ square (compare with the example of the $7\times 7$ square in \cref{fig:3-square}).}
  \label{fig:fault-line}
\end{figure}

Such a tiling can tile the whole plane, a half-plane or a quarter-plane, depending on how we sequentially choose the orientation of the central crosses. Lines between valid half or quarter planes must consist only of arms. If the tiling divides the plane into two half-planes, the patterns in these half-planes may be shifted relative to one another by an arbitrary even number of cells. In this case, the line separating the two half-planes is called a \keyword{fault} (see \cref{fig:fault-line}). However, if a half-plane is further divided into two quarter-planes, no further shift is possible between these quarter-planes.

\subsection{Rigidity of the Robinson tiling}\label{sec:rigidity}
We have just seen that with the original Robinson tiles, the quasi-periodic pattern of borders does not necessarily extend throughout the entire plane; there can be a fault line between two half-planes, with ``slippage'' between the patterns on either side of the fault (see \cref{sec:Robinson} and \cite[\S~8]{Robinson} and \cref{fig:fault-line}).

\begin{figure}[hbtp]
  \centering
  \includegraphics[width=.8\columnwidth]{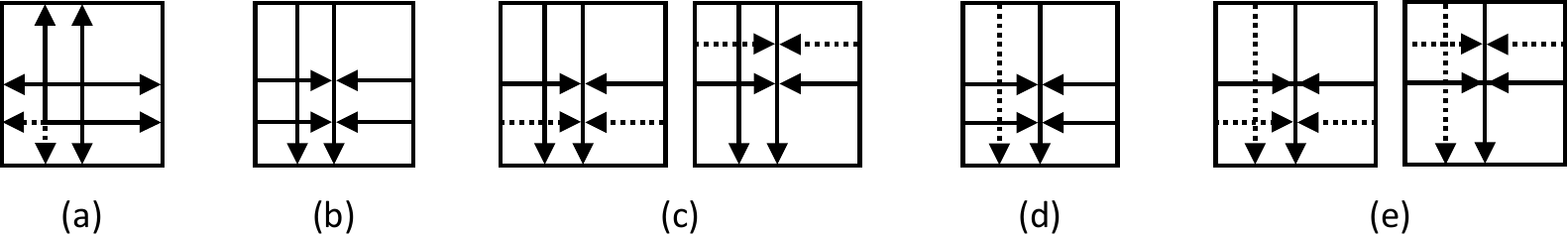}
  \caption{The basic tiles of the modified Robinson tiling.}
  \label{fig:modified_Robinson_tiles}
\end{figure}

It will be convenient to slightly modify the Robinson tiles in order to prevent fault lines. To this end, wherever one of the five basic tiles does not have side-arrows, we add dashed side-arrows parallel to the solid central arrows (see \cref{fig:modified_Robinson_tiles}). We then extend the five basic tiles to the full tile set by rotation and reflection, adding parity markings, and adding red and green colourings to the solid side arrows, exactly as in \cref{sec:Robinson}. We refer to these as the \keyword{modified Robinson tiles}. (Similar modifications to the Robinson tile set appear elsewhere in the literature, e.g.\ in \cite{Miekisz}.)

Since the all original tile markings are still present, any tiling using the modified tiles must also be a valid tiling using the original tiles if the dashed side-arrows are ignored. It is easy to verify that, for any tiling in which the $2^n$\nobreakdash-borders repeat periodically throughout the plane for all $n$ (i.e.\ any tiling that does not contain a fault line), dashed side-arrows can be added to make it into a valid tiling using the modified tiles.

However, if the tiling contains a fault line, there is no way to consistently add dashed side-arrows. To see this, assume without loss of generality that the fault line is vertical. Then there must exist two back-to-back crosses on either side of the fault that are not aligned, so that one points up and the other down (see \cref{fig:fault-line}). Consider the cross to the left of the fault line. The dashed side-arrow on this cross that points to the right forces a dashed arm to extend in that direction, towards the other cross. Similarly, the left-pointing dashed side-arrow on the cross to the right of the fault forces a dashed arm to extend to the left. These dashed arms must meet somewhere between the two crosses. But, since one cross points up and the other down, one of these arms has the dashed side-arrow on top, the other has it on the bottom, so there is no way to consistently join up the arms.

The crucial property we will use is that the Robinson tiling has a very ``rigid'' structure. Even if we start from a tile configuration that contains a defect (a pair of non-matching adjacent tiles), outside of a ball around the defect, the defect has no effect on the pattern of borders that appears in the rest of the tiling. The following \namecrefs{Robinson_rigidity} make this rigorous.\footnote{Our ``rigidity'' results for the Robinson tiling could certainly be strengthened, but they suffice for our purposes.}

\begin{lemma}[Cross rigidity]\label{crosses}
  In any tiling of a connected region with Robinson tiles, crosses must
  occur in alternate columns and in alternate rows.
\end{lemma}

\begin{proof}
  Immediate from the observation that, for each ``parity tile'' in
  \cref{fig:parity-markings}, its neighbour in each direction is uniquely
  defined.
\end{proof}

\begin{lemma}[Robinson rigidity]\label{Robinson_rigidity}
  Consider a connected region of a 2D square grid made up of square blocks of size $2^{n+1}\times 2^{n+1}$. Any tiling of such a region with modified Robinson tiles must contain a periodic pattern of $2^n$\nobreakdash-borders (some of which may be incomplete due to boundaries), repeating horizontally and vertically with period $2^{n+1}$. That is, the tiling must contain the same periodic pattern of $2^n$\nobreakdash-borders as a section of the tiling of the infinite plane, up to translation and defects that do not affect the $2^n$ borders.
\end{lemma}

\begin{proof}
  By \cref{crosses}, crosses must occur in alternate rows and columns throughout the region being tiled. Following \textcite{Robinson},  we refer to these crosses as ``1\nobreakdash-squares''. (Note that there may be additional crosses in the tiling that do not constitute 1\nobreakdash-squares.) Pick an orientation for one of these 1\nobreakdash-square crosses. Its solid side-arrows force a solid arm to extend in the directions it faces. Unless there is insufficient space to the boundary of the region, there must be a cross two cells away in the direction of each arm. Just as in \textcite[\S 3]{Robinson}, these solid arms force the orientation of any two facing crosses to be mirror images.

  Meanwhile, the dashed side-arrows force a dashed arm to extend in the directions that the cross faces \emph{away} from. Again, these arms force the orientations of any back-to-back crosses to be mirror images. Thus, as soon as we pick the orientation of a single 1\nobreakdash-square cross, the orientations of all adjacent crosses are forced, and hence the orientations of all 1\nobreakdash-square crosses throughout the connected region.

  \paragraph{Base case} Consider first the case $n=1$ with $4\times 4$ blocks. Since orientations of adjacent pairs of 1\nobreakdash-square crosses are mirror images of each other, they are forced to form groups of four facing towards the centre of a $3\times 3$ block. As in \textcite[\S 3]{Robinson}, these blocks must be completed to form ``3\nobreakdash-squares'' with 2\nobreakdash-borders running around the edges and crosses in the centre. (The orientation of the central crosses is not fixed.) In our modified Robinson tiles, the dashed side-arrows also force the orientations of \emph{back-to-back} crosses to be mirror images, which forces adjacent 3\nobreakdash-squares to be aligned. Thus the pattern of complete 2\nobreakdash-borders must repeat horizontally and vertically with period $4$ throughout the region.

  However, in general some of the 1\nobreakdash-squares will be too close to the boundaries of the region to form complete 3\nobreakdash-squares. We need to show that these still force partial 2\nobreakdash-borders. If a 1\nobreakdash-square faces a boundary and is located at that boundary, then there is no further section of 2\nobreakdash-border in that direction in any case, and we have nothing to show. If a 1\nobreakdash-square faces a boundary and is more than two cells away from it, there must be another 1\nobreakdash-square facing it before reaching the boundary, and the section of 2\nobreakdash-border between them must be completed as usual. The only interesting case is if a 1\nobreakdash-square faces a boundary one cell away from it. In this case, the solid side-arrows still force the cell between the cross and the boundary form an arm extending to the boundary. The only difference is that there is no facing cross in that direction to force the side-arrows of the arm to point inwards; they could instead point towards the boundary. Thus any of the arm tiles with solid side-arrows (in a suitable orientation) can be used to form this piece of the 2\nobreakdash-border. For our purposes, we consider any of these to still constitute a section of the 2\nobreakdash-border. So partial 2\nobreakdash-borders are still forced at the boundaries.

  This completes the proof for the base case $n=1$. It will be important for later to note that, if a partial 2\nobreakdash-border is large enough to contain the cell that would normally be at the centre of the border, then the central cross is still forced in that cell. To see this, note that because our region is made up of adjacent square blocks, there must be arms on at least two sides of the central cell (not necessarily opposite sides, if the partial 2\nobreakdash-square is located in a corner of the region). Regardless of which tile is used to form these arms, the central arrows perpendicular to the arms must face outwards as usual, because the solid side-arrows on the arms must be on the side away from the central cell. But the only tile that has outward-facing central arrows on adjacent sides is the cross, so it is forced.

  \paragraph{$\mathbf{n=2}$ case} Before proving the general case, it is instructive to consider the next case $n=2$ (with $8\times 8$ blocks). We can always divide each $8\times 8$ block into four $4\times 4$ blocks and apply the previous argument, so the 2\nobreakdash-borders must repeat horizontally and vertically with period~4, throughout the region, with a cross at the centre of each 2\nobreakdash-border (whether partial or complete). Each $8\times 8$ block must therefore contain at least one complete 2\nobreakdash-border, and all 2\nobreakdash-borders must be aligned (hence their central crosses are also aligned).

  Pick an orientation for one of these central crosses. The side-arrows of this cross force solid arms extending in the directions that the cross faces, and dashed arms extending in the directions it faces away from. Now, the only tile at which an arm can terminate is the cross. But there can be no crosses between adjacent 2\nobreakdash-borders. Thus the arms must extend all the way from a central cross to its neighbours, again forcing the orientation of both facing and back-to-back crosses to be mirror images of each other, so that they form groups of four facing towards the centre of a $5\times 5$ block. As in \textcite[\S 3]{Robinson}, these blocks must be completed to form ``5\nobreakdash-squares'' with 4\nobreakdash-borders running around the edges and crosses in the centre. As before, the dashed arms force adjacent 5\nobreakdash-squares to be aligned. So the pattern of complete 4\nobreakdash-borders repeats horizontally and vertically throughout the region, with period~8.

  It remains to consider crosses that are too close to the boundaries to form complete 4\nobreakdash-borders. If the central cross of a 2-border faces a boundary (rather than another cross), its solid side-arrows still force an arm that extends towards the boundary. The only tile that can terminate an arm is a cross. However, we cannot place a cross anywhere within the 3-square surrounding the central cross, as it would break the tiling within the 3-square. The only remaining possibility is to place a cross in the one-cell-wide corridor that runs along each side of a 3\nobreakdash-square. However, where an arm intersects a 2-border it must point outwards, which prevents placing a cross in the corridor. Thus the arm must be extended all the way to the boundary to form a partial 4-border (with some additional freedom in which arm tiles are used to form this section of 4-border).

  This proves the \namecref{Robinson_rigidity} for $n=2$. As in \cite[\S~3]{Robinson}, a cross is forced at the centre of all complete 4-borders. If a partial 4-border is large enough to surround the cell that would be at the centre of the complete border, a cross is forced there too. To see this, observe that for partial borders of this size, the centre cell has partial 4-borders on at least two sides. The central arrows perpendicular to a border necessarily have arrow tails on the inside of the border. These force an arm extending back towards the central cell. That cell therefore has arrow tails on at least two sides, and the central cross is forced.

  \paragraph{General case} The argument for general $n$ is very similar to the argument for the $n=2$ case, and proceeds by induction on $n$. Assume for induction that the pattern of $2^{n-1}$\nobreakdash-borders repeats horizontally and vertically with period $2^n$ in any connected region made up of $2^n\times 2^n$ blocks, and that a cross is forced at the centre of each of those $2^{n-1}$\nobreakdash-borders (whether partial or complete).

  Now consider a connected region made up of $2^{n+1}\times 2^{n+1}$ blocks. We can always divide each block into four $2^n\times 2^n$ blocks. So, by assumption, the pattern of $2^{n-1}$\nobreakdash-borders must repeat horizontally and vertically with period $2^n$ throughout the region, with a cross at the centre of each (partial or complete) $2^{n-1}$\nobreakdash-border. Each $2^{n+1}\times 2^{n+1}$ block therefore contains at least one complete $2^{n-1}$\nobreakdash-border, and adjacent borders (hence also their central crosses) are aligned. The solid and dashed side-arrows of these central crosses force arms extending between adjacent crosses, which in turn force the orientations of adjacent crosses to be mirror images. Thus the arms extending between facing crosses form $2^n$\nobreakdash-borders, and adjacent $2^n$\nobreakdash-borders are aligned. The pattern of $2^n$\nobreakdash-borders therefore repeats horizontally and vertically throughout the region, with period $2^{n+1}$.

  For central crosses that face a boundary instead of an adjacent cross, the solid side-arrows on the cross still force an arm extending towards the boundary. To show that this arm necessarily extends all the way to the boundary, we must show that it is impossible to place a cross anywhere along the arm. Within the $(2^n-1)$\nobreakdash-square surrounding the central cross, we cannot place a cross along the arm without breaking the tiling within  $2^n-1$\nobreakdash-square (cf.\ \cite[\S~3]{Robinson}). Thus the arm must extend to the boundary of this $(2^n-1)$\nobreakdash-square. Similarly, we cannot place a cross along the arm anywhere within  the adjacent $(2^n-1)$\nobreakdash-square. So, if the arm reaches the adjacent $(2^n-1)$\nobreakdash-square, it must continue onwards to the boundary. The only remaining possibility is to place a cross in the one-cell-wide corridor between the adjacent $(2^n-1)$\nobreakdash-squares. However, the arm necessarily points out of the $(2^n-1)$\nobreakdash-square, which prevents placing a cross in the corridor.

  Finally, for the induction step we must show that a cross is forced at the centre of each (complete or partial) $2^n$\nobreakdash-border. The argument applies to any border that extends sufficiently far to surround its centre cell (which includes complete borders). The central cell has (complete or partial) $2^n$\nobreakdash-borders on at least two sides. The central arrows perpendicular to a border necessarily have arrow tails pointing inwards, which force arms extending back towards the central cell. That cell therefore has arrow tails on at least two sides, forcing the central cross.

  Together with the base case $n=1$, this completes the proof.
\end{proof}

We refer to the top edge of a $2^n$\nobreakdash-border as a \emph{$2^n$\nobreakdash-segment}.\footnote{It is on these $2^n$\nobreakdash-segments that the QTMs will ultimately ``run'' in the final construction.}

\begin{lemma}[Segment bound]\label{defect-free}
  The  number of $2^n$\nobreakdash-segments in a tiling of an $L\times H$ rectangle (width $L$, height $H$) using modified Robinson tiles is $\ge \lfloor H/2^{n+1}\rfloor \bigl(\lfloor L/2^{n+1}\rfloor -1\bigr)$ and $\le \bigl(\lfloor H/2^{n+1}\rfloor +1 \bigr)\lfloor L/2^{n+1}\rfloor$ for all $n$.
\end{lemma}

\begin{proof}
  \Cref{Robinson_rigidity} implies that the tiling must contain the usual periodic pattern of $2^n$\nobreakdash-borders, up to translation. The result then is a trivial consequence of the fact that borders repeat vertically and horizontally with period $2^{n+1}$.
 \end{proof}

The following \namecref{segment_bound} is the key rigidity result that we will need later. It shows that a defect in the tiling (i.e.\ a non-matching pair of adjacent tiles) cannot affect the pattern of $2^n$\nobreakdash-borders in the tiling outside a ball of size $2^{n+1}$ centred on the defect.

\begin{lemma}[Segment rigidity]\label{segment_bound}
  In any tiling of an $L\times H$ rectangle (width $L$, height $H$) with $d$ defects using modified Robinson tiles, the total number of $2^n$\nobreakdash-segments is at least $\lfloor H/2^{n+1}\rfloor \bigl(\lfloor L/2^{n+1}\rfloor -1\bigr) - 2d$.
\end{lemma}

\begin{proof}
  Divide the $L\times H$ rectangle into contiguous square blocks of size $2^{n+1} \times 2^{n+1}$, with $\lfloor L/2^{n+1}\rfloor$ blocks in each row and $\lfloor H/2^{n+1}\rfloor$ in each column. (Any leftover cells play no further role in the argument.) Delete any block that contains at least one defect (arbitrarily assigning defects that occur at block boundaries to the bottom/left block). Consider any horizontal row of blocks. If the row contains $d_i$ deleted blocks, those missing blocks divide it into at most $d_i+1$ strips of height $2^{n+1}$.

  Since each strip is a connected, defect-free region made up of $2^{n+1} \times 2^{n+1}$ blocks, \cref{Robinson_rigidity} implies that it must be tiled with the usual periodic pattern of $2^n$\nobreakdash-borders, up to translations. The strip may be connected to strips in adjacent rows, in which case we are not free to translate the tiling in each strip independently. Nonetheless, we can lower-bound the number of $2^n$\nobreakdash-segments by allowing the tiling within each strip to be translated independently, and minimising over all such translations.

  Each strip has height $2^{n+1}$, so, regardless of how the pattern is translated vertically, it must contain exactly one row of $2^n$\nobreakdash-segments. The minimum number of segments within a strip of block-length $l$ is $l-1$ (obtained by translating the pattern horizontally so that there are incomplete segments at the beginning and end of the strip). In other words, each block in the strip effectively contributes one segment to the total, except for the block at the right end of the strip, which does not contribute (we assign the segment between adjacent blocks to the left one).

  If the row contains $d_i$ deleted blocks, those missing blocks divide it into at most $d_i+1$ strips.\footnote{For $d_i>n/2$, this substantially over-counts the number of strips, but it is sufficient for our purposes.} Thus the row contains $\lfloor L/2^{n+1}\rfloor - d_i$ non-deleted blocks, with at most $d_i+1$ of them located at ends of the strips. The whole row therefore contains at least $\lfloor L/2^{n+1}\rfloor - d_i - \bigl(d_i+1\bigr) = \lfloor L/2^{n+1}\rfloor - 2d_i - 1$ segments.

  Summing over all $\lfloor H/2^{n+1}\rfloor$ rows, and noting that the total number of deleted blocks cannot be more than the total number of defects, we obtain the claimed lower bound on the total number of $2^n$\nobreakdash-segments.
\end{proof}


\clearpage
\section{Putting it all together}
\label{sec:put-together}

\subsection{Undecidability of the g.s.\ energy density}
\label{sec:gs-energy}
To prove our first main result, \cref{thm:gs_density}, we will prove a sequence of lemmas which allow us to combine together the Hamiltonian constructions from the previous sections, and progressively build up the final Hamiltonian.

We will repeatedly need to refer to 1D translationally-invariant Hamiltonians with a particular set of properties. For conciseness, we will call these ``Gottesman-Irani Hamiltonians'', captured in the following definition:

\begin{definition}[Gottesman-Irani Hamiltonian]\label{def:G-I_Hamiltonian}
  Let $\C^Q$ be a finite-dimensional Hilbert space with two distinguished orthogonal states labelled $\ket{\sleftend}$, $\ket{\srightend}$. A \emph{Gottesman-Irani Hamiltonian} is a 1D, translationally-invariant, nearest-neighbour Hamiltonian $H_q(r)$ on a chain of length $r\geq 2$ described by Hilbert space $(\C^Q)^{\ox r}$ with local interaction $h_q\in\cB(\C^Q\ox\C^Q)$, which satisfies the following properties:
  \begin{enumerate}
  \item $h_q\geq 0$.
  \item $\comm{h_q}{\proj{\sleftend}\ox\proj{\sleftend}} = \comm{h_q}{\proj{\sleftend}\ox\proj{\srightend}} = \comm{h_q}{\proj{\srightend}\ox\proj{\sleftend}} = \comm{h_q}{\proj{\srightend}\ox\proj{\srightend}} = 0$.
  \item $\lambda_0(r) := \lambda_0(H_q(r)|_\Sbr) < 1$, where $\Sbr$ is the subspace of states with fixed boundary conditions $\ket{\sleftend},\ket{\srightend}$ at the left and right ends of the chain, respectively. \label[condition]{G-I_Hamiltonian:bracketed}
  \item $\forall n\in\N: \lambda_0(4^n)\geq 0$ and $\sum_{n=1}^\infty\lambda_0(4^n) < 1/2$. \label[condition]{G-I_Hamiltonian:upper-bound}
  \item $\lambda_0(H_q(r)|_{S_<}) = \lambda_0(H_q(r)|_{S_>}) = 0$, where $S_<$ and $S_>$ are the subspaces of states with, respectively, a $\ket{\sleftend}$ at the left end of the chain or a $\ket{\srightend}$ at the right end of the chain. \label[condition]{G-I_Hamiltonian:0-energy}
  \end{enumerate}
\end{definition}

\begin{lemma}[Tiling + quantum layers]\label{tiling+quantum}
  Let $h_c^{\mathrm{row}},h_c^{\mathrm{col}}\in\cB(\C^C\ox\C^C)$ be the local interactions of a 2D tiling Hamiltonian $H_c$, with two distinguished states (tiles) $\ket{L},\ket{R}\in\C^C$. Let $h_q\in\cB(\C^Q\ox\C^Q)$ be the local interaction of a Gottesman-Irani Hamiltonian $H_q(r)$, as in \cref{def:G-I_Hamiltonian}. Then there is a Hamiltonian on a 2D square lattice with nearest-neighbour interactions $h^\mathrm{row},h^\mathrm{col}\in\cB(\C^{C(Q+1)}\ox\C^{C(Q+1)})$ with the following properties: For any region of the lattice, the restriction of the Hamiltonian to that region has an eigenbasis of the form $\ket[c]{T}\ox\ket[q]{\psi}$, where $\ket[c]{T}$ is a \emph{product} state representing a classical configuration of tiles. Furthermore, for any given $\ket[c]{T}$, the lowest energy choice for $\ket[q]{\psi}$ consists of ground states of $H_q(r)$ on segments between sites in which $\ket[c]{T}$ contains an $\ket{L}$ and an $\ket{R}$, a 0-energy eigenstate on segments between an $\ket{L}$ or $\ket{R}$ and the boundary of the region, and $\ket{0}$'s everywhere else.
\end{lemma}

\begin{proof}
  The idea is to sandwich the two Hamiltonians $H_c$ and $H_q$ together in two ``layers'', so that the overall Hamiltonian acts as $H_c$ on the $c$-layer, with constraints between the layers that force low-energy configurations of the $q$-layer to be in the auxiliary $\ket{0}$ ``blank'' state, \emph{except} between pairs of $\ket{L}$ and $\ket{R}$ states appearing in the same row of the $c$-layer, where the $q$-layer acts like $H_q$ on that line segment.

  To this end, define the local Hilbert space to be $\HS := \HS_c\ox(\HS_e\oplus\HS_q) \simeq \C^C\ox(\ket{0}\oplus\C^Q)$. The Hamiltonian $H$ is defined in terms of the two-body interactions as follows:
  \begin{subequations}\label{TQ:overall_H}
  \begin{align}
    h^{\mathrm{col}}_{j,j+1} =
      \label[term]{TQ:cols} &h_c^{\mathrm{col}}\ox \id_{eq}^{(j)} \ox \id_{eq}^{(j+1)}\\
    h^{\mathrm{row}}_{i,i+1} =
      \label[term]{TQ:Hc}   &h_c^{\mathrm{row}}\ox\id_{eq}^{(i)}\ox\id_{eq}^{(i+1)}\\
      \label[term]{TQ:Hq}   &+\id_c^{(i)}\ox\id_c^{(i+1)}\ox h_q\\
      \label[term]{TQ:<L}   &+\proj[c]{L}^{(i)} \ox (\id_{eq}-\proj{\sleftend})^{(i)}
                              \ox \id_{ceq}^{(i+1)}\\
      \label[term]{TQ:L<}   &+(\id_c- \proj[c]{L})^{(i)} \ox \proj{\sleftend}^{(i)}
                              \ox \id_{ceq}^{(i+1)}\\
      \label[term]{TQ:>R}   &+\id_{ceq}^{(i)} \ox \proj[c]{R}^{(i+1)}
                              \ox (\id_{eq} - \proj{\srightend})^{(i+1)}\\
      \label[term]{TQ:R>}   &+\id_{ceq}^{(i)}
                              \ox (\id_c-\proj[c]{R})^{(i+1)}
                              \ox\proj{\srightend}^{(i+1)}\\
      \label[term]{TQ:1R}   &+\id_c^{(i)} \ox \proj[e]{0}^{(i)}
                              \ox \proj[c]{R}^{(i+1)} \ox \id_{eq}^{(i+1)}\\
      \label[term]{TQ:L1}   &+\proj[c]{L}^{(i)} \ox \id_{eq}^{(i)}
                             \ox\id_c^{(i+1)} \ox \proj[e]{0}^{(i+1)}\\
      \label[term]{TQ:10-a} &+\id_c^{(i)} \ox \proj[e]{0}^{(i)}
                              \ox (\id_c-\proj[c]{L})^{(i+1)}
                              \ox (\id_{eq}-\proj[e]{0})^{(i+1)}\\
      \label[term]{TQ:10-b} &+(\id_c-\proj[c]{R})^{(i)}
                              \ox (\id_{eq}-\proj[e]{0})^{(i)}
                              \ox\id_c^{(i+1)} \ox \proj[e]{0}^{(i+1)},
  \end{align}
  \end{subequations}
  where $\id_c$, $\id_{eq}$ and $\id_{ceq}$ are the identity operators on the corresponding Hilbert spaces. Operators living on $\HS_q$, or $\HS_q\otimes \HS_q$, such as $h_q$ in \cref{TQ:Hq}, are assumed to be extended in the trivial way (they are defined as zero in the extra sectors) to the larger space $\HS_q \oplus \HS_e$, or $(\HS_q \oplus \HS_e)\otimes (\HS_q \oplus \HS_e)$ .

  The Hamiltonian can be understood as follows. \Cref{TQ:L<,TQ:<L} force a $\ket{\sleftend}$ in the $q$-layer whenever there is an $\ket{L}$ in the $c$-layer. \Cref{TQ:>R,TQ:R>} do the same with $\ket{\srightend}$ and $\ket{R}$. \Cref{TQ:1R,TQ:L1} force non-blank to the left and right of an $\ket{R}$ or $\ket{L}$, respectively. Finally, \cref{TQ:10-a,TQ:10-b} force a non-blank to the left and right of any other non-blank in the $q$-layer, except when a non-blank coincides with an $\ket{L}$ or $\ket{R}$ in the $c$-layer.

  Since $h_c$ is a tiling Hamiltonian, \cref{TQ:Hc} is by assumption diagonal in the canonical product basis on $\HS_c^{\ox 2}$ (and acts trivially on $(\HS_q\oplus\HS_e)^{\ox 2}$). Meanwhile, \cref{TQ:Hq} is block-diagonal with respect to the four one-dimensional subspaces of $\HS_q^{\ox 2}$ spanned by products of $\ket{\sleftend}$ and $\ket{\srightend}$, together with the orthogonal complement thereof. (\Cref{TQ:Hq} acts trivially on $\HS_c^{\ox 2}$.) The remaining terms are manifestly block-diagonal with respect to both of these decompositions simultaneously. The overall Hamiltonian is therefore block-diagonal w.r.t.\ the product basis on the $c$-layer. Hence there is a basis of eigenstates of $H$ of the form $\ket[c]{T}\ket[q]{\psi}$, where $\ket[c]{T}$ is a product state in the canonical basis of the $c$-layer. This proves the first claim of the \namecref{tiling+quantum}.

  For a given classical tile configuration $\ket[c]{T}$ on the $c$-layer, let $\mathcal{L}$ denote the set of all horizontal line segments $\ell$ that lie between an $\ket{L}$ and an $\ket{R}$ (inclusive) in the classical configuration $\ket[c]{T}$. Let $\mathcal{L}_L$ denote the set of all horizontal line segments between an $L$ and the right boundary of the region, and similarly $\mathcal{L}_R$ the segments between the left boundary and an $R$.

  Consider first a state $\ket[q]{\psi_0}$ consisting of the ground state of $H_q(\ell)$ in the $q$-layer for each $\ell\in \mathcal{L}$, a 0-energy eigenstate of $H_q(\ell)$ in the $q$-layer for each $\ell\in\mathcal{L}_L\cup\mathcal{L}_R$, and $\ket{0}$ everywhere else in the $q$-layer. The associated energy $\bra[c]{T}\bra[q]{\psi_0}|H|\ket[c]{T}\ket[q]{\psi_0}$ is clearly $\braket{T|H_c|T} + \sum_\mathrm{\ell\in\mathcal{L}} \lambda_0(\abs{\ell})$. Indeed $\braket{T|H_c|T}$ is the contribution of \cref{TQ:cols,TQ:Hc}, and $\sum_\mathrm{\ell\in\mathcal{L}} \lambda_0(\abs{\ell})$ the contribution of \cref{TQ:Hq}, the rest of the terms being~$0$ as $\ket[c]{T}\ket[q]{\psi_0}$ satisfies all constraints imposed by \crefrange{TQ:<L}{TQ:10-b}.

  To see that this is the lowest energy state for a given classical configuration on the $c$-layer, we define a signature $\sigma$ for each state in the canonical basis of the $q$-layer. The local Hilbert space at site $i$ in the $q$-layer is $\HS_e\oplus\HS_q$, so we assign $\sigma_i=0$ for states in $\HS_e$ and $\sigma_i=1$ for states in $\HS_q$. By collecting computational basis states with the same signature, we decompose $\HS_q$ into a direct sum of subspaces with given signature,
  \begin{equation}
    \bigotimes_{i\in\Lambda}\HS_q^{(i)}\simeq\bigoplus_{\sigma}\HS_\sigma.
  \end{equation}
  The overall Hamiltonian $H$ is block-diagonal with respect to this decomposition, so all eigenstates $\ket[c]{T}\ket[q]{\psi}$ have a $\ket[q]{\psi}$ part with well-defined signature. We distinguish two cases:

  \paragraph{Case 1:} $\sigma_i= 1$ for all $i\in \ell\in \mathcal{L}$. Consider a $q$-layer state $\ket[q]{\psi_\sigma}$ with this signature; that is, a state supported on the corresponding Hilbert space $H_\sigma$. Since all Hamiltonian terms \crefrange{TQ:<L}{TQ:10-b} are $\ge 0$, and for all $\ell\in \mathcal{L}$ the \cref{TQ:Hq} contribution is $\ge \lambda_0(\abs{\ell})$, we easily get that $\bra[c]{T}\bra[q]{\psi_\sigma} H \ket[c]{T}\ket[q]{\psi_\sigma} \ge \braket{T|H_c|T} + \sum_\mathrm{\ell\in\mathcal{L}} \lambda_0(\abs{\ell})$, with equality iff $\ket[q]{\psi_\sigma} = \ket[q]{\psi_0}$.

  \paragraph{Case 2:} There exists $i\in \ell\in \mathcal{L}$ such that $\sigma_i=0$. Let $j$ be the rightmost position in $\ell$ such that $\sigma_j=0$. If this is the right or left end of $\ell$, then the state picks up an energy contribution of~$1$ from \cref{TQ:<L} or \cref{TQ:>R}. If it is the position next to the right or left end, it picks up a contribution of~$1$ from from \cref{TQ:1R} or \cref{TQ:L1}. Finally, if it is not one of the above cases, it acquires a contribution of~$1$ from \cref{TQ:10-b}. In all cases, the contribution is $\geq \lambda_0(\abs{\ell})$ since $\lambda_0(\abs{\ell}) < 1$ by assumption for a Gottesman-Irani Hamiltonian (see \cref{def:G-I_Hamiltonian}). As all terms in the Hamiltonian apart from \cref{TQ:cols,TQ:Hc} are positive-semidefinite, and the contribution from \cref{TQ:cols,TQ:Hc} is $\braket{T|H_c|T}$, we have that $\bra[c]{T}\bra[q]{\psi_\sigma} H \ket[c]{T}\ket[q]{\psi_\sigma} \ge \braket{T|H_c|T} + \sum_\mathrm{\ell\in\mathcal{L}} \lambda_0(\abs{\ell})$, which completes the proof of the \namecref{tiling+quantum}.
\end{proof}

With this, we can now prove the following:
\begin{lemma}[Robinson + Gottesman-Irani Hamiltonian]
  \label{put-promise-together} \hfill\newline
  Let $h_c^{\mathrm{row}},h_c^{\mathrm{col}}\in\cB(\C^C\ox\C^C)$ be the local interactions of the tiling Hamiltonian associated with the modified Robinson tiles. For a given ground state configuration (tiling) of $H_c$, let $\mathcal{L}$ denote the set of all horizontal line segments of the lattice that lie between down/right-facing and down/left-facing red crosses (inclusive).
  Let $h_q\in\cB(\C^Q\ox\C^Q)$ be the local interaction of a Gottesman-Irani Hamiltonian $H_q(r)$, as in \cref{def:G-I_Hamiltonian}.

  Then there is a Hamiltonian on a 2D square lattice of width $L$ and height $H$ with nearest-neighbour interactions $h^\mathrm{row},h^\mathrm{col}\in\cB(\C^{C(Q+1)}\ox\C^{C(Q+1)})$ such that, for any $L,H$, the ground state energy $\lambda_0(H^{\Lambda(L\times H)})$ on a lattice of size $L\times H$ is contained in the interval
  \begin{equation}\label{eq:interval-0-defects}
\left[  \sum_{n=1}^{\lfloor\log_4(L/2)\rfloor}
      \left(
        \left\lfloor\frac{H}{2^{2n+1}}\right\rfloor
        \left(\left\lfloor\frac{L}{2^{2n+1}}\right\rfloor -1\right)
      \right)
      \lambda_0(4^n)\; ,\;   \sum_{n=1}^{\lfloor\log_4(L/2)\rfloor}
      \left(\left(
        \left\lfloor\frac{H}{2^{2n+1}}\right\rfloor +1\right)
        \left\lfloor\frac{L}{2^{2n+1}}\right\rfloor
      \right)
      \lambda_0(4^n)\right]
  \end{equation}
  \end{lemma}

\begin{proof}
  Construct the Hamiltonian $H$ as in \cref{tiling+quantum}, with the red down/right- and down/left-facing crosses from the modified Robinson tile set as the designated $\ket{L}$ and $\ket{R}$ states.

  The tiling and QTM Hamiltonians satisfy the requirements of \cref{tiling+quantum}, implying that the lowest energy for a given $c$-layer configuration is attained by putting $\ket{0}$'s in the $q$-layer everywhere except in the segments between an $\ket{L}$ and an $\ket{R}$, inclusive. In the modified Robinson tiling, these are exactly the $4^n$\nobreakdash-segments. Therefore, to bound the ground state energy, we can restrict our attention to classical configurations of Robinson tiles (not necessarily valid tilings) with an eigenstate of $h_q$ along each $4^n$\nobreakdash-segment.

  By \cref{defect-free}, the minimum energy of an eigenstate with a defect-free Robinson tiling on the $L\times H$ rectangle, $E(0 \text{ defects}) = \sum_{\ell\in\mathcal{L}}\lambda_0(\abs{\ell})$, is contained in the interval \cref{eq:interval-0-defects}. (Recall that we only use the red $4^n$\nobreakdash-segments to define line\nobreakdash-segments on which $h_q$ acts non-trivially in the ground state.)

  On the other hand, since each defect in the classical tile configuration contributes energy at least~1 from the $h_c$ term, \cref{segment_bound} implies that the energy of an eigenstate with $d$ defects on the $L\times H$ rectangle is at least
  \begin{equation}
    E(d \text{ defects})
    = d + \sum_{\ell\in\mathcal{L}} \lambda_0(\abs{\ell})
    \geq d + \mspace{-20mu} \sum_{n=1}^{\lfloor\log_4(L/2)\rfloor}
         \Biggl(
           \left\lfloor\frac{H}{2^{2n+1}}\right\rfloor
           \left(\left\lfloor\frac{L}{2^{2n+1}}\right\rfloor -1\right)
           - 2d
         \Biggr) \lambda_0(4^n).
  \end{equation}
  Since $\sum_{n=1}^\infty\lambda_0(4^n) < 1/2$ by assumption for a Gottesman-Irani Hamiltonian (see \cref{def:G-I_Hamiltonian}), for all $d>0$  this is in turn lower-bounded by
  \begin{equation}
    E(d \text{ defects}) \geq
    \sum_{n=1}^{\lfloor\log_4(L/2)\rfloor}
      \left(
        \left\lfloor\frac{H}{2^{2n+1}}\right\rfloor
        \left(\left\lfloor\frac{L}{2^{2n+1}}\right\rfloor -1\right)
      \right)
      \lambda_0(4^n)\;.
  \end{equation}
  The \nameCref{put-promise-together} follows.
\end{proof}

We can now apply this \namecref{put-promise-together} to construct a Hamiltonian $h_u$ with ground state energy that is undecidable even with a constant promise on the energy gap.

\begin{proposition}[Diverging g.s.\ energy]
  \label{diverging_Hu} \hfill\newline
  Choose any $\beta\in \Q$, as small as desired. There exists a family of interactions $h_u^{row}(n), h_u^{col}(n) \in \cB(\C^U\otimes \C^U)$ and $h_u^{(1)}(n) \in \cB(\C^U)$ with operator norm $\le 1/2$ and algebraic (hence computable) matrix entries, and there exist strictly positive (uncomputable) functions $\delta_1(n), \delta_2(n)$ such that either $\lambda_0(H_u^{\Lambda(L)}(n)) \le -L\; \beta/2$ for all $L$, or $\lambda_0(H_u^\Lambda(n)) \ge L^2\; \delta_2(n) - L\; \delta_1(n)$ for all $L\ge L_0(n)$ ($L_0(n)$ uncomputable), but determining which is undecidable.

  Moreover, the interactions can be taken to have the following form: $h^{(1)}_u(n)=-\beta(1+\alpha_2(n))\1$ where $\alpha_2(n)$ is an algebraic number, $h^{col}_u(n)$ is $\{0,\beta\}$-valued and independent of $n$ and
  \begin{equation}
    h^{row}_u(n)=\beta \left(A+e^{i\pi\varphi} B+e^{i\pi2^{-\abs{\varphi}}} C\right) + h.c.
  \end{equation}
  where $A\in \cB(\C^U\otimes \C^U)$ is independent of $n$ and has coefficients in $\Z+\frac{1}{\sqrt{2}}\Z$, and $B,C\in \cB(\C^U\otimes \C^U)$ are independent of $n$ and have coefficients in $\Z$. (Recall that $\varphi$ is defined as the rational number whose binary fraction expansion contains the digits of $n$ in reverse order after the decimal.)
\end{proposition}

\begin{proof}
  Let $h_{q\mathit{0}}$ be the Hamiltonian obtained by applying \cref{QTM_in_local_Hamiltonian} with $K=3$ to the QTM from \cref{phase-estimation_QTM} with a proper reversible universal TM dovetailed after it. The Hamiltonian $h_q(n)$ in \cref{put-promise-together} will then be $h_q(n) = h_{q\mathit{0}}(n) + \proj{\top}\ox\id + \id\ox\proj{\top}$, where $\ket{\top}$ is the halting state of the universal TM. Clearly, $h_q$ has the form given in \cref{QTM_in_local_Hamiltonian:explicit-form} of \cref{QTM_in_local_Hamiltonian}. We claim that this Hamiltonian is a Gottesman-Irani Hamiltonian according to \cref{def:G-I_Hamiltonian}.

  The only requirements in \cref{def:G-I_Hamiltonian} that are not immediate are those concerning the minimum eigenvalues restricted to various subspaces. Recall the construction of the computational history state Hamiltonian from \cref{QTM_in_local_Hamiltonian}. In the initialisation sweep, the $\arrRzero$ normally sweeps once from left-to-right along the chain, turns around at the $\rightend$ to become a $\arrLzero$, which sweeps once right-to-left back along the chain. The superpositions over the standard basis states in these sequences contain no illegal pairs as long as the other tracks are initialised correctly.
  However, if the $\rightend$ is missing, then when the $\arrRzero$ reaches the right end of the chain, there is no forward transition out of the resulting state. Thus the uniform superposition over the first part of the initialisation sweep, involving just the left-to-right sweep, is an eigenstate of $h_q$. Furthermore, it has 0~energy if the rest of the tracks are correctly initialised. Similarly, if the $\leftend$ is missing, there is no further forward transition once the $\arrLzero$ reaches the left end of the chain, and the superposition over the full initialisation sweep is a 0-energy eigenstate of $h_q$. Thus \cref{G-I_Hamiltonian:0-energy} of \cref{def:G-I_Hamiltonian} is satisfied.

  Let $\ket{\psi} = \frac{1}{\sqrt{T}}\sum_{t=1}^T\ket{\phi_t}\ket{\psi_t}$ be the normalised computational history state for the QTM, where $T = \Omega(\abs{\Sigma\times Q}^r)$ and $\ket{\psi_t}$ is the state encoding the $t$'th step of the computation. Note that $\ket{\psi}$ is a zero-energy eigenstate of $H_{q\mathit{0}}$, and at most one $\ket{\psi_t}$ can have support on the state $\ket{\top}$ that represents the halting state of the universal TM, by \cref{QTM_in_local_Hamiltonian}. For $r>2$, we have
  \begin{equation}
    \begin{split}
      \lambda_0(r) &\leq \braket{\psi|H_q(r)|\psi}
      = \braket{\psi | \left( \sum_i h_{q\mathit{0}}^{(i,i+1)}(n) + \proj[i]{\top}\ox\id_{i+1}
                       + \id_i\ox\proj[i+1]{\top} \right) | \psi}\\
      &= \sum_{t=1}^T \frac{1}{T}
         \braket{\psi_t | \left( \sum_i \proj[i]{\top}\ox\id_{i+1} + \id_i\ox\proj[i+1]{\top} \right)
           | \psi_t}
      \leq O\left(\frac{1}{\abs{\Sigma\times Q}^r}\right),
    \end{split}
  \end{equation}
  thus $\sum_{m=1}^\infty \lambda_0(4^m) < 1/2$. The remaining \cref{G-I_Hamiltonian:bracketed,G-I_Hamiltonian:upper-bound} of \cref{def:G-I_Hamiltonian} are therefore also satisfied. Hence $h_q(n)$ is a Gottesman-Irani Hamiltonian, as claimed.

  Let $\tilde{h}_u^{\text{row}}(n),\tilde{h}_u^{\text{column}}(n)$ be the local interactions obtained by applying \cref{put-promise-together} to $h_q(n)$. Let $N(n) := \max\{\norm{\tilde{h}^{row}_u(n)},\norm{\tilde{h}^{\text{column}}_u(n)}\}$, and fix any rational number $\beta \le \frac{1}{N(n)}$ for all $n$. Such a $\beta$ exists by the form of $h_q$ guaranteed by \cref{QTM_in_local_Hamiltonian:explicit-form} in \cref{QTM_in_local_Hamiltonian} and the definition of $\tilde{h}_u^{\text{row}}(n),\tilde{h}_u^{\text{column}}(n)$ based on $h_q$. Define the normalised local interactions $h_u^{\text{row}}(n) := \beta \tilde{h}_u^{\text{row}}(n)$, $h_u^{\text{column}}(n) := \beta \tilde{h}_u^{\text{column}}(n)$.

  For any $r\geq\abs{n}+6$, the QTM from \cref{phase-estimation_QTM} has sufficient tape and time to finish, and we can be sure that the reversible universal TM starts. Consider first the case in which the universal TM does \emph{not} halt on input $n$. In that case, for all $r\geq\abs{n}+6$ we have that $\lambda_0(r)=0$. Let $L_0(n)$ denote the minimal $L$ such that the modified Robinson tiling of $\Lambda(L)$ necessarily contains a $4^m$-segment of size $4^m \ge \abs{n}+6$. If we take $L\ge L_0(n)$, thanks to \cref{put-promise-together} we can bound the ground state energy for $H^{\Lambda(L)}(n)$:
  \begin{subequations}
  \allowdisplaybreaks
  \begin{align}
    \lambda_0(H^{\Lambda(L)})
    &\le  \beta\sum_{\substack{1\le r \le \abs{n}+6\\ r=4^m,\; m\in\N}}
        \left\lfloor\frac{L}{2r}\right\rfloor
        \left(\left\lfloor \frac{L}{2r}\right\rfloor+1\right) \lambda_0(r)\\[1em]
    &=\beta L^2\Biggl[
          \sum_{\substack{1\le r \le \abs{n}+6\\ r=4^m,\; m\in\N}}
          \frac{\lambda_0(r)}{4r^2}
        \Biggr]\;\;
        +\;\; \beta L\Biggl[
          \sum_{\substack{1\le r \le \abs{n}+6\\ r=4^m,\; m\in\N}}
          \frac{\lambda_0(r)}{2r}
          \left(1 - 2\rem\left(\frac{L}{2r}\right)\right)
        \Biggr] \notag\\
        &\mspace{50mu}
        -\beta \Biggl[
          \sum_{\substack{1\le r \le \abs{n}+6\\ r=4^m,\; m\in\N}}
          \lambda_0(r) \rem\left(\frac{L}{2r}\right)
          \left(1 - \rem\left(\frac{L}{2r}\right)\right)
        \Biggr]\\[1em]
    &=: \beta\;\left(L^2\alpha_2(n) + L\alpha_1(n,L) - \alpha_0(n,L)\right),
  \end{align}
  \end{subequations}
  where $\rem(x):=x-\lfloor x\rfloor$ denotes the fractional part of $x$. Since the number of terms in the sum is bounded by $\abs{n}$ and for any fixed $r$ the quantity $\lambda_0(r)$ is an eigenvalue of a finite-dimensional matrix, $\alpha_2(n)$ is an algebraic number. Moreover, $\alpha_2(n)\le \beta$.

  We now shift the ground state energy by adding $\beta \1\otimes \1$ to $\hrow$ and adding a 1-body term $-\beta\left(1+\alpha_2(n)\right)\1$. Overall, this adds the Hamiltonian $\sum_{\mathrm{rows}}\sum_{c\in\mathrm{cols}} (\beta \1_{c}\otimes \1_{c+1}) - \sum_{i\in \Lambda(L)} \beta(1+\alpha_2(n))\1_i = -\beta(L^2 \alpha_2(n) + L) \1_{\Lambda(L)}$. Note that this commutes with all the other terms in the Hamiltonian. After this shift, and using the fact that $\alpha_0(n,L)\ge 0$, the ground state energy in the non-halting case becomes
  \begin{equation}
    \lambda_0(H^{\Lambda(L)}) \le \beta L\left(\alpha_1(n,L) - 1\right) \le -\frac{\beta}{2}L\,,
  \end{equation}
  since
  \begin{equation}
    \alpha_1(n,L) =
      \sum_{\substack{1\le r \le \abs{n}+6\\ r=4^m,\; m\in\N}}
      \frac{\lambda_0(r)}{2r}
      \left(1 - 2\rem\left(\frac{L}{2r}\right)\right)
    \le \sum_{\substack{1\le r \le \abs{n}+6\\ r=4^m,\; m\in\N}}
      \frac{\lambda_0(r)}{2r}
    \le \frac{1}{2}.
  \end{equation}

  Consider now the case in which the universal TM \emph{does} halt on input $n$. Take $r$ of the form $4^m$ large enough so that the TM has sufficient time and tape to halt. Let $r_1(n)$ denote the minimal such $r$. The  computational history state encoding the evolution necessarily has support on $\ket{\top}$ and is the unique ground state of $h_{q\mathit{0}}\ge 0$. Thus $\lambda_0(r)>0$ and by \cref{put-promise-together} the ground state energy $\lambda_0(H^{\Lambda(L)})$ is, after the energy shift, lower-bounded by
  \begin{subequations}
  \allowdisplaybreaks
  \begin{align}
    \begin{split}
    \lambda_0(H^{\Lambda(L)}) &\geq
    -\beta L^2 \alpha_2(n) - \beta L + \beta\!\sum_{\substack{1\le r \leq \abs{n}+6\\ r=4^m,\; m\in\N}}
        \left\lfloor\frac{L}{2r}\right\rfloor
        \left(\left\lfloor \frac{L}{2r}\right\rfloor-1\right)
        \lambda_0(r) \\
      &\mspace{50mu}
      + \beta\!\sum_{\substack{r \geq r_1(n)\\ r=4^m,\; m\in\N}}
        \left\lfloor\frac{L}{2r}\right\rfloor
        \left(\left\lfloor \frac{L}{2r}\right\rfloor-1\right)
        \lambda_0(r)
    \end{split} \\
    \begin{split}
    &= \beta L^2 \left(\sum_{\substack{r \geq r_1(n)\\ r=4^m,\; m\in\N}}
          \frac{\lambda_0(r)}{4r^2}
      \right)\;\;
      - \;\; \beta L \left(1 + \!\!\sum_{\substack{|r|\le n+6 \text{ or } r \geq r_1(n)\\ r=4^m,\; m\in\N}}
          \frac{\lambda_0(r)}{2r}
          \left(2\rem\left(\frac{L}{2r}\right) + 1\right)
      \right) \\
      &\mspace{50mu}
      + \beta\left(
        \sum_{\substack{|r|\le n+6 \text{ or } r \geq r_1(n)\\ r=4^m,\; m\in\N}}\!\!
          \lambda_0(r) \rem\left(\frac{L}{2r}\right)
          \left(\rem\left(\frac{L}{2r}\right) + 1\right)
      \right)
    \end{split} \\[1em]
    &=: \beta\left(L^2\; \delta_2(n) - L \; \delta_1(n,L) + \delta_0(n,L)\right)\\
    &\ge \beta\left(L^2\; \delta_2(n) - L \; \delta_1(n)\right),
  \end{align}
  \end{subequations}
  where
  \begin{equation}
  \delta_1(n):=\left(1+ \sum_{\substack{|r|\le n+6 \text{ or } r \geq r_1(n)\\ r=4^m,\; m\in\N}}
        \frac{3\lambda_0(r)}{2r}\right)\ge  \delta_1(n,L)\;,
  \end{equation}
  and $\delta_2(n)>0$, since in the halting case $\lambda_0(r)>0$ for all $r\ge r_1(n)$.

  Summarising, the ground state energy of $H^{\Lambda(L)}$ in the non-halting case is bounded by
  \begin{equation}\label{eq:alpha}
    \lambda_0(H^{\Lambda(L)})\leq  -L\; \frac{\beta}{2}
  \end{equation}
  whereas in the halting case we have the bound
  \begin{equation}\label{eq:delta}
      \lambda_0(H^{\Lambda(L)}) \geq L^2\; \delta_2(n) - L\; \delta_1(n).
  \end{equation}
  The \namecref{diverging_Hu} follows from undecidability of the Halting Problem.
\end{proof}

The following corollary follows immediately by letting $L(n)$ be the minimal $L$ in \cref{diverging_Hu} such that, in the halting case, $L^2\delta_2(n) - L \delta_1(n) \geq 1$. This will be useful shortly.

\begin{corollary}[Undecidability of g.s.\ energy with promise]\label{promise_Hu}
  There exists a family of interactions $h_u^{\text{row}}(n), h_u^{\text{col}}(n)\in \cB(\C^U\ox\C^U)$ and $h_u^{(1)}(n)\in\cB(\C^U)$ with  operator norm $\le 1/2$ and algebraic (hence computable) matrix entries, and an (uncomputable) function $L(n)$, such that either $\lambda_0(H_u^{\Lambda(L)}(n)) \leq 0$ for all $L$, or $\lambda_0(H_u^\Lambda(n)) \ge 1$ for all $L\ge L(n)$, but determining which is undecidable. Moreover, the interactions can be taken to have the same form as in \cref{diverging_Hu}.
\end{corollary}

Undecidability of the ground state energy density is now immediate from \cref{diverging_Hu} by the definition $E_\rho := \lim_{L\rightarrow\infty}\lambda_0(H^{\Lambda(L)})/L^2$ of the ground state energy density. We restate this result here for convenience:

\makeatletter
\@ifundefined{r@thm:gs_density@cref}{}{
  \setcounter{tmpcounter}{\value{theorem}}
  \cref@getlabel{thm:gs_density}{\@tempa}
  \def\@tempb{\setcounter{theorem}}
  \expandafter\@tempb\expandafter{\@tempa}
  \addtheoremline{theorem}{Undecidability of g.s.\ energy density -- restated}
  \setcounter{theorem}{\value{tmpcounter}}}
  \addtocounter{theorem}{-1}
\makeatother

\begin{reptheorem}{thm:gs_density}
  Let $d\in\mathbb{N}$ be sufficiently large but fixed, and consider translationally-invariant nearest-neighbour Hamiltonians on a 2D square lattice with open boundary conditions, local Hilbert space dimension $d$, algebraic (hence computable) matrix entries, and local interaction strengths bounded by~1. Then determining whether $E_\rho=0$ or $E_\rho>0$ is an undecidable problem.
\end{reptheorem}

\subsection{From g.s.\ energy density to spectral gap}
\label{sec:spectral-gap}
\begin{figure}[hbtp]
  \centering
  \includegraphics[width=1.0\columnwidth]{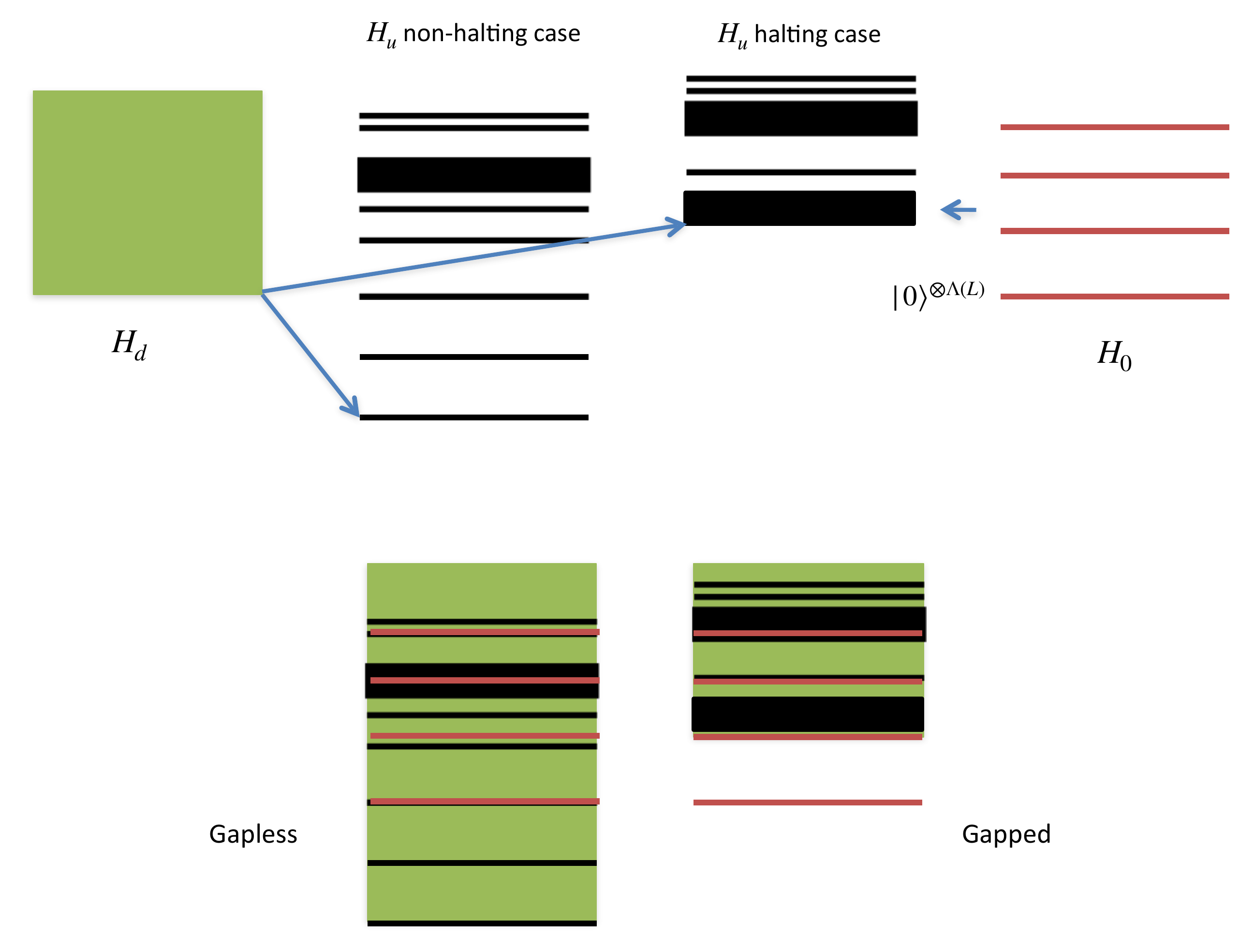}\\
  \caption{Starting with the Hamiltonian $H_u$ constructed in \cref{diverging_Hu}, we consider two additional Hamiltonians, $H_d$ and $H_0$, with dense and trivial spectrum respectively. These are combined into a final Hamiltonian $H$ such that the different spectra get combined as indicated by the arrows in the figure. This results in an overall Hamiltonian $H$ with gapped or gapless behaviour, as shown in the bottom figure, depending on whether the TM encoded in Hamiltonian $H_u$ halts or not. }
  \label{fig:gs-energy-to-spectral-gap}
\end{figure}

We are finally in a position to prove our main result: undecidability of the spectral gap, which we restate here for convenience.
The overall intuition behind the proof is illustrated in \cref{fig:gs-energy-to-spectral-gap}.

Recall that for each natural number $n$, we define $\varphi=\varphi(n)$ to be the rational number whose binary fraction expansion contains the digits of $n$ in reverse order after the decimal.\footnote{The reverse order being an unimportant artefact of the way we constructed the phase-estimation QTM in \cref{sec:phase-estimation}.}

\makeatletter
\@ifundefined{r@thm:promise@cref}{}{
  \setcounter{tmpcounter}{\value{theorem}}
  \cref@getlabel{thm:promise}{\@tempa}
  \def\@tempb{\setcounter{theorem}}
  \expandafter\@tempb\expandafter{\@tempa}
  \addtheoremline{theorem}{Undecidability of the spectral gap (restated)}
  \setcounter{theorem}{\value{tmpcounter}}}
  \addtocounter{theorem}{-1}
\makeatother

\begin{reptheorem}{thm:promise}\label{repthm:promise}
  For any given universal Turing Machine UTM, we can construct explicitly a dimension $d$, $d^2\times d^2$ matrices $A,A',B,C,D,D'$, a $d\times d$ diagonal projector $\Pi$ and a rational number $\beta$ which can be as small as desired, with the following properties:
  \begin{enumerate}
  \item $A$ is diagonal with entries in $\Z$.
  \item $A'$ is Hermitian with entries in $ \Z+ \frac{1}{\sqrt{2}}\Z$,
  \item $B,C$ have integer entries,
  \item $D$ is diagonal with entries in $\Z$,
  \item $D'$ is hermitian with entries in $\Z$.
  \end{enumerate}
  For each natural number $n$, define:
  \begin{align*}
    &\begin{aligned}
      &h_1(n)=\alpha(n)\Pi, &\qquad&  \text{$\alpha(n)\le 2\;\beta$ an algebraic number}\\
      &h_{\text{col}}(n)=D +\beta D', &\qquad& \text{independent of $n$}
    \end{aligned}\\
    &h_{\text{row}}(n)=A + \beta\left(A'+e^{i\pi\varphi} B + e^{-i\pi\varphi} B^\dg + e^{i\pi2^{-\abs{\varphi}}} C + e^{-i\pi2^{-\abs{\varphi}}} C^\dg\right).
  \end{align*}
  Then:
  \begin{enumerate}
  \item  The local interaction strength is bounded by~1, i.e.\ \linebreak $\max(\norm{h_1(n)}, \norm{h_{\text{row}}(n)}, \norm{h_{\text{col}}(n)}) \leq 1$.
  \item If UTM halts on input $n$, then the associated family of Hamiltonians $\{H^{\Lambda(L)}(n)\}$ is gapped in the strong sense of \cref{def:gapped} and, moreover, the gap $\gamma\ge 1$.
  \item If UTM does not halt on input $n$, then the associated family of Hamiltonians $\{H^{\Lambda(L)}(n)\}$ is gapless in the strong sense of \cref{def:gapless}.
  \end{enumerate}
\end{reptheorem}

\begin{proof}
  We assign a Hilbert space $\HS^{(i)} := \ket{0}\oplus\HS_u\ox\HS_d \simeq \C\oplus\C^U\ox\C^D$ to each site $i\in\Lambda$, where $\ket{0}$ denotes the generating element of $\C$. Let $h_u^{(i,j)}$ be the two-body interactions obtained in \cref{promise_Hu} (keeping the one-body term $h^{(i)}_u(n)$ separate, for later convenience).

  Let $h_d$ be the two-body interaction of any nearest-neighbour Hamiltonian $H_d$ with 0~ground state energy whose spectrum becomes dense in the thermodynamic limit (cf.\ \cref{def:gapless}): $\lim_{L\rightarrow\infty}\spec H_d^{\Lambda(L)} \rightarrow [0,\infty)$. (For example~\cite{LiebSchultzMattis}, we can take $D=2$ and $h_d$ to be the critical XY-model $h_{\text{row}} = \sigma_x\ox\sigma_x + \sigma_y\ox\sigma_y + \sigma_z\ox\id + \id\ox\sigma_z$ along the rows, where $\sigma_{x,y,z}$ are the Pauli matrices, with no interactions along the columns.) We normalise the interaction strength such that $\norm{h_d}\le \frac{1}{2}$.

  Define the Hamiltonian $H(n)$ in terms of its two-body and one-body interactions $h(n)$ as follows (with $\alpha(n) = \alpha_2(n)$, $\Pi = \Pi_{ud}$):
  \begin{subequations}\label{eq:promise_H}
  \begin{align}
    h^{(i,j)}(n) :=
      &\proj{0}^{(i)}\ox\Pi_{ud}^{(j)}+ \Pi_{ud}^{(i)}\ox\proj{0}^{(j)}
        \label[term]{eq:promise_boundary}\\
          &+ \1_{\vphantom{d}u}^{(i)}\ox\1_u^{(j)}\ox h_d^{(i,j)}\;,
        \label[term]{eq:promise_hq}\\
          &+ h_{\vphantom{d}u}^{(i,j)}(n)\ox\1_d^{(i)}\ox\1_d^{(j)}
        \label[term]{eq:promise_hu}\\
    h^{(i)}(n) := &-\beta(1+\alpha_2(n))\Pi_{ud}^{(i)}.
      \label[term]{eq:promise_alpha}
  \end{align}
  \end{subequations}
  Clearly, both have norm bounded by $1$. Note that we can also rescale $h_d$ so that $\norm{h_d}\le \beta$.

  As in \cref{tiling+quantum}, $\1_u,\1_d$ denote the identity operators on $\HS_u,\HS_d$, and $\Pi_{ud}$ denotes the projection of $\HS$ onto its $\HS_u\ox\HS_d$ subspace. We decompose the global Hamiltonian in the square $\Lambda(L)$ as $H^{\Lambda(L)}=:\tilde{H}_0 + \tilde{H}_d + \tilde{H}_u$, where $\tilde{H}_0,\tilde{H}_d,\tilde{H}_u$ are defined by taking the sum over sites separately for the expressions in \cref{eq:promise_boundary}, \cref{eq:promise_hq}, and \cref{eq:promise_hu} + \cref{eq:promise_alpha}, respectively. Note that the three terms commute with each other and
 \begin{equation}\label{eq:spectra-final-theorem}
 \spec \tilde{H}_d =\spec H_d\; , \;\; \spec \tilde{H}_u=\{0\} \cup \spec H_u\; , \;\; \spec \tilde{H}_0\subset \Z_{\ge 0}
 \end{equation}

 Let us analyse the spectrum of $H^{(\Lambda(L)}$ in both the halting and non-halting cases. First consider the halting case, and take $L\ge L(n)$ as defined in \cref{promise_Hu}. In that case, $\tilde{H}_d, \tilde{H}_u\ge 0$ and hence $H^{\Lambda(L)}\ge \tilde{H}_0$. Since $|0\rangle^{\otimes \Lambda(L)}$ is the unique ground state of $\tilde{H_0}$, with energy $0$, and is also a 0-energy state for $H^{\Lambda(L)}$, we have that the spectral gap of $H^{\Lambda(L)}$ is at least as large as the spectral gap of $\tilde{H}_0$, which is $1$.

 In the non-halting case, it is clear from the structure of the Hamiltonians \cref{eq:promise_H} that
 \begin{equation}
 \spec \tilde{H}_u + \spec \tilde{H}_d\subset \spec H^{\Lambda(L)}\subset \spec \tilde{H}_u + \spec \tilde{H}_d + \spec \tilde{H}_0\;.
  \end{equation}
  Since $\spec \tilde{H}_0$ is contained in the set of non-negative integers, $\spec \tilde{H}_d\subset [0,+\infty)$ and converges in the thermodynamic limit to $[0,+\infty)$, and $\lambda_0(\tilde{H}_u)\le 0$ for all $L$ by \cref{eq:spectra-final-theorem}, statement (iii) in the Theorem follows.
\end{proof}

\subsection{Periodic boundary conditions}
\label{sec:periodic_bc}
The previous section proves \cref{thm:promise} for open boundary conditions. This is arguably the most important type of boundary conditions in the context of the spectral gap problem. Both physically, because periodic boundary conditions rarely occur in real physical systems, and mathematically, because the thermodynamic limit is better behaved. Nonetheless, our result can also be extended to other types of boundary condition. Extending it to fixed boundary conditions is trivial, so we omit the argument. In this section, we consider the more interesting case of periodic boundary conditions.

\begin{figure}[hbtp]
  \centering
  \includegraphics{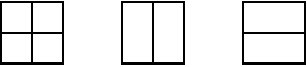}\\
  \caption{Boundary tiles for periodic boundary condition construction.}
  \label{fig:boundary_tiles}
\end{figure}

To extend our result to periodic boundary conditions, we add to the modified Robinson tiles of \cref{sec:quasi-periodic} the three new tile types shown in \cref{fig:boundary_tiles}.

Let $h_c$ be the standard tiling Hamiltonian of the modified Robinson tiles. We consider the Hamiltonian given by $3h_c$ plus the weighted terms given by the following tables (where $\wctile$ stands for any modified Robinson tile):

\begin{equation}\label{eq:boundary_tile_weights}
  {\def\arraystretch{1.2}
  \begin{array}{cc|cccc}
    & & \multicolumn{3}{|c}{\text{Right tile}} \\
    & & \wctile  & \horizline & \vertline & \cross \\
    \hline
    \multirow{4}{*}{\rotatebox{90}{\text{Left tile}}}
    & \wctile &  0   & 5 & 0 & 5 \\
    & \horizline &  5   & 0 & 5 & 1 \\
    & \vertline  &  0   & 5 & 5 & 5 \\
    & \cross     &  5   & 1 & 5 & 5
  \end{array}
  \qquad
  \begin{array}{cc|cccc}
    & & \multicolumn{3}{|c}{\text{Tile above}} \\
    & & \wctile  & \horizline & \vertline & \cross \\
    \hline
    \multirow{4}{*}{\rotatebox{90}{\text{Tile below}}}
    & \wctile &  0   & 0 & 5 & 5 \\
    & \horizline &  0   & 5 & 5 & 5 \\
    & \vertline  &  5   & 5 & 0 & 1 \\
    & \cross     &  5   & 5 & 1 & 5
  \end{array}}
\end{equation}

The resulting weighted tiling Hamiltonian $\tilde{h}_c$ effectively turns the weighted tiling problem for an $L\times L$ square region with periodic boundary conditions (a torus), into the standard tiling problem for the modified Robinson tile set on a square region with open boundary conditions of size $(L-1)\times(L-1)$:

\begin{lemma}[Periodic to open b.c.]
  \label{periodic-to-open-bc}
  For any $L\in\N$, all minimum weight tilings of an $L\times L$ torus using the tile set described above consist of a single $\cross$ tile at an arbitrary location, a complete row of $\horizline$ tiles extending horizontally from the $\cross$, a complete column of $\vertline$ tiles extending vertically from the $\cross$, and a valid modified Robinson tiling of the remaining $(L-1)\times(L-1)$ region.
\end{lemma}

\begin{proof}
  The tilings described in the statement of the \namecref{periodic-to-open-bc} have total weight $+4$ (coming from the $\vertline$ and $\horizline$ tiles adjacent to the $\cross$). We must show that all other tilings have higher weight.
  First, consider the case in which only modified Robinson tiles are used. By the aperiodicity of Robinson tiling, there are at least two mismatching tiles in an $L\times L$ torus, one in the vertical and one in the horizontal direction. Since we multiplied the original modified Robinson tiling Hamiltonian $H_c$ by a factor of~3, each of these mismatches has weight $3$ and we get a total weight of $6$. Thus any weight $<5$ tiling must contain at least one $\horizline$, $\vertline$ or $\cross$.

  If the tiling contains any tile other than $\horizline$ or $\cross$ to the left or right of a $\horizline$, it has weight $\geq 5$. Thus, in any weight $\leq 5$ tiling, $\horizline$ tiles can only appear in complete rows of $\horizline$ and $\cross$ tiles. The analogous argument for columns implies that $\vertline$ tiles can only appear in complete rows of $\vertline$ and $\cross$ tiles. A $\cross$ adjacent to a modified Robinson tile has weight $+5$, so $\cross$ tiles can only appear where such rows and columns meet. Moreover, adjacent $\horizline$ and $\vertline$ tiles have weight $+5$, so \emph{all} such rows and columns must meet at $\cross$ tiles.

  Therefore, any tiling with weight $<5$ must consist of some non-zero number of rows of $\horizline$ tiles and columns of $\vertline$ tiles meeting at $\cross$ tiles, with modified Robinson tiles everywhere else. Each intersection of a $\horizline$ row with a $\vertline$ column at a $\cross$ contributes weight $+4$, so the weight is minimised by having just one such row and column.

  Finally, the remaining region to be filled with modified Robinson tiles is a square of size $(L-1)\times (L-1)$. It contributes weight~0 iff, in addition, they form a valid tiling.
\end{proof}

Using this weighted tile set in \cref{tiling+quantum}, the proofs of \cref{tiling+quantum,put-promise-together,diverging_Hu,promise_Hu} go through unchanged for square lattices with periodic boundary conditions. Undecidability of the ground state energy density (\cref{thm:gs_density}), and undecidability of the spectral gap (\cref{thm:promise}) follow easily.



\section{Acknowledgements}
TSC would like to thank IBM T.\ J.\ Watson Laboratory for their hospitality during many visits, and Charlie Bennett in particular for insightful discussions about aperiodic tilings and reversible Turing Machines. Also, for pointing out that he [Charlie] had wrestled with implementation details of reversible Turing Machines thirty years ago, and it was time a younger generation took over! TSC also thanks Ashley and Catherine Montanaro for their kind hospitality in lodging him for two months in their house in Cambridge, where part of this work was carried out.

The authors are very grateful to Norbert Schuch for pointing out that the aperiodic tiling itself can be exploited to both simplify and at the same time strengthen our original version of the periodic boundary condition results, and to James Watson for pointing out a number of errors in the draft manuscript.

TSC, DPG and MMW thank the Isaac Newton Institute for Mathematical Sciences, Cambridge for their hospitality during the programme ``Mathematical Challenges in Quantum Information'', where part of this work was carried out; also the organisers of the 2014 Seefeld Quantum Information Workshop, where another part of this work was carried out. TSC and DPG are both very grateful to MMW and the Technische Universit\"at M\"unchen for their hospitality over summer 2014, when some of this work was completed.

TSC is supported by the Royal Society.

DPG acknowledges support from MINECO (grants MTM2014-54240-P, MTM2017-88385-P and ICMAT Severo Ochoa projects SEV-2015-0554 and CEX2019-000904-S), from Comunidad de Madrid (grants QUITEMAD, ref.
S2013/ICE-2801 and S2018/TCS-4342), and the European Research Council (ERC) under the European Union's Horizon 2020 research and innovation programme (grant agreement GAPS No 648913).

DPG and MMW acknowledge support from the European CHIST-ERA project CQC (funded partially by MINECO grant PRI-PIMCHI-2011-1071).

MMW acknowledges funding by the Deutsche Forschungsgemeinschaft (DFG, German Research Foundation) under Germany's Excellence Strategy -- EXC-2111 -- 390814868.

This work was made possible through the support of grant \#48322 from the John Templeton Foundation. The opinions expressed in this publication are those of the authors and do not necessarily reflect the views of the John Templeton Foundation.



\printbibliography

\end{document}